# Euclid

## Mapping the geometry of the dark Universe

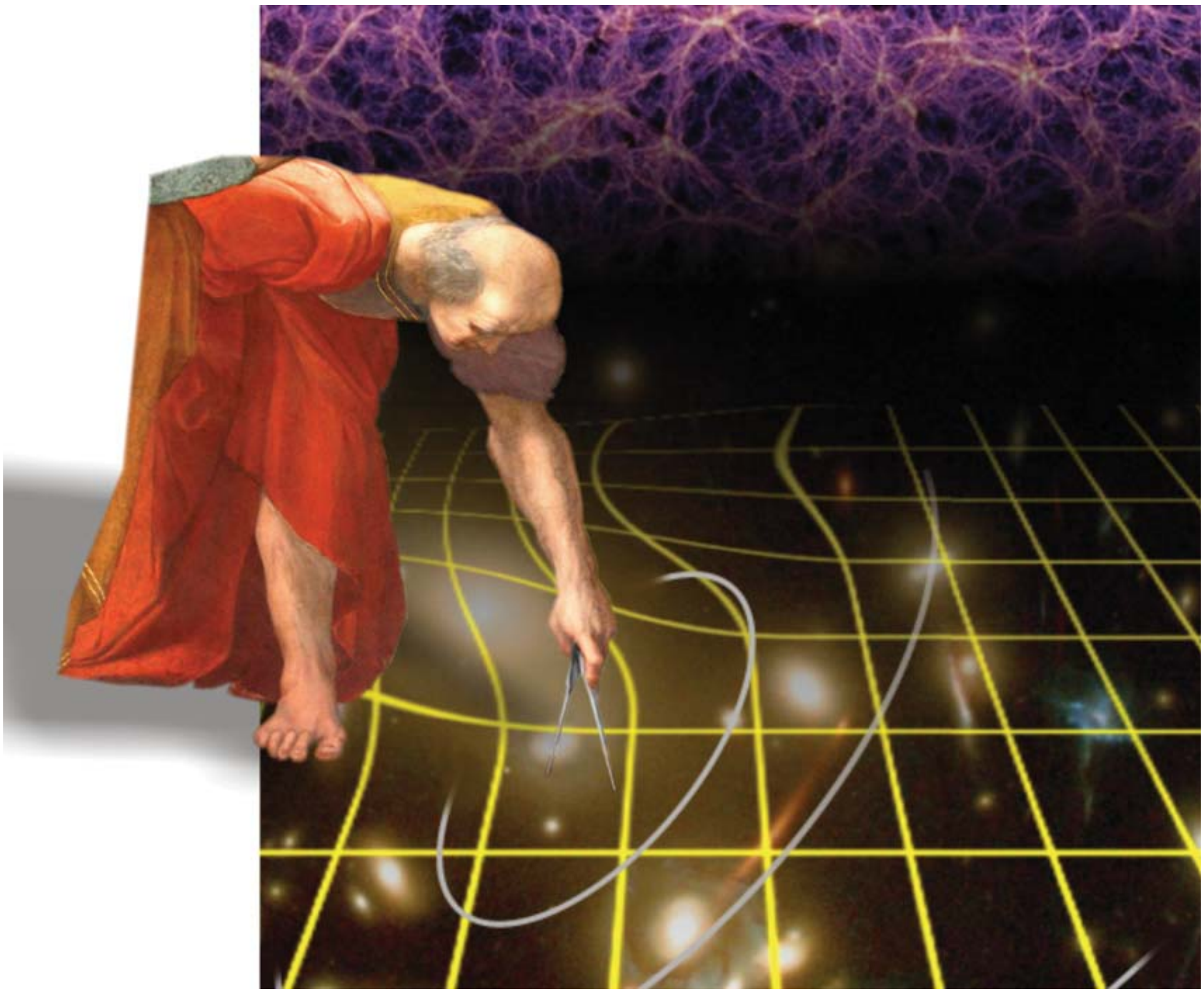

Definition Study Report

# Euclid

## Mapping the geometry of the dark Universe

Definition Study Report

***On the front cover***: composite of a fragment from Raphael's fresco "The School of Athens" in the Stanza della Segnatura of the Vatican Palace depicting the Greek mathematician Euclid of Alexandria, a simulation of the cosmic web by Springel et al, and an image of Abell 1689; the composition is made by Remy van Haarlem (ESA/ESTEC).

# Euclid Mission Summary

| Main Scientific Objectives |
|---|
| ***Understand the nature of Dark Energy and Dark Matter by:***<br>• Reach a dark energy *FoM* > 400 using only weak lensing and galaxy clustering; this roughly corresponds to 1 sigma errors on $w_p$ and $w_a$ of 0.02 and 0.1, respectively.<br>• Measure $\gamma$, the exponent of the growth factor, with a 1 sigma precision of < 0.02, sufficient to distinguish General Relativity and a wide range of modified-gravity theories<br>• Test the Cold Dark Matter paradigm for hierarchical structure formation, and measure the sum of the neutrino masses with a 1 sigma precision better than 0.03eV.<br>• Constrain $n_s$, the spectral index of primordial power spectrum, to percent accuracy when combined with Planck, and to probe inflation models by measuring the non-Gaussianity of initial conditions parameterised by $f_{NL}$ to a 1 sigma precision of ~2. |

| SURVEYS | | |
|---|---|---|
| | Area (deg2) | Description |
| Wide Survey | 15,000 (required)<br>20,000 (goal) | Step and stare with 4 dither pointings per step. |
| Deep Survey | 40 | In at least 2 patches of > 10 $deg^2$<br>2 magnitudes deeper than wide survey |

| PAYLOAD | | | | | |
|---|---|---|---|---|---|
| Telescope | 1.2 m Korsch, 3 mirror anastigmat, f=24.5 m | | | | |
| Instrument | VIS | NISP | | | |
| Field-of-View | 0.787×0.709 $deg^2$ | 0.763×0.722 $deg^2$ | | | |
| Capability | Visual Imaging | NIR Imaging Photometry | | | NIR Spectroscopy |
| Wavelength range | 550– 900 nm | Y (920-1146nm), | J (1146-1372 nm) | H (1372-2000nm) | 1100-2000 nm |
| Sensitivity | 24.5 mag<br>10σ extended source | 24 mag<br>5σ point source | 24 mag<br>5σ point source | 24 mag<br>5σ point source | 3 $10^{-16}$ erg cm-2 s-1<br>3.5σ unresolved line flux |
| Detector Technology | 36 arrays<br>4k×4k CCD | 16 arrays<br>2k×2k NIR sensitive HgCdTe detectors | | | |
| Pixel Size<br>Spectral resolution | 0.1 arcsec | 0.3 arcsec | | | 0.3 arcsec<br>R=250 |

| SPACECRAFT | |
|---|---|
| Launcher | Soyuz ST-2.1 B from Kourou |
| Orbit | Large Sun-Earth Lagrange point 2 (SEL2), free insertion orbit |
| Pointing | 25 mas relative pointing error over one dither duration<br>30 arcsec absolute pointing error |
| Observation mode | Step and stare, 4 dither frames per field, VIS and NISP common FoV = 0.54 $deg^2$ |
| Lifetime | 7 years |
| Operations | 4 hours per day contact, more than one ground station to cope with seasonal visibility variations; |
| Communications | maximum science data rate of 850 Gbit/day downlink in K band (26GHz), steerable HGA |

| Budgets and Performance | | | | |
|---|---|---|---|---|
| | *Mass (kg)* | | *Nominal Power (W)* | |
| industry | TAS | Astrium | TAS | Astrium |
| Payload Module | 897 | 696 | 410 | 496 |
| Service Module | 786 | 835 | 647 | 692 |
| Propellant | 148 | 232 | | |
| Adapter mass/ Harness and PDCU losses power | 70 | 90 | 65 | 108 |
| **Total (including margin)** | **2160** | | **1368** | **1690** |

# Foreword

Euclid is a Medium Class mission of the ESA Cosmic Vision 2015-2025 programme, and competes for one of the two foreseen launch slots in 2017 and 2018. This report (the *Euclid Red Book*) describes the outcome of the Phase A study, which started with the industrial kick-off in July 2010 and ended with the completion of the Preliminary Requirement Review in July 2011. Phase A is the first part of the Mission Definition Phase in which the mission is prepared for the Implementation Phase where an industrial prime contractor will lead the system development and the Euclid Mission Consortium will provide elements of payload and science ground segment.

The Assessment Phase of Euclid, which preceded Phase A, showed that the Euclid science and mission concept are feasible and could meet the M Class mission requirements. However, the study also showed that a serious optimisation of the mission was necessary to stay within the programmatic constraints with sufficient margin. In particular, the payload mass had to be reduced to create margin for a Soyuz launcher, and a lower number of detectors was recommended, since procurement of many detectors has a high schedule risk beyond the foreseen launch in 2018.

The Euclid study team considered several options to optimise the mission, but only a few cases could be investigated in detail. The most viable solution was the merging of the near-infrared photometric and spectroscopic channels into one instrument with a single focal plane array and with a reduced number of near-infrared detectors. This concept provides a comfortable system mass margin due to a payload mass reduction as a result of a more compact payload design. The optimisation comes with a price: the near-infrared spectroscopic and photometric observations are now done sequentially instead of simultaneously, thereby significantly increasing the observation time for a single field.

A team of 6 scientists, the Euclid Optimisation Advisory Team (EOAT), was selected to provide re-commendations on the set of mission parameters that can meet the Euclid science objectives and top level requirements, taking into account the optimised payload. It was decided that a payload configuration with 16 NIR detectors for NISP, would best meet the Euclid science requirements. The alternative configuration with a smaller FoV and consequently fewer detectors was rejected as baseline because of its inability to meet the sufficient survey area during the nominal mission.

The new payload and mission concept was used to prepare the invitation to tender (ITT) for two competing Definition Phase studies by two independent industrial consortia. Two industrial contractors were selected: Thales Alenia Space Italy (Turin) and Astrium GmbH Germany (Friedrichshafen). Both industrial teams already performed the Assessment Phase studies, and carry a lot of expertise about the Euclid mission.

In June 2010 the Euclid Science Management Plan was accepted by the Science Programme Committee. The science management plan provides the programmatic framework for the issue of the Euclid Announcement of Opportunity for a single consortium - the Euclid Mission Consortium - to provide elements of the payload and the science ground segment. A European wide consortium consisting of 7 lead countries submitted their proposal in October 2010 and was selected by the Science Programme Committee in February 2011.

The Euclid Science Management Plan was updated in June 2011 and reflects the latest programmatic developments as well as the data release scheme for Euclid.

This report provides a description of the Euclid mission as emerged from the phase A study. Section 1 gives the executive summary. The science case is provided in Section 2. The first level science requirements and the top level instrument and mission requirements are provided in Section 3. Payload and mission are described in Sections 4 and 5, respectively. An assessment of the system and instruments performance in view of the science requirements is provided in Section 6. The description of the ground segment is given in Section 7, and the management and programmatics are described in Section 8.

*The Euclid Study Team, July 2011*

# Authorship and Acknowledgements

### Euclid Red Book Editorial Team

| | |
|---|---|
| R. Laureijs | ESA/ESTEC, The Netherlands (Chair; ESA Study Team) |
| J. Amiaux | CEA /IRFU Saclay, France |
| S. Arduini | IAP UPMC Paris, France |
| J.-L. Auguères | CEA/IRFU Saclay, France |
| J. Brinchmann | Leiden Univ., The Netherlands |
| R. Cole | UCL-MSSL, London, UK |
| M. Cropper | UCL-MSSL, London, UK (ECB, EST member)) |
| C. Dabin | CNES, Toulouse, France |
| L. Duvet | ESA/ESTEC, The Netherlands (ESA Study Team) |
| A. Ealet | CPPM and LAM, Marseille, France |
| B. Garilli | INAF IASFMI, Milan, Italy |
| P. Gondoin | ESA/ESTEC, The Netherlands (ESA Study Team) |
| L. Guzzo | INAF OA Brera, Italy |
| J. Hoar | ESA/ESAC, Spain (ESA Study Team) |
| H. Hoekstra | Leiden Univ., The Netherlands |
| R. Holmes | MPIA, Heidelberg, Germany |
| T. Kitching | ROE, Edinburgh, UK |
| T. Maciaszek | CNES, Toulouse, France |
| Y. Mellier | IAP UPMC, Paris, France (ECL) |
| F. Pasian | INAF OA Trieste, Italy |
| W. Percival | Univ. Portsmouth, UK |
| J. Rhodes | JPL, Pasadena CA, USA (ECB) |
| G. Saavedra Criado | ESA/ESTEC, The Netherlands (ESA Study Team) |
| M. Sauvage | CEA/IRFU Saclay, France |
| R. Scaramella | INAF OA Roma, Italy (ECB) |
| L. Valenziano | INAF IASFBO Bologna, Italy |
| S. Warren | Imperial College London, UK |

### ESA Study Team

| | |
|---|---|
| L. Duvet | Study Payload Manager |
| P. Gondoin | Study Manager |
| J. Hoar | Science Operations |
| R. Laureijs | Study Scientist |
| G. Saavedra Criado | System Engineer |

### Euclid Consortium Board (ECB)

| | |
|---|---|
| R. Bender | MPE Garching and USM Muenchen, Germany |
| F. Castander | IEEC Barcelona, Spain (EST member) |
| A. Cimatti | Univ. of Bologna, Italy (EST member) |
| M. Cropper | UCL-MSSL London, UK (EST member) |
| O. Le Fèvre | LAM Marseille, France |
| H. Kurki-Suonio | Univ. Helsinki, Finland |
| M. Levi | LBL Berkeley CA, USA |
| P. Lilje | ITA Univ. Oslo, Norway |
| Y. Mellier | IAP UMPC Paris, France (ECL) |
| G. Meylan | EPFL Lausanne, Switzerland |
| R. Nichol | Univ. Portsmouth, UK |
| K. Pedersen | SSC Univ. Copenhagen, Denmark |
| V. Popa | ISS Bucharest, Romania |
| R. Rebolo Lopez | IAC, Spain |
| J. Rhodes | JPL, Pasadena CA, USA |
| H.-W. Rix | MPIA, Heidelberg, Germany (EST member) |
| H. Rottgering | Leiden Univ, The Netherlands |
| R. Scaramella | INAF OA Roma, Italy |
| W. Zeilinger | IfA Univ. Wien, Austria |

## Other Euclid Tiger Team (TT) contributors (not in previous lists)

| | |
|---|---|
| F. Grupp | MPE Garching and USM Muenchen, Germany |
| P. Hudelot | IAP UPMC, France |
| R. Massey | ROE Edinburgh, UK |
| M. Meneghetti | INAF OA Bologna, Italy |
| L. Miller | Oxford Univ., UK |
| S. Paltani | Univ. Genève, Switzerland |
| S. Paulin-Henriksson | CEA/IRFU Saclay, France |
| S. Pires | CEA/IRFU Saclay, France |
| C. Saxton | UCL-MSSL London, UK |
| T. Schrabback | SLAC CA, USA and USM Muenchen, Germany |
| G. Seidel | MPIA Heidelberg, Germany |
| J. Walsh | ESO Garching, Germany |

## Contributions from the Euclid Science Working Group (not in previous lists):

N. Aghanim (IAS, Orsay), L. Amendola (MPIA, Heidelberg), J. Bartlett (APC, Paris), C. Baccigalupi (SISSA Trieste), J. -P. Beaulieu (IAP, Paris), K. Benabed (IAP, Paris), J. -G. Cuby (LAM, Marseille), D. Elbaz (CEA, Saclay), P. Fosalba (IEEC, Barcelona), G. Gavazzi (U. Milano Bicocca), A. Helmi (U. Groningen), I. Hook (U. Oxford, and OA Roma), M. Irwin (U. Cambridge), J.-P. Kneib (LAM, Marseille), M. Kunz (U. Genève), F. Mannucci (OA Arcetri), L. Moscardini (U. Bologna), C. Tao (CPPM, Marseille), R. Teyssier (CEA Saclay and U. Zurich), J. Weller (USM München, MPE Garching), G. Zamorani (OA Bologna), M.R. Zapatero Osorio (IAC, Spain)

## Contributions from the Euclid VIS instrument (not in previous lists):

O. Boulade (CEA Saclay), J. J. Foumond (IAS, Orsay), A. Di Giorgio (IFSI Roma), P. Guttridge (UCL-MSSL, London), A. James (UCL-MSSL, London), M. Kemp (UCL-MSSL, London), J. Martignac (CEA, Saclay), A. Spencer (UCL-MSSL London), D. Walton (UCL-MSSL London)

## Contributions from the Euclid NISP instrument (not in previous lists):

T. Blümchen (MPIA Heidelberg), C. Bonoli (OA Padova), F. Bortoletto (OA Padova), C. Cerna (CPPM, Marseille), L. Corcione (OA Torino), C. Fabron (LAM, Marseille), K. Jahnke (MPIA Heidelberg), S. Ligori (OA Torino), F. Madrid (IEEC Barcelona), L. Martin (LAM Marseille), G. Morgante (IASFBO Bologna), T. Pamplona (LAM, Marseille), E. Prieto (LAM Marseille), M. Riva (OA Brera), R. Toledo (UPCT, Cartagena), M. Trifoglio (IASFBO Bologna), F. Zerbi (OA Brera),

## Contributions from the Euclid Science Ground Segment (not in previous lists):

F. Abdalla (UCL London), M. Douspis (IAS, Orsay), C. Grenet (IAP, Paris), S. Borgani (U. Trieste), R. Bouwens (U. Leiden), F. Courbin (EPFL, Lausanne), J.-M. Delouis (IAP, Paris), P. Dubath (U. Genève), A. Fontana (OA Roma), M. Frailis (OA Trieste), A. Grazian (OA Roma), J. Koppenhöfer (MPE Garching), O. Mansutti (OA Trieste), M. Melchior (FHNW Windisch), M. Mignoli (OA Bologna), J. Mohr (USM München), C. Neissner (PIC, Barcelona), K. Noddle (ROE Edinburgh), M. Poncet (CNES, Toulouse), M. Scodeggio (IASFMI, Milan), S. Serrano (IEEC Barcelona), N. Shane (UCL-MSSL London), J.-L. Starck (CEA Saclay), C. Surace (LAM, Marseille), A. Taylor (ROE Edinburgh), G. Verdoes-Kleijn (U. Groningen), C. Vuerli (OA Trieste), O. R. Williams (U. Groningen), A. Zacchei (OA Trieste)

## Contributions from ESA

B. Altieri (ESAC), I. Escudero Sanz (ESTEC), R. Kohley (ESAC), T. Oosterbroek (ESTEC)

## Other contributors:

P. Astier (LPNHE, Paris), D. Bacon (ICG, Portsmouth), S. Bardelli (OA Bologna), C. Baugh (U. Durham), F. Bellagamba (U. Bologna), C. Benoist (OCA, Nice), D. Bianchi (Univ. Milan), A. Biviano (OA Trieste), E. Branchini (U. Roma), C. Carbone (U. Bologna), V. Cardone (OA Roma), D. Clements (Imperial, London), S. Colombi (IAP Paris), C. Conselice (U. Nottingham), G. Cresci (OA Arcetri), N. Deacon (UH Honolulu), J. Dunlop (IFA Edinburgh), C. Fedeli (U. Florida), F. Fontanot (OA Trieste), P. Franzetti (IASFMI Milan), C. Giocoli (U. Bologna), J. Garcia-Bellido (UAM Madrid), J. Gow (Open U.), A. Heavens (IFA Edinburgh), P. Hewett (U. Cambridge), C. Heymans (IFA Edinburgh), A. Holland (Open U.), Z. Huang (OA Roma), O. Ilbert (LAM, Marseille), B. Joachimi (ROE Edinburgh), E. Jennins (U. Durham), E. Kerins (U. Manchester), A. Kiessling (ROE, Edinburgh), D. Kirk (UCL, London), R. Kotak (QUB Belfast), O. Krause (MPIA Heidelberg), O. Lahav (UCL), F. van Leeuwen (U. Cambridge), J. Lesgourgues (CERN Geneva), M. Lombardi (U. Milano), M. Magliocchetti (IFSI), K. Maguire (U. Oxford), E. Majerotto (OA Brera), R. Maoli (U. Roma), F. Marulli (U. Bologna), S. Maurogordato (OCA Nice), H. McCracken (IAP, Paris), R. McLure (IFA Edinburgh), A. Melchiorri (U. Roma), A. Merson (U. Durham), M. Moresco (U. Bologna), M. Nonino (OA Trieste), P. Norberg (ROE Edinburgh), J. Peacock (IfA Edinburgh), R. Pello (IRAP Toulouse), M. Penny (U. Manchester), V. Pettorino (SISSA, Trieste), C. Di Porto (U. Roma), L. Pozzetti (OA Bologna), C. Quercellini (OA Roma), M. Radovich (OA Padova), A. Rassat (EPFL, Lausanne), N. Roche (OA Bologna), S. Ronayette (CEA Saclay), E. Rossetti (U. Bologna), B. Sartoris (U. Trieste), P. Schneider (U. Bonn), E. Semboloni (U. Leiden), S. Serjeant (Open U.), F. Simpson (ROE Edinburgh), C. Skordis (U. Nottingham), G. Smadja (IPNL, Lyon), S. Smartt (QUB Belfast), P. Spano (OA Brera), S. Spiro (OA Roma), M. Sullivan (U. Oxford), A. Tilquin (CPPM Marseille), R. Trotta (Imperial London), L. Verde (ICC Barcelona), Y. Wang (U. Oklahoma), G. Williger (U. Louisville), G. Zhao (U. Portsmouth), J. Zoubian (LAM, Marseille), E. Zucca (OA Bologna)

# 1 Executive Summary

Understanding the acceleration of the expansion of the Universe is one of the most compelling challenges of cosmology and fundamental physics. The Euclid surveys will show how cosmic acceleration modifies the expansion history and the 3-dimensional distribution of matter in the Universe. To achieve this, Euclid will measure the shapes over a billion galaxies and accurate redshifts of tens of millions of galaxies for weak gravitational lensing and galaxy clustering studies. The Phase A study of Euclid has led to a payload and a mission design concept able to support the requirements imposed by the need of high visible image quality, and of near-infrared imaging photometry and slitless spectroscopy over a very large sky area. With the expected mission performance and exquisite control of systematics, the Euclid space mission puts Europe in a leading position to address a most fascinating question that may revolutionise physics.

**Primary Science Objectives**: Our understanding of cosmology is that of the Universe evolving from a homogeneous state after the Big Bang to a hierarchical assembly of galaxies, clusters and superclusters at our epoch. However this view relies on two untested assumptions, about the initial conditions of the Universe and the nature of gravity itself, and the existence of two dominant components whose nature is entirely unknown. Of these unknown components, 76% of the energy density is in the form of *dark energy*, which is causing the Universe expansion to accelerate. Dark energy is in conflict with our knowledge of fundamental physics, if it behaves as predicted from the cosmological constant (introduced by Einstein), then its value is $10^{60}$ times smaller than that predicted from theory; this is the largest discrepancy between theory and observation ever encountered in modern physics. Another 20% of the energy density of the Universe is in the form of *dark matter*, which exerts a gravitational attraction as normal matter, but does not emit or absorb light. While several candidates for dark matter exist in particle physics, its nature remains unknown. Another possibility to explain one or both of these puzzles is that Einstein's General Relativity, and thus our understanding of gravity, needs to be revised. This diversity of theoretical ideas shows our current ignorance but also defines the need for future observations. Based on our present-day knowledge, the existing plausible models will only change observational signatures by tiny amounts that can only be decisively distinguished by using high-precision astronomical surveys covering a major fraction of the sky.

Euclid is a survey mission designed to understand the origin of the Universe's accelerating expansion. It will use *cosmological probes* to investigate the nature of dark energy, dark matter and gravity by tracking their observational signatures on the geometry of the universe and on the cosmic history of structure formation. Euclid will map large-scale structure over a cosmic time covering the last 10 billion years, more than 75% the age of the Universe. The mission is optimised for two independent primary cosmological probes: Weak gravitational Lensing (WL) and Baryonic Acoustic Oscillations (BAO). WL is a technique to map dark matter and measure dark energy by quantifying the apparent distortions of galaxy images, a change in a galaxy's observed ellipticity, caused by mass inhomogeneities along the line-of-sight. The lensing signal is derived from the measurement of shape and distance of galaxies. BAO are wiggle patterns imprinted in the clustering of galaxies that provide a standard ruler to measure the expansion of the Universe. The properties of the wiggles are derived from accurate distance measurements of galaxies. Surveyed in the same cosmic volume, these two probes provide necessary cross-checks on systematic errors. They also provide a measurement of large scale structure via different physical fields (potential, density and velocity), which are required for testing dark energy and gravity at all scales. In addition, the Euclid surveys yield data of several important complementary cosmological probes such as galaxy clusters, redshift space distortions and the integrated Sachs Wolfe effect.

WL requires a high image quality on sub-arcsec scales for the galaxy shape measurements, and photometry at visible and infrared wavelengths to measure the photometric distances of each lensed galaxy out to $z \geq 2$. BAO requires near-infrared spectroscopic capabilities to measure accurate redshifts of galaxies out to $z \geq 0.7$. Both probes require a very high degree of system stability to minimise systematic effects, and the ability to survey a major fraction of the extra-galactic sky. Such a combination of requirements cannot be met from the ground, and demands a wide-field-of-view space mission. Euclid is designed for that purpose.

To understand the nature of dark energy its equation of state needs to be determined. Euclid uses WL and BAO to measure the constant and time varying terms of the dark energy equation of state to a 1-sigma precision of 0.02 and 0.1, sufficient to make a decisive statement on the nature of dark energy. Euclid tests the validity of General Relativity by measuring the rate of cosmic structure growth to a 1-sigma precision of < 0.02, sufficient to distinguish General Relativity from a wide range of modified-gravity theories. As Euclid maps the dark matter distribution with unprecedented accuracy, subtle features produced by neutrinos are measured, providing constraints on the sum of the neutrino masses with a 1-sigma precision better than 0.03 eV. Likewise, the initial conditions of the seeds of cosmic structure growth are unveiled by determining the power spectrum of density perturbations to one percent accuracy. Euclid and Planck together measure deviations to a Gaussian distribution of initial perturbations with a precision one order of magnitude better than current constraints, allowing Euclid to test a broad range of inflation models. Euclid is therefore poised to uncover new physics by challenging all sectors of the cosmological model.

**Legacy Science:** The Euclid wide and a deep surveys yield a treasure-trove with unique legacy science in various fields of astrophysics and a primary data base for next generation multi-wavelength surveys.

Euclid produces a legacy dataset with images and photometry of more than a billion galaxies and several million spectra, out to high redshifts $z$>2. At low redshift, Euclid resolves the stellar population of all galaxies within ~5 Mpc, providing a complete census of all morphological and spectral types of galaxies in our neighborhood. It also delivers morphologies, masses, and star-formation rates out to $z$~2 with a 4 times better resolution, and 3 NIR magnitudes deeper, than possible from ground. Euclid derives the mass function of galaxy clusters (in combination with eROSITA, Planck and SZ telescopes), and finds over $10^5$ strong lensing systems. Gravitational lensing together with near infrared photometry of lensing sources explores the relationship between light, baryons and dark matter between galaxy and super cluster scales as function of look-back time and environment.

The Euclid deep survey will be *the* primary target for follow-up observations. Deep data contain thousands of objects at $z$>6 and several tens of $z$>8 galaxy or quasar candidates that will be critical targets for JWST and E-ELT. As the deep survey fields are visited repeatedly over a time span of several years they are a unique baseline for the discovery of variable sources.

Euclid probes the formation and evolution of our Galaxy. Euclid augments the Gaia survey, taking it several magnitudes deeper, and provides complementary information, adding infrared colours and spectra for every Gaia stars it observes. This breaks the age-metallicity degeneracy, which is critical for the chemical enrichment history of our Galaxy. Also, the wide sky coverage of Euclid will detect nearby extremely low surface brightness tidal streams of stars.

**Payload**: The Euclid payload consists of a 1.2 m Korsch telescope designed to provide a large field of view. The telescope directs the light to two instruments via a dichroic filter in the exit pupil. The reflected light is led to the visual instrument (VIS) and the transmitted light from the dichroic feeds the near infrared instrument (NISP) which contains a slitless spectrometer and a three bands photometer. Both instruments cover a large common field-of-view of ~0.54 deg$^2$.

VIS is equipped with 36 CCDs. It measures the shapes of galaxies with a resolution better than 0.2 arcsec (PSF FWHM) with 0.1 arcsec pixels in one wide visible band (R+I+Z). The NISP photometer contains three NIR bands (Y, J, H), employing 16 HgCdTe NIR detectors with 0.3 arcsec pixels. The spectroscopic channel of NISP operates in the wavelength range 1.1-2.0 micron at a mean spectral resolution $\lambda/\Delta\lambda$ ~ 250, employing 0.3 arcsec pixels. While the VIS and NISP operate in parallel, the NISP performs the spectroscopy and photometry measurements in sequence by selecting a grism wheel in case of spectroscopy and a filter wheel in case of photometry.

**Mission:** Euclid will be launched in 2018 on a Soyuz ST-2.1B rocket, with an all-year round launch window. A direct transfer of ~30 days is targeted to a large-amplitude free-insertion orbit at the 2$^{nd}$ Lagrange Point of the Sun-Earth System. It takes 6 years to complete a wide survey with the deep survey interspersed. Spacecraft commissioning, performance verification and initial calibration require an additional 3-6 months. The sky mapping mode is step and stare. Image stability is maintained by scanning the sky along circles of constant solar aspect angle. Possible variations in the solar aspect angle between fields are kept to a maximum of 5 deg to minimise overheads for thermal stabilisation. At least one ground station is available to receive the science data from the spacecraft at a rate of at most 850 Gbit over a daily pass time of 4 hours.

**Surveys:** The wide survey covers 15,000 deg$^2$ of the extragalactic sky and is complemented by two 20 deg$^2$ deep fields observed on a monthly basis. For WL, Euclid measures the shapes of 30 resolved galaxies per

arcmin$^2$ in one broad visible R+I+Z band (550-920 nm) down to AB mag 24.5 (10 $\sigma$). The photometric redshifts for these galaxies reaches a precision of $\sigma_z/(1+z) < 0.05$. They are derived from three additional Euclid NIR bands (Y, J, H in the range 0.92-2.0 micron) reaching AB mag 24 (5 $\sigma$) in each, complemented by ground based photometry in visible bands derived from public data or through engaged collaborations with projects such as DES, KiDS, and Pan-STARRS. To measure the shear from galaxy ellipticities, requirements are imposed on the PSF such that it can be reconstructed with an error $\leq 2\text{x}10^{-4}$ in ellipticity and its dimension varies by less than $2\text{x}10^{-3}$ across the FoV. The BAO are determined from a spectroscopic survey with a redshift accuracy $\sigma_z/(1+z) \leq 0.001$. The slitless spectrometer, with $\lambda/\Delta\lambda$~250, predominantly detects H$\alpha$ emission line galaxies. The limiting line flux is $3\text{x}10^{-16}$ erg $s^{-1}cm^{-2}$ (1 arcsec extended source, 3.5 sigma at 1.6 micron), yielding over 50 million galaxy redshifts with a completeness higher than 45%.

The deep survey is two magnitudes deeper than the wide survey. This is needed for calibration of the slitless spectroscopy and is also unique as a self-standing survey. The deep survey monitors the stability of the spacecraft and payload through repeated visits of the same regions.

**Expected performance**: The end-to-end simulations carried out to assess the performance of VIS demonstrate that the PSF of the VIS channel can be known to a high level of accuracy. The spatio-temporal variations of the PSF which are most critical for WL are sampled and modeled with sufficient accuracy by using the ≥1,800 stars suitable for calibration spread over each Euclid field. The PSF smearing produced by the CCD charge transfer inefficiency is corrected using the standard method routinely applied to HST images, provided the readout noise of the Euclid detectors is lower than 4.5 electrons. Overall, the residuals in the knowledge of the size and the ellipticity of the PSF will be lower than the upper limit requirements during the whole life of the Euclid mission. Furthermore, the throughput of the VIS channel guarantees that the galaxy number density of ~30 gal./arcmin$^2$ down to AB 24.5 mag can be achieved in less than 4000 s.

Similar simulations carried out for the NISP imaging mode show detection limits of $Y_{AB}$, $J_{AB}$ and $H_{AB}$ = 24 mag (5$\sigma$) in less than 6 minutes per filter, while preserving the required encircled energy in the three near-infrared bands. The NISP imaging mode provides Y, J and H photometric data for 95% of VIS sources suitable for WL analysis. The limited number of near-infrared detectors results in an under-sampled design, with a pixel scale of 0.3 arcsec, which will be compensated by multiple dither observing sequences.

End to end simulations of NISP in spectroscopic mode show that Euclid can measure 3,000 redshifts/deg$^2$ with the required S/N, completeness (fraction of spectra measured above a given line flux limit) and purity (fraction of spectra for which the measured redshift is correct) over the whole range of redshifts explored by the Euclid BAO sample and down to the expected H-alpha flux limit. The redshift completeness for BAO is predicted to be significantly higher than the required 35%.

**Mission management:** For the space segment, ESA provides the spacecraft and telescope through a selected industrial contractor, as well as the CCD and NIR detectors. The Euclid Mission Consortium (EMC), funded by national agencies, has been selected to provide the VIS and NISP instruments, and elements of the science ground segment (SGS) related to the scientific pipelines generating the data products and the instrument in-orbit maintenance and operations.

The EMC is organised to support the instrument development, assessment of scientific requirements and performance, and the SGS. Together with ESA, the consortium has worked out the SGS operations concept, which has led to an agreed set of science implementation requirements for the SGS. The implementation encompasses the definition of tasks and interfaces of the science data centres (SDCs) and the architecture of the SGS that includes operation of the mission and the legacy archives.

**Data release:** Scientifically validated data are released via the Euclid Legacy Archive on an annual basis. The scientific products are categorised in three data levels. The first data level consists of the raw decompressed telemetry frames, the second level consists of calibrated data with instrument signatures removed, and the third level consists of extracted scientific information such as catalogues. In addition, Euclid provides quick-release data, Level Q, representing transient products suitable for most purposes in astronomy, except for the core cosmology objectives. The earliest public data release takes place 14 months after the start of the routine operations, and contains Level Q products of the first year of routine operations. The other associated data levels are released 12 months later, together with the Level Q products of the second year, and so on.

# Table of Contents

# 2 Euclid Science Objectives

## 2.1 Cosmology today

Our view of the Universe has changed dramatically over the past century: less than a hundred years ago it was believed to consist of only our own Galaxy. The discovery of the expansion of the Universe and the subsequent realisation that the Universe is very old, but had a beginning, are major triumphs of astronomy that have changed our view of humanity's place in the Universe. At the turn of the Millennium, further crucial observational progress led to the emergence of the *concordance* cosmological model. This provides a remarkably accurate description of a wide range of independent observations through a fully self-consistent theoretical framework with a small number of parameters. However, it relies on two untested assumptions about the initial conditions of the Universe and the nature of gravity itself, as well as the existence of two dominant components whose nature is entirely unknown. Of these unknown components, about 76% of the overall mass-energy density is in the form of *dark energy*, which is causing the Universe to accelerate its expansion at the current epoch. Another 20% is in the form of non-baryonic *dark matter*, which exerts a gravitational attraction as normal matter, but cannot emit or absorb light. The biggest of all these puzzles, one that raises a potential crisis in fundamental physics, is the nature of dark energy. There are a plethora of ideas, ranging from including a "cosmological constant" in the Einstein field equations of General Relativity (GR), to the inclusion of additional fields, or even a revision of our theory of gravity. More than an order of magnitude improvement in the quality and quantity of observational data is needed to address this problem.

Euclid has been conceived to make this step forward by measuring the expansion history and growth of large-scale structure with a precision that will allow us to distinguish time-evolving dark energy models from a cosmological constant, and to test the theory of gravity on cosmological scales. The same measurements also constrain the initial conditions in the very early Universe, by determining the statistical distribution of the primordial density fluctuations with high precision, on scales that cannot be probed using observations of the cosmic microwave background (CMB). As such, Euclid will not only provide insights into the ultimate fate of the Universe, but also into how it began. Euclid will achieve this by taking the concept of galaxy surveys into a new regime in terms of size and control of systematics. Surveys such as the Sloan Digital Sky Survey (SDSS) have provided one of the primary pillars upon which the concordance model has been built. Euclid will take this to a new level by surveying 15,000 $deg^2$ of the extra-galactic sky. It will directly map the dark matter distribution in the Universe through weak gravitational lensing by imaging 1.5 billion galaxies with HST-like resolution and providing near infrared (NIR) photometry. At the same time, it will carry out a spectroscopic redshift survey of 50 million galaxies over a volume 500 times larger than the SDSS, observing galaxies over 75% of the lifetime of the Universe.

The scientific impact of Euclid, however, is not limited to cosmology and in fact spans most of extra-galactic astronomy. The unique combination of high-resolution optical imaging, multi-band NIR imaging and spectroscopy up to $z\sim2$ over most of the extra-galactic sky, will result in a vast range of non-cosmology science. It is only possible to provide a selection of scientific highlights in this document. The legacy of Euclid will last for many decades and will have an impact across all of physics and astronomy. Finally, when we make the leap forward that Euclid enables us to do, important serendipitous discoveries will be almost a certainty.

### 2.1.1 A dark Universe

Of the two disturbing ingredients of the concordance model, dark matter is the most familiar. Evidence for its existence goes back to the 1930s, when Fritz Zwicky realised that the dynamical mass of the Coma cluster exceeded that expected from the luminosities of its member galaxies, suggesting a dominant non-luminous (dark) component. In the 1970s further support came from the measurements of flat rotation curves in spiral galaxies. Finally, a significant dark matter contribution was found to be necessary to reconcile the low level of anisotropy in the CMB with the very existence of galaxies and structures today. This, together with early observations of large-scale structures by the first extensive *galaxy redshift surveys* during the 1980s, brought forth the emergence of the *Cold Dark Matter* (CDM) paradigm. The "standard" CDM scenario was based on a flat geometry on large scales with hierarchical structure formation on small-scales induced by the dominant dark-matter component. Although observations did support the existence of some dark matter, the estimated density fell short by more than half of the density needed to ensure flatness, ruling out the "standard" scenario.

A fully coherent picture emerged only in the late 1990s thanks to observations of distant Type-Ia supernovae and the CMB: the Universe is indeed close to being spatially flat, but cold dark matter and ordinary matter make up only 24% of the energy density today. Various lines of evidence strongly support a non-baryonic nature for the dark matter particle. Several plausible candidates arise naturally in extensions of the standard model of particle physics.

The remaining 76% is provided by a component that is causing the expansion of the Universe to accelerate, named *dark energy* to reflect its mysterious nature. It represents one of the biggest puzzles in modern physics because our current theories cannot explain its magnitude or its origin. In General Relativity, the cosmological constant, a simple additive term to the Einstein field equations, appears as a natural candidate that could cause the accelerated expansion. Current observations are consistent with this explanation of dark energy. The cosmological constant, however, could be interpreted as the 'vacuum energy of empty space' and if this is the case its observed value is not consistent with current theories: in the context of a quantum-field theory, the value of a cosmological constant should be set by an upper cut-off in the vacuum energy which, considering current experimental bounds from particle physics, would imply a cosmological constant at least $10^{60}$ times larger than observed. This is the largest discrepancy between theory and observation ever encountered in modern physics. It implies that either the cosmological constant is not the correct description of dark energy, leaving the possibility open for more exotic models, or that a radical change in our most fundamental theories of physics, such as quantum mechanics and GR, is in order.

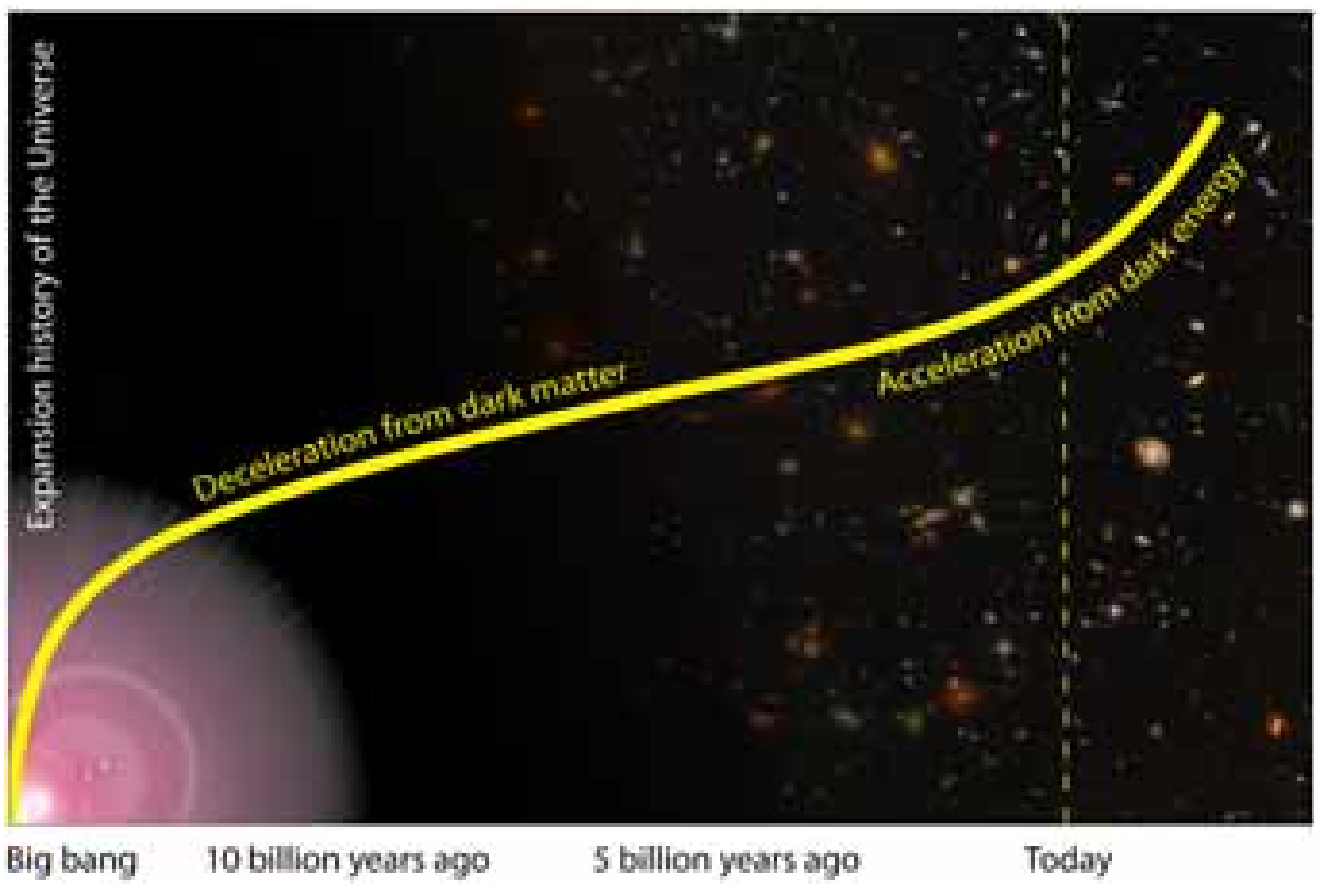

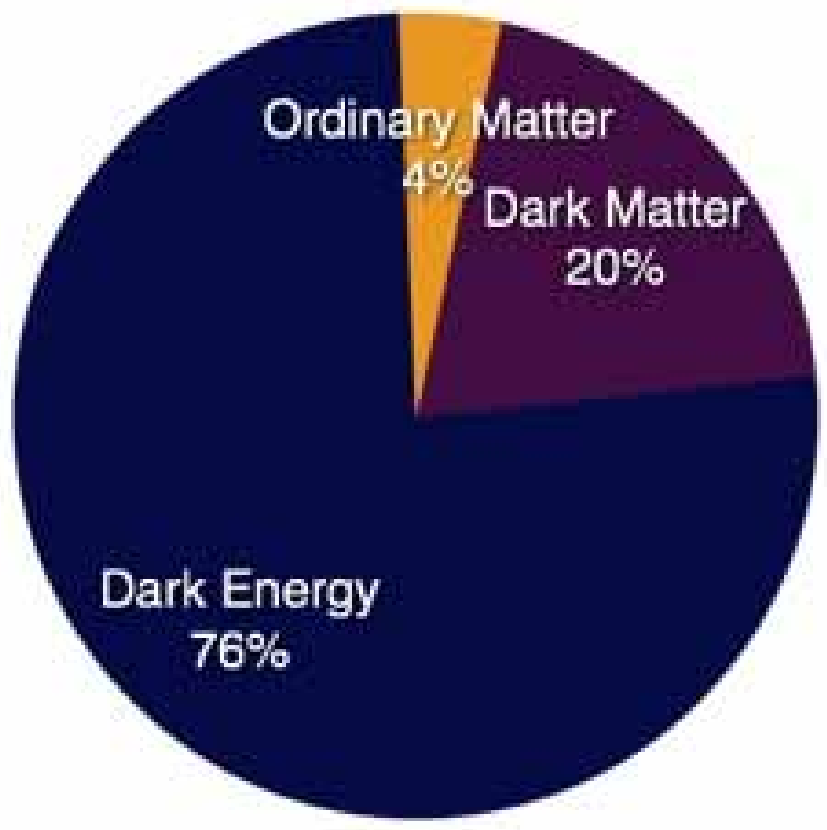

*Figure 2.1: Left panel: The Universe evolves from a homogeneous state after the big bang through cooling and expansion. The small initial inhomogeneities grow through gravity to produce the large-scale structures that are observed today. Right panel: The mass-energy budget at our cosmological epoch. The Universe appears to be dominated by ingredients, dark energy and dark matter, whose nature poses some of the most pressing questions in fundamental physics.*

### 2.1.2 The key questions

Determining the nature of dark energy and dark matter represents no less than an endeavour to understand the majority of the mass-energy content of the Universe. These are the biggest mysteries in contemporary cosmology, and credibly in all of physics. Substantial progress has been made by dedicated space missions that have measured the temperature fluctuations of the CMB with high precision. They provide a snapshot of how density fluctuations on different scales behaved at the time of hydrogen recombination and represent a formidable confirmation of the CDM scenario for structure formation. While ESA's Planck experiment will provide by far the best view of the CMB sky, a number of fundamental questions will remain unanswered, as it can only weakly probe the subsequent 13 billion years of evolution of structure in the Universe. For example, the relative heights of the peaks in the CMB power spectrum have ruled out a baryonic origin for the dark matter, but cannot provide us with more information about the nature of dark matter. Similarly, the CMB power spectrum provides strong evidence that dark energy exists, but its sensitivity to the actual nature of this component is limited.

In this proposal we argue that progress requires high precision measurements that can only be obtained by a dedicated space-based mission. Euclid is designed to accurately measure the expansion history of the Universe and the growth of cosmic structures using a large area optical and NIR imaging survey and a massive spectroscopic survey in the NIR. Euclid will create the largest and most comprehensive maps of the Universe around us, simultaneously tracing the distribution of both its luminous and dark components over more than one-third of the sky and out to epochs when the Universe was less than 3 billion years old. The statistical properties of these distributions will also constrain several properties of dark matter, including the contribution from neutrinos. Euclid is unique in the combination of its two primary probes, namely galaxy clustering and weak gravitational lensing: they not only allow the experiment to reach unprecedented statistical precision, but provide a crucial crosscheck of systematic effects, which become dominant at these levels of precision. Consequently, Euclid will be able to address the following key questions about the dark side of the Universe:

1. **Dynamical Dark Energy:** Is the dark energy simply a cosmological constant, or is it a field that evolves dynamically with the expansion of the Universe?
2. **Modification of Gravity:** Alternatively, is the apparent acceleration instead a manifestation of a breakdown of General Relativity on the largest scales, or a failure of the cosmological assumptions of homogeneity and isotropy?
3. **Dark Matter:** What is dark matter? What is the absolute neutrino mass scale and what is the number of relativistic species in the Universe?
4. **Initial Conditions:** What is the power spectrum of primordial density fluctuations, which seeded large-scale structure, and are they described by a Gaussian probability distribution?

The fourth question will help improve our understanding of the physics that caused inflation, the first period of accelerated expansion when the Universe was only a fraction of a second old. Euclid will complement the high-redshift picture of cosmology from the ESA Planck experiment by completing our census of the Universe at lower redshifts; together these two regimes will allow us to track the evolution of the Universe with unprecedented accuracy over cosmic time.

We now explain in more detail each of the primary science goals, and explain the level of precision needed to resolve each of the questions that were posed. To quantify the level of precision we will use parameterisations that are summarised in Table 2.1.

**Dark Energy:** The physical nature of the dark energy can be characterised at the most basic level by the relation between its average pressure $p(a)$ and energy density $\rho(a)c^2$, i.e. the equation-of-state $w(a) = p(a)/\rho(a)c^2$, where $a = 1/(1+z)$ is the scale factor of the Universe. In this formulation, the cosmological constant corresponds to $w(a) = -1$, and any deviation from this would imply a dynamical dark energy. A key goal of Euclid is to place tight observational bounds on any such deviation. The functional form of $w(a)$ is unknown, but to first order it can be approximated by a constant term and one that captures the dynamical nature of the dark energy:

$$w(a)= w_p+ w_a(a_p-a) \ ,$$

where the 'pivot' scale factor $a_p$ is chosen such that the statistical errors on these parameters, $\Delta w_p$ and $\Delta w_a$, are not correlated. Alternatively $a_p$ is set to 1 (present day) so that $w(a)= w_0+w_a(1-a)$, where $w_0$ and $w_a$ are correlated parameters. This simple parameterisation is convenient for the purposes of this document, as it allows us to compare the performance of current and future experiments. This parameterisation is commonly used to define the dark energy figure of merit $FoM = 1/(\Delta w_p \times \Delta w_a)$, the reciprocal ratio of the product of the $1\sigma$ error on these parameters, where a larger $FoM$ implies a more precise measurement of the dark energy properties. Note that this definition differs from the one proposed by the Dark Energy Task Force (Albrecht et al., 2006) who used the reciprocal of the area of the error ellipse enclosing the 2-sigma confidence limit. More sophisticated approaches will be employed to fully exploit the statistical power of Euclid.

An important question for the design of Euclid is how well $w(a)$ should be determined. Unfortunately, we lack clear theoretical guidance or indications from current observations, but we can present a statistical argument that the value of $w$ needs to be measured with a precision of ~1% to robustly test the cosmological constant model. In fact this line of reasoning is the most pessimistic case: any detection that $w$ deviates from $-1$ would be a scientific sensation, providing the first constraints on the nature of the dark energy. One can imagine comparing two models of dark energy, a cosmological constant and a model in which the two parameters $w_p$ and $w_a$ vary within a reasonable range. For very small variations the cosmological constant model is simpler and would be favoured because of Occam's razor; for very large deviations the more complex two-parameter model would be favoured. This is quantified using a Bayesian evidence calculation which shows that if the data are consistent with a cosmological constant, and the *FoM* > 400 (e.g. $\Delta w_p \sim 0.016$ and $\Delta w_a \sim 0.16$) then the cosmological constant would be favoured with odds of more than 100:1, which is considered "decisive" statistical evidence. A less precise experiment will not have enough statistical power to decisively favour a cosmological constant over the more complex alternative, and thus could not determine the nature of dark energy.

In Section 2.3 we show that the constraints from the two primary cosmological probes, weak lensing and galaxy clustering, provide complementary constraints on $w_p$ and $w_a$. Using these primary dark energy probes alone Euclid will meet its dark energy science goal of a *FoM*≥400, with subdominant systematic uncertainties. The impact of the primary probes, however, is not limited to learning more about the nature of dark energy.

*Table 2.1. An illustrative summary of the key parameterisations discussed throughout for each of the four primary science goals. These parameters are simply indicative of the potential of Euclid; in practice, a significantly more extended set of physical quantities will be constrained.*

| Goal | Parameter | Parameter Details |
|---|---|---|
| **Dark Energy** | $w(a)$ | The *dark energy equation of state* is the ratio of the pressure to density of dark energy $p(a) = w(a)\times\rho(a)c^2$. Euclid will determine the redshift dependence of this function with high precision. This dependence can be parameterised using a first order Taylor expansion with respect to the scale factor $a=1/(1+z)$, $w(a)= w_p+(a_p-a)w_a$. Detecting a $w(a)\neq-1$ at any redshift would demonstrate that dark energy is not a cosmological constant, but rather a dynamical field. |
| **Modified Gravity** | $\gamma$ | The *growth factor* [or its derivative, the growth rate $f(z)$] quantifies the efficiency with which structure is built up in the Universe as a function of redshift. It is directly sensitive to the nature of gravity. The growth rate is well described using a simple parametric function $f(z)=\Omega_m(z)^\gamma$. A detection of $\gamma\neq0.55$ would indicate a deviation from General Relativity, and thus a completely different origin of cosmic acceleration, rather than dark energy. |
| **Dark Matter** | $m_\nu$ | The total *neutrino mass* is the sum of the masses of the three known species (electron, muon and tau neutrinos). Massive neutrinos damp structure growth on small scales. The larger the mass, the more damping occurs, leaving a clear signature in the matter power spectrum observed by Euclid. |
| **Initial Conditions** | $f_{NL}$ | The concordance cosmological model assumes an initial Gaussian random field of perturbations, from which large-scale structure grows. A detection of *non-Gaussianity* would signify a departure from this central assumption of the current standard model. The $f_{NL}$ parameter is a way to quantify the amplitude of this effect. |

**Modified Gravity:** An alternative exciting possibility to explain cosmic acceleration is that Einstein's theory of General Relativity (GR), and thus our understanding of gravity, needs to be revised on cosmological scales. Models that modify GR in a generic way change both the expansion history and the evolution of the perturbations, and thus the growth of structure in the Universe. Although not the most general modification, in the simplest models the deviation from GR can be captured as a change in the growth of structure with respect to canonical dark energy models. This is often expressed by an additional parameter $\gamma$, which parameterises the rate of growth of a matter density fluctuation: $f(z)=\Omega_m(z)^\gamma$, where a deviation from $\gamma=0.55$ indicates a breakdown of GR. The goal of Euclid is to measure the exponent $\gamma$ with a $1\sigma$ uncertainty of 0.02, thus testing GR to an exquisite level of precision. Moreover, the measurements are precise enough to reconstruct the growth history in several redshift bins (i.e., going beyond the simple $\gamma$-parameterisation) and to constrain the form of the metric gravitational potentials.

In Newtonian gravity the acceleration is given by the gradient of the Newtonian gravitational potential. In General Relativity the acceleration is due to potential gradients in an analogous way, except that two potentials are needed. These are usually called Ψ and Φ, and describe the time-time and space-space parts of the (scalar) metric perturbations respectively. The difference between these potentials is called *gravitational slip*. Non-relativistic (slowly moving) objects such as galaxies experience a Newton-like acceleration proportional to the gradient of Ψ (which is then the close analogue of Newton's potential), while the path of a ray of light is affected by the sum, Ψ+Φ. Models with a cosmological constant or scalar field dark energy models tend to influence both potentials in the same way, while modified gravity models generically affect them differently. Hence the measurement of both, which Euclid will perform through the combination of its primary probes, can rule out whole classes of dark energy and modified gravity models.

**Neutrinos and Dark Matter:** We already know that some non-baryonic dark matter exists: oscillation experiments have shown that at least two of the three neutrino species have non-zero mass, but other large scale structure and CMB measurements suggest they make up only a small fraction of the dark matter in the Universe. Although particle physics experiments have established that the larger mass difference is of the order of 0.06 eV, it is extremely difficult for laboratory experiments to determine the total absolute neutrino masses. This is important, because such a measurement could answer the question whether the neutrino masses obey a normal (two light neutrinos, one massive neutrino) or inverted (two massive neutrinos, one light neutrino) hierarchy; understanding this will give indications about the mechanism that gave neutrinos their mass. To distinguish the two cases, it is necessary to measure the sum of the neutrino masses to a precision comparable to the larger mass difference. As neutrinos are light (or hot), they move large distances in the early Universe resulting in a tiny suppression of structure formation, but over a large range of scales. Euclid will measure this small, but observable effect on the matter power spectrum and reach a precision of $\Delta m_\nu < 0.03$eV. This is sufficient to determine the neutrino mass hierarchy, if the total mass turns out to be small, $m_\nu<0.1$ eV.

If the dark matter particle can interact with itself, the formation of structures on relatively small scales is suppressed. Current observations suggest that dark matter particles have a low self-interaction cross- section and a relatively high mass (eliminating hot dark matter). Euclid will measure the density profiles of dark matter halos and the matter power spectrum with very high precision, improving constraints on the dark matter self-interaction cross section by three orders of magnitude over current limits: this will either show that dark matter is warm or else put a lower limit on the dark matter mass of 2 keV. While the Large Hadron Collider at CERN and direct detection experiments probe dark matter particles directly, Euclid will test their macroscopic distribution in galaxies and clusters. This provides an important consistency check on their properties as well as constraints on interactions within the dark sector.

**The Early Universe:** Euclid will also provide new insights into the physical properties of the Universe when it was only a small fraction of a second old. This is possible because the large-scale structure that Euclid will observe arises from quantum fluctuations that grew to cosmological scales during *inflation*. This early period of accelerated expansion has been proposed to solve a number of problems in Big Bang cosmology, and could be related to the physics of dark energy.

Simple models of inflation predict a primordial power spectrum that is nearly scale invariant with power law spectral index $n_s$ close to unity. In more complicated models the spectral index $n_s$ may also depend on scale, often referred to as a running spectral index. Euclid will measure the spectral index and its scale dependence with a precision similar to that of Planck, thus providing independent confirmation of the CMB results and extending them to much smaller scales. By combining measurements, Euclid will improve constraints on the power spectrum of initial density fluctuations by a factor 2 over Planck alone.

The characterisation of the primordial power spectrum provides an important test of inflation models, but Euclid can do more: even the simplest inflationary models predict small deviations from a Gaussian probability distribution for the primordial density fluctuations, with more complicated models predicting higher levels of non-Gaussian fluctuations. The $f_{NL}$ parameter, which gives the amplitude of a quadratic term in the potential, is often used to quantify amplitude of non-Gaussianity. Planck is expected to determine $f_{NL}$ to an accuracy of ~5, while Euclid will measure determine $f_{NL}$ to an accuracy of ~2, a significant improvement using a complementary approach.

## 2.2 Mapping the luminous and dark Universe

We now describe the primary scientific techniques for which Euclid has been optimised, as well as the additional cosmological probes that do not drive requirements, but are made possible by the Euclid data. The combination of the primary probes challenges all aspects of the current concordance model of cosmology, whereas the impact of the secondary probes is mostly limited to improving our understanding of dark energy (see Section 2.3 for more details). The primary probes are:

**Weak gravitational lensing:** By measuring the correlations in the shapes of the 1.5 billion galaxies that Euclid will image, the expansion and growth history of the Universe can be determined with high precision. This is possible because the gravitational potential of intervening structures perturbs the paths of photons emitted by distant galaxies: it is as if these galaxies are viewed through a piece of glass with a varying index of refraction. As a result the images of the galaxies appear slightly distorted. The amplitude of the distortion provides us with a direct measure of the gravitational tidal field, which in turn can be used to 'map' the distribution of (dark and luminous) matter *directly*. This phenomenon is called weak gravitational lensing and is commonly referred to as 'cosmic shear' when applied to determining the statistical properties of the large-scale structure (see Figure 2.2).

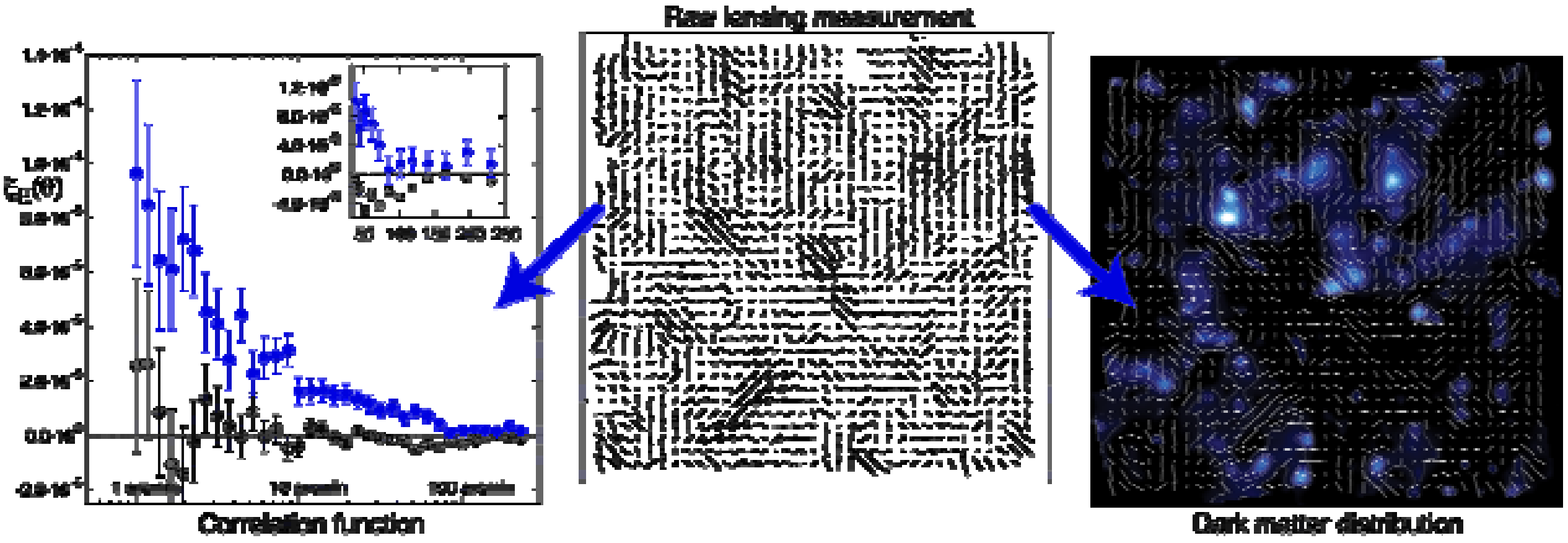

*Figure 2.2: (Middle panel) Raw gravitational lensing data obtained from HST observations (Massey et al., 2007). Each line/stick shows the average ellipticity of about 200 galaxies. Lines of sight without coherent lensing produce circular average galaxies, with zero ellipticity, represented by a dot; the line length indicates the magnitude of the lensing effect, and the orientation shows the major axis. (Right panel) Foreground dark matter lenses are mapped by filtering the observed shear field for the circular patterns. (Left panel) The clumpiness of dark matter on different physical scales can be quantified statistically via the correlation between shear along different lines of sight (ground-based results from Fu et al., 2008). An independent estimate of systematics is indicated by the black open symbols in the left panel; it should be consistent with zero but is not quite because of residual systematics, inherent to ground-based weak lensing studies*

Euclid will accurately measure the coherent pattern in galaxy ellipticities that gravitational lensing by large-scale structure imprints, which provides the most direct measurement of the (projected) matter power spectrum. Importantly, the availability of photometric redshifts for the sources means that is possible to study the mass distribution in *three dimensions*, called weak lensing tomography.

**Galaxy clustering:** By measuring the spectroscopic redshifts of 50 million galaxies in the redshift range $0.7<z<2.1$ the *three dimensional* galaxy distribution of the Universe can be mapped to high precision, and quantified in terms of its power spectrum (or correlation function) within several redshift bins over the time interval when dark energy becomes dynamically important. The amplitude, shape and anisotropy of these statistics contain the crucial information on the expansion and structure growth histories of the Universe.

We choose these primary probes for the following reasons.

- Weak lensing and galaxy clustering are the most sensitive probes of dark energy *and* of the theory of gravity on cosmological scales (Peacock et al., 2006 and Albrecht et al., 2006).
- Their combination simultaneously probes the cosmological expansion, the growth of structure and the relation between dark and luminous matter, providing fundamental cross-checks and cross-calibration of the systematic effects of each probe.

- They involve a minimum of astrophysics to interpret the results, compared to other approaches. At the precision that Euclid will achieve no probe is perfectly "clean" of astrophysical effects but fortunately, as discussed in Section 2.4, Euclid will also significantly improve our understanding of the astrophysics of galaxy formation and evolution.

In summary, weak lensing and galaxy clustering are two powerful independent probes of both expansion and growth, which complement each other ideally while having a natural synergy in the observing strategy and thus in their practical implementation.

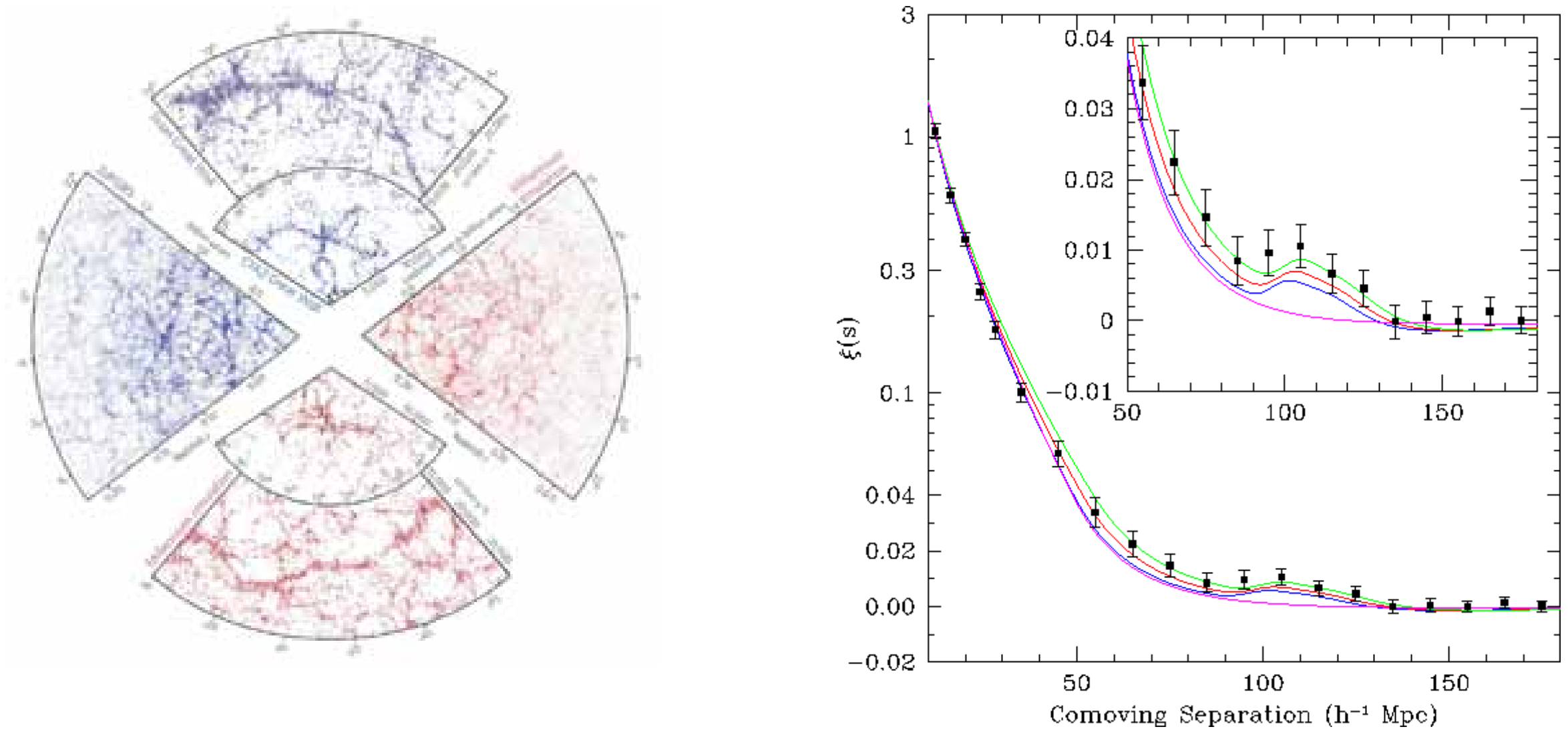

*Figure 2.3: (Left panel) The galaxy distribution in the largest surveys of the local Universe, compared to simulated distributions from the Millennium Run (Springel et al., 2005); (Right panel) The two-point correlation function of SDSS "luminous red galaxies", in which the BAO peak at ~105 $h^{-1}$ Mpc has been clearly detected (Eisenstein et al., 2005).*

### 2.2.1 Mapping the Expansion History of the Universe

The signature of sound waves in the photon-baryon plasma in the early Universe has now been measured with high precision in CMB data. These *Baryon Acoustic Oscillations* (BAO) are not limited to CMB observations: they have been seen in the galaxy distribution as a preferred co-moving separation of galaxies of ~150 Mpc (Figure 2.3; Eisenstein et al., 2005), or, equivalently, as a series of oscillations in the galaxy power spectrum (Cole et al., 2005). These are the low redshift equivalent of the wiggles in the CMB power spectrum and as such provide a uniquely precise and robust standard ruler to probe the expansion history of the Universe. What makes this a key method for measuring $w(z)$ is its low level of systematic uncertainties.

The physical size $s$ of the "ruler" is known, and the measurement of its angular size on the sky constrains $s/D_A(z)$, where $D_A(z)$ is the angular diameter distance, which provides a direct measurement of the expansion history. Thanks to the spectroscopic data, Euclid also makes sub-percent level measurements of $sH(z)$, where $H(z)$ is the Hubble parameter. The combination of radial and transverse distance measurements is used to determine the curvature at different redshifts to test for changes in geometry over cosmological time-scales.

Weak gravitational lensing also constrains the expansion history. The amount of distortion caused by any gravitational lensing is directly dependent on the observer-lens-source geometry. On cosmological scales this means that weak lensing is dependent on the geometry of Universe through a combination of angular diameter distances $D_A(z)$, to the weakly lensed galaxies. The correlation function of the weak lensing signal contains this geometric information (Hu, 1999), and using this Euclid will map the expansion history of the Universe to percent precision. Further techniques can extract this geometric signal, by using the weak lensing signal around galaxy clusters, and taking a ratio of quantities to cancel out any cluster mass dependence (Jain & Taylor, 2003; Taylor et al., 2006).

### 2.2.2 Mapping the growth of large scale structure in the Universe

The unique feature of Euclid is that it will measure *both* the power spectrum of the galaxy distribution, through its redshift survey, and that of the underlying total matter distribution through its weak lensing survey.

Weak lensing directly probes the total (dark plus luminous) matter power spectrum, as photons are sensitive to the integrated gravitational effect of all the mass along the path, while they travel towards us. The Euclid redshift survey has been optimised to fully exploit the BAO signature on scales around 150 Mpc, but by design it will also provide an unprecedented measurement of galaxy clustering over the maximum possible range of separations given by the full sky. The full power spectrum provides significantly more cosmological information, boosting the *FoM* of dark energy by a factor of a few beyond using BAO-only constraints. This comes at the price of requiring us to model the relation between the galaxy clustering signal and the predicted linear matter power spectrum, i.e. how galaxies trace mass (see Section 2.3 for more details). However, as remarked above, this relationship (also known as *galaxy bias*), will be determined by Euclid with unprecedented accuracy, by matching the maps of luminous and dark matter produced by its two main probes (see next section).

By determining the dark matter distribution at different redshifts, weak lensing maps will directly measure the growth rate of structure. At the same time, Euclid will be the first experiment to exploit another effect to measure the same quantity from its redshift survey: the maps of the galaxy distribution display a measurable anisotropy in *redshift space*, introduced by the additional contribution of galaxy peculiar velocities to the smooth Hubble flow. The effect is maximal along the line of sight, and is easily understood if one considers galaxies as test particles that partake in the overall growth of structure. As a result, on large scales peculiar velocities take the form of coherent bulk flows towards filaments and superclusters, and away from voids. Standard two-point clustering estimators can be generalised to account for this angular dependence and extract the anisotropic signal through specific modelling (Kaiser 1987). As such, *redshift-space distortions* provide us with an independent measurement of the growth rate $f(z)$, which, together with that from lensing tomography, will be Euclid's key measurement to detect the gravitational slip and break the dark energy versus modified gravity degeneracy (e.g. Guzzo et al., 2008, Percival & White, 2009).

### 2.2.3 Combining the primary probes

What makes Euclid a unique experiment is that it allows for the combination of the results from the two probes. The key benefit comes from the full combination of a deep imaging survey and an extensive redshift survey *over the same area of sky*. As a result the relationship between the distribution of luminous objects and the mass density field can be reconstructed, which in turn enables new science. For instance, Reyes et al. (2010) used the combination of the clustering amplitude, redshift distortions and weak lensing measurements from the SDSS to test gravity on large scales. In addition, Bernstein & Cai (2011) have recently shown that complementing a redshift survey with weak lensing data increases information on the growth rate by an amount equivalent to a 10-fold increase in the volume of a standard redshift-space distortion measurement, such that the latter can be measured *nearly free of sample variance*. Thus, a combined lensing and redshift survey over a common volume like Euclid is a more powerful test of GR than isolated redshift or imaging surveys, boosting the achievable precision by an order of magnitude. This is what is required to test gravity on cosmological scales, as the potentials are small (of the order of $1{:}10^5$, similar to the CMB $\Delta T/T$), requiring the statistical power and exquisite control of systematics that only the combined Euclid can provide.

### 2.2.4 Additional cosmological probes

Its two primary probes make Euclid a unique and powerful experiment to tackle key questions in cosmology. It will, however, also allow researchers to extract a number of ancillary measurements, which will further constrain cosmological models, which are briefly discussed here.

**Clusters of galaxies:** The extreme positive peaks of the matter density field are traced by galaxy clusters and provide complementary information, which is nearly independent from that of the primary cosmological probes. The cluster population bears the imprints of the statistical distribution of initial fluctuations, their subsequent growth and the dynamics of the collapse of dark matter halos. This threefold dependence makes clusters an excellent probe of the growth of structure in the Universe: the observed redshift distribution of galaxy clusters already provides the most stringent limit on models of modified gravity (Lombriser et al., 2010). In addition, the number of high mass clusters at high redshift is extremely sensitive to any primordial non-Gaussianity and deviations from standard dark energy models, which can also be constrained by exploiting the predicted scale-dependence of the bias.

The most efficient method to detect clusters with Euclid relies on analysing the photometric data, an approach that has been clearly demonstrated by the SDSS at low redshifts (e.g., Koester et al., 2007). Euclid will be able to push towards much higher redshifts over a large area, thanks to its unique capabilities in the infrared, which cannot be matched from the ground. Conservative estimates, based on simulated mock catalogues, indicate that Euclid will find of order 60,000 clusters with an $S/N$ of better than 3, between $z$=0.2 and $z$=2.0, with 10,000 having $z$>1. With such good statistics, our cluster-based constraints on cosmological parameters will be limited by our understanding of the catalogue selection function, systematic errors and, most important, the cluster mass determinations and their uncertainty. This is where the strength of Euclid lies: it is able to calibrate the important mass-observable relations and their scatter through lensing measurements. The exquisite image quality and high number density of sources will enable Euclid to measure masses of clusters much more accurately and out to higher redshifts than is possible from the ground.

The combination of Euclid data with other surveys, such as eROSITA or Planck, enables the cross-calibration of non-lensing mass-observable relations, which are currently limited to low redshifts and small samples. For example, Euclid will provide mass proxies via the stacking of clusters according to their X-ray or Sunyaev-Zel'dovich signals.

**Dark matter density profiles:** Numerical simulations of cold dark matter dominated universes make strong predictions about the average density profiles of collapsed structures (or halos) and the amount of substructure within them. For example, the *concentration* of a halo is believed to correlate with its formation history (e.g. Navarro, Frenk & White, 1996). On galaxy scales these predictions are difficult to test because most observations probe the regions where baryons (rather than the simulated CDM) dominate the gravitational potential. Clusters of galaxies are better targets, allowing us to probe a physical regime where the fraction of baryons trace the CDM in a more simple way, and where complementary X-ray or Sunyaev-Zel'dovich observations provide additional insight.

Euclid will determine the average weak lensing signal around the clusters with unprecedented precision, providing unique constraints on the density profiles on scales larger than ~100 kpc. This will improve on current results (Leauthaud et al., 2010), by a factor ~50 out to $z$~1 and extend this work to higher redshifts. Importantly, the large number of clusters and the high precision of the measurements make it for the first time possible to correlate with other observables (e.g., luminosity of the central galaxy or X-ray properties), such that their impact on the density profile can be studied.

The mass distribution in the central regions can be studied best by modelling strong lensing features that Euclid will discover thanks to its high image quality. In particular the rare radial arcs constrain the local slope of the density profile, while tangential arcs place tight limits on the enclosed projected mass. With more modelling the morphology and distribution of the multiply-lensed images can provide direct constraints on the presence of substructure or constrain the density profile with high precision (e.g. Smith et al., 2009; Jullo & Kneib, 2009; Meneghetti et al., 2010).

**Complementing ESA Planck data:** Constraints on cosmological parameters are significantly tightened when Euclid measurements are combined with the Planck observations of the cosmic microwave background. Euclid data themselves also enable new science using the Planck observations, by providing information on the matter distribution in front of the surface of last scattering. For instance, the integrated Sachs-Wolfe effect (ISW; Sachs & Wolfe, 1967) is a secondary anisotropy of the CMB and a direct signature of dark energy in a spatially nearly flat Universe. The effect is caused by the decaying gravitational potentials of the large-scale structure at low redshift. As such it is sensitive to the derivative of the growth factor, which in combination with the primary probes can place constraints on alternative models of gravity. The ISW effect is a weak signal that can be detected only through correlations of large-scale structure with the CMB. Euclid represents a near ideal survey for the ISW effect (Douspis et al., 2008; Dupe et al., 2011).

Another exciting new area is the measurement of the lensed CMB sky. Euclid will provide a map of lenses over the redshift range $0.5<z<1.5$, where the weak lensing cross-section for CMB anisotropies is maximal.

This will allow the determination of the lensing distortions on CMB anisotropies, allowing us to put constraints on the high redshift behaviour of the dark energy, as well as helping to subtract the lensing foreground in the context of CMB B-mode measurements.

**Type Ia Supernovae in the Deep Survey:** The use of Type Ia supernovae (SNe) as calibrated light sources to determine the distance-redshift relation led to the discovery of dark energy and is therefore a well-established technique (Riess et al., 1998; Perlmutter et al., 1999). It has been recently demonstrated that the intrinsic scatter in the Hubble diagram is lower in the rest-frame *I*-band (observed in the Euclid NIR bands at $z\sim 0.2-1.2$) than in the "traditional" rest-frame *B*-band (Freedman et al., 2009). Repeated observations by Euclid will be used to calibrate the wide survey and, depending on the cadence, would provide NIR light-curves and colours for a few thousand Type Ia SNe. Furthermore, the Euclid spectroscopy would provide spectroscopic redshifts for many of the host galaxies, although ground-based spectroscopic redshifts would still be required for at least a subset of the SNe to confirm the classifications. Euclid will be the first large-scale NIR search for SNe from space.

With a cadence of approximately 1 observation per week, the feasibility of which will be considered during the optimisation phase of the survey strategy, Euclid will find about 3,000 Type Ia SNe to $z\sim 1.2$ with NIR light-curves and colours. A further 6,000 type Ia SNe and ~10,000 core-collapse SNe would be discovered with at least one detection. The latter sample will contain several tens of well-observed Type-IIP SNe out to $z\sim0.3$. These detections will be used as an additional probe to measure the expansion history of the Universe (see Section 2.2.4), whereas the measurement of SNe rates provides information on the star formation history of the Universe and on stellar evolution.

## 2.3 Precise and accurate cosmology with Euclid

The science goals of Euclid are compelling, addressing some of the most fundamental questions about the Universe and its earliest stages. Achieving the science objectives is challenging, but feasible only because of the stable observing conditions in space that Euclid provides. Although a number of ground-based projects have similar objectives, only Euclid can measure cosmological parameters to high precision, while simultaneously keeping residual systematics to a sub-dominant level of uncertainty. A detailed discussion of the requirements that Euclid needs to meet to provide both accurate and precise measurements is presented in Section 3. Nonetheless it is useful to briefly review the main benefits that Euclid provides over ground-based observations for the primary cosmology probes.

**A small and stable PSF:** The cosmic shear survey requires accurate measurements of the shapes of many galaxies, which is made possible thanks to the stability of Euclid and its exquisite image quality. The latter minimises any corrections for the inevitable blurring caused by the PSF and the stability allows the PSF to be known accurately as a function of time and across the field of view.

**Deep NIR photometry:** Weak lensing tomography not only requires exquisite shape measurements, but also information about the redshifts of the source galaxies. Euclid will obtain *photometric redshifts* from multi-band photometry. The optical colours will be obtained using ground-based observations, but to achieve the required small scatter and outlier rate NIR observations are required. To be of use for Euclid's science goals, this can only be done from space because of the much lower background. Euclid will therefore provide NIR photometry down to $m_{\rm AB}\sim24$, which is three magnitudes deeper than can be achieved from the ground over such a large area.

**Deep NIR spectroscopy:** The precision measurements of the clustering of galaxies require an unprecedented number of redshifts for galaxies over most of the extra-galactic sky in the redshift range $0.7<z<2.1$ in order to probe the epoch when dark energy became important. This requires measuring the Hα line using deep NIR spectroscopy that can only be provided by space observations, because ground-based efforts suffer from absorption and emission lines in the atmosphere.

Consequently Euclid can constrain dark energy and modified gravity parameters to percent-level accuracy, *only because* any biases caused by systematic effects are subdominant compared to the cosmological signal. This leads to realistic predictions of the performance, which are detailed in the following section.

### 2.3.1 Cosmological performance

Given the systematic control that is explained in detail in Section 3, here we will quantify the expected performance of Euclid instead. We explain the performance calculations in some technical detail; for a summary see Section 2.3.3 and Table 2.2.

The concordance model of cosmology ($\Lambda$CDM) can be parameterised by a simple set of values that describe the two dominant components, which are a static cosmological constant $\Lambda$ and cold dark matter (CDM). We define the concordance model with fiducial values as $\{\Omega_m$: 0.25, $\Omega_\Lambda$: 0.75, $\Omega_b$: 0.0445, $\sigma_8$: 0.8, $n_s$: 1.0, $h$: 0.7$\}$. This model assumes General Relativity and Gaussian initial conditions and contains a dominant cosmological constant, $w(a)=-1$, and dark matter components. To address the four questions described in Section 2.1, we extend this concordance model in four ways, as summarised in Table 2.1. Each extension tests a fundamental aspect of our Universe.

- **Dynamical Dark Energy.** Euclid will test the concordance model of cosmology by determining the dynamical nature of dark energy. To capture this we extend the parameter set with $w_p$ and $w_a$, and assume values of -0.95 and 0.0 respectively.
- **Modifications to Gravity.** To include modified gravity we parameterise the growth of structure using $f(z)=\Omega_m(z)^\gamma$. For General Relativity there is an approximate prediction that $\gamma$=0.55 and that any deviation from this value would suggest a break down of the concordance model.
- **Dark Matter.** We extend the concordance model to include massive neutrinos, a sub-dominant type of dark matter that is not cold. We add the sum of the neutrino masses $m_\nu$ as an extra parameter with an assumed value of 0.25eV.
- **Initial Conditions.** In the concordance model the initial perturbations from which large-scale structure grows are assumed to have a Gaussian distribution. We parameterise the amplitude of the non-Gaussianity by $f_{NL}$ where any non-zero value would signify a departure from the concordance model.

These extensions are shown as an illustrative example of the performance of Euclid, which will scrutinise every facet of the $\Lambda$CDM model to unprecedented accuracy: if the concordance model is refuted in any way this will pave the way for a new understanding of our Universe.

We now proceed to describe some of the details used in the cosmological forecasting calculations. As a baseline approach we use the standard error-prediction Fisher matrix formalism (Tegmark, Taylor & Heavens, 1997) that allows the expected performance of an experiment in terms of forecasted parameter errors to be determined in a well-understood way. All errors shown are the one-parameter 1$\sigma$ confidence limits marginalised over all other parameters in the set. When performing these calculations we do not assume flatness, unless specified (i.e. we *do not* assume that $\Omega_m + \Omega_\Lambda = 1$).

**Weak Lensing Forecasts:** The expected performance for the Euclid weak lensing probe was calculated using the weak lensing tomography technique (Hu et al., 1999), in which the galaxies are divided into a number of redshift bins. The calculations assume a survey covering 15,000 deg$^2$, a realistic source redshift distribution (with a mean of $z$~0.9) and a number density of 30 galaxies per square arcminute, resulting in a total of 1.5 billion galaxies. The survey is split into 10 tomographic redshift bins such that each bin contains the same number of galaxies. This technique uses the two-point correlation of the weak lensing signal, the photometric redshifts and the shape information from galaxies in the wide survey.

The results shown here are significantly more realistic than those in the Euclid Assessment Phase Study Report (*Yellow Book*, Laureijs et al., 2009). A restricted range of scales is used, such that the effects of baryonic feedback on the lensing power spectrum are minimised (a maximum $l_{max}$=5000, see Kitching & Taylor, 2011; Semboloni et al., 2011). Tidal interactions can lead to “intrinsic alignments” of galaxies with the surrounding density field that can mimic a lensing signal. We account for this primary astrophysical systematic by extending the cosmological parameter set to allow for unknown variation in the cosmic shear power spectrum caused by intrinsic alignments. This self-calibration is performed following Joachimi & Bridle (2010); to account for the uncertainty in the intrinsic alignment power spectra, a grid parameterisation over the relevant ranges in redshift and wave-number is introduced, where the values on the grid enter the analysis as additional parameters, which should be conservative. In addition to galaxy ellipticity correlations, this approach incorporates information from number density-ellipticity correlations and number density correlations (including lensing magnification).

The galaxy bias entering these signals is modelled with the same grid parameterisation as the intrinsic alignment contribution. The intrinsic alignment signal, though diluting the signal due to the increase in the parameters to be constrained, also helps to constrain the cosmological parameters. Nonetheless the constraints are clearly dominated by the weak lensing signal. We anticipate that ground based experiments such as the Dark Energy Survey (DES), Pan-STARRS and the Kilo Degree Survey (KiDS) will have constrained the intrinsic alignment signal to ~10% as a function of redshift and scale by the time Euclid launches.

**Galaxy Clustering Forecasts:** To forecast the dark energy constraints from galaxy clustering we use the Power Spectrum *P(k)* method developed by Seo & Eisenstein (2007), which utilises the measured clustering signal in its entirety, with a cut-off at intermediate scales to avoid non-linear effects (Rassat et al., 2008). The galaxy clustering constraints use the spectroscopic information of galaxies in the Euclid wide survey.

The implementation used to forecast the constraints on dark energy and growth rate from galaxy clustering is described in Wang et al. (2010). We account for the primary astrophysical systematic in galaxy clustering, an unknown galaxy bias, by marginalizing over variable amplitude. To minimise non-linear effects, wavenumbers are restricted to the quasi-linear regime.

The weak lensing intrinsic alignment signal has a negligible contribution from galaxy clustering because in these calculations only the photometric redshift information for the weak-lensing measurement is used (see Joachimi & Bridle, 2010). Hence combining the weak lensing predictions with the galaxy clustering results is justifiable. This leads to an improvement of the predicted performances compared to the Assessment Phase Study Report, despite the increased realism in the accounting of systematics: thanks to recent work, it is now possible to include more information from both the primary probes.

**Forecasts from Clusters and ISW:** Constraints on cosmological parameters can be improved by augmenting the two primary probes with other measurements. When making forecasts, we include the expected constraints from two of these that are only possible thanks to Euclid data: measurements of clusters of galaxies and the ISW effect (see Section 2.2.4 for details).

To predict cosmological constraints from the expected sample of galaxy clusters, first the survey selection function is calibrated using two approaches based on photometric richness as a cluster mass proxy. The first uses a mock galaxy catalogue constructed from the COSMOS survey, while the second is an analytic calculation of the confidence level of detection of a cluster identified as a spike in the photo-z galaxy distribution. Only *H*-band information was used and hence our estimates are conservative. A cluster finding algorithm was applied to the mock galaxy catalogue and the resulting cluster catalogues were then compared to predictions from the analytical approach. The two were found to give comparable results. For the analysis presented below, the conservative analytic estimate of the limiting mass with an $S/N=3$ was used. We use a survey limiting mass and a mass function given by Tinker et al. (2008), and account for systematics in the richness relation (using the approach of Lima & Hu, 2005).

For the ISW the spherical harmonic cross-correlation power spectrum of galaxy and CMB temperature maps is used as an observable. This is calculated using linear theory (on non-linear scales the Rees-Sciama effect occurs) and the calculation uses a tomographic approach, where the redshift distribution of each bin is modified to account for photometric redshift errors. As a by-product, the method produces a reconstructed ISW temperature map, which can then be used to investigate anomalies in the primordial CMB on large scales.

## 2.3.2 Unprecedented cosmology constraints

The accuracy that will be achieved by Euclid is impressive, as is demonstrated by the predictions presented in Table 2.2. Euclid will dramatically improve over current constraints and will provide a fundamentally new level of precision. Euclid will test each aspect of the concordance model, improving by a factor of 10 or more each unknown aspect of our understanding of the Universe. This is a result of the power of the individual probes, their complementarily *and* the control of systematics that Euclid provides.

Euclid will determine the dark energy equation of state to 1% alone, with no additional assumptions, and will reduce the uncertainty in $\gamma$ by a factor of 30. The uncertainty in the neutrino mass is reduced by a factor of 30. Finally, our understanding of dark energy, quantified by the *FoM* will improve by a factor of >300 over the current *FoM*. The current constraints on the neutrino mass and $f_{NL}$ measurements are from Komatsu et al. (2010) and the current dark energy constraints from Suzuki et al. (2011) who include supernovae, BAO and CMB constraints. Current γ constraints are taken from Rapetti et al. (2009) who make a measurement under the assumption of flatness; we do not make this assumption, so the improvement derived for this parameter should be considered a conservative estimate.

*Table 2.2: A summary of the forecasted cosmology constraints from Euclid. The figure of merit (FoM) is listed in the last column. Note that a larger FoM is better.* ***Euclid Primary:*** *Combined constraints from Euclid weak lensing tomography and galaxy clustering.* ***Euclid All:*** *Constraints from primary probes combined with galaxy clusters and ISW.* ***Current*** *constraints from Rapetti et al. (2009), Komatsu et al. (2010) and Suzuki et al. (2011).* ***Improvement Factor:*** *improvement over the current constraints compared to the Euclid+Planck case. For modified gravity a simple parameterisation of the growth factor $f(z)=\Omega_m^{\gamma}$ is used. The neutrino mass $m_\nu/eV$ is the total mass summed over all species, assuming a degenerate hierarchy. All constraints are 1σ predicted errors marginalised over all other parameters ($\Omega_m$: 0.25, $\Omega_\Lambda$: 0.75, $\Omega_b$: 0.0445, $\sigma_8$: 0.8, $n_s$: 1.0, h: 0.7). Here we use expected 2-point (TT, ET, EE, BB) correlations from Planck, and do not include CMB lensing.*

| | **Modified Gravity** | **Dark Matter** | **Initial Conditions** | **Dark Energy** | | |
|---|---|---|---|---|---|---|
| **Parameter** | $\gamma$ | $m_\nu/eV$ | $f_{NL}$ | $w_p$ | $w_a$ | *FoM* |
| Euclid Primary | 0.010 | 0.027 | 5.5 | 0.015 | 0.150 | 430 |
| Euclid All | 0.009 | 0.020 | 2.0 | 0.013 | 0.048 | 1540 |
| Euclid+Planck | 0.007 | 0.019 | 2.0 | 0.007 | 0.035 | 4020 |
| Current | 0.200 | 0.580 | 100 | 0.100 | 1.500 | ~10 |
| **Improvement Factor** | **30** | **30** | **50** | **>10** | **>50** | **>300** |

The *FoM* provides a convenient way to assess the statistical power of a combination of measurements, but does not take into account the detrimental effects of systematic errors. Hence a means to assess the influence of such biases is critical: the FOM only makes sense if systematic errors are negligible. In this particular respect, the Euclid mission can be compared to HST Key Project on the Hubble constant $H_0$, which primarily focused on reducing the systematics on absolute calibration of a few highly resolved Cepheids (Freedman et al., 2001). The primary strength of Euclid is its control of biases produced by systematics and on the use of several methods jointly, applied to the same survey. The primary probes are individually sufficiently precise to test for consistency between results. This ability is critical given the profound implications of an observed deviation from the concordance model and is lost if the statistical uncertainty of any individual probe is large compared to the objective. Although a *FoM*~400 may appear achievable if current constraints are combined with future data from the Dark Energy Survey (DES[1]), the Baryon Oscillation Spectroscopic Survey (BOSS[2]), and Planck, the relatively large uncertainties of the individual ground-based probes prevents their internal consistency to be determined. The debate about the value of the $H_0$ provides a well-known example: both sides claimed small statistical uncertainties (i.e. large *FoM*), yet the actual values were different.

Our forecast results are an improvement over the numbers presented in the Yellow Book (Assessment Phase Study Report) because we now include the full galaxy power spectrum. Previously only the localised BAO peak position was used, which contains less information. We also include realistic secondary dark energy probes for the "Euclid All" scenario in Table 2.2. By themselves the secondary probes constrain the dark energy properties to $\Delta w_p$=0.05 and *FoM*=55; however in combination with the weak lensing and clustering results, the sum is much more than the individual parts leading to a substantially improvement *FoM*>1500. The results presented here are consistent with the findings of the ESA-ESO working group on fundamental cosmology (Peacock et al., 2006), the NASA Dark Energy Task Force (Albrecht et al., 2006) as well as numerous papers available on the predicted constraints obtainable for the Euclid cosmological probes.

---

[1] http://www.darkenergysurvey.org/reports/proposal-standalone.pdf
[2] http://www.sdss3.org/collaboration/description.pdf and Eisenstein et al. (2011)

**Constraints on Dark Energy:** Table 2.2 and Figure 2.4 show that the Euclid primary probes alone will determine the dark energy equation of state with a *FoM*>400. In combination with the secondary dark energy probes, clustering and ISW Euclid will surpass the science requirement of *FoM*=400 by a factor of 3. In combination with Planck results, Euclid can surpass the primary dark energy science goal by a factor of 10, improving upon current constraints by over a factor of 100. These constraints will allow each of the broad classes of dark energy models to be tested: freezing models where $w$ tends to -1 at low redshift, thawing models where $w$ deviates from -1 at low redshift, and phantom models where $w$ is less than -1 at any redshift.

A deviation from $w$=−1 at any redshift would signify that dark energy is not a cosmological constant. Expressing the constraints in the ($w_p$, $w_a$) plane as a constraint on the redshift evolution of $w(z)$ it is clear that the functional form of $w(z)$ will be constrained to percent accuracy over the redshift range $0<z\leq 2$. Figure 2.4 shows that the Euclid primary probes can constrain $w(z)$ around $z$~0.5 to percent accuracy, which by itself could provide evidence for a departure from a cosmological constant. In combination with the secondary probes and the CMB the redshift dependence can be constrained to percent level over a wide redshift range.

**Constraints on Modified Gravity:** Euclid will test the theory of General Relativity on cosmological scales. One way to do so is to examine the growth of structure using the γ-parameter described earlier. Our results suggest that Euclid can constrain this parameter to 0.01 (where ΛCDM corresponds to γ=0.55). Figure 2.5 shows the expected constraints on γ, which are consistent with other studies (e.g. Heavens, Kitching & Verde, 2007). As discussed in Section 2.1, the γ-parameterisation is merely an example. In general at least two parameters should be used in order to have a sufficiently flexible model to capture general modifications to gravity (e.g. Amendola, Kunz & Sapone, 2008; Ferreira & Skordis, 2010) and it has been shown (e.g. Daniel et al., 2010; Amendola et al., 2010) that a Euclid-like survey could measure these parameters to high precision.

**Constraints on Neutrino Mass:** Euclid will be sensitive to the properties of weakly interacting particles in the eV mass range, such as massive neutrinos. Table 2 shows that Euclid will constrain the sum of the neutrino masses with a precision of 0.019 eV. Here we assume that the mass is 0.25 eV; if the mass is larger (0.5 eV) then a Euclid combined constraint of 0.022 eV is found, and if the mass if smaller (0.1eV) the Euclid combined constraint is 0.060 eV. If the neutrino mass is the smaller of these possible outcomes then the neutrino hierarchy could also be constrained. These are conservative estimates because the expected signal from weak lensing of the CMB itself is not included, which can also be used to place limits on the neutrino mass.

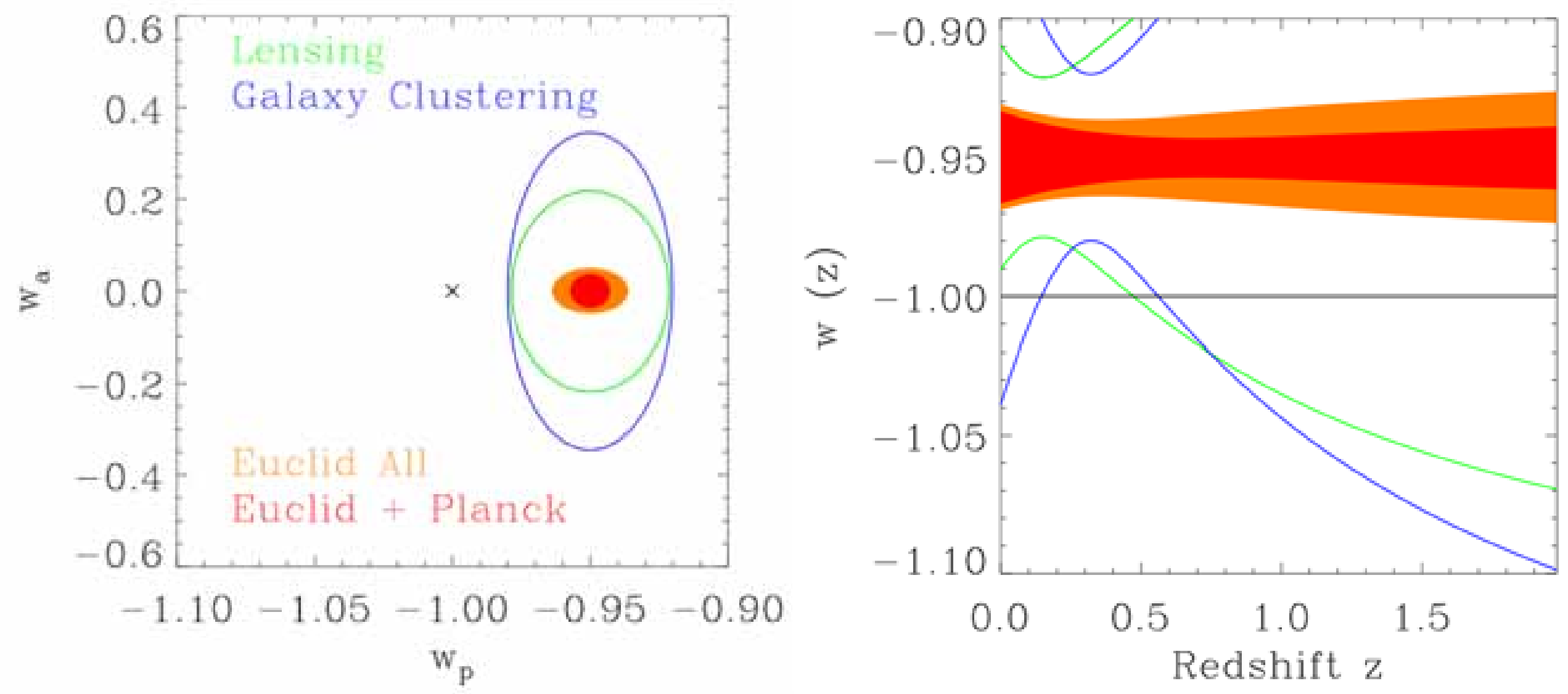

*Figure 2.4: The expected constraints from Euclid in the dynamical dark energy parameter space. We show lensing only (green), galaxy clustering only (blue), all the Euclid probes (lensing+galaxy clustering+clusters+ISW; orange) and all Euclid with Planck CMB constraints (red). The cross shows a cosmological constant model. Left panel: the expected 68% confidence contours in the ($w_p$, $w_a$). Right panel: the 1σ constraints on the function w(z) parameterised by ($w_p$, $w_a$) as a function of redshift (green-lensing alone, blue-galaxy clustering alone, orange-all of the Euclid probes, red-Euclid combined with Planck).*

**Constraints on Initial Conditions:** As shown in Figure 2.5, Euclid will constrain the shape of the primordial power spectrum parameterised by the spectral index $n_s$ to percent accuracy when combined with Planck results. If the assumption of a Gaussian random field is relaxed then Euclid can constrain the amplitude of the non-Gaussianity $f_{NL}$ through 3-point statistics of the weak lensing and galaxy clustering signals and through the correlation function of clusters of galaxies. We find agreement with previous results (e.g. Fedeli et al., 2011), where the combination of the galaxy power spectrum with the cluster-galaxy cross spectrum can decrease the error on the determination of $f_{NL}$ by up to a factor of 2 relative to either probe individually. Through the combination of lensing, galaxy clustering and clusters we find that Euclid can constrain $\Delta f_{NL}$~2, competitive and possibly superior to future CMB experiments.

In fact, if the simplest inflationary scenario holds, Euclid is expected to detect a non-Gaussian signal due to large-scale corrections needed in the Poisson equation from general relativistic effects, while no such imprint should be detectable in the CMB. Here the unique combination of the two primary cosmological probes again enables the discrimination among models for the origin of cosmological structures.

To conclude, we have presented the primary science goals of Euclid, and shown that these laudable objectives can be met by the experiment that we present. Euclid provides a major step forward, reducing the uncertainties of a number of key cosmological parameters by impressive factors. It will either confirm the concordance model with unprecedented accuracy, or else lead the way to exciting alterations of it, signalling the need for a revision of fundamental physics.

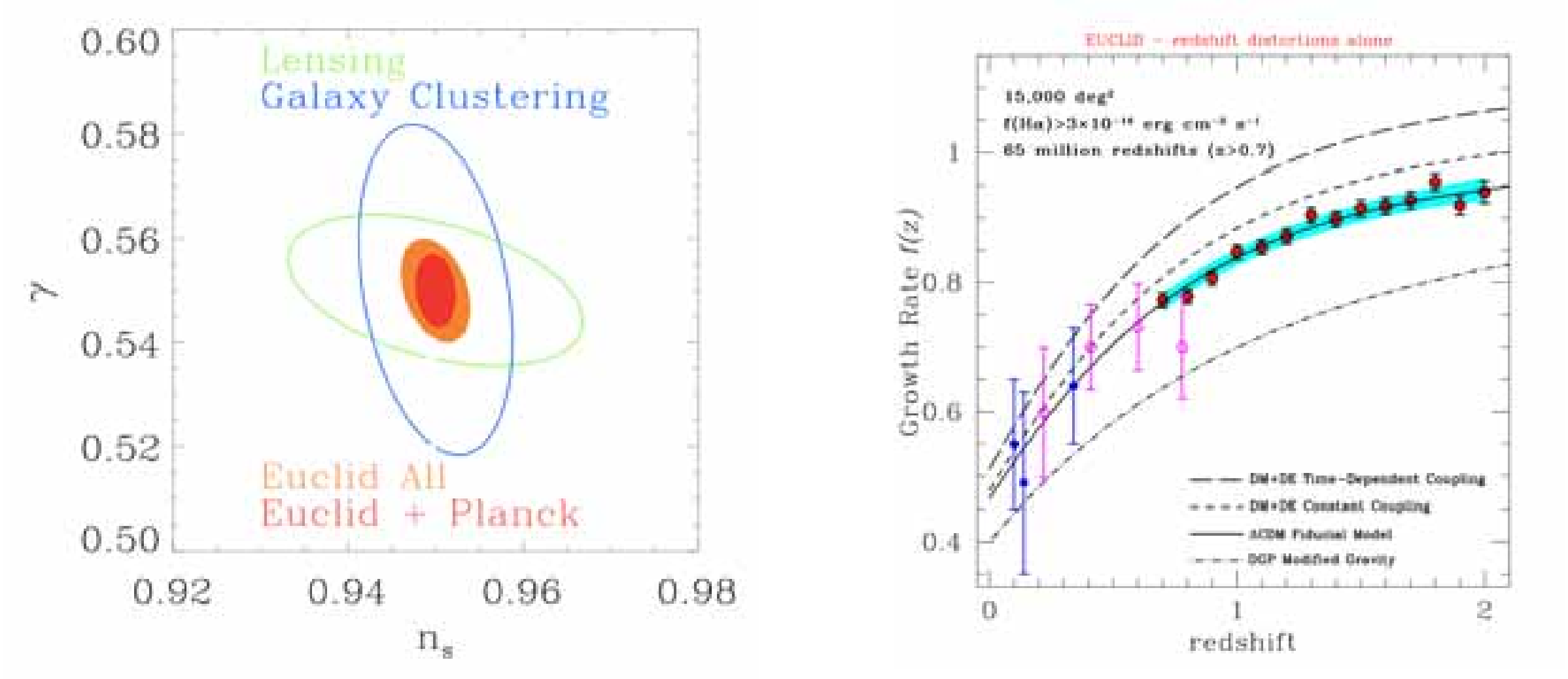

*Figure 2.5: In the left panel we show the parameter space constraints on the γ parameter describing the growth factor and the scalar spectral index. Green is lensing, blue galaxy clustering, orange includes the primary and secondary Euclid probes and red is combined with Planck. These errors are marginalised over all other parameters. Right panel: Predicted Euclid measurements of the growth rate of structure f(z) using redshift-space distortions alone. The cyan (shaded) area gives the expected 1σ error, with the red points illustrating a corresponding simulated observation. Current state-of-the-art measurements by the SDSS (filled pentagons), 2dF (filled square, Hawkins et al., 2003) and Wigglez (open hexagons, Blake et al. 2011) are also shown. The lines show predictions for f(z) by the concordance model and by three alternative models in which DE couples with DM (Di Porto & Amendola, 2007) or gravity is generalised to a 5-dimensional brane-world (DGP, Dvali et al., 2000).*

## 2.4 Legacy science

The design of Euclid is driven by our desire to study some of the most fundamental problems in cosmology, but the survey that is needed to achieve these goals will provide a dataset that will be of immense value for astrophysics as well: it will be important for understanding the formation and evolution of structures in the Universe at all scales, from galaxy clusters to brown dwarfs. The Euclid wide survey required to achieve the cosmological goals (see Section 3) will image 15,000 $deg^2$ of extra-galactic sky in the optical with a spatial resolution approaching that of HST, and to a depth in the near-IR at which only an area 1000 times smaller can feasibly be surveyed from the ground.

While Euclid will not go as deep as next-generation telescopes such as JWST or E-ELT, it will be far superior for surveying a large volume of space, detecting brown dwarfs, faint halo stars, low-redshift dwarf galaxies, luminous giant galaxies to $z\sim3$ and ultra-luminous galaxies and quasars at $z>7$. This offers enormous gains in studies using population statistics and for discovering rare objects. In fact rare sources often offer the most stringent constraints on models: getting the mean right might be feasible, but explaining extreme objects is usually very challenging. Rare sources often require several beneficial coincidences to take place. An example of this is the effect of strong lensing which occasionally magnifies distant sources by large amounts.

The wide survey will be complemented by two deep survey fields, which are located near the Ecliptic Poles. These fields, each covering ~20 deg$^2$, will be visited repeatedly for calibration purposes (see Section 5). This allows for the discovery of variable sources, such as supernovae. Furthermore the added depth will make these fields prime targets for follow-up observations. Hence Euclid will produce a legacy dataset with images of more than 1.5 billion galaxies and more than 50 million spectra, considerably larger than the SDSS and reaching well beyond $z\sim1$. The number of papers produced by scientists not in the SDSS consortium outnumbers the number of consortium papers by a factor of several. One can anticipate that Euclid legacy science papers will similarly come to dominate over the cosmology papers after sufficient time. In the following we provide a quantitative summary of the science that will be uniquely possible with the Euclid data, but note once more that many other areas of astronomy will also benefit.

## 2.4.1 Euclid legacy in numbers

At high redshift, Euclid is the only planned survey capable of finding the brightest $z>7$ galaxies and quasars in statistically significant numbers. The deep survey will detect hundreds of galaxies at redshift $z>7$, brighter than $3L^*$ (where $L^*$ indicates the luminosity of the knee of the luminosity function), none of which has been found to date, with this brightness. Furthermore, Euclid will detect tens of quasars at $z>8$, brighter than $J$=22, suitable for detailed spectroscopic study of the inter-galactic medium (IGM), and in the epoch when the Universe was re-ionised.

Another important redshift range where Euclid will have a significant impact is $1<z<3$, which is the period when the star formation and Active Galactic Nuclei (AGN) activity was at its peak. Based on existing surveys, Euclid is expected to observe $\sim2\times10^8$ galaxies with $H<22.5$ with good photometric redshifts and stellar mass estimates in this redshift range. About $\sim8\times10^6$ of these will be quiescent galaxies with stellar masses exceeding $4\times10^{10}$ M$_\odot$ and with z>1.5, and a few thousand of them will be very massive, but rare, quiescent galaxies ($M_{star}>4\times10^{11}$M$_\odot$) for which spectra will be available as well; no on-going or planned survey will even approach these numbers. The properties of galaxies will be studied across a wide range in environment with superb statistics. For instance, Euclid is expected to observe a total of ~18,000 galaxy cluster members (brighter than $H$=22.5) in massive clusters ($M>4\times10^{14}$ M$_\odot$,) beyond $z\sim1$. Measurements of Hα luminosities for $\sim4\times10^7$ sources are expected in the wide survey, going below the knee of the luminosity function out to $z\sim1.8$. Of these, $\sim5\times10^5$ ($1\text{x}10^5$) sources with $1.26<z<2$ in the wide (deep) survey will also have Hα and [OIII]5007. By combining Hα, [OIII] and [OII] measurements accurate metallicities for $\sim5\times10^3$ ($\sim2\times10^4$) galaxies with $1.95<z<3$ in the wide (deep) survey can be determined.

Strong gravitational lensing observations, which require high-resolution images such as those provided by Euclid, allow uniquely precise studies of the matter distribution in galaxies. At the same time the lensed galaxies provide a magnified view of the high-redshift Universe. Euclid can discover an estimated ~300,000 galaxy-scale strongly lensed systems, ~1000 multiply-imaged quasars, and ~5,000 clusters containing strongly distorted arcs. These numbers represent an increase by several orders of magnitude relative to the total of all other surveys preceding it.

If the deep fields can be observed with a suitable cadence, Euclid could measure NIR light-curves and colours of ~3000 Type Ia supernovae (SNe) out to $z\sim$ 1.2, an order of magnitude more than what will be achieved up to the launch of Euclid. To study SNe rates, single detections of a further 6000 Type Ia SNe and over 10,000 core-collapse SNe with $z<1.2$ can be used as well.

Closer to home Euclid will resolve the stellar populations of galaxies within ~5 Mpc of our Galaxy, a volume sufficiently large to include statistically significant samples of most morphological types. In addition, Euclid will be able to detect more than $10^5$ dwarf galaxies with luminosities down to $10^6$ solar luminosities out to 100 Mpc, thus probing a cosmologically representative volume for this dominant-by-number galaxy population.

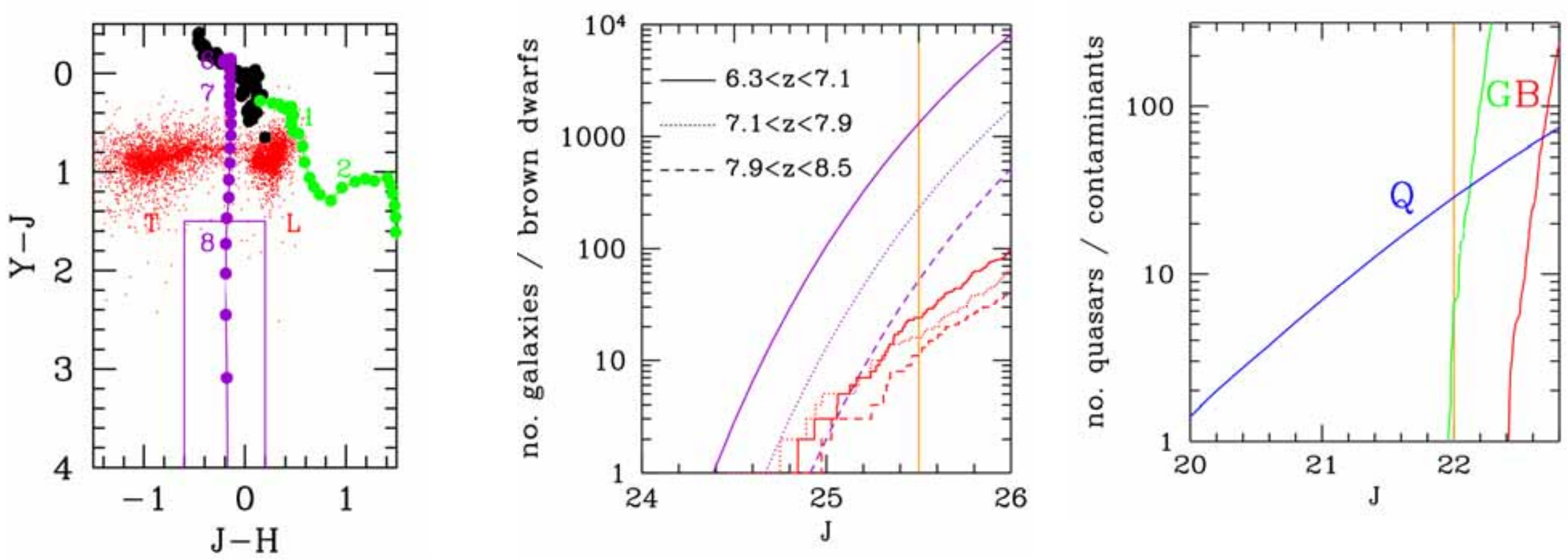

*Figure 2.6: Selection of very high-redshift galaxies and quasars, and quantification of contamination. Left: YJH two-colour diagram showing selection of Lyman Break Galaxies (LBGs) at 7.9<z<8.5. Purple dots show LBGs with an interval Δ(z)=0.1 from redshifts 6 to 8; Black dots show O-M stars; Green dots show early-type galaxies (formed at z=10), Δ(z)=0.1, redshifts 1, 2 marked; Small red dots L and T brown dwarfs, simulated using the model of Deacon & Hambly (2006), updated to match the survey of Burningham et al (2010), and assuming a disk scale height of 300pc. Middle: Cumulative number counts of LBGs, purple lines, in three redshift slices, calculated using the Luminosity Function of McLure et al (2010), and compared against the predicted contamination by brown dwarfs, (red lines). The nominal sample depth of J=25.5 is marked, fainter than the magnitude at which the LBG samples becomes significantly incomplete. Right: Predicted number counts of quasars at z>8.1 (Q), and contaminants; early-type galaxies with 1.5<z<1.8 (G) and brown dwarfs (B). The nominal depth of J=22 is marked, but clean quasar samples down to J=22.5 should be selectable as z=1.6 galaxies can be distinguished from point sources at this S/N of 20.*

### 2.4.2 The high redshift Universe

**Galaxies at z>7:** At high redshifts galaxies can be selected using the sharp decrease in flux blue-ward of the Lyman-α break. To reflect the way they are selected, these galaxies are commonly referred to as Lyman Break Galaxies (LBG). Using this technique, galaxies out to redshifts $z$~8.5 with $H$~28-29 have now been successfully identified with HST, but only over tiny areas of sky (a few arcmin$^2$). At these redshifts the break has shifted into NIR wavelengths, requiring space-based NIR imaging. Recent studies have started to revolutionise our view of the Universe at z > 7 (e.g. McLure et al., 2010; Bouwens et al., 2010), providing the first measurements of the evolution of the galaxy UV luminosity function (LF) from $z$ = 6.5 out to $z$ = 8, and enabling the first meaningful estimates of the contribution of star-forming galaxies to cosmic re-ionisation

The bright end of the LF at $z$>6.5 remains very poorly determined, because of the small field of view of WFC3/IR, or alternatively the difficulty of reaching $J$>25 from the ground. An accurate measurement of the bright end is, however, of vital importance for determining the form of the LF, and testing theories of galaxy formation. No survey is able to match the combination of depth and area that Euclid provides, although some progress in this direction will be made by two ambitious surveys that will be completed around 2015. The deep HST/WFC3 CANDELS will cover 800 arcmin$^2$ to $H$~27, while the shallower, wider *YJHK* UltraVISTA survey will reach $H$~25.5 over an area of 0.75 deg$^2$.

In contrast, Euclid will detect hundreds of $z$>7 galaxies brighter than $J$~26. The results of a detailed feasibility study of bright galaxies, $J$<26, at redshifts 6.3<$z$<8.5 with Euclid is illustrated in Figure 2.6. For such a survey the main contaminants are brown dwarfs (see, however, Section 2.4.5). The large number of $z$>7 galaxies that Euclid will find allows studies of the clustering on large scales, and detailed spectroscopic follow-up with E-ELT and JWST, including measurement of the UV slope, Ly-α equivalent widths, and the strength of HeII λ1640 emission, a signature of Pop III stars, to characterise the metallicity and initial mass function of the stellar populations. The evolution of Ly-α strength and the difference in clustering strength between strong and weak Ly-α emitters will provide a test of models of re-ionisation (McQuinn et al., 2007).

The area covered by the deep survey is also well matched to 21 cm studies with e.g. LOFAR and SKA, and will allow the measurement of the 21 cm galaxy cross-power spectrum (Lidz et al., 2009), quantifying bubble growth in the epoch of re-ionisation.

**Quasars at z>7:** When the epoch of re-ionisation occurred is currently constrained by two observations. Measurements of the polarisation of the CMB indicate that the half-way point of the re-ionisation epoch was near $z$=10 (Jarosik et al., 2011). The discovery of Gunn-Peterson troughs in the spectra of $z$~6 quasars found in SDSS (Fan et al., 2006) shows that the neutral fraction in the IGM is increasing rapidly towards higher redshift, and has reached $f \sim 10^{-3}$ by $z$=6.4, the limit of optical surveys. More can be learned about the period of re-ionisation by finding bright sources at redshifts $z$>6.5. Progress in the redshift range 6.5<$z$<7.2 will come from UKIDSS, VISTA, and Pan-STARRS, which will have discovered tens of bright quasars in this redshift interval by 2018 so that the condition of the IGM back to $z$=7.2 will be well established by then.

Quasars in the redshift interval 7.2<$z$<8.1, cannot be discriminated from L dwarfs on the basis of broadband *zYJH* (or similar) colours, without very deep imaging in the $z$-band. Simulations indicate that the combination of Euclid colours and spectra will allow the detection of a few tens of the brightest quasars in this redshift range. Then at $z$>8.1, quasars may be selected with *YJH* photometry, as very red objects in *Y-J*. Euclid is by a large margin the most powerful of any planned mission in this redshift range. Because quasars are redder in *Y-J* than LBGs early-type galaxies are the dominant source of contamination (rather than brown dwarfs) but they are easily eliminated on the basis of morphology at $J$=22, and clean selection of quasars down to $J$=22.5 should be possible.

The predicted numbers of quasars $z$>8.1 based on the Willott et al. (2010) luminosity function are plotted in Figure 2.6 (right panel) as the solid blue line. Euclid is expected to find 30 quasars $z$>8.1 brighter than $J$=22, and 55 quasars brighter than $J$=22.5 in the wide survey. Such a sample will transform our knowledge of the process of cosmic re-ionisation. These quasars can be used to measure the neutral fraction of the IGM in the vicinity of the quasar through the detection of the red damping wing (Miralda-Escude, 1998) for large neutral fractions, $f$>0.1, and several lines of sight must be probed as re-ionisation is predicted to be a very inhomogeneous process (Mesinger & Furlanetto, 2008). Following up these quasars with other facilities, it becomes possible to make a census of both high and low ionisation metal absorption lines (e.g. CIV, OI) which provides additional information on re-ionisation, as well as on the history of early star formation, and the importance of galactic winds at these very early phases in the Universe.

### 2.4.3 The cosmic co-evolution of galaxies and active galactic nuclei

Euclid's huge legacy dataset will do for galaxy evolution studies at 1<$z$<3 what the SDSS did for the $z$<0.2 Universe. This will allow some of the most urgent issues in galaxy evolution theory to be addressed: what is the relationship between the luminous baryon content of the Universe and the dark matter that dominates the gravitational potential; what is the internal structure of dark matter halos; what is the role of the central black hole in the evolution of galaxies; where and when does star formation take place and is quenched; what is the origin of galaxy scaling relations and how do they depend on environment? For most of these questions, the period 1<$z$<3 is of utmost importance because that is the period when the star formation and AGN activity peaked.

**Correlation and distribution functions:** The history of galaxy evolution is encoded in the spatial distribution of galaxies and in the distribution functions of their properties (e.g., luminosity and mass functions). Galaxies, however, cannot be characterised by a single parameter and real progress comes when large samples can be used to study evolution as a function of a variety of indicator quantities, as demonstrated by the SDSS (e.g. Blanton & Moustakas, 2009). Euclid can study clustering statistics and luminosity functions over a wide range of galaxy properties such as morphology (where Euclid is uniquely suited), colours, environment and stellar masses. Much work has been done out to $z$~1, but the peak of star formation activity and assembly of passive galaxies is at $z$>1 (Hopkins & Beacom, 2006; Ilbert et al., 2010) where samples are small and often selected on star formation activity rather than stellar mass. By combining the deep and wide data Euclid will be the first survey to enable these kinds of studies with sufficient numbers of galaxies at $z$>1.

The rest-frame optical luminosity function and stellar mass function can be studied ~2 mag fainter than $L^*$ for $z$<2 and the evolution of the luminosity and mass functions of different galaxy types in different environments can be followed, including under-dense regions, filaments and the cores of virialised structures.

In addition, the large volume sampled minimises any uncertainties due to cosmic variance. These will be crucial observations to help understand how star formation proceeds and is quenched in galaxies and how the efficiency of star formation depends on the properties of their host halo.

The spatial distribution of galaxies can be used to estimate the masses of their parent halos, and to place stringent constraints on galaxy evolution models. For distribution functions, clustering analyses can be carried out split by physical parameters. At higher redshift, which is the most active period of galaxy assembly, but where ground-based spectroscopy is inefficient, Euclid will truly revolutionise the field and will be a unique facility for galaxy correlation analysis.

**Tracing the star formation and enrichment history of the Universe:** The cosmic star formation rate represents a key constraint for the state-of-the-art theoretical models: the redshift evolution of the star formation rate (SFR) distribution function, both at the bright and faint end, is an essential ingredient to discriminate between different models for star formation and feedback. The stars also return heavy elements to the interstellar medium and thus the evolution of the metal abundance and dust content is tightly linked to the star formation history of galaxies: studying their interplay and the mechanisms that regulate them at different cosmic epochs is fundamental for understanding galaxy evolution.

Due to this importance there are many surveys pursuing this goal, but the combination of NIR slitless spectroscopy (i.e. no pre-selection on broad-band flux/colours) and large volume sampled means that Euclid is truly unique. It will increase the number of sources by two orders of magnitude relative to current work and likely also relative to JWST as Euclid will be most sensitive to the most luminous sources with densities of ~few/deg$^2$ of which JWST will find very few. For other distribution functions, the sample size will allow the study of the inter-relationship between the SFR and physical parameters of galaxies, such as the stellar mass and the metallicity (e.g. Mannucci et al., 2010), and trace their evolution with redshift.

**Mergers:** Galaxy mergers are an essential part of the evolution of galaxies and large-scale structure in any hierarchical cosmological model. Yet theoretical predictions for merger rates differ by an order of magnitude (Hopkins et al., 2010). Despite these large theoretical uncertainties, current observational constraints on the merger rate cannot distinguish between models due to the small existing samples. The current best surveys of galaxy mergers (e.g. de Ravel et al., 2009; Lin et al., 2008; Conselice et al., 2009) use sample sizes of a few thousand galaxies out to $z$~1, and vastly smaller numbers of a few hundred at $z$>1 (Conselice et al., 2008). Euclid will multiply those numbers by 4 orders of magnitude for the spectroscopic redshift sample, and well over 5 orders of magnitude for the photometric redshift sample. Using the combination of the wide and deep surveys, spectroscopy and multi-band photometry, Euclid will enable studies of merger rates as a function of galaxy and dark matter halo properties, to move from today's limit of $z$~1 out to $z$~6.

**Evolution of galaxies in dense environments:** The physical properties of galaxies depend on their environment. For example, a key characteristic of galaxy clusters in the low-$z$ Universe is that galaxies fall on a tight colour-magnitude relation, the red sequence. The assembly of this sequence can now be traced back to $z$~0.9, and these results show that the massive end of the red sequence, where most of the stellar mass resides, was in place already at $z$~1 and must have assembled at an earlier time. To make significant progress at $z$>1 requires deep NIR imaging over wide areas. The Euclid data can be used to probe the assembly of the red sequence two magnitudes past the knee of the $H$-band luminosity function out to $z$~2.

The assembly of galaxies in galaxy clusters will be accompanied by a range of processes that inject energy and metals into the intra-cluster medium and the combination of Euclid with X-ray data from eROSITA will be very powerful for studying scaling relations in galaxy clusters and will allow this work to push past $z$~0.9, improving the accuracy of the constraints by an order of magnitude. Euclid will also extend such studies to smaller galaxy groups, which is where most galaxies reside, thanks to the large number of spectroscopic redshifts. This extends the study of scaling relations to lower mass halos and provides the way forward to study feedback in groups and star formation efficiency as a function of halo mass.

**Quiescent galaxies:** Extrapolations from existing surveys indicate that Euclid will increase the sample of $z$>1.5 quiescent galaxies by an order of magnitude to ~8x10$^6$. This large number of galaxies can be used to follow in detail the growth of this fundamental class of rare galaxies that remain a challenge for models of galaxy formation. Euclid will be truly unique for finding and identifying spectroscopically the rarest and most massive quiescent galaxies (>4x10$^{11}$M$_\odot$) at $z$>1.8, since their number density (~1 deg$^{-2}$) makes their identification in smaller surveys difficult. The abundance and distribution of these sources place strong constraints on galaxy formation models, but they require deep NIR spectroscopy. Only JWST can compete with Euclid in this area, and then only for very small samples. Thus the number of galaxies of this class that can be studied spectroscopically is expected to increase by at least two orders of magnitude.

**The co-evolution of stars and black holes:** In addition to stellar feedback, AGN feedback plays an important role in current models of galaxy formation and its impact may even be relevant on cosmological scales (e.g. Van Daalen et al., 2011; Semboloni et al., 2011). Yet, the adopted models are necessarily simplified and rather different prescriptions have been proposed in the literature. One key area where information is lacking is in the interplay between star formation and black hole accretion at different redshifts and the related growth (or co-evolution) between galaxies and black holes at the peak of AGN activity ($z$~2). Euclid spectroscopy will detect more than $10^6$ broad line (i.e. Type 1) AGNs over the entire redshift range $0.7<z<9$. This sample will be more than one order of magnitude larger and a few magnitudes fainter than the current SDSS sample of such AGN.

In addition, Type 2 AGNs, where the accretion disk is thought to be obscured by a dusty torus, can be identified using the [OIII]/Hβ and [NII]/Hα emission-line ratios (e.g. Baldwin et al., 1981). Based on end-to-end spectroscopic simulations Euclid is able to measure the [NII]/Hα flux ratio with small enough errors to reliably distinguish narrow-line AGN and star-forming galaxies down to Hα-fluxes of $\sim 1.5\times10^{-15}$ erg cm$^{-2}$s$^{-1}$. Using a range of models, the number of these types of galaxies is estimated to be in the range 6,000 – 25,000 over the entire Euclid Wide Survey at $0.7<z<2.0$. At redshifts $1.26<z<2.0$ both line ratios are measureable, leading to a more secure classification of these objects.

The identification and characterisation of such a large sample of Type 2 AGN, more than one order of magnitude larger than anything available at the time Euclid flies, will provide a great amount of information about the evolution of narrow-line AGN and the clustering and environment of their host galaxies out to $z$~2, much beyond existing and future ground-based surveys. In particular, it will be possible to investigate fundamental open issues, such as the demography of obscured AGN, i.e. the ratio between type 2 and type 1 AGN (and therefore the covering factor of the obscuring medium) and its evolution with redshift and luminosity, tracing the growth of super-massive black holes up to $z$~2 and providing an important constraint on galaxy evolution models.

### 2.4.4 The relationship between dark and baryonic matter

A fundamental question in astrophysics is how galaxies trace the underlying dark matter distribution, both as a function of scale and redshift. The evolution of baryonic and dark matter is intimately interconnected. For instance, theoretical models indicate that the mass of the dark halo in which a galaxy resides is a crucial ingredient for predicting its evolution. Furthermore, it is becoming clear from observations at $z<1$ that only a relatively small fraction of all baryons are converted into stars and that the conversion efficiency is a strong function of halo mass. Understanding this relation in detail, particularly at $z>1$ where much of the galaxy assembly occurs, is challenging. The Euclid data enable a major step forward to study the relationship between dark and baryonic matter over half of the Hubble time through a number of complementary probes.

**Galaxy-mass correlation function:** The distribution of matter within galaxies can be probed using a range of dynamical tracers, but the outskirts of galaxies are more difficult to study. Weak gravitational lensing, however, provides a powerful way to probe the properties of dark matter halos well out to the virial radius (e.g., Hoekstra et al., 2004; Mandelbaum et al., 2005; Leauthaud et al., 2011). Such measurements therefore provide a direct link between predictions from N-body simulations and what is observed. With Euclid it will be possible to study the relation between baryons and dark matter halos using weak lensing out to $z>1$ for significant samples of galaxies and study the evolution in the build-up of both the baryonic and dark matter content of galaxies.

The weak lensing signal can also be measured to smaller radii, thanks to the high resolution imaging data. This is particularly useful in the case of galaxies that reside in clusters of galaxies. These are believed to have been stripped of most of their dark matter and the study of their density profile requires measurements of the lensing signal on scales of ~20 kpc, which are inaccessible from the ground, especially at high redshift. As a result Euclid can test predictions from numerical simulations of the stripping of dark matter halos over a wide range of environments.

Another important prediction from CDM simulations is that dark matter halos are tri-axial. If there is an alignment between the stars and the dark matter halo, this will cause an azimuthal variation in the weak lensing signal, which can be measured with high precision as a function of radius by Euclid. This measurement is intimately related to the study of intrinsic alignments that are an astrophysical source of bias for cosmic shear studies. Comparison with predictions of the alignments of dark matter halos will reveal possible misalignments between shapes/spins of baryonic and dark matter in galaxies, and shed light on the formation of galaxies within dark matter haloes. Euclid will measure gravitational shear-intrinsic shape alignments for both early- and late-type galaxies as a function of galaxy environment, luminosity, and other properties (see Hirata et al., 2007, Mandelbaum et al., 2011, Joachimi et al., 2011) with unprecedented accuracy, yielding insight into the structure of large-scale gravitational fields and their influence on galaxy evolution.

**Strong lensing:** Strong gravitational lensing provides precise measurements of the mass for individual lenses, as opposed to ensembles in the case of weak lensing. Euclid is a perfect survey for the intrinsically rare strong lensing events because of its combination of large area and high quality optical images. Based on past experience, several orders of magnitude more galaxy-scale lenses, giant arcs and multiply-imaged quasars relative to the total of all other surveys preceding Euclid are expected to be discovered. This enormous increase will enable the systematic use of strong lensing in galaxy evolution studies.

The precise mass modelling can be used to probe the balance between dark and luminous matter as a function of radial distance and for different galaxy types (e.g. Treu & Koopmans, 2004; Auger et al., 2010; Treu et al., 2011). Furthermore, by combining weak and strong lensing it is possible to extend the studies mentioned above over two decades in size (e.g. Gavazzi et al., 2007). Since lenses will mainly be found up to $z\sim1$ the measurements cover the stellar-to-dark matter evolution over half of the Hubble time (Treu & Koopmans, 2004). This dramatic increase of the number of strong galaxy-galaxy lens systems means that surface brightness anomalies (e.g. Koopmans, 2005) might become the main mode to detect mass-substructure in galaxies at cosmological distances.

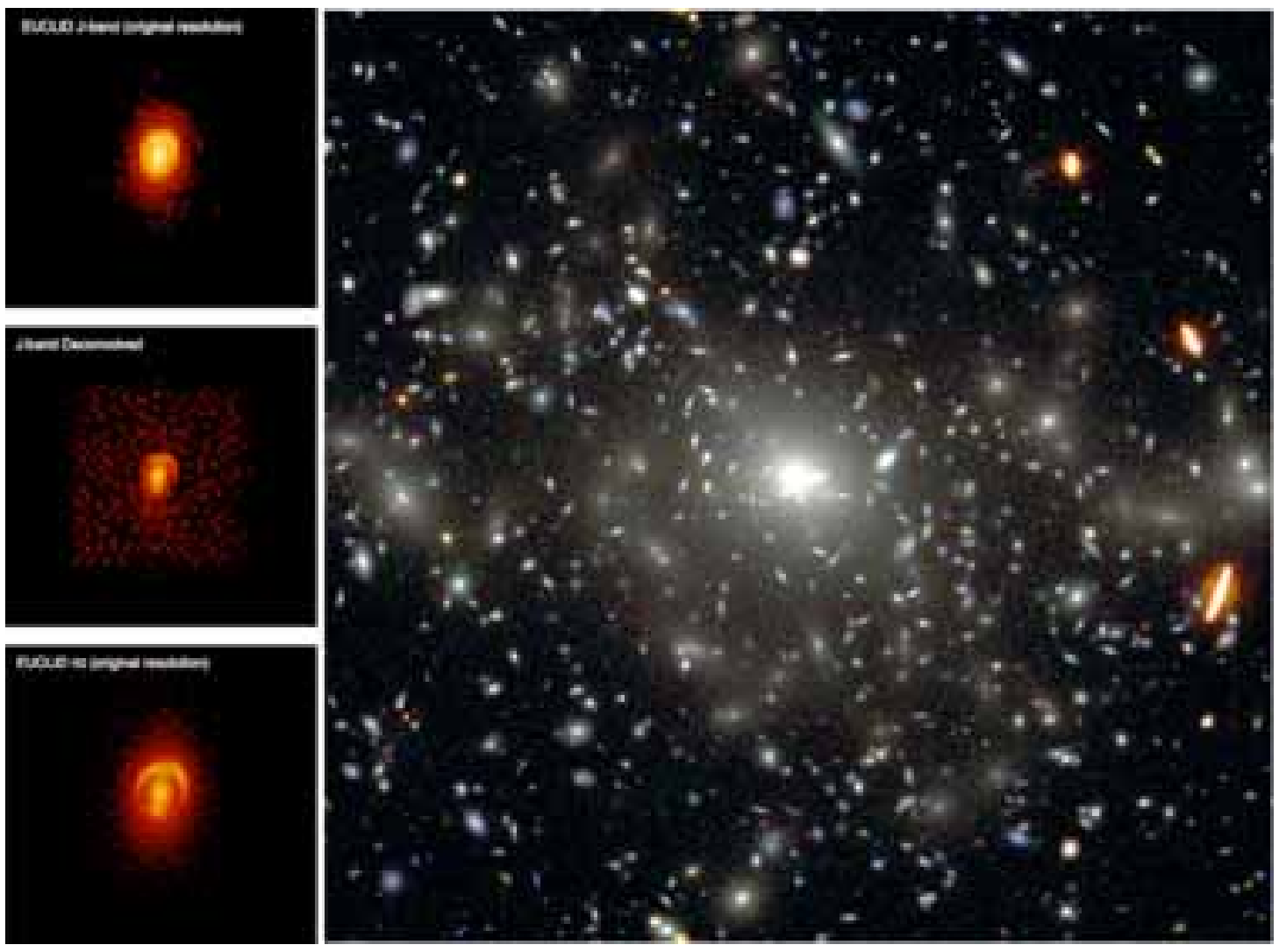

*Figure 2.7: Simulated examples of Euclid images of strong lenses. Left panel: Upper and lower panels are the Euclid J and VIS images of a strong lensing galaxy, with a strong arc visible in the VIS image. The middle panel shows the upper panel after deconvolution, revealing the arc structure. Right panel: Multicolour VIS+YJ image of a galaxy cluster, including a prominent strongly magnified multiply-imaged high-redshift source.*

The large number of lensing events provides not only a unique view of the mass distribution of the lenses, but they also act as gravitational telescopes permitting very high S/N observations of intrinsically faint sources, yielding a unique view of low-mass sources at high redshift and offering the possibility to study high redshift galaxies at ~100-500 pc scales (see Figure 2.7). The increase in sample size by several orders of magnitude provides the exciting possibility to proceed from studies of single objects to population studies of these sources.

### 2.4.5 Near-field cosmology and astrophysics

Much can be learned from a detailed study of the Milky Way and its nearest neighbours. Euclid NIR imaging is able to resolve the luminous Giant Branch stars in galaxies out to 5 Mpc, a volume sufficiently large to include statistically significant samples of most morphological types. The fine structure of stellar distributions constrains the nature of dark matter and defines the shapes and total mass of the halos of galaxies such as the Milky Way and M31. The detailed properties of their stellar components define the evolutionary history of their assembly. Of particular interest are the luminous Asymptotic Giant Branch (AGB) stars. These not only give a window into recent star formation processes, but are also major contributors to the chemical enrichment of the interstellar medium in these systems. The spatial distribution of Giant Branch stars will probe substructure and detect the more luminous stellar streams in and around these galaxies and thereby shed light on their recent merger history. The depth and resolution of the imaging data will also enable a census and characterisation of the globular cluster systems throughout these groups and address the question of free floating star clusters in galaxy groups, the variation of star cluster luminosity functions, physical size, and metallicities as a function of environment.

The number, nature and baryonic content of low mass dark matter haloes in the vicinity of Milky-Way-like galaxies is a strong function of the nature of the dark matter, and the recent discovery of ultra-faint satellites of the Milky Way has revived interest in this faint end of the galaxy luminosity function. Within the Milky Way halo and the Local Group, satellite galaxies and substructure in the form of streams and stellar clusters are known to exist over a large volume. However, fully characterising the distribution and nature of these structures using ground-based imaging is constrained by lack of deep large area NIR surveys. Deep wide area spectroscopic studies are impractical but much of the age-metallicity degeneracy, and hence access to the evolutionary history, is broken by having access to a wide baseline of deep optical through NIR imaging.

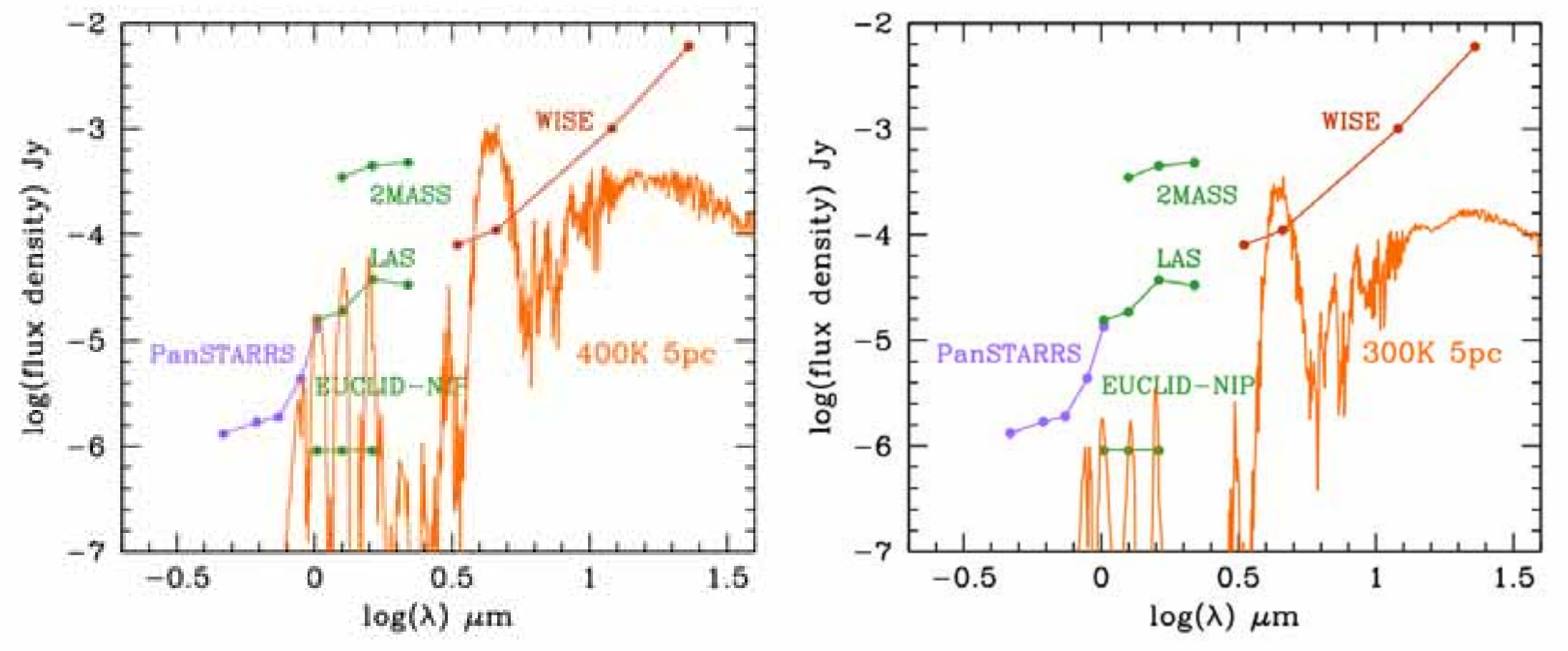

*Figure 2.8: 5σ point-source depths of a number of surveys as a function of wavelength: PanSTARRS, 2MASS, UKIDSS-LAS, Euclid-NIP, and WISE. Overplotted are Burrows' model spectra for Y dwarfs of the quoted effective $T_{eff}$ and distance. Only Euclid has the depth to confirm WISE detections of the coolest brown dwarfs with $T_{eff}$~300K.*

In combination with WISE, Euclid will complete the census of the coldest brown dwarfs, finally closing the gap between the coolest brown dwarfs currently known, $T_{eff}$= 500 K, and the giant planets in the solar system, $T_{eff}$=150K, and in the process answering the question of whether any remaining spectral types remain to be discovered i.e. the hypothesised Y dwarfs. As illustrated in Figure 2.8, at an effective temperature as cool as 400K, most of the flux emerges in the mid-infrared, particularly in the second WISE band at 4.6μm, so that in WISE many of the coolest brown dwarfs will only be detected in this band. Therefore confirmation of a candidate will require detection at a second wavelength. Euclid will probe a volume 100 times larger than UKIDSS+VISTA. WISE with UKIDSS+VISTA could discover a handful of objects as cool as 400K, whereas Euclid is expected to identify a few hundred such sources, and also reach as cool as 300K out to a distance of 5pc. Euclid therefore will open up the study of the properties of the Y dwarfs in detail.

### 2.4.6 Euclid synergies with multi-wavelength surveys

Euclid provides a deep NIR imaging and spectroscopic survey that complements other wide-field surveys such as Planck, eROSITA, LSST, PanSTARRS, LOFAR and MWA. The Euclid wide survey becomes the primary near-infrared resource for large area multi-wavelength studies of different classes of sources, and provides the principal means of identifying sources detected in wide surveys at longer wavelengths. Euclid's ability to find rare objects leads to natural synergies with facilities that have a large collecting area, but a small field-of-view, such as JWST, E-ELT, ALMA and current 8 m class telescopes.

In this section we provide some illustrative examples of new science that can be done by combining the Euclid data with results from other surveys. For instance, Planck can find proto-clusters of dusty star forming galaxies at high-$z$. Euclid's sensitivity and angular resolution are powerful tools to examine members of such clusters. Planck also provides Cosmic Infrared Background (CIB) maps, where correlated fluctuations have been measured. These can be used to search for large high-$z$ structures that Euclid can identify and characterise. Furthermore SKA precursors will trace HI evolution *and* the star-formation rate density over 90% of cosmic time, free of dust obscuration, while Euclid can trace stellar mass assembly. The combination of these data will show how quickly stars form as a function of galaxy mass and redshift. Euclid H$\alpha$ spectra can also test the calibration of star-formation rate in the radio continuum surveys, as well as detect AGN, measuring the cosmic evolution of accretion activity.

The Euclid deep fields are located close to the ecliptic poles, where extensive data are available from other surveys. Deep fields already exist with AKARI, Herschel, Spitzer, and BLAST data, and shortly LOFAR. Also the eROSITA mission will include a deep component near the ecliptic poles. In turn, because of Euclid, the ecliptic poles are likely to become key fields for future missions such as SPICA. AKARI's multi-wavelength coverage in the near-IR and mid-IR covers the peak bolometric output of hot dust in AGN and redshifted PAH/silicate features. Euclid will identify and morphologically classify all the host galaxies of AGN in this region, selected via the AGN-heated host dust emission. Euclid will also obtain redshifts for most mid-IR star-forming galaxies, constraining evolving PAH dust contributions and link H$\alpha$-based star formation rate estimators with the mid-IR PAH and continuum, the far-IR bolometric output, the decimetric radio flux, etc.

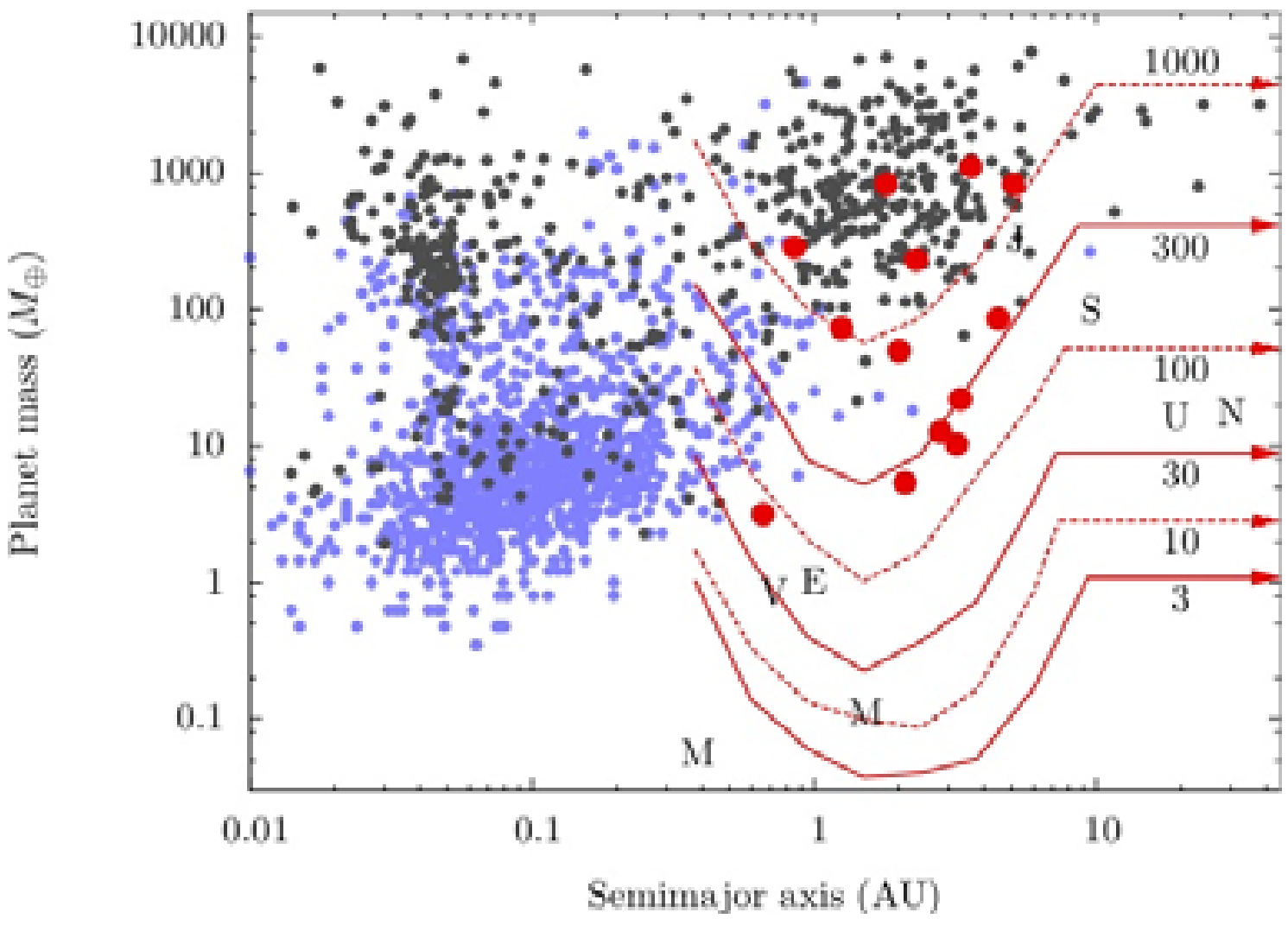

*Figure 2.9: Euclid planet detection capability for a NISP H-band survey observing for a total of 10 months (not necessarily continuously). The curves indicate the number of detections for the various Solar System planets (which are indicated by their initials). For instance a Mars analogue lies on the contour of 10 discoveries. The red dots are published planets detections by microlensing, dark grey dots are published planets detections with other techniques and blue dots are the published Kepler candidates. A Euclid programme lasting 10 months in total, will have sensitivity down to all Solar System analogues except for Mercury and probe planets unreachable by any other technique.*

## 2.4.7 Prospects for additional surveys

Current estimates are that the primary surveys can be completed in less than 6 years although additional optimisation to improve the survey efficiency might be possible (see Section 5). Once completed, one could decide to extend the cosmology surveys. Another option, however, is to survey new areas on the sky, in particular in the NIR (because of data rate constraints and VIS degradation). For example, a survey of the Galactic Plane would provide the most detailed structural study of the thin and thick disk of our Galaxy, tracing the spiral arms, the extent of the Galactic bar and would observe obscured young thin-disk regions in the disk and bulge inaccessible to the Gaia mission.

Another compelling science case is the search for extra-solar planets through observations of microlensing of stars towards the Galactic Bulge. If such a survey can be implemented, Euclid will measure the distribution of the cold exoplanet population down to Earth masses. This will provide a perfect complement to the study of the hot and warm populations currently studied by Kepler. The parameter space probed by Euclid coincides with the region where theories predict the bulk of planet formation to take place (Ida & Lin, 2004). Some theories predict that planets are ejected during the formation process and Euclid is sensitive to such "free-floating" exoplanets. Recent ground-based observations suggest that a significant population of such systems may exist (Sumi et al., 2011).

To reach the sensitivity to detect Earth-mass planets requires the combination of wide-field NIR observations with a well-behaved PSF, which only Euclid can provide. For instance, a NISP *H*-band survey of three fields centred at (*l,b*)=(1.1,-1.7) with a cadence of 20 minutes with a total survey time of ~10 months (not necessarily continuous) is sensitive to exoplanets with host separations of 0.5AU and masses above an Earth mass (see Figure 2.9). When combined with observations in the other bands the colour information can be used to determine the source star type, and distance, which will enable exoplanet mass measurements. If observations can be dispersed over the mission lifetime, the proper motion of the lens relative to the source provides constraints on the distance and mass of the planets for a substantial fraction of the events.

# 3 Scientific Requirements

In Section 2 we saw that there is a compelling case for a space-based dark-energy mission, encompassing a number of observational techniques, and the high-level science goals (level 0) for the survey were presented. In this Section we describe the design of a mission that can fulfil this science remit, presenting the lower level (level 1 and level 2) science requirements and their rationale. A table summarising the formal requirements is provided after each subsection.

## 3.1 Choice of techniques

The primary driver for the mission is to understand the expansion of the Universe and the rate of structure growth through observations of weak gravitational lensing and galaxy clustering. Only these two probes were considered when designing the science requirements for the mission. In particular, we considered:

- For **weak lensing**, using two-point 3D cosmic shear measurements.
- For **galaxy clustering**, using two-point 3D position measurements.

Other weak lensing and galaxy clustering statistics will be measured, but these are used to set further *goals* on the mission, rather than *requirements*. The focus on these two techniques was chosen because they yield the best science return for dark energy; they are particularly complementary for modified gravity science, and they represent the best-understood cosmological probes of the low-redshift Universe (see Section 2). The combination of these primary probes is crucial for the top-level science objectives. The power of Euclid is the ideal combination and mutual benefit of obtaining visible imaging, near infrared imaging, and spectroscopic data that enable both of these probes. The limitation of only considering these two probes was also designed to minimise mission creep and complexity. The only exception to this focus is that considerations of the legacy science achievable with the Euclid data were used to set additional requirements for a short-duration deep survey covering a smaller area, alongside calibration requirements for the primary probes as measured using a wide survey.

## 3.2 Designing an accurate dark energy experiment

### 3.2.1 Survey area

As argued in Section 2.1.2, it is necessary to measure the dark energy equation-of-state with a precision of $\Delta w_p$~0.016 and $\Delta w_a$~0.16, or to reach a Figure-of-Merit (*FoM*) for the dark energy of ~400, with sub-dominant systematic control, in order to decisively distinguish a cosmological constant model from a dynamic model. These requirements place important constraints on the survey that needs to be carried out by Euclid. We have performed an optimisation assuming a fixed total time for the wide survey S=5 years, which limits the product of the area A (in units of the field-of-view) and exposure time E, S=A×E. A plot of the resulting trade-off between area and *FoM* for different probes and their combination is provided in Figure 3.1 (see Taylor et al., 2006 for the scaling used for the lensing survey). We find that, for the combined weak lensing and galaxy clustering measurements, a *FoM*>400 is expected and that an optimal combined weak lensing and galaxy-clustering survey should cover 15,000deg$^2$.

The weak lensing optimisation alone favours smaller, deeper surveys. In the absence of systematic errors, in particular by ignoring intrinsic alignments, the optimal survey would be an increasing function of survey area, the number density of sources for which shapes can be determined, and the mean source redshift (Amara & Refregier 2007). Here we include a flexible model for intrinsic alignments with 5 parameters in scale and redshift (Joachimi & Bridle, 2009, Section 2), and a modest prior on these parameters (10%) from external data. Galaxy clustering alone favours larger, shallower surveys to maximise the volume observed (Wang et al., 2010). In combination, galaxy clustering provides more information than weak lensing for very large area surveys, whilst the opposite is true for smaller area surveys.

A further scientific argument is based on the synergy of the probes: to ensure that equal reliance is placed on each, the cosmological information content (parameterised by FoM) should be approximately equal. All of these arguments suggest a survey covering ~15,000deg$^2$ is optimal. Note that we define the survey area as the region of sky that Euclid has observed at least once. This includes masked areas due to the presence of bright stars, cosmic rays, etc., because their impact on *FoM* is minor (see section 6).

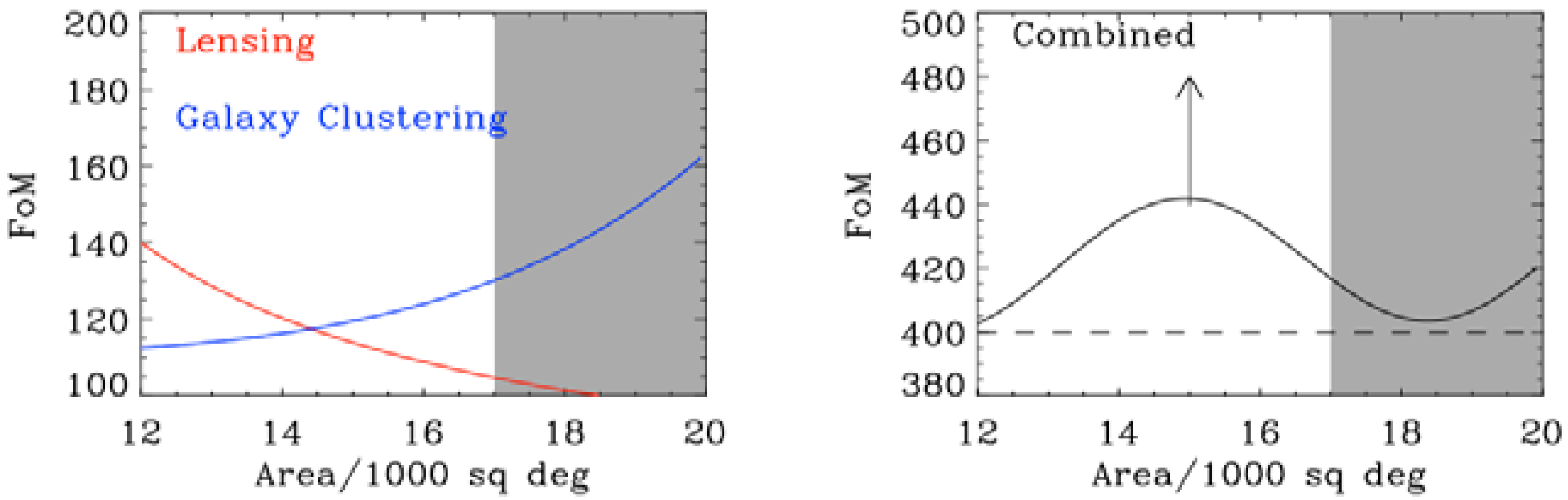

*Figure 3.1: Variation of Figure-of-Merit for Galaxy Clustering and Weak Lensing (left), and the combination (right) as a function of area for an assumed total survey time of 5 years. The sweet spot shown by the arrow provides the optimal approach for a single survey. The grey shaded region shows where sky coverage starts to become less efficient due to having to target areas with higher Zodiacal-light background. For presentation purposes here we fit a smooth function to discrete points in area and FoM.*

The *FoM*s shown here are *lower limits* on the performance of Euclid, where the mutual reduction in systematics (intrinsic alignments for weak lensing and bias for galaxy clustering) from the combination of the probes, has *not* been taken into account (this is a conservative assumption). As shown in Section 2, the combination of the primary and secondary dark energy probes with Planck data results in a *FoM*>4000. Note that the above arguments are only correct if we consider a survey with a fixed total survey time. For a survey with fixed exposure time, both probes would provide better results for larger areas. Similarly for a fixed area, both probes would provide more information given longer exposure times. It is only when we trade-off area against exposure time that we find an optimal finite solution. Consequently, any extension to the survey to cover more area with the same exposure time would be highly beneficial in terms of scientific return.

*Table 3.1 : The wide survey area requirement for weak lensing and galaxy clustering.*

| Req. ID | Parameter | Requirement | Goal |
|---|---|---|---|
| WL.1-1 & GC.1-1 | Survey Area | >15,000 deg$^2$ | >20,000 deg$^2$ |

### 3.2.2 Galaxy density from visible imaging

In a tomographic weak lensing analysis, where the lensing signal is binned in several redshift slices/shells, the observed galaxy shapes probe fluctuations in the gravitational potential caused by matter fluctuations transverse to the line-of-sight. The gravitational lensing signal from large-scale structure represents a ~1% change in a galaxy ellipticity so, to extract a measurable signal, a large number of galaxies are required. On average, the mean additional induced ellipticity is zero, therefore the correlation function or power spectrum of the ellipticity contains the cosmological information (the power spectrum and correlation function form a Fourier pair and so contain the same information). The amplitude of the lensing signal depends on the mass of the lens, and on the source-lens-observer geometry. To use lensing to measure the large scale structure at scales useful for understanding dark energy, and for a combined *FoM*>400, the surface density of galaxies needs to be at least 30 per square arcmin: at this density the lensing signal of group-scale massive objects and larger contribute to the lensing signal. Galaxies that are useful for weak lensing are those that are distant (to maximise the lensing effect) and resolved (to measure the shapes). The galaxy shape measurement requires the radius of the galaxy to be larger than the point-spread function (PSF); also, as the signal-to-noise decreases the ability of algorithms to measure galaxy shapes decreases too; methods are unbiased to a signal-to-noise of 10. To set requirements we therefore restrict the galaxy sample to have a signal-to-noise ratio of at least 10, and a size larger than 1.25 times the PSF FWHM. Once the survey area has been fixed, *FoM* is optimal when the number density of galaxies is maximised. However, fainter galaxies are also smaller, which leads to an increase in the residual systematics if the correction for instrumental effects (PSF or CTI) is imperfect. For a number density of 30 galaxies per arcmin$^2$ almost all source galaxies are useful for weak lensing, meeting the signal-to-noise and size criteria.

Image simulations, based on a Euclid representation of the Hubble Space Telescope Ultra Deep Field (UDF), show that with a small PSF (less than 0.2 arcsec) and a depth of 24.5 mag in a broad RIZ filter (discussed in Meneghetti et al., 2008 and Section 6.4.1) Euclid surveys slightly more than 30 galaxies/arcmin$^2$ with a median redshift greater than 0.9, that can be used for weak lensing analyses.

Once galaxies have been identified we also require their shapes to be measured, to extract the lensing signal. Shape measurement is commonly done by a forward convolution of a galaxy model (see Miller et al., 2007, Kitching et al., 2008) or by measuring the moments from galaxy images. Shape measurement of galaxies for weak lensing is complicated by two factors, whose dominance can only be suppressed to a level required for sub-percent dark energy science, using a space-based platform. A pixel scale of 0.1 arcsec means that the smallest galaxies used in the weak lensing science are well sampled, allowing sufficient information to be available. A small and stable PSF ensures that even the smallest galaxies used in the weak lensing science can be resolved, galaxies that are unresolved have ambiguous shapes that are dominated by the characteristics of the PSF. The Euclid PSF introduces no systematic errors as a result of under-sampling of the galaxies, or poor resolution. In addition the stability requirement means that the PSF can be modelled to the required accuracy using the stars detected in each exposure. This is discussed in Section 3.3.1 and Section 6.

The stability and knowledge of the PSF are vital for weak lensing studies and drive the requirements for the visible side of the Euclid mission. Sections 3.3.1 and 6.4.1 show that we can reliably predict the performance of Euclid with regard to these concerns.

The raw statistical power of a cosmic shear survey increases with the number density and mean redshift of the galaxies that can be used in the analysis. Since the first weak-lensing detections, based on areas of a few square degrees, the survey area has steadily increased. The coming years will see another major step forward with the Kilo Degree Survey (KiDS) and the Dark Energy Survey (DES), which will survey 1500 deg$^2$ and 5000 deg$^2$ respectively. In fact the gain in the measured cosmic shear signal-to-noise of the weak-lensing measurement over the past decade is consistent with an exponential increase in precision over time, with Euclid providing a breakthrough that can only be achieved with a space-based platform (discussed in more detail in Section 3.4), with approximately an order of magnitude improvement.

*Table 3.2: The top level image quality requirements from weak lensing. The Level 1 requirements on the number density and redshift of the galaxies used for weak lensing propagate into Level 2 requirements on the image quality and wavelength coverage.*

| Req. ID | Parameter | Requirement | Goal |
|---|---|---|---|
| WL.1-2 | Density of galaxies | ≥30 gals/arcmin$^2$ | >40 gals/arcmin$^2$ |
| WL.1-3 | Median redshift | >0.8 | |
| WL.2.1-1 | Wavelength Coverage | 550nm–900nm | |
| WL.2.1-2 | Number of visible Filters: NF | ≥1 | ≥2 |
| WL.2.1-3 | Vis PSF Size: FWHM | ≤0.18 arcsec | |

### 3.2.3 Photometric data requirements

Estimates of galaxy redshifts are needed for the majority of sources used in the weak lensing analysis, in order to split the Universe into redshift slices for lensing tomography, and to correct for contamination by galaxy intrinsic alignments. In order to obtain sufficient galaxies for the WL experiments, we need to include all types of galaxies at $0<z<2$ (star-forming with a wide range of luminosities, weakly star-forming, passive etc); it would be impractical to obtain a spectroscopic redshift for all of these galaxies, given the integration times that would be required. Thus, we need photometric redshifts for the galaxies for which we cannot easily obtain spectroscopic redshifts. Note that we do not require photometric redshifts for *all* galaxies, but do need these data for a subsample covering the full Euclid survey area, which would then constitute the galaxy sample used in the analysis. Photometric redshift derivation relies on finding the best galaxy spectral-template match to the broad-band colours. The accuracy of photometric redshift measurements is dictated by the number of filters and the signal-to-nose ratio of the observations, leading to dispersion between the photometric redshifts and the true (spectroscopic) redshifts.

To reach the full potential of the weak lensing technique, galaxy redshifts need to be estimated using photometry such that the standard deviation with respect to the true redshifts is $\sigma_z/(1+z) \leq 0.05$(required)-0.03 (goal) (see Abdalla et al., 2008, Kitching et al., 2008). As this error increases, the information available in the redshift direction is diluted. We require this dilution to occur on scales that are less than or equal to scales that are useable for dark energy science (cluster and group scales). This level of precision can only be achieved by the combination of deep NIR photometry ($Y_{AB}$ = 24, $J_{AB}$ = 24 and $H_{AB}$ =24) with visible data. While visible photometry is foreseen to be provided by ground-based facilities, Euclid needs to provide imaging in the three NIR bands, covering the wavelength range 0.92-2.0 micron. This deep NIR imaging is very challenging from the ground and unfeasible over a significant area of sky. To maximise the photometric results in the NIR, a sampling of about 1 detector pixel (or slightly higher) per system PSF will be applied.

For the photo-z measurement it is important that the photometry is uniform within a field and between fields on the sky. As a consequence, we put a strong requirement that the relative photometric accuracy for all galaxies within the survey should be better than 1.5%, after full data processing. At a later stage the data can be scaled to adjust the photometry to a consistent absolute calibration based on standard stars or other data sets. To minimise systematics from external light sources, the straylight level should be below the sky background level.

The broad-band Euclid data alone are not sufficient to achieve the required photometric redshift accuracy and precision, which means that additional ground-based data are required. The Euclid survey area (covering 15,000 deg$^2$) needs to be imaged from the ground using at least 4 filters, covering at least the full wavelength range 420−930 nm, with an overlap between the filters less than ~10%. The commonly used *griz* filter set meets these requirements. The photometric data themselves can come from different telescopes. The Euclid Consortium plan for obtaining these data is discussed in Section 3.2.4. Simulations of a wide range of filters and depths demonstrate the need for ground-based photometry, with limiting magnitudes (10σ for extended sources in the AB system) of $g$=24.4, $r$=24.1, $i$=24.1 and $z$=23.7. Note that a slight reduction in depth in some of the filters can be compensated with increased depth in other filters. These numbers represent a minimum requirement on the depth of the ground-based photometry and increased depth or extension of the wavelength coverage to bluer wavelengths will improve the photometric redshifts further.

The mean redshift needs to be known to better than $\sigma(<z>)<0.002(1+z)$ in each redshift bin to be used (see Bordoloi et al., 2009; Kitching et al., 2008). To be unbiased, the set of templates used to generate the photometric redshifts needs to be complete and a fair representation of the true galaxy spectra. We can obtain these templates from observed spectra, and this translates into a (conservative) requirement of having $10^5$ accurate spectroscopic redshifts, with a fraction of incorrect redshifts less than $10^{-4}$, from external sources. The Euclid Mission Consortium plan for obtaining these data is discussed in Section 3.2.5

*Table 3.3: The top level photometric requirements from weak lensing. The Level 1 requirements on the fidelity of photometric redshifts propagate into requirements on the NIR image quality and photometry.*

| Req. ID | Parameter | Requirement | Goal |
|---|---|---|---|
| WL.1-5 | Redshifts error ($\sigma(z)/(1+z)$) | $\leq$ 0.05 | $\leq$ 0.03 |
| WL.1-6 | Catastrophic failures | 10% | 5% |
| WL.1-7 | Error in mean redshift in bin | <0.002 | |
| WL.2.1-17 | NIR wavelength range | 920 to ≥1600nm | |
| WL.2.1-18 | NIR number of filters: | ≥3 | |
| WL.2.1-19 | NIR PSF size: | EE50 and EE80<br>Y: (<0.30”, <0.62”)<br>J: (<0.30”, <0.63”)<br>H: (<0.33”, <0.70”) | |
| WL.2.1-20 | NIR Pixel scale: | 0.3±0.03 arcsec | |
| WL.2.1-21 | Relative Photometric Accuracy | <1.5% | |

### 3.2.4 Ground based imaging data

**Wide-field imaging data**: The Euclid survey area covers much of the extra-galactic sky, which will also be covered by a range of planned and ongoing ground based imaging surveys. The Euclid consortium has explored several options to obtain data that comply with the required depth and wavelength coverage. The options are described below, but we note that a number of other projects are also underway. For example, HyperSuprimeCam will image about 2,000 $deg^2$ in *grizy* using Subaru to depths that exceed our requirements, starting in ~2012.

**1. DES**: The Dark Energy Survey (DES) is scheduled to start in late 2011 using the Blanco 4m telescope at CTIO. Its baseline survey will cover 5,000 $deg^2$ of the Southern sky in 5 years with at least $g$, $r$, $i$ and $z$ photometry at the depth needed for the Euclid photometric redshifts. The DES and Euclid consortia are planning for collaboration.

**2. Pan-STARRS**: PS1, the first of the planned Pan-STARRS telescopes, has been surveying the sky from Hawai'i since May 2010. PS2 is a planned extension, which would provide an additional clone of the PS1 telescope, doubling the power of the experiment. The PS1 3π survey (which covers most area) is too shallow to meet our requirement. However, this survey covers more (Northern) sky than is needed for Euclid photometric redshifts (as the Southern sky can be covered by DES, LSST or VST). The maximum amount of Northern sky coverage that the Euclid consortium would request PS to cover is therefore 7,500 $deg^2$ (and most likely less than that). This option is under discussion with the Pan-STARRS consortium. Both parties understand the benefit of creating a joined PS-Euclid Northern sky survey and reaching the required depth is feasible if the PS1 and PS2 systems spent 70% of the observing time over 5 years on a Euclid-optimised survey over 7,500 $deg^2$ in the North Galactic cap. In this scenario, based on the current performance of PS1 the resulting survey could reach 10σ AB flux limits of $g = 24.7$, $r = 24.3$, $i = 24.1$, $z = 23.6$ in 400 visits (this assumes an equal exposure time per filter, which could be optimised).

The etendue of PS2 is comparable to that of DES thanks to its larger field-of-view that compensates for the smaller mirrors. The funding of PS2 is almost secured and we consider this the best option to date to cover the Northern sky for Euclid. The PS and Euclid consortia are planning for collaboration.

The data from the current DES+PS2 scenario would cover 12,500 $deg^2$ assuming no overlap, representing 83% of the sky covered by the Euclid wide survey. This leaves an additional 2,500 $deg^2$ that are not covered by this baseline. There are, however, a number of viable options that are under discussion in the Euclid consortium:

**Option 1:** The baseline DES survey will be completed before the launch of Euclid. An increase of the DES survey area has already been discussed by the DES collaboration internally if DES is successful. These data could be obtained within the duration of the Euclid mission. This option requires no development in terms of instrumentation and software.

**Option 2:** The VLT Survey Telescope (VST) is currently being commissioned. It will be used to carry out the Kilo Degree Survey (KiDS), one of the ESO public surveys. The survey will commence October 2011 and will cover 1,500 $deg^2$ to $u$=24, $g$=24.6, $r$=24.4, $i$=23.1 (10σ extended), and $Z$ = 23.1 and $Y$ = 22.1 (5σ) with VISTA in 3 years. The survey covers parts of the southern and northern skies that do not overlap with DES or with the northern PS Euclid-survey under discussion. The nominal survey will be completed before the Euclid launch and an extension of the survey to cover more area and increase depth is a realistic option and should be well received at ESO, if Euclid is selected. As is the case for DES, this option does not require developments in instrumentation and software.

**Option 3:** The Large Synoptic Survey Telescope (LSST) would cover ~20,000 $deg^2$ to depths that are much deeper than our requirements. This project has been ranked highly by the US Decadal Survey and is anticipated to commence in 2020. As the telescope diameter is 8.4m, and will cover the sky every 3 nights using a multi-visit observing strategy, LSST will get the depth and the bands needed for Euclid photometric redshifts on a short time scale, likely in 3 years after the beginning of the survey. The LSST and Euclid consortia are discussing possible collaboration plans.

### 3.2.5 Ground based spectroscopic data

To verify the parameters of the photometric redshift sample, and the consequent definition of the redshift bins in which weak-lensing tomography is performed, we require a sample of at least 100,000 spectroscopic redshifts to a VIS limiting magnitude of $RIZ_{AB}$~24.5.

The need for ground-based data for this task is somewhat limited: in the crucial range $0.7<z<2.0$, which contains sources carrying most of the weak lensing signal, part of the redshift information can be provided by a deep slitless spectroscopic survey performed by Euclid. A Euclid Deep Survey with an expected flux limit $\sim5\times10^{-17}$ erg $s^{-1}$ $cm^{-2}$, provides a sample where a large fraction of the galaxies have $23<H_{AB}<25$ (Shim et al., 2009). The spectroscopic redshifts correspond to flat-spectrum star-forming galaxies (i.e. exactly those for which photometric redshifts are notoriously difficult to estimate and have the largest errors) and peer deep through the "redshift desert" of ground-based observations ($z>1.3$). The synergy between the spectroscopic and imaging channels of Euclid therefore fulfils some of the need for spectroscopic calibrators of the WL photometric redshifts.

The remaining redshifts required, including those for galaxies in the Euclid sample at $z<0.7$, and the early-type and weakly star-forming populations, (for which in any case Euclid photometric redshifts are intrinsically more accurate) are covered by completed, ongoing, or planned surveys.

For example, BOSS is targeting specifically the bright-end of the early-type population out to z~0.7. Surveys with the ESO/VLT already collected several thousands spectra for faint galaxies: the VVDS surveys (Le Fevre et al., 2005) include more than 20,000 redshifts to $i$=22.5 (z<1.2) and ~10,000 down to $i$=24 (z up to 4.5); zCOSMOS (Lilly et al., 2007) has collected spectra for 25,000 galaxies with $i$<22.5 and 12,500 galaxies with $1.3<z<2.5$ and B<25. Also, the DEEP2 survey at Keck obtained spectra for ~50,000 galaxies down to R~23.5, while the new VIPERS survey, currently ongoing at the VLT, will collect 100,000 redshifts for galaxies with iAB<22.5 and $0.5<z<1.2$.

For the faint, red galaxy population (passive early-types, weakly star-forming galaxies, dusty starbursts, ...), several surveys will provide useful data, such as the ESO VLT UDSz (PI O.Almaini) targeting ~3500 IR-selected galaxies to a limit of KAB=23, the VIMOS Ultra-Deep Spectroscopic survey in the UltraVISTA field (PI: O.Le Fevre), which expects to collect ~5000 redshifts to $i$=24.75, and the Magellan-COSMOS survey (PI: C. Impey), as well as other samples from already completed surveys of near-IR-selected galaxies at $z>1$ (e.g. K20 and GMASS, Cimatti et al., 2008; GDDS, McCarthy et al., 2004; ESO-GOODS-South, Vanzella et al., 2008, Balestra et al., 2010).

This implies that a significant fraction of the calibration sample is already or will be soon available. Finally, many very deep redshift surveys are planned with the current very large telescopes and with the next generation of Extremely Large Telescopes and radio arrays. For example, the Australian SKA Pathfinder (ASKAP) is planning to measure HI redshifts for over 100,000 late-type galaxies (again, those harder for photometric redshifts measurements), out to $z$=0.4. It is also worth mentioning that JWST is expected to provide spectra and redshifts for the galaxies for which it is not possible to derive redshifts from the ground due to their faintness, colours, or lack of spectral features in the observed spectral range. These samples will be available before the end of the Euclid mission.

### 3.2.6 Galaxy density from NIR spectroscopy

Using a sample of galaxies with accurate spectroscopic redshifts, significant science can be undertaken based on the direct clustering signal measured (see Section 2). The galaxy-clustering signal increases strongly with galaxy density, saturating at the cosmic (sample) variance limit at which point the errors are dominated by the limited volume covered. By using slitless spectroscopy and targeting Hα emitting galaxies, Euclid represents a unique opportunity to quickly and uniformly obtain redshifts for a large number of galaxies that are strongly clustered (Blake et al., 2011). In order to have a combined *FoM*>400, we require that Euclid obtain correct redshifts for at least 3,500 galaxies per square degree, giving a sample of $52.5\times10^{6}$ galaxies in total over 15,000 deg$^{2}$. The capabilities of Euclid go far beyond those achievable from the ground, as shown in Figure 3.2. Euclid-spectroscopy is also complementary to other surveys, both in redshift range and in terms of galaxy type (luminous star forming galaxies selected with Hα, vs. [OII] emitters for BigBOSS, luminous red galaxies for BOSS (and BigBOSS) and Ly-alpha emitters for HETDEX. Note that, of these projects, only BOSS is fully funded and ongoing: BigBOSS and HETDEX are still at the proposal stage. The recent Blake et al., (2011) analysis of the WiggleZ survey demonstrated the feasibility of using star-forming galaxies to make BAO measurements, compared with all the other current BAO analyses, which use Luminous Red Galaxies.

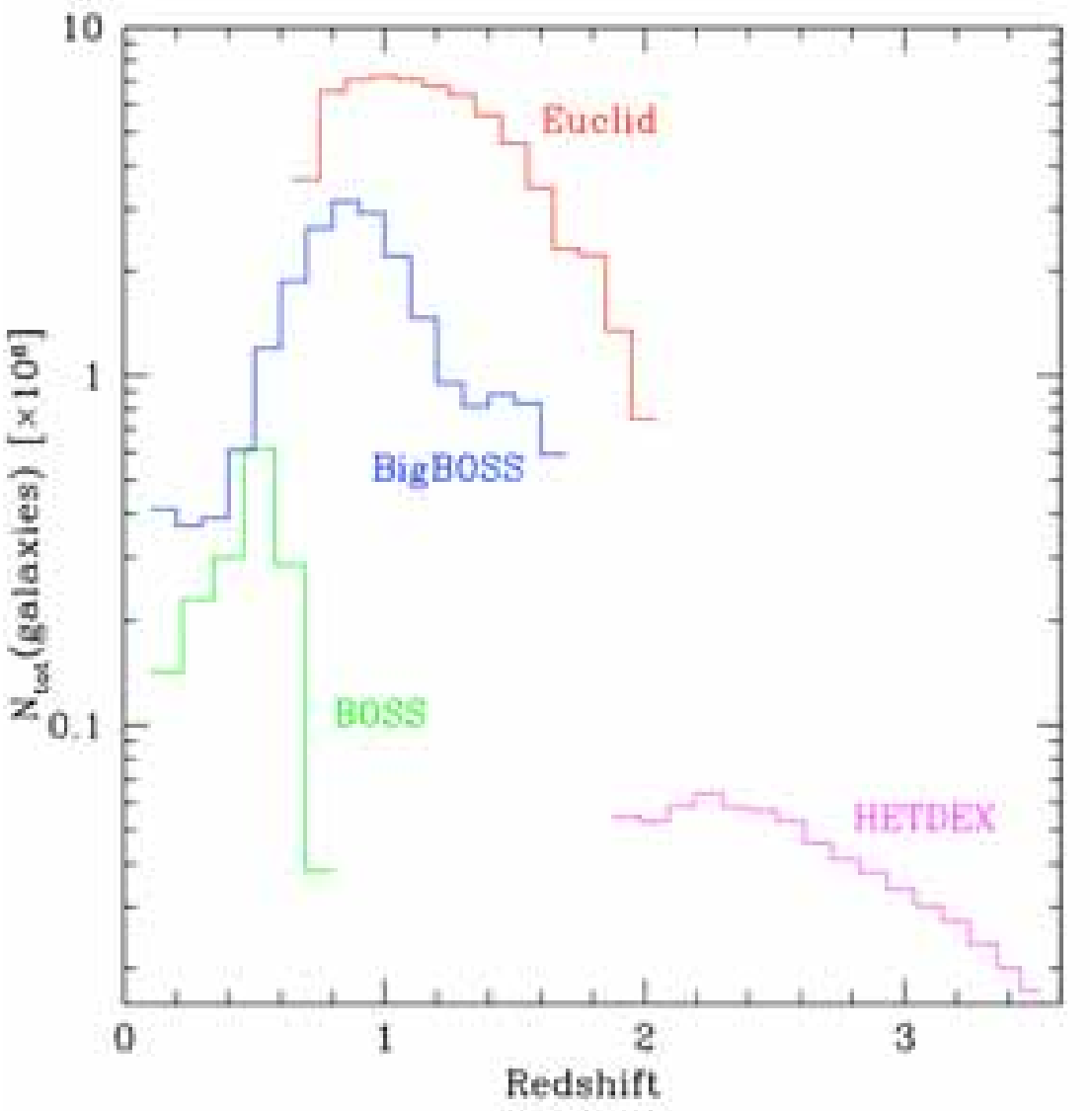

*Figure 3.2: comparison of number of galaxies in redshift bins of width Δz=0.1, for the Euclid spectroscopic survey compared with various ongoing and proposed ground-based surveys, including the Baryon Oscillation Spectroscopic Survey (BOSS; Eisenstein et al., 2011), the proposed BigBOSS project (Schlegel et al., 2009) and the proposed Hobby-Eberly Telescope Dark Energy Experiment (HETDEX; http://hetdex.org).*

The Hα line will be the main spectral feature for the determination of the redshift. Given the high number of detectable spectra in the field of view, and the size of the spectra covering the full wavelength range, a large number of spectra are contaminated (or confused) by spectra from other galaxies. Based on the top-level science requirement on the number density, the average effective Hα line flux limit from a 1-arcsec diameter source shall be lower than or equal to $3 \times 10^{-16}$ erg cm$^{-2}$ s$^{-1}$ at 1600nm. The flux limit is defined as the line flux for which the signal-to-noise is > 3.5. A slitless spectroscopic survey has a success rate in measuring redshifts that is a function of the emission line flux. As such, the Euclid survey cannot be characterised by a single flux limit, as in conventional slit spectroscopy. Given the steep log N – log S number counts of Hα emitters in this flux range, the number of galaxies with a measured redshift below the nominal flux limit is significant. At this stage these galaxies are not included in our derivation of the *FoM*; their inclusion would enhance the *FoM* due to the sample covering a broader redshift range.

To fulfil the top-level scientific requirements, the redshift completeness $N_{meas}$ / $N_{tot}$ should be in excess of 45%. Here $N_{meas}$ is the number of galaxies with measured redshifts, and $N_{tot}$ is the total number that can be detected at this flux limit, based on the current knowledge of the spatial density and luminosity function of Hα emitting galaxies (Geach et al., 2009). Given the confusion due to overlapping spectra, which is intrinsic in the slitless technique, to achieve such a high fraction one has to obtain multiple observations of the same field of view at different orientations. Additionally, confusion can be further diminished by splitting the total spectral range over different exposures of the same field through the use of filters, which make the spectra shorter and reduce the sky background while keeping a nearly constant signal-to-noise ratio on the emission lines. Euclid will adopt a combination of these two techniques, which has been shown to provide optimal results through end-to-end simulations. The same simulations show that any residual density-dependent incompleteness is not expected to be a significant effect (see Section 6).

It is necessary to obtain a NIR image of the same field as covered by the slitless spectrograph, sufficiently deep as to always allow association between an emission line detected in the dispersed image with a counterpart in the field image. The NIR image will provide positions of the objects, allowing the zero-point in the wavelength scale to be measured for each object, and remove ambiguities with zero order spectra contamination. Equally important is the fact that the NIR image will give the object sizes and orientations enabling the correct definition of the best extraction aperture of the spectra, as well as the flagging of contaminated spectra. Within the Euclid concept, the NIR photometry channel can be used to obtain the field image: an image exposure in the H-band to a depth of $H_{AB} < 24$ mag, would be sufficient to meet the needs of the slitless spectroscopy. It would also be possible to produce a composite image using all three NIR imaging bands.

The imaging quality of the spectrometer is important to minimise the loss of sensitivity. For the imaging mode the size of the PSF at 80% encircled energy must be less than 1 arcsec. In the case of the spectroscopic mode, the size of the PSF at 80% encircled energy must be better than 1 detector pixel in cross dispersion mode at all wavelengths.

*Table 3.4: The top level requirements for galaxy clustering. The Level 1 requirement on the galaxy redshift distribution and fidelity propagate into Level 2 requirements on flux limits and completeness.*

| Req. ID | Parameter | Requirement | Goal |
|---|---|---|---|
| GC.1-2 | Galaxy sky density | 3,500 / deg$^2$ | 5,000 / deg$^2$ |
| GC.1-8 | Bias of all galaxies | >1 | |
| GC.1-9 | Bias, upper quartile in redshift | >1.3 | |
| GC.2.1-1 | Flux limit | $\leq 3\times10^{-16}$ erg cm$^{-2}$ s$^{-1}$ | |
| GC.2.1-2 | Completeness | >45% | |
| GC.2.1-3 | Flux limit at all wavelengths | <120% of GC.2.1-1 | |

## 3.3 Control of systematics

Here we discuss the requirements that result from having to achieve a tight control of the observations, and our understanding of them.

### 3.3.1 Image quality

For weak lensing, the primary requirement is on how accurately we can measure the cosmic shear power spectrum. Any systematic effect that acts to change this will bias the final cosmological parameter constraints. In particular, un-modelled image quality variations, such as unknown spatial variation in the PSF, have a direct impact as a redshift and scale-dependent change on the cosmic shear power spectrum.

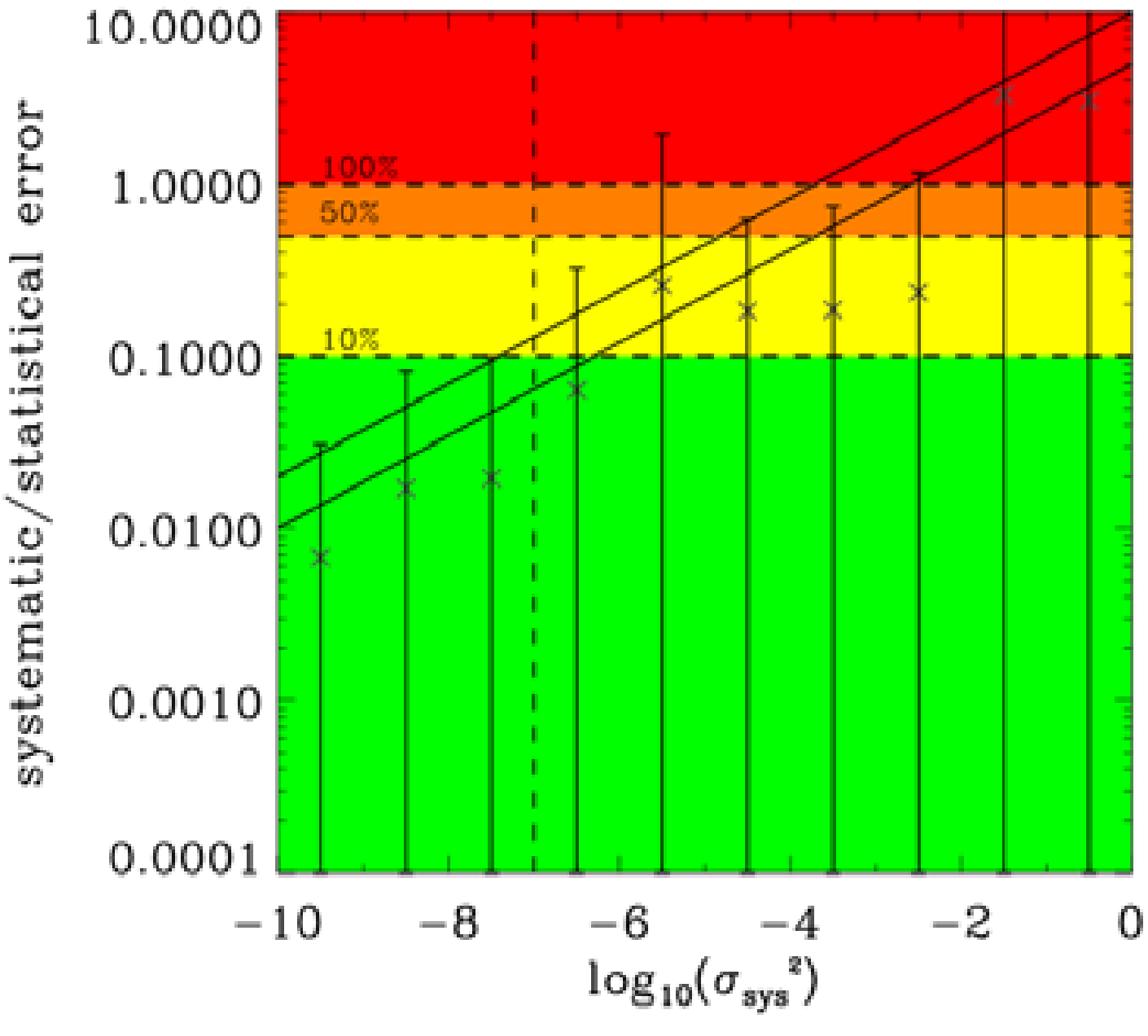

*Figure 3.3: The relative bias divided by the expected marginalised error for $w_p$ and $w_a$ (lower and upper solid lines) calculated from the mean (averaged over redshift and scale) additive systematic $\sigma^2_{sys}$, generated by a Monte Carlo suite of simulations. The points show the mean relative bias over the Monte Carlo simulations and the errors are 95% limits, the solid lines show a linear fit to these points for $w_0$ and similarly for $w_a$. Imposing an upper limit of $10^{-7}$ on the systematic error guarantees that the relative bias will be less than 10% of the statistical error budget.*

To achieve the science objectives, the amplitude of the lensing signal needs to be measured accurately as a function of redshift. Inability to do this leads to limitations for our ability to measure the shapes of faint galaxies that have been convolved by the Euclid PSF. In general there are two classes of systematic effect with which we are concerned; a multiplicative bias *m* proportional to the signal and, primarily, an additive bias $\sigma^2_{sys}$ that is constant with respect to the signal. An additive bias would be introduced if the assumed shape of the PSF were wrong, while a multiplicative bias would result if the assumed size were wrong. These biases propagate into the cosmic shear correlation function (and similarly into the power spectrum) such that

$$\hat{C}_{ij} = (1+m)^2 C_{ij} + \sigma^2_{sys}$$

where $C_{ij}$ are the shear correlation functions for redshift bins $i$ and $j$; the lensing signal is determined by correlating the shapes of the galaxies in different tomographic redshift bins.

To place a requirement on the magnitude of $\sigma^2_{sys}$ that could be tolerated, we show in Figure 3.3 the result of a Monte Carlo suite of simulations where we create arbitrary systematic errors on the additive bias $\sigma^2_{sys}$ as a function of spatial scale and redshift, and propagate these to expected biases on the dark energy equation of state parameters $w_p$ and $w_a$ (as defined in Section 2). We show the relative bias, with respect to the expected error from the wide survey as a function of $\sigma^2_{sys}$. The requirement on $\sigma^2_{sys}$ is $10^{-7}$, which means we want $w_p$ and $w_a$ biases to be less than 10% the expected error. To calculate these we make the same conservative assumptions described in Section 2.3.1. These represent the top-level requirements that can be translated to lower level constraints on the allowed PSF characteristics, but also place constraints on detector-induced biases. Furthermore the multiplicative bias should be $m<2\times10^{-3}$ (Kitching et al., 2008).

We now explain the level of image quality requirements that this implies. We place requirements on the *absolute* size and ellipticity of the PSF and requirements *knowledge* of the PSF, as a function of space and time over all exposures. The requirement on the multiplicative biases leads to a constraint on the PSF size expressed in terms of $R^2=Q_{11}+Q_{22}$ where $Q_{ij}$ are the quadrupole moments as defined in Paulin-Henriksson et al. (2008). For a given population of galaxies with sizes $R^2_{gal}$ the multiplicative bias $m$ is given by

$$m = \frac{1}{P^{sh}}\left(\mu + \frac{\sigma[R^2_{PSF}]}{R^2_{PSF}}\right)\left\langle\left(\frac{R_{PSF}}{R_{gal}}\right)^2\right\rangle,$$

where $P^{sh}$~2 accounts for the fact that we measure shapes using weighted moments, $\mu$ is a measure of the accuracy of the algorithm used to measure galaxy shapes and this term depends on the fractional uncertainty in the knowledge of the PSF size $\sigma(R^2_{PSF})/R^2_{PSF}$. Similarly for the same population of galaxies the additive bias is given by

$$\sigma^2_{sys} = \left(\frac{1}{P^{\gamma}}\right)^2\left\langle\left(\frac{R_{PSF}}{R_{gal}}\right)^4\right\rangle\left[\left(\frac{\sigma[R^2_{PSF}]}{R^2_{PSF}}+c\right)^2|e_{PSF}|^2 + 2\sigma^2[e_{PSF}]\right].$$

where $P^{\gamma}$~1.84 is the shear polarisability. This term also depends on the absolute PSF ellipticity $e_{PSF}$ and the uncertainty in the knowledge of the ellipticity $\sigma^2(e_{PSF})$. The parameter $c$ is a measure of how well the algorithm used to measure galaxy shapes can correct for PSF anisotropy. These definitions enable us to relate the PSF properties directly to systematic biases: for example the larger $R^2_{PSF}$ the larger the correction for the PSF blurring required.

The terms that depend on the shape measurement algorithms are not the driving terms. Tests on simulated data, such as those presented in Bridle et al. (2009), indicate that state-of-the-art correction schemes can currently achieve a shape measurement accuracy of $\mu\sim2\times10^{-3}$. Extrapolating progress in the past years to the future indicates that $\mu<<10^{-3}$ is likely by the time Euclid is launched, but we find that $\mu=2\times10^{-3}$ would be sufficient. Tests indicate $c\sim10^{-3}$ for current methods.

We can now place constraints on the absolute size and ellipticity of the PSF required. To correct for the blurring caused by the PSF we need to measure the PSF properties from the individual exposures. To ensure no significant loss of information, the PSF needs to be sampled sufficiently well. With a choice of 0.1 arcsec-/pixel we require that the FWHM is less than 0.18 arcsec at a reference wavelength of 800nm. Furthermore the ellipticity of the PSF should be $e_{PSF}<0.15$, significantly smaller than the average intrinsic galaxy ellipticity, to minimise selection biases and reduce the overall correction itself. Similarly shear introduced by the geometric distortion should be <0.03 i.e. less than the contribution from the PSF for the population of galaxies that Euclid will observe.

It is clear that we need model the PSF size variation sufficiently well and we require that $\sigma[R^2_{PSF}]/R^2_{PSF} \leq 10^{-3}$. We also find that the wings of the VIS PSF should be low, such that $(R_{PSF}/R_{ref})^2<4$, where the reference value is for a Gaussian profile with a 0.2 arcsec FWHM. The PSF size is measured from bright stars in the images, but we need to know the "effective PSF" size for galaxies. The galaxies have lower signal-to-noise ratios, so this extrapolation also leads to a requirement on how well the linearity of the detector is known.

The PSF ellipticity also needs to be modelled sufficiently well. In the above expressions because the *c* factor is very small the first term in the additive bias expression is sub-dominant and the additive bias is dominated by uncertainties in the model for the PSF anisotropy $\sigma[e_{PSF}]$. This leads to the requirement that the variation of the PSF, both spatially and in time, should be such that $\sigma[e_{PSF}]<2\times10^{-4}$ for each ellipticity component. This can be achieved if the system PSF has a finite number of degrees of freedom for its dynamic variation, because it allows us to combine measurements from a sequence of exposures.

Finally, the PSF is diffraction limited, and therefore we need to account for the wavelength dependence of the PSF. Cypriano et al. (2010) have shown that the Euclid ground-based data are sufficient for this. Furthermore the colours of galaxies may vary spatially. Such colour gradients need to be accounted for, but fortunately this can be done statistically (Voigt et al., 2011). To calibrate this correction 1% of the survey will be imaged in a narrow filter so that the colour gradient of galaxies used for weak lensing can be determined.

*Table 3.5: The top level systematic requirements from weak lensing. The Level 1 requirements on the systematic contamination for weak lensing propagate into Level 2 requirements on the knowledge of the PSF, image distortion and image defects.*

| Req. ID | Parameter | Requirement | Goal |
|---|---|---|---|
| WL.1-4 | Systematics ($\sigma^2_{sys}$) | $10^{-7}$ | |
| WL.2.1-4 | Vis pixel scale | 0.1'' ± 0.01'' | |
| WL.2.1-5 | PSF ellipticity | <0.15 | |
| WL.2.1-6 | PSF profile: | $(R_{PSF}/R_{REF})^2<4$ where $R_{REF}$=0.2'' for Gaussian profile | $(R_{PSF}/R_{REF})^2<3$ |
| WL.2.1-8 | PSF ellipticity Stability: $\sigma(\mathbf{e}_{res})$ | $<2x10^{-4}$ | |
| WL.2.1-9 | PSF size Stability: $\sigma(R^2)/<R^2>$ | $<10^{-3}$ | |
| WL.2.1-10 | PSF wavelength dependence | <0.9 | |
| WL.2.1-12 | Stray light | <20% of eclip zod | |
| WL.2.1-13 | Lost pixels: | <3%, VIS<br><10%, NIR | <2% , VIS<br><5%, NIR |
| WL.2.1-15 | Pre-calibration distortion | <3% | |
| WL.2.1-16 | Post-calibration distortion | <0.03% | <0.01% |
| WL.2.1-23 | Contrast ratio of ghost images | $\leq10^{-4}$ | |

## 3.3.2 Redshift measurement and contaminants

In the Euclid Assessment Phase Study Report, galaxy redshifts are required to cover the range $0.5 < z < 2.05$: we now identify an optimal mission where the redshift range is limited to $0.7<z<2.05$, corresponding to a longer-wavelength blue cut-off for the blue grism. Ongoing ground-based surveys such as BOSS will cover the low redshift region, and it is the high redshift galaxies for which Euclid has the unique ability to obtain significant numbers of spectra. By focusing our efforts on this area where Euclid is particularly well suited, we can boost the signal-to-noise of the remaining observations, as the observations are background limited (and so a reduction in wavelength range cause an increase in signal-to-noise at the remaining wavelengths), the Zodiacal light is strongest in the blue, and the shorter spectra give rise to less confusion. This does mean that the FoM from Euclid alone is reduced as the FoM is biased towards measurements at low redshifts. This is a known limitation of the FoM in quantifying the overall total value of a dark energy experiment, which should be always kept in mind when comparing different experiments based only on one value (see e.g. the findings of the Joint Dark Energy Mission Figure-of-Merit Science Working Group, Albrecht et al., 2009). The end-to-end simulations described in Section 6 show that the total number of galaxies with redshifts is actually boosted by this change by 20%.

Extensive simulations show that the statistical reconstruction of large-scale structure and the accurate measurement of the galaxy clustering power spectrum analysis require a redshift accuracy of $\sigma_z \leq 0.001(1+z)$. In case of larger values of $\sigma_z$, the galaxy clustering FoM will decrease significantly. A similar accuracy is required also for the redshift distortion analysis. This is demonstrated in Figure 3.4, taken from Wang et al. (2010), which shows how degradation in the redshift error degrades the recovered FoM for a galaxy clustering analysis.

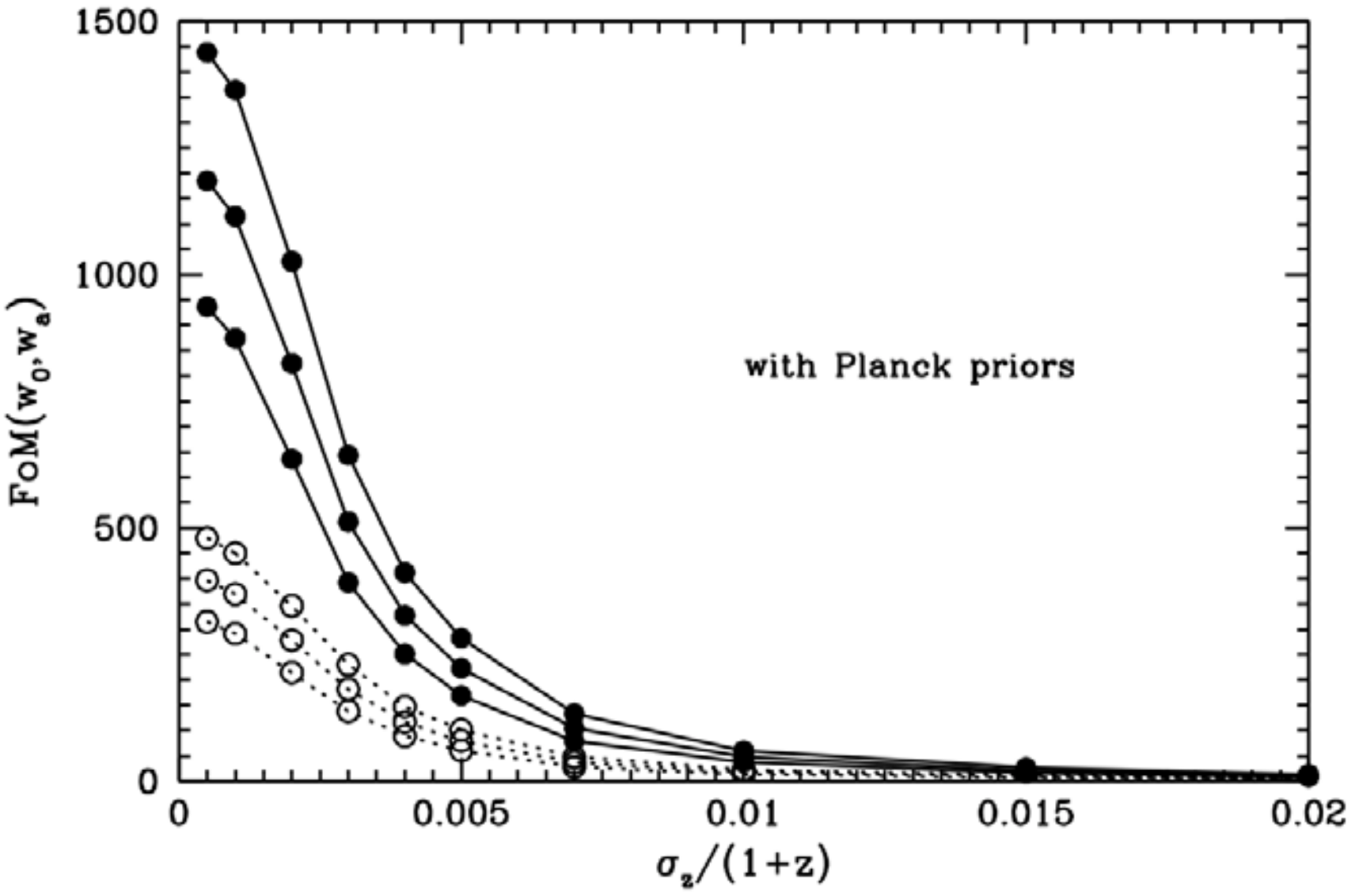

*Figure 3.4: Combined FoM for a fiducial galaxy survey, including predicted Planck data, as a function of average redshift error. The solid and dotted lines in each panel are the FoM for ($w_0$,$w_a$) with growth information included and marginalised over respectively. The different lines show different expected number densities for the galaxy survey. Figure taken from Wang et al. (2010). Photometric redshifts give errors larger than $\sigma_z/(1+z)=0.02$, while the slitless redshift accuracy is $\sigma_z/(1+z)=0.001$.*

Slitless spectroscopy by its very nature provides a spectrum for every object within the field-of-view. We need to subsample the spectra that can be matched to objects within the NIR image, in order to obtain the set of galaxies to be analysed. When selecting these galaxies, we want to retain as many correct redshifts ($N_{\rm correct}$) as possible, while also maximising the fraction $p$ of measured redshifts that are correct out of those measured ($N_{\rm meas}$), we call $p$ the purity of the sample,

$$p = \frac{N_{\rm correct}}{N_{\rm meas}}$$

.

If the fraction of redshift failures is denoted $f$, then $p$=1-$f$. In section 3.2.4 we defined the completeness of the sample. We see that the

$$\text{success rate} = \frac{N_{\rm correct}}{N_{\rm total}} = p \times \text{completeness}$$

is the fraction of correct redshifts divided by the total number of observed galaxies, and is equal to the completeness multiplied by the purity. Contamination degrades the clustering signal (for random contaminants the clustering signal is reduced by $p^2$), reducing our ability to measure BAO. We set a constraint on the contamination of 20%: i.e. the purity has to be >80%. Knowing this contamination level is crucial for understanding the recovered clustering measurement, and we require that this number is known to 1%.

By observing BAO in different redshift slices we will measure the projected distances within those slices to high accuracy. However, in order to use these measurements to constrain cosmological models, the average redshift within each bin must be known more precisely, or this will enter as a systematic in the analysis. We have identified this as the primary and most challenging potential systematic for the Euclid galaxy clustering analysis. To show the impact of this systematic, Figure 3.5 shows the relative bias in the dark energy para-meters, with respect to the statistical error, using the standard Fisher prescription (e.g. Kitching et al., 2008). In order to be conservative in our calculation, we have assumed that the sign of the redshift errors is correlated.

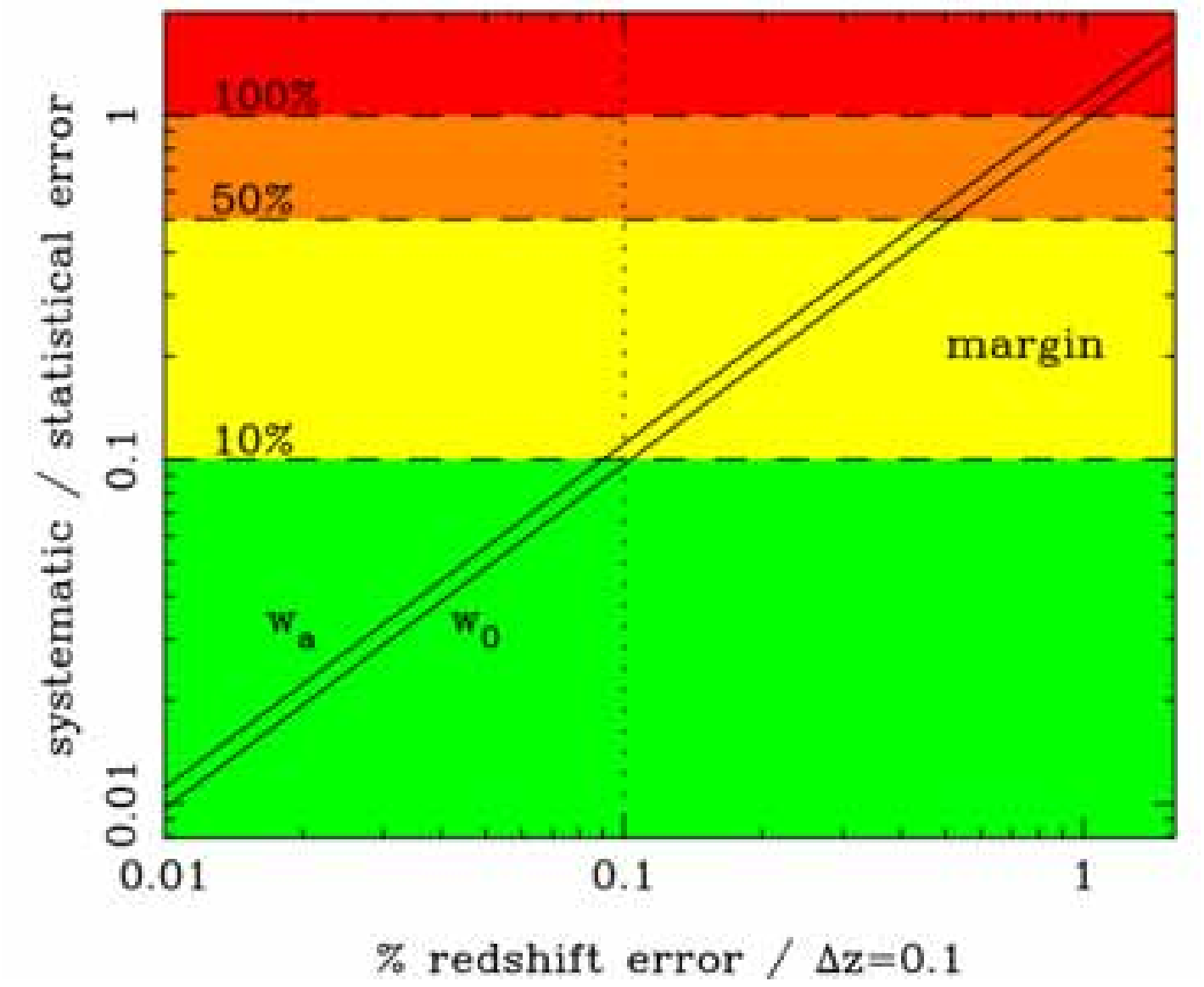

*Figure 3.5: The predicted systematic error divided by the marginalised statistical error for parameters $w_0$ and $w_a$, from Euclid BAO measurements, (including the Planck data), required to normalise the BAO scale (solid lines). These are calculated for different values of the systematic error placed on the mean redshift in a bin of width $\Delta z$=0.1. The current requirement has been put at 0.1 % (shows by the vertical dotted line) and can be verified using a sample of 140,000 galaxies with purity >99%.*

Our requirement that the mean systematic error in all redshift bins of width 0.1 is less than 0.1% (shown by the vertical dotted line), sets an error on $w_0$ and $w_a$ that is approximately 10% of the statistical error. The fractional error on $w_a$ is only slightly higher. There is margin available for this level 1 requirement, by accepting that a larger fraction of the final error budget will arise from this systematic.

Although challenging, we believe that we can achieve the required level of accuracy using the deep survey: simulations show that obtaining a sample of 140,000 galaxies with the same statistical properties as the wide spectroscopic survey sample, over 40 $deg^2$ region with 99% purity is sufficient to achieve this requirement. Provided that individual observations of the deep survey are taken using the same strategy as the wide, and that the field is rotated between observations such that spectra are projected along axes that are evenly spread over a large number of angles, then we can achieve this (see Section 6 for more details). This will be confirmed using follow-up spectroscopy using ground-based instrumentation.

*Table 3.6: The top level systematic requirements from galaxy clustering. The Level 1 requirements on the systematic contamination for galaxy clustering propagate into Level 2 on spectral resolution, PSF, stray light and purity.*

| Req. ID | Parameter | Requirement | Goal |
|---|---|---|---|
| GC.1-3 | Redshift accuracy | $\sigma(z)<0.001(1+z)$ | |
| GC.1-4 | Systematic offset in redshift | <1/5 redshift accuracy | |
| GC.1-5 | Redshift range | $0.7<z<2.05$ | also gals $z<0.7$ |
| GC.1-6 | Median of redshift distribution | >1 | >1.1 |
| GC.1-7 | Upper quartile of redshifts | >1.35 | |
| GC.1-10 | fraction of catastrophic failures | f<20% | |
| GC.1-11 | fraction of catastrophic failures | known to 1% | |
| GC.1-12 | mean redshift in 0.1 redshift bin | known to 0.1% | |
| GC.2.1-4 | Spectral range: lower limit<br>Spectral range: upper limit | less than 1.1 micron<br>greater than 2.0 micron | |
| GC.2.1-5 | Spectral resolution | >250 | |
| GC.2.1-6 | Resolution element | sampled by > 2 pixels | |
| GC.2.1-7 | Wavelength error | line sampling f < 0.25 | |

| Req. ID | Parameter | Requirement | Goal |
|---|---|---|---|
| GC.2.1-8 | PSF size and shape in spectroscopic mode | FWHM<0.6" and rEE80 radius <0.6" | |
| GC.2.1-9 | Stray light | <20% of Zodiacal light at ecliptic poles | |
| GC.2.1-10 | Subsample of galaxies | >140,000 gals, with >99% purity | |

## 3.4 The need for a space mission

The top-level science requirements described in the previous section imply the following important features must be met: high image quality, accessibility to near infrared wavelengths, and a homogeneous survey of the extragalactic sky with a minimum of sources of potential systematic effects. In the following we will argue that a space mission is the only option to meet these conditions.

Weak lensing measurements rely on accurate galaxy shape measurements, which in turn rely on a stable, small and well-known PSF. This can only be met in space, because observing from space provides a constant diffraction limited resolution, a well-controlled and stable environment, and avoids sources of systematic errors caused by the Earth's atmosphere and thermal variations, which seriously limit similar observations from ground. Modelling the spatio-temporal behaviour of the random turbulent atmosphere presents a serious challenge for weak lensing that requires the PSF properties to be known to sub-percent accuracy, and even if this is done then the size of the PSF renders the majority of useful weak lensing galaxies unresolved. A $\sigma^2_{sys} \sim 10^{-4}$ is possible from the ground, which for a 15,000 square degree survey with a median redshift of ~1.0, would mean dark energy parameters could not be measured with a systematic bias less than the statistical error. To control atmospheric systematics below statistical errors is feasible for near-term experiments, such as KiDS and DES, but becomes unfeasible for more ambitious experiments. Furthermore, the lack of deep NIR data and spectroscopy also leads to larger systematic biases for the photometric redshifts.

To meet the required redshift accuracies for the weak lensing and galaxy clustering analyses, photometry and spectroscopy must be performed in the near-infrared at wavelengths beyond ~1 micron. The top level science requirements aim at imaging galaxies and obtaining photometric redshifts in the range at $0<z<2$ with a median at $z\sim1$. At these redshifts the galaxy spectral energy distributions must be probed in the near infrared. To obtain spectroscopic redshifts, the prime diagnostic emission line (Hα) lies between 1.1 and 2.0 micron for galaxies at $0.7 < z < 2.05$, requiring NIR data. About 30% of the wavelength range between 0.8 and 2 micron is invisible from ground and, for the remaining fraction, bright night sky lines completely dominate the background. At low spectral resolution ($R\sim350$), the filling factor of the sky lines is virtually 100%. These emission lines, variable in intensity on time scales of minutes, add background noise, and make cosmic line and redshift determination of faint sources from the ground extremely difficult, no matter how long one integrates. A decent sky subtraction of ground-based NIR spectra can then be achieved only using fairly high spectral resolution, which results in very long spectra on the detector, something that directly clashes with the high multiplexing required by large-scale cosmological surveys. The absence of the Earth's atmosphere is therefore the unique advantage of infrared spectroscopy and photometry from space, and cannot be duplicated using current or future ground-based telescopes.

Thanks to the conspiracy of the cosmic evolution of the star formation density of the Universe, strongly increasing from $z$=0 to $z$=2+, the Euclid spectroscopic survey will provide > 50 million secure galaxy redshifts in a relatively short time. These achievements are not possible from the ground as indicated in Figure 3.2.

By observing from space, a single experiment should be able to provide the ideal combination and mutual benefit of stable visible and NIR imaging, and NIR spectroscopic data. In addition, the required integration time for NIR spectroscopy closely matches that required for the visible imaging, leading to an efficient mission design.

Last but not least, a space-based spectroscopic survey such as that provided by Euclid will give a homogeneous, uniform and well-controlled data-set which will maximise the robustness to systematic errors required for the high-precision cosmology experiments enabled by Euclid.

To summarise, only from space is:

- the environment stable enough that a visible diffraction-limited PSF can be modelled to high degree of accuracy.
- the NIR background low enough to enable the resolution of faint and small objects needed to measure the statistical signal required.
- the NIR PSF sufficiently small and stable due to the absence of terrestrial atmospheric effects.
- the resulting data-set homogeneous, uniform and well-controlled on large-scales, which is fundamental to minimise the sources of potential systematic error
- it possible to perform a survey of 15,000 deg$^2$ extended to $z \geq 2$ in $\leq 6$ years, including both imaging and spectroscopic observations.

## 3.5 Legacy science requirements for the deep survey

The dark energy science requirements described above suggest that a mission should be undertaken including both a wide survey over 15,000deg$^2$, and a deeper survey over 40deg$^2$, both with imaging and spectroscopic components. While both surveys will provide data sets that can be used for a variety of legacy science, the deep survey data would form a unique resource. We have therefore included requirements on these data based on legacy science (as described in Section 3.1), to ensure its strong legacy value. We describe these requirements in this section.

The deep field will be 2 magnitudes deeper than the wide-survey and this, combined with an area of 40 deg$^2$ will give a truly unique survey, 50 times larger than the NIR UltraVista survey and 3 times larger and 2 magnitudes fainter than the NIR VIDEO survey. A comparison with current and ongoing NIR surveys is provided in Figure 3.6. The Euclid deep survey will allow us to detect high-redshift star forming galaxies at $z$>7, measure the faint end slope of the Hα luminosity function at all redshifts for which it is detectable, and allow us to relate galaxies and their dark matter halos for normal galaxies at $z$~2.

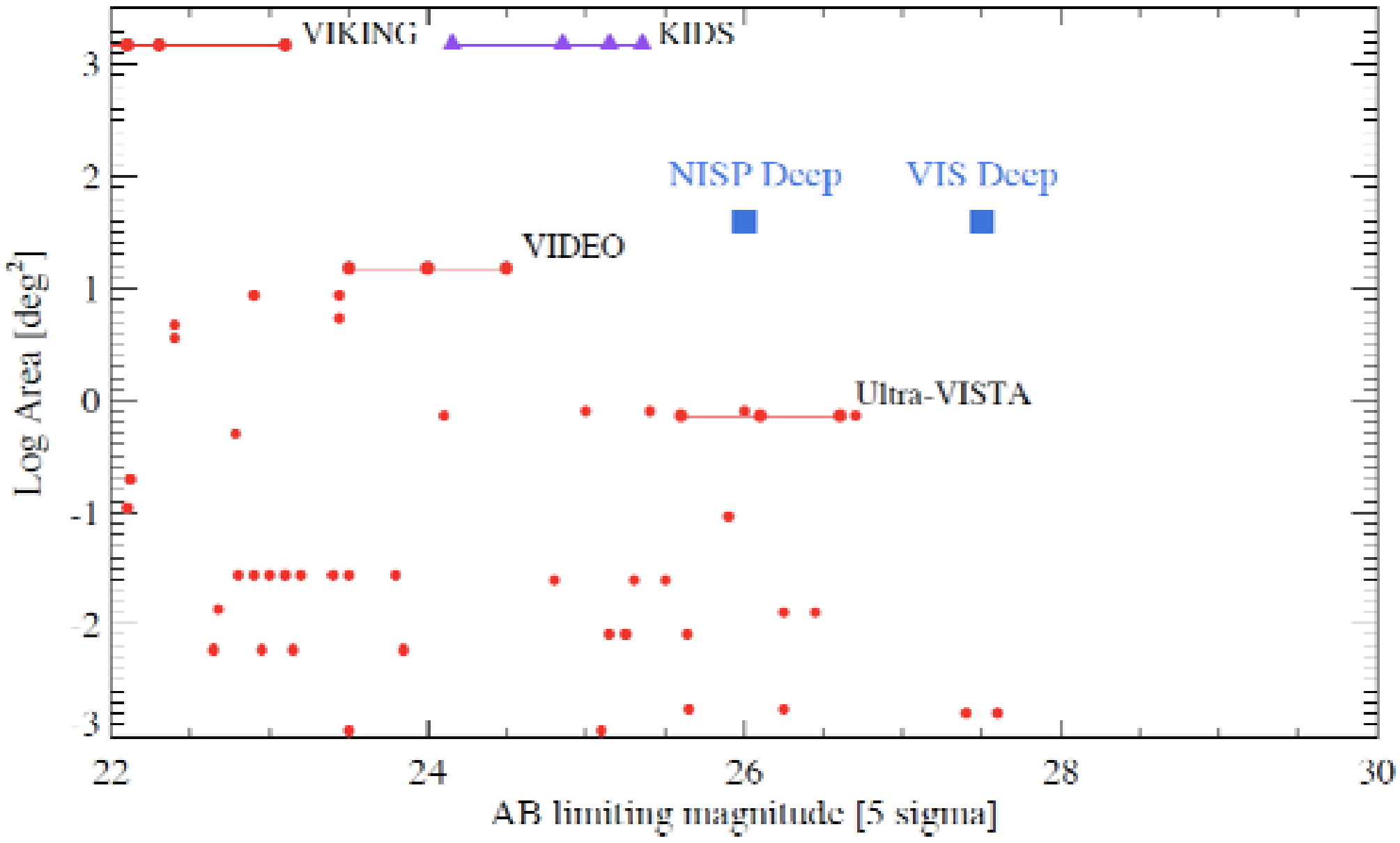

*Figure 3.6: The depth of current and on-going NIR surveys (red points), and the optical KIDS survey (purple triangles), compared to the Euclid deep surveys in visible and NIR wavelengths (blue squares).*

The depth and area of the deep survey are set by the requirements of identifying a substantial sample of the brightest galaxies at $z$~8 for detailed spectroscopic study, while providing a significant detection of the 21cm-galaxy cross-power spectrum, for the study of the epoch of re-ionisation. The proposed survey detects galaxies in three different redshift intervals, 6.3-7.1, 7.1-7.9, 7.9-8.5, using different colour combinations. There are currently only two galaxies at $z$>7 with robust spectroscopic redshifts (Vanzella et al., 2011), and both have J>26.

The choice of the combination of area and depth is a compromise between maximising the sample size (go deeper), and finding the brightest galaxies (go wider) to allow detailed spectroscopic study. For a given total survey time, the number of galaxies in a survey is maximised when the depth reached corresponds to the luminosity at which the slope of the luminosity function is $\alpha$=-3. But this occurs at J>26 at z>7, and the faintest galaxies will be difficult to follow up spectroscopically, so a wider shallower survey is better. Much brighter than J=25.5 the surface density of high-redshift galaxies drops steeply and the survey becomes inefficient. Therefore J=25.5 is the brightest flux at which it is practical to find substantial samples of galaxies at z>7. To find galaxies at this brightness with small contamination requires detection at 8σ (see below), meaning a survey reaching J=26 at 5σ.

The required sample size is based on statistical considerations. A commonly used criterion for defining a 'statistical sample', when computing the mean of a quantity (e.g. the galaxy UV slope here) is 20, since then the error on the measured mean is less than a quarter of the standard deviation of the sample $\sigma(\mu) < \sigma/4$. To investigate correlations between quantities, such as between Ly equivalent width and galaxy UV slope, a larger sample size is required. Also given the uncertainty in the galaxy luminosity function a margin for error needs to be included. For these reasons the sample size is set at 50.

To discover 50 galaxies in the highest redshift interval 7.9<*z*<8.5 (using YJH selection) brighter than J=25.5 requires a survey over 40 $deg^2$, based on the luminosity function of McLure et al. (2010). The predicted contamination for detection at 8 σ at this depth, illustrated in Figure 2.6, is 20%, which is acceptable.

A survey similar in scope is indicated to provide a high S/N detection of the 21cm galaxy cross-power spectrum, in the lowest redshift interval 6.3-7.1, which has the largest numbers of galaxies. Lidz et al. (2009) describe a proposed survey within this redshift interval over 3 $deg^2$, with Subaru, that will detect the predicted 21cm galaxy cross-spectrum at S/N=2-3. In considering how to achieve higher S/N, they note, "sizeable Lyman-break galaxy catalogues at very high redshift likely await a wide-field near-infrared instrument in space". We now consider the requirements for such a survey. Scaling the proposed Subaru survey, and matching the galaxy space density, to achieve a detection of the cross-spectrum at S/N=10, requires a survey reaching J=25.4 over 35 $deg^2$, again at 8σ to minimise contamination. These requirements, J=25.9, 5σ, 35 $deg^2$, are similar to the above requirements for spectroscopic study of the brightest highest redshift galaxies, J=26, 5 σ, 40 $deg^2$, which are therefore adopted as the baseline for the deep survey.

The top-level deep survey requirements can be fulfilled in a much shorter observing time than the wide survey. Deep survey areas of several tens of square degrees can be obtained in a few months observing time. Since the deep survey observations will be used to monitor the stability of the system, it is necessary that the observing mode is identical to that of the wide survey. This condition can be fulfilled by repeating the wide survey measurement on the same sky position until the required depth (2 magnitudes deeper) has been achieved. It is also required that the rotation angles at which the spectroscopic images are taken should be spread in angular orientation so that the deep field is suited for understanding the spectroscopic confusion in the wide survey.

With repeated visits of the same field a deep survey would open up the possibility of time-domain science, which otherwise would be impossible. In particular it would enable a systematic search of Supernovae Type Ia. We have set a goal that the cadence of revisits should be every 4 to 6 days to fulfil this science remit. Longer cadences up to 10 days are acceptable when combined with periods of faster cadences twice per month. This is not critical for the Euclid experiment, and will only be accommodated in the final deep survey strategy if feasible while meeting the other requirements.

The location of the deep survey fields in the sky requires a balance between orbital constraints and the availability of existing data. Stability of the spacecraft argues for choosing deep survey areas close to the ecliptic poles, which have good visibility throughout the year (see Section 5). This is therefore the baseline, but we retain a goal to target equatorial fields both because of the wealth of existing complementary data in this region, as well as their accessibility for ground-based follow-up observations.

## 3.6 Hardware and mission requirements

### 3.6.1 Technological readiness

To minimise risks, a number of technology development activities (TDAs) are in progress to bring all subsystems to a technology readiness level TRL ≥ 5 by the end of 2011. Details are provided in Section 5.6.

### 3.6.2 Telescope aperture

The required sensitivity, wavelength coverage, and spectral resolution for Euclid are modest compared to other astronomy space missions. In order to meet the sensitivity and angular resolution a 1 meter class telescope is sufficient. An early assessment showed that an aperture of maximum 1.2 m diameter is sufficient to meet the science objectives.

### 3.6.3 Mission duration

The current assessment is that the primary surveys can be completed in less than 6 years (see Section 5), with the deep field observations interspersed between those of the wide survey. Further optimisation of the mission operations concept may reduce this duration. However, the exact duration of the survey depends on many parameters and details of the satellite design - such as slew speed and stabilisation times - which are still uncertain at this point. The time required to commission the spacecraft and its payload and to verify and measure their in-orbit performances - currently budgeted at between 3 and 6 months - is also uncertain at this stage. For these reasons, a mission duration of up to 7 years has been assumed.

### 3.6.4 Daily telemetry rate

A constraining condition is the daily telemetry rate for the science data that are collected by the experiments. For K-band communications, the maximum daily downlink telemetry rate is 0.85 Tbits/day for a daily telemetry communication period of 4 hours. Due to the large information content of the Euclid data, no specific scientific data processing will be done on board to extract the science content. The data rate provides an upper limit for the number of detectors (although other considerations may provide tighter constraints) and hence the maximum field of view which can be fitted for a given pixel size. It also limits the maximum number of independent exposures (e.g. by using filters or by dithering), which can be collected during one day.

### 3.6.5 Orbit

In order to meet the stability requirements, a large amplitude free-insertion libration orbit around the Sun-Earth second Lagrange point (SEL2 or L2) was selected as a baseline, similar to that used by Herschel and Planck. This orbit has the great advantage that there are no disturbances by the Earth magnetic field or Moon, no thermal perturbations, and no significant gravity gradient as is the case for Earth-bound orbits. The Euclid wide survey can be performed completely without occultation or illumination of the payload by Sun, Earth or Moon, which is advantageous for obtaining a homogeneous data-set. Finally, L2 has a benign radiation environment compared to an Earth orbit, which can be a limiting factor in terms of exposure time and total lifetime of the detectors selected.

### 3.6.6 Instruments

The scientific requirements described above suggest a payload solution with two instruments, one providing wide-band visible images with exquisite control of systematic errors (the VIS instrument), and one capable of both NIR imaging and slitless spectroscopy (NISP). By using a dichroic to split the incident light, these instruments can be operated simultaneously, benefiting from the similar exposure times required to reach the required sensitivities (see Section 5). The design of these instruments is outlined in the following section.

# 4 Payload

The Euclid payload consists of a 1.2 m aperture telescope with two instruments: the visual imager (VIS) and the near-infrared spectrometer and photometer (NISP). Both instruments share a large common field of view. VIS provides high quality images to carry out the weak lensing galaxy shear measurements. NISP performs imaging photometry to provide NIR photometric measurements for photometric redshifts, and also carries out slitless spectroscopy to obtain spectroscopic redshifts. The telescope and instrument designs are challenging, but proven heritage from e.g. Gaia and JWST ensures high confidence to deliver the performance required. This section describes the features of the payload module.

## 4.1 Optical design and telescope description

The telescope has to provide both excellent visible channel imaging quality and a simultaneous field for near-IR spectroscopy and imaging. The optical design was defined by ESA after iteration and refinement with industry and the Euclid Mission Consortium.

The demanding image quality requirements in the VIS channel for the point spread function's ellipticity, full width at half maximum, and encircled energy, are the driving parameters in the telescope design. Although theoretically the ellipticity requirement does not imply the need for diffraction limited performance, in practice this requirement can not be met if the telescope images depart significantly from the diffraction limit. Two mirror telescopes do not yield good enough image quality for Euclid large field of view. The alternative is a three mirror configuration, either an off axis three mirror anastigmat (thus, without central obstruction) or a Korsch configuration. With three mirrors, there are enough degrees of freedom (three curvatures, three conic constants and two distances between mirrors) to achieve a good level of aberration correction (quasi diffraction limited images) and the required image scale and low distortion.

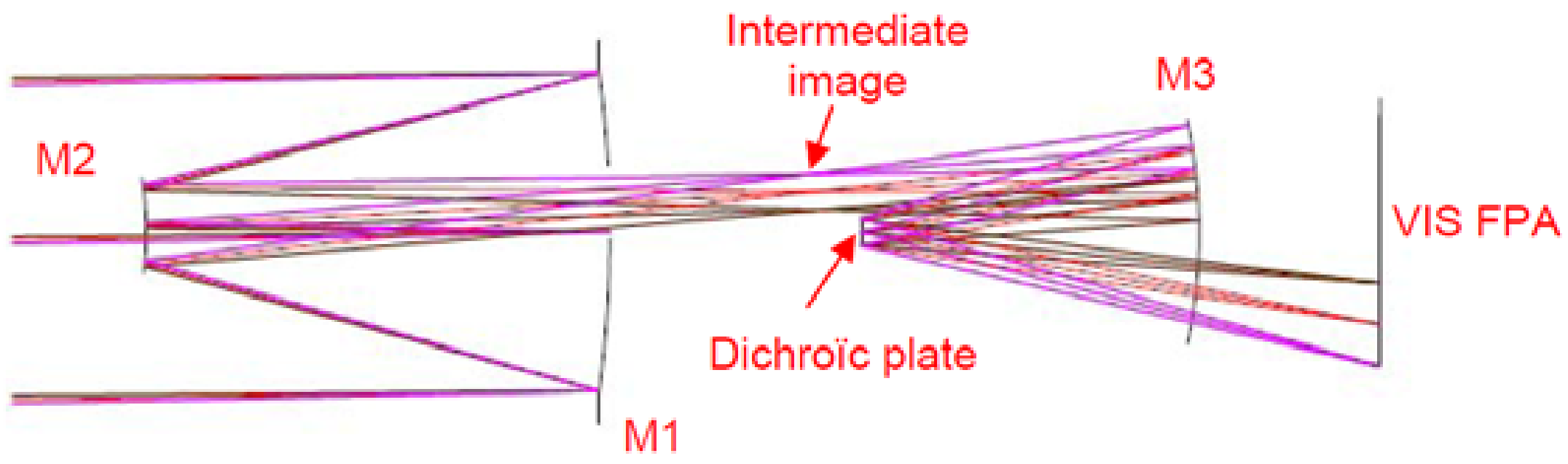

*Figure 4.1: scaled view of unfolded Euclid telescope unfolded design. NISP instrument, which is placed in the transmitted beam of the dichroic, is not shown.*

A schematic of the resulting optical design for Euclid is given in Figure 4.1. The Euclid telescope is a three mirror Korsch configuration with a 0.45 deg off-axis field and an aperture stop at the primary mirror. The entrance pupil diameter is 1.2 m, the optically corrected and unvigneted field of view is 0.79×1.16 deg$^2$, and the focal length is 24.5 m. The dichroic beam-splitter is located at the telescope exit pupil for separating the VIS and NISP optical channels with the reflected output from the dichroic beam-splitter going to the VIS instrument and the transmitted output to the NISP. The interface between the VIS instrument and the telescope is the focal plane, while the interface between the telescope and the NISP instrument is the dichroic beam-splitter.

To meet scientific performance objectives, such as an internal background well below the zodiacal sky background, the telescope and payload module have to operate at a reduced temperature. To minimise dark current noise, the maximum telescope temperature was determined to be ~240K.

The telescope is built on a truss hexapod concept (Figure 4.2): 6 struts are connecting the secondary mirror (M2), mounted on a frame through spiders to the primary mirror (M1) optical bench. The upper part of the optical bench supports M1 and the M2 structure, the lower part supports the other telescope optics and both VIS and NISP instruments. The optical bench provides also interface points to the service module.

For the M1/M2 sub-systems, industry has concentrated on two options that reflect their respective expertise base: Silicon Carbide (SiC) for Astrium and Zerodur for TAS. M1 is supported via 3 isostatic bipods and is on the upper side of optical bench supporting the instruments. In the SiC option, the payload bench and the mirrors are made of lightweight SiC, designed for a temperature of 150 K, which is achieved by passive means. M1 is a monolithic SiC mirror with CVD SiC coating. The use of an all-SiC design allows an athermal homothetic approach to guarantee PSF stability. TAS uses light-weighted Zerodur class 0 mirrors with very low coefficient of thermal expansion (CTE) kept at a temperature below 240 K, which eases the thermal stability aspects of the primary figure control. Further thermal stability is provided by application of stable materials: a ceramic ($Si_3N_4$) for the strut structure and carbon fibre reinforced plastic (CRFP) for the optical bench. The Zerodur/ceramic approach requires active thermal control while being less sensitive in principle to the payload module power dissipation and solar aspect angle variations.

The secondary mirror is supported by a spider. The telescope optical performance is most sensitive to the M1/M2 separation. The secondary mirror is integrated on a mechanism which may be used after cool-down to correct for misalignment after launch, and potentially for any relaxation and thermal changes to focus, especially in the case of large excursions of the Solar aspect angle from a nominal 90 deg up to 120 deg. Thales uses the mechanism to compensate for long term drifts in the ceramic structure. Astrium uses it to compensate misalignment errors at integration and for refocusing in orbit.

A number of measures have been taken to suppress straylight and ghosts. Illumination from the Sun is blocked by a sunshield on which the Solar cells are mounted. A telescope baffle further minimises scattered light from outside the field of view entering the telescope.

The required image quality and the stability of the point spread function depends on misalignments of the optical system, optical surface manufacturing errors, thermo-elastic deformations and residual pointing jitter of the attitude and orbit control system (AOCS). The budgeting between these factors is ongoing, and the assessment requires an end to end approach. Specific telescope properties will be different for the different industrial designs. The performance calculations are described in Section 6

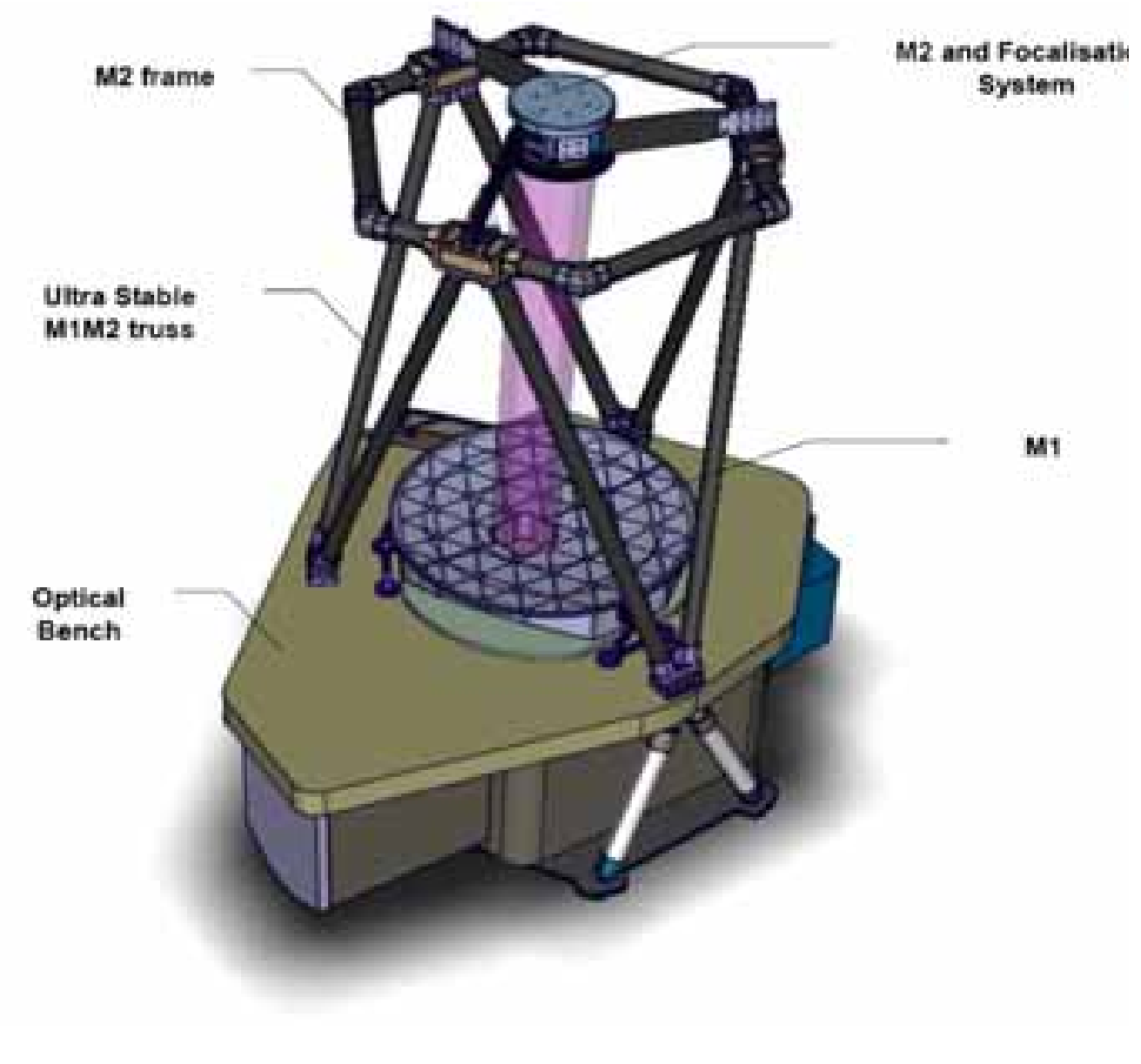

*Figure 4.2: Typical telescope mechanical architecture (drawing kindly provided by TAS). The optical bench supports M1 and the truss hexapod with M2 on one side and the other telescope optics and the instruments on the other side.*

# 4.2 Visible instrument (VIS)

## 4.2.1 Visible Channel description

*Table 4.1: VIS functional description*

| Name | UNIT | Function |
|---|---|---|
| **VI-FPA** | VIS Focal Plane Assembly | Detection of visible light for imaging |
| **VI-SU** | VIS Shutter | Close VIS optical path for read out<br>Close VIS optical path for dark calibration |
| **VI-CU** | VIS Calibration Unit | Illuminate the FPA with Flat Field for calibration |
| **VI-CDPU** | Control and Data Processing Unit | Control Instrument<br>Perform data processing<br>Interface with Spacecraft for data handling |
| **VI-PMCU** | Power and Mechanism Control Unit | Control Units |
| **VI-FH** | Flight Harness | Connection of units |

The Visible Imaging Channel (VIS) is an assembly of the units listed in Table 4.1. The layout of these units (excluding the harness) is shown in Figure 4.3.

The VI-FPA is a thermal-mechanical structure that supports the 6×6 CCDs and its associated Read Out Electronics units (ROE) which together comprise the visible focal plane array. In addition to providing structural support of the massive focal plane array, the FPA structure ensures mechanical and thermal stability of the CCD array and accurate location of the CCD array with respect to the optical beam image plane.

The VI-CU calibration unit is designed to allow flat fields of the visible channel. This structure encloses a 12-LEDs panel illuminating a diffusing panel inside an integrating sphere.

A shutter VI-SU stops the light beam before entering the VIS focal plane assembly in order to prevent trails in the images during visible detector array readout in normal operations and flat field calibration.

Two electronics units are associated with the instrument: the Control and Data Processing Unit (VI-CDPU) and the Power and Mechanism Control Unit (VI-PMCU). The VI-CDPU controls the instrument and compresses the scientific data before transfer to the payload mass memory. The VI-PMCU controls the instrument mechanisms and calibration units.

These are now described in more detail.

***Focal Plane Assembly (VI-FPA)*** is composed of two main parts, the detector subassembly and the electronics subassembly.

The detector subassembly is the 6×6 CCD matrix composing the visible focal plane array together with a thermo-mechanical structure to support them. It provides a thermal path for the power dissipated by the CCDs to the radiator associated with the visible focal plane. It also ensures stability of the temperature on the whole extent of the focal plane array to operational temperature 153 ±5 K with a gradient lower than ±3 K.

The electronics subassembly consists of the Read Out Electronics (ROE) and associated Power Supply Units (RPSU) associated with the CCDs plus a mechanical support structure. It is also linked to a dedicated radiator to provide the operational temperature environment to the electronics. This structure provides thermal shielding to prevent thermal radiative coupling between the ROE warm electronics and the cold CCDs in the focal plane.

The two subassemblies (detector and electronics) are not mechanically linked: they are linked only electrically by the CCD harnesses. This is in order to prevent mechanical perturbation at launch and thermal perturbation of the cold stabilised CCD structure by variations in the power dissipation in the warm ROE and RPSU. The CCD harnesses have sufficient degrees of freedom to accommodate the small relative movement under vibration.

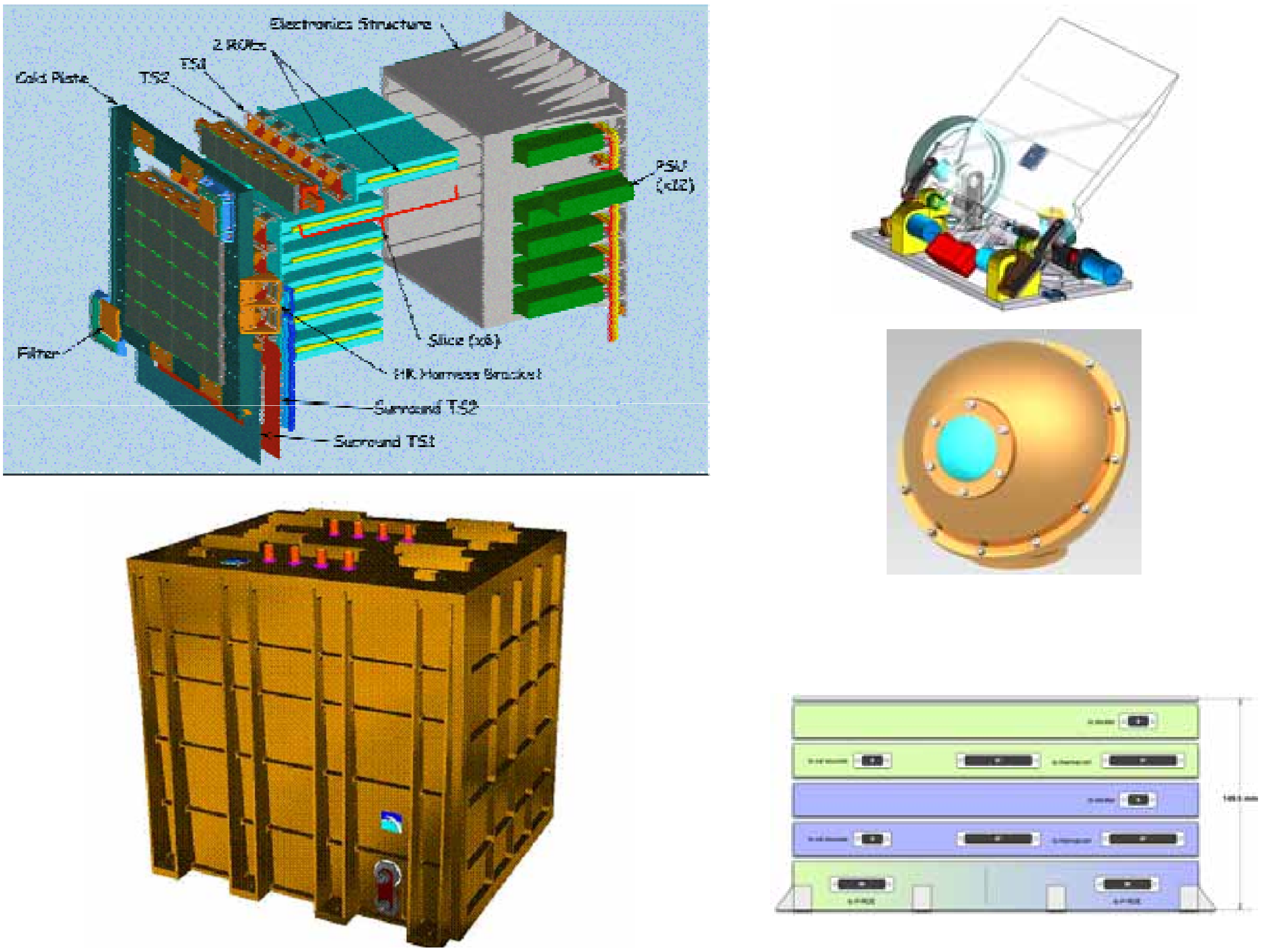

*Figure 4.3: The units comprising the VIS instrument. Clockwise from top left: the VIS-FPA, the VIS shutter, a calibration unit, and, in the service module, the Mechanism Control Unit and the Control and Data Processing Unit.*

The CCDs are supported on an Aluminium bar (6 CCDs per bar) through the standard mechanical interface on the CCD package. Each of the 6 bars is integrated on a common Aluminium structure. This is then interfaced to the payload module.

To provide the second VIS optical passband on 2 corner CCDs, two filters subassemblies have been implemented in two corners of the focal plane. The filter housing has been designed to avoid vignetting on nearby CCDs. For this reason the optical filter will be interfaced with its mechanical housing on only two sides.

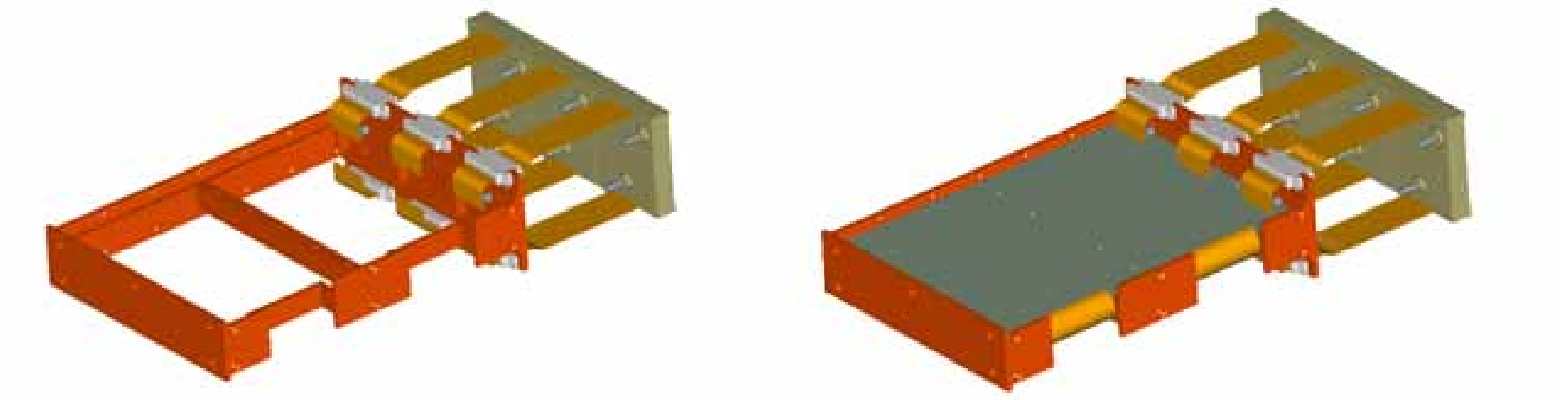

*Figure 4.4: Read-out electronics (ROE) unit with CCDs*

The size of the CCD package dictates the dimensions of the ROE unit in two axes. A tradeoff for redundancy and system resources has led to a design where each ROE supports three CCDs (12 video chains) (Figure 4.2). Access to assemble or replace any one CCD, ROE or RPSU unit is also a key driver to the system design which in turn drives sub-system design. The interconnection between the CCDs and the ROEs is optimised to minimise thermal coupling between the cold CCDs and the warm electronics whilst not degrading critical low level analogue video signals.

The ROE electronics will be built on one folded flex-rigid printed circuit board which will accommodate the 12 readout chains and the circuitry that is common to the three CCDs – most of the latter contained within a single FPGA. The PCB mounting technique is chosen for good thermal conductivity to the chassis which is fabricated as a monolithic structure. Special provision is made to allow the CCD interface connectors to float slightly (Figure 4.4), so they can be locked from the rear of the ROE after the integration of the CCDs.

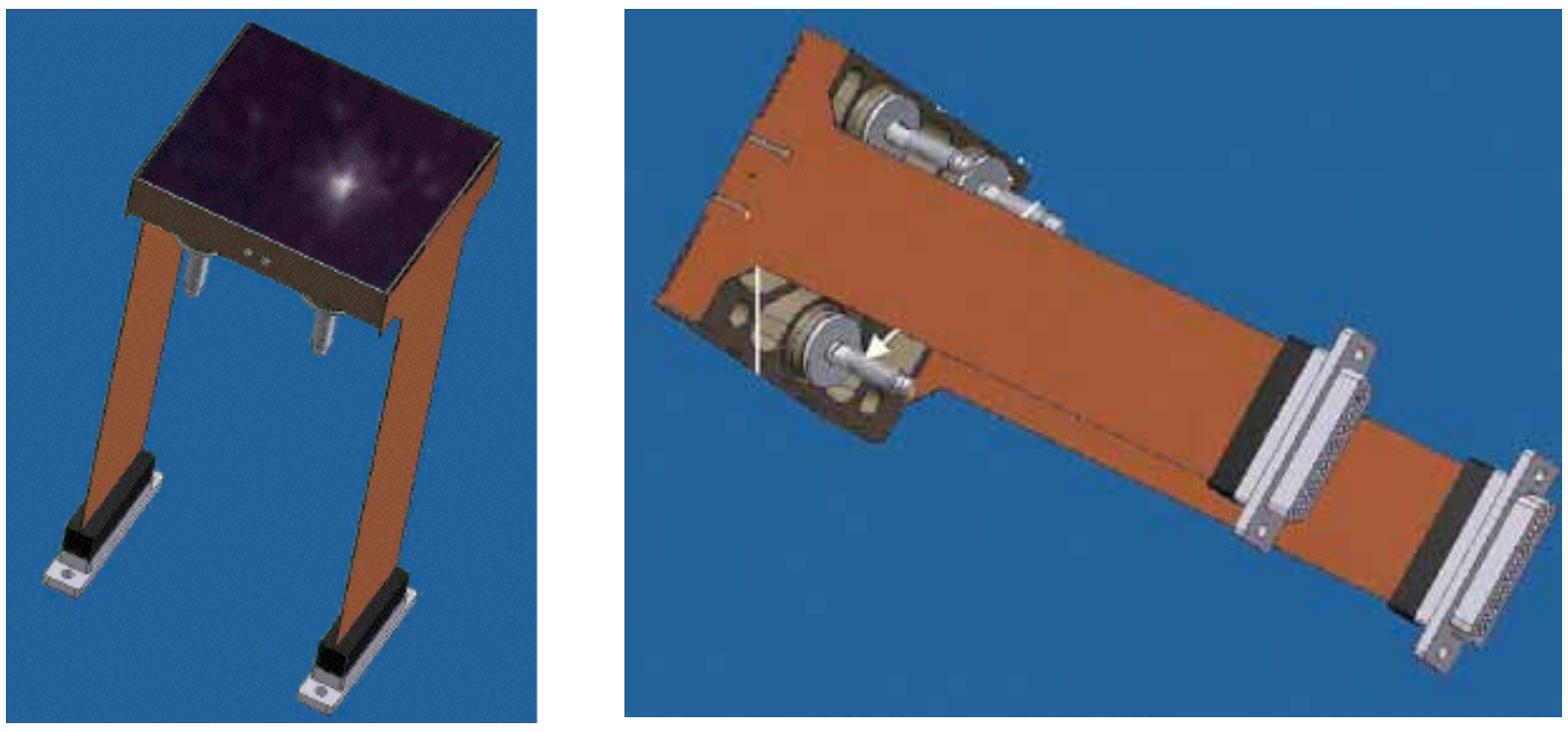

*Figure 4.5: The CCD273 seen from the front and the back.*

New CCDs, designated CCD273, have been developed especially for Euclid VIS, to optimise its radiation performance. Each provides a 4k×4k pixel format, each pixel 12 μm square. The package is shown in Figure 4.5. In the CCD array, the areas of active Si are separated by 7.8 mm between readout registers, and 1.6 mm in the other dimension.

***VIS Shutter:*** The shutter (VI-RSU Figure 4.3, top right) prevents direct light from falling onto the CCDs during the closed phase while allowing the fine guidance sensors to be exposed to light continuously. During the open phase it must avoid any interference with the light beams.

A single hatch concept has been adopted after extensive tradeoff. The rotation axis is parallel to the longer side of the FPA. The hatch closes away from the detector and opens towards the detector.

The hatch is closed and opened by redundant stepper motors (via a drive arm) and a torsion spring working against the retreating motor head (Figure 4.6). The drive arm allows for an easy means of decoupling the motor from the hatch in case of failure, and of introducing the redundant motor. A counter mass and a flywheel have been implemented to counter the disturbance loads. The counter mass of the hatch eliminates linear momentum while the flywheel is optimised in terms of moment of inertia to sufficiently reduce the angular disturbances. Each stepper motor has a nominal and redundant winding.

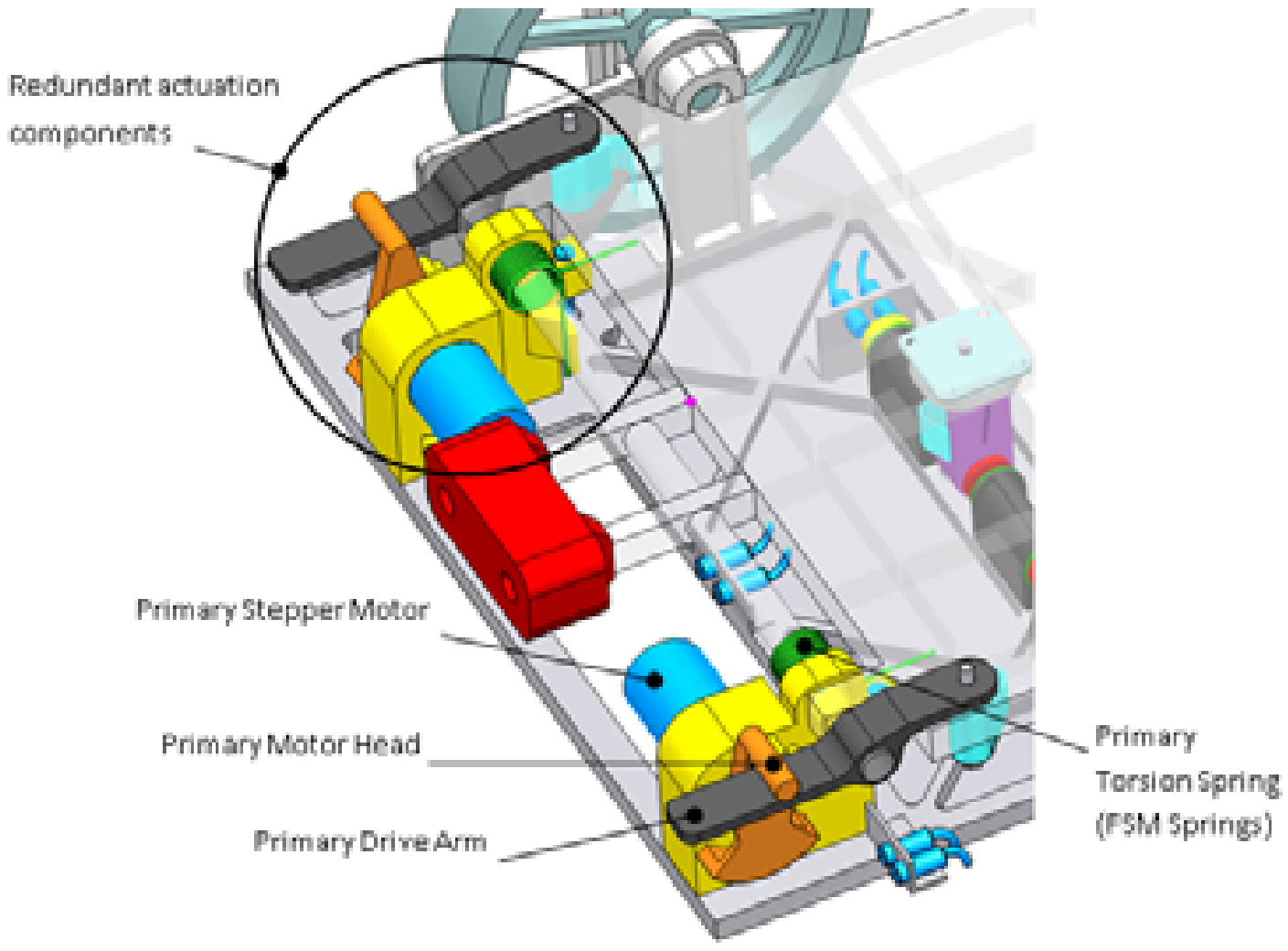

*Figure 4.6: Shutter Actuator System*

Between the baffle and the hatch, labyrinth seals will be implemented to reduce the light as required. Because the hatch opens towards the VI-FPA, it can be the cause of unwanted reflections from the incoming beam. Geometrical protections can be implemented on the shutter to prevent this reflection from hitting the CCDs. The first possibility is to place a series of teeth on the shutter, always presenting an angle > 45 deg to the beam. Another possibility is to equip the top side of the shutter with light baffles. The shutter is driven by the VI-PMCU.

***VIS Calibration Unit Design (VI-CU):*** The calibration unit allows flat fields of the focal plane array to be obtained. The concept of the unit is based on an integrating sphere (Figure 4.3). The inner optical surfaces of the sphere are coated with Spectralon, which has a high reflectivity over the wavelength of interest for the VIS instrument (from 600 to 900nm). Two optical ports are foreseen, one for the illuminating sources and another one for the output light which illuminates the VI-FPA.

The illuminating sources port supports both main and the redundant channel. Each channel provides three wavelengths from 600 nm to 900 nm, each of them using two LEDs. A minimum of twelve LEDs are used for the whole CU. The output light port is closed by a BK7 window in order to maintain the inside of the CU extremely clean. This glass has been chosen for its high transmission from 600 to 900 nm.

The CU will be driven by the VI-PMCU.

***Control and Data Processing Unit (VI-CDPU):*** Mechanically the CDPU consists of a six panel Aluminium box with ribs on external side and feet with cantilever on two opposite sides (Figure 4.3). It is located within the Service Module. The VI-CDPU is responsible for the following main activities:

1. Telemetry and telecommand exchange with the spacecraft control and data management unit
2. Instrument commanding, based on the received and interpreted telecommands
3. Instrument monitoring and control, based on the housekeeping data acquired from the other instrument units
4. Synchronisation of all the instrument activities
5. Data acquisition from the ROEs, pre-processing and formatting according to the selected telemetry protocol

For a fully redundant unit all the functions are implemented on 16 electronic boards.

The software running on the VI-CDPU is composed of startup and application modules, including the loss-less compression, resident on data processing and science processing boards. This software is implemented as a real-time multitasking application.

***Power and Mechanism Control Unit (VI-PMCU)*** The Power and Mechanism Control Unit (Figure 4.3) encompasses all the functions required to control VIS mechanisms as well as the calibration sources. For a fully redundant unit all the functions are implemented on 5 electronic boards.

### 4.2.2 Thermal architecture

The VI-FPA thermal interface internal and external toward the PLM is summarised in Figure 4. There are two radiators located outside VIS: a cold radiator for the CCDs and a warm one for the ROEs.

The VI-FPA detector subassembly is attached to the payload via a thermo-mechanical interface located on the mechanical plate supporting the CCD. A thermal and mechanical structure provided by the spacecraft provides the thermal link to the payload cold radiator bay that ensures passive cool-down to 153 K.

The VI-FPA electronics subassembly is located on the payload without stringent mechanical location requirements. It is directly mounted on the payload warm radiator in order to optimise thermal coupling.

The packaging of the ROE and PSU electronics is designed to optimise conductive thermal paths. A monolithic machined chassis based structure has been chosen for this reason. The thermal and mechanical interfaces for the ROE units are at the front (towards the CCDs) where there is a connector panel and at the rear where a matrix of rails is used to transfer heat and dynamic forces from the ROE units to the edges of the electronics subassembly.

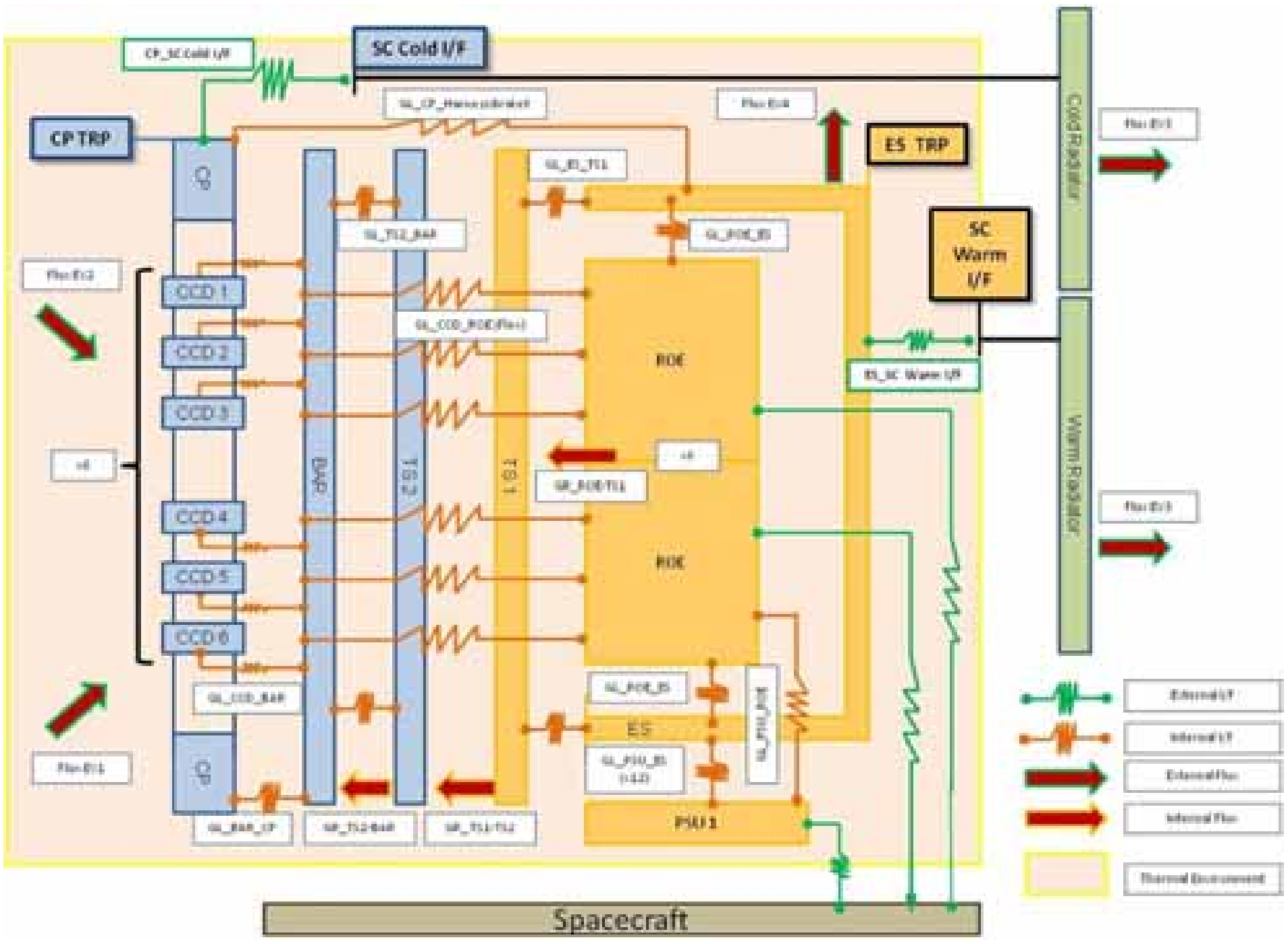

*Figure 4.7: VIS FPA I/F description (red link internal interface, green link external interface)*

In order to achieve the thermal stability of ±0.3K at CCD level, heaters and temperature sensors are implemented for each pair of ROE (i.e. each CCD row). A full cold redundancy is adopted, so the focal plane will be equipped with 12 heaters and 12 temperature sensors in total.

## 4.2.3 Electronics architecture

Figure 4.8 shows the electrical architecture diagram between the VIS subsystems, and from the instrument to the platform. This architecture is split in several temperature stages ranging from cold operation (153 K for VIS detectors) to ambient temperature operation (<300 K).

The architecture of the proposed electronics to control and read out the 36 CCDs has been developed based on experience of previous projects, most notably Gaia. Three CCD273 interface to each ROE and all VIS CCDs are all operated in synchronism. This simplifies the sub-system design and minimises the danger of coupled noise sources. To minimise read out time, all four CCD ports are used concurrently to read out each Euclid VIS CCD. The ROE units digitise with high resolution (16-bit) and provide low noise, high precision video processing in order to sample the telescope PSF accurately. Great care is taken in the design of mixed signal circuitry which handles very low level analogue inputs, to prevent cross-talk and other noise pick-up. Multi-layer printed circuit boards are used with separate ground planes for analogue and digital functions. System grounding and decoupling is carefully planned to prevent circulating currents in ground lines from introducing noise sources. Figure 4.9 shows prototype ROE developed for Euclid VIS; a second-generation board is also currently in use for the CCD test programme. It has already been demonstrated at engineering model quality that the electronics to support three CCDs fit on the ROE board size using only surface mount devices (i.e. without any hybrid microcircuits).

To preserve the redundancy concept, each ROE is provided with its own Power Supply Unit (RPSU). The performance of the RPSU is critical to overall instrument performance because supply line noise can easily be the factor that limits the achievable signal to noise ratio. Also, analogue housekeeping parameters are acquired internally to these units especially to avoid noise propagation from the noisy digital units (typically the payload digital electronics) toward the sensitive analogue electronics in the ROE.

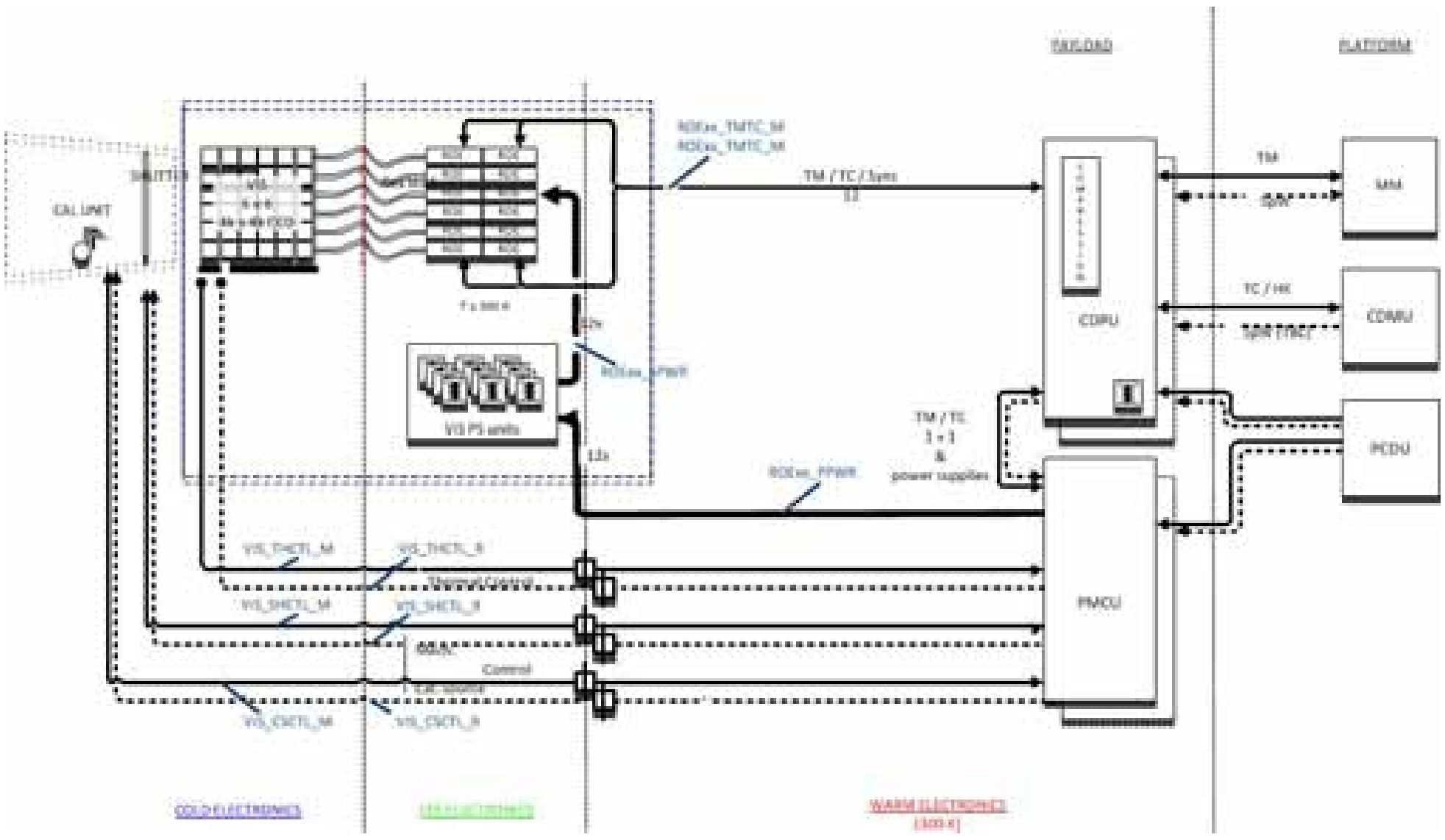

*Figure 4.8: VIS instrument electrical system architecture*

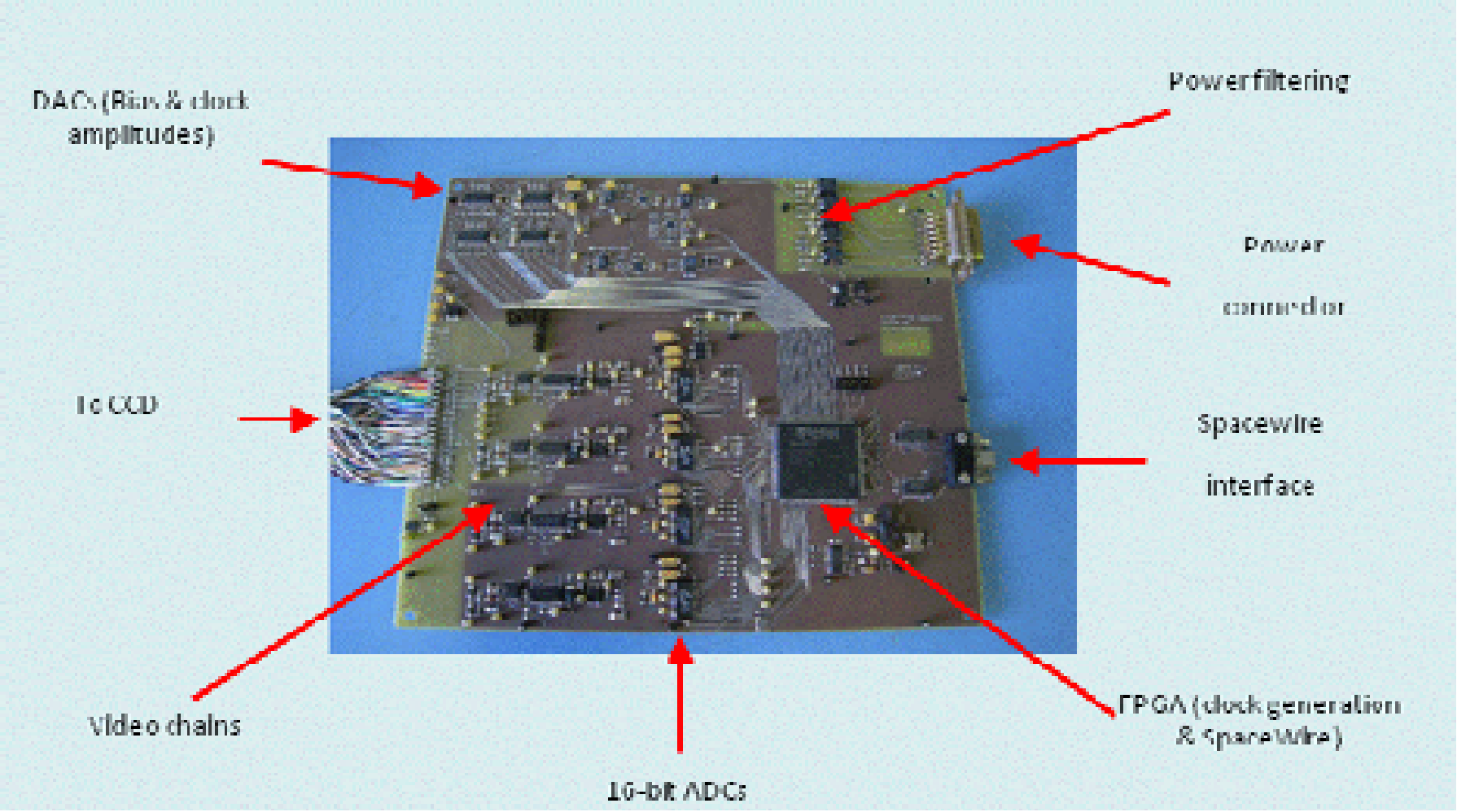

*Figure 4.9: The Electrical Evaluation Model (EVM) read-out electronics*

The digital electronics is located in the VI-CDPU unit. The VI-CDPU controls all the sub-systems of the instrument by distributing low level commands, collecting and monitoring housekeeping parameters. It is also responsible for collecting all the uncompressed data streams transferred from the sub-systems and performs lossless compression in order to match instrument data rates and spacecraft telemetry rate. The data are then downloaded to the spacecraft mass memory through a dedicated SpaceWire high-speed data link. The VI-CDPU interfaces the spacecraft command and data management unit through a MIL STB 1553B link for telecommands and housekeeping telemetry. The VI-CDPU internal interfaces (to VI-PMCU and to the ROEs) are implemented through dedicated SpaceWire high-speed data links. The VI-CDPU is a fully redundant unit.

The VI-PMCU includes control electronics for instrument mechanisms; these are the drive electronics for motor control and position sensors. It also includes drive electronics to bias the calibration source LEDs. It includes temperature monitoring lines for focal plane temperatures acquisition. Finally it includes the main-bus power switching needed for the ROE power supply units. It is also redundant.

For the VI-CDPU and VI-PMCU the secondary power supplies are derived from the spacecraft power bus by mean of DC/DC converters located inside the CDPU.

### 4.2.4 Mass and power budget

*Table 4.2: VIS instrument Mass Breakdown (includes 20% margin)*

| Unit | Mass (kg) |
|---|---|
| VI-FPA | 66 |
| VI-SU | 15 |
| VI-CDPU | 17 |
| VI-PMCU | 14 |
| VI-CU | 1 |
| VIS Total | 113 |

*Table 4.3: VIS power breakdown*

| | Stand by (W) | Observation (W) | | | | | Calibration (W) | | | | |
|---|---|---|---|---|---|---|---|---|---|---|---|
| | | Mean | Imaging[1] | Read-out | Shutter | Idle | Mean | Imaging[1] | Read-out | Shutter | Idle |
| **Time (s)** | | **803** | **544** | **100** | **20** | **139** | **803** | **544** | **100** | **20** | **139** |
| Detector | 0.0 | 1.8 | 1.7 | 2.0 | 1.7 | 2.0 | 1.8 | 1.7 | 2.0 | 1.7 | 2.0 |
| ROE (12x) | 0.0 | 78.0 | 76.7 | 86.3 | 76.7 | 86.3 | 78.0 | 76.7 | 86.3 | 76.7 | 86.3 |
| PSU (12x) | 0.0 | 35.2 | 34.7 | 38.7 | 34.7 | 38.7 | 35.2 | 34.7 | 38.7 | 34.7 | 38.7 |
| CDPU | 53.8 | 47.5 | 53.8 | 53.8 | 53.8 | 53.8 | 53.8 | 53.8 | 53.8 | 53.8 | 53.8 |
| PMCU | 7.1 | 24.7 | 27.6 | 27.6 | 41.8 | 27.6 | 28.9 | 33.3 | 27.6 | 41.8 | 27.6 |
| **Total Primary** | **60.9** | **187.2** | **194.5** | **208.4** | **208.7** | **208.4** | **197.6** | **200.2** | **208.4** | **208.7** | **208.4** |
| **Inc 20%** | **73.1** | **224.7** | **233.4** | **250.0** | **250.5** | **250.0** | **237.2** | **240.3** | **250.0** | **250.5** | **250.0** |

### 4.2.5 VIS Critical items

The following critical items and the mitigating activities undertaken can be identified for VIS:

***The CCD***. With the very high imaging quality required for VIS to achieve the weak lensing performance, there are many aspects of the CCDs which require deep understanding and characterisation. Fortunately, there is substantial TRL associated with these devices. The Euclid CCD273 has been developed especially for Euclid VIS: it optimises the CCD203/204 family design, which itself has space heritage. Radiation damage effects are a particular concern. The criticality of the CCD has been addressed through the close cooperation of the Euclid Consortium, ESA and the manufacturer e2v, with an active CCD Working Group. Flight standards for the operation of the device throughout the test programme (clocking schemes, bias voltages) to ensure inter-comparison and fidelity of the test programme data have been established early. e2v have already provided the first engineering CCD273s for evaluation, and these are operational in the testing. Prior to this, many CCD203/204 devices have been radiation tested and optically characterised, to provide inputs for the performance evaluations.

***The ROEs***. The Gaia programme showed the importance of early development of the detection chain, in order for the required performance to be reached, and for test results to be representative. The ROEs are required to provide similar performance to the Gaia equivalents, but with half of the power budget and with truly radiation-hardened 16-bit analog-digital converters. As noted above, already in the Assessment Phase, flight-like ROEs were developed and improved, reducing risk and providing standards for the test programme. A further development has been undertaken in the Definition Phase to examine the detailed noise characteristics arising from simultaneous operation of 3 CCDs.

***The FPA assembly and metrology***. The Euclid VIS focal plane will be the second largest flown (after Gaia), and, being tightly packed, has particular assembly/disassembly and metrology requirements, all to be exe-cuted within a clean environment. Detailed procedures have been developed, and the requirements for the associated jigs and other mechanical ground support equipment has been analysed.

***The Shutter***. This large shutter will need to make a large number of actuations in orbit over its lifetime, and a shutter failure will seriously impact VIS. Great care has been taken in the failure analysis and in the redundancy planning.

## 4.3 Near IR spectrometer and imaging photometer (NISP)

The NISP Instrument is the near-infrared Spectrometer and Photometer operating in the 0.9-2.0 micron range at a temperature lower than 140K, except for detectors, cooled to ~120 K. The warm electronics will be located in the service module, at a temperature higher than 240 K.

The NISP instrument has two main observing modes: the photometric mode, for the acquisition of images with broad band filters, and the spectroscopic mode, for the acquisition of slitless dispersed images on the detectors. The main elements of the NISP instrument are listed in Table 4.4.

*Table 4.4: NISP main elements*

| Name | UNIT | Function |
|---|---|---|
| **NI-OMA** | NISP Opto-Mechanical Assembly | This holds the optical elements and the Focal Plane Array |
| **NI-GWA** | NISP Grism Wheel Assembly | This holds the four dispersing elements for the spectroscopic mode and it allows them to be placed in the optical beam. |
| **NI-FWA** | NISP Filter Wheel Assembly | This holds the three filters for the photometric mode and it allows them to be placed in the optical beam. It provides also a closed and open position. |
| **NI-CU** | NISP Calibration Unit | This injects calibration signal in the optical beam for calibration purposes |
| **NI-DS** | NISP Detector system | This system provides detection of the NIR signal in photometric and spectroscopic mode |
| **NI-WE** | NISP Warm Electronics | This is composed of the NI-DCU, NI-DPU and NI-ICU (see below). |
| **NI-DCU** | NISP Detector Control Unit | This provides the data and command interface to NI-DS and also detector acquisition and cosmic ray identification. |
| **NI-DPU** | NISP Data Processing Units | This provides data compression and packeting as well as the interface to S/C Mass Memory and to the NI-DCU |
| **NI-ICU** | NISP Instrument Control Unit | This controls the instrument, powers and controls mechanisms, provides instrument thermal control, and the command interface with NI-DPU and NI-DCU. |

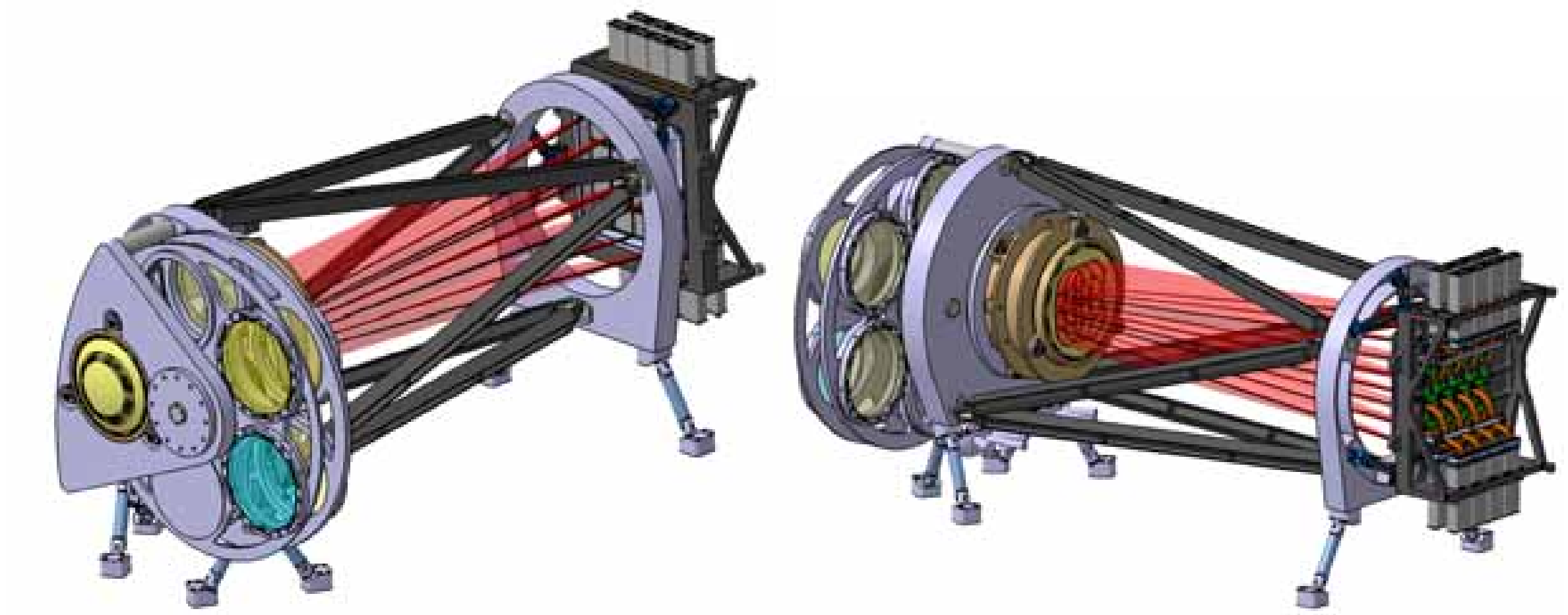

*Figure 4.10: NI-OMA plus NI-DS CAD view*

### 4.3.1 Opto-mechanical design

The NISP Opto-Mechanical Assembly (NI-OMA, Figure 4.10) holds the optical elements of the instrument at cryogenic temperature. It is located in the payload cavity. Two mechanisms mounted on the NI-OMA select the two main observing modes.

In the photometer mode the NISP instrument images in the wavelength range from 920 nm to 2000 nm (Y, J, H bands) on an array of 16 detectors. The spatial resolution is required to be 0.30±0.03 arcsec per pixel in all three bands. The FoV of the instrument is 0.55 deg$^2$ having a rectangular shape of 0.763 deg×0.722 deg. In the spectrometer mode the light of the observed target is dispersed by means of grisms in wavelength range of 1.1 – 2μm. In order to provide a flat resolution over the specified wavelength range two sets of two grisms each are used, selected via a filter wheel.

The NI-OMA is is made up from two sub-assemblies: the Corrector Lens Assembly(NI-CoLA) which corrects residual abberation after the telescope pupil, and the Camera Lens Assembly (NI-CaLA), which images the field of view onto the focal plane.

The 2D model of the NISP optical design is presented in Figure 4.11, which shows:

- the corrector lens (CL) made from fused silica with its holding structure.
- the filter is also made from fused silica with mildly powered spherical entrance surface and flat exit surface.
- the grism (fused silica) with mildly powered spherical entrance surface and binary optic (curved line) grating exit surface.
- the three spherical-aspherical meniscus lenses L1-L3 which, together with their holding structure called the Camera Lens Assembly (CaLA).L1 is made from CaF2, while L2 and L3 are made of LF5G15.

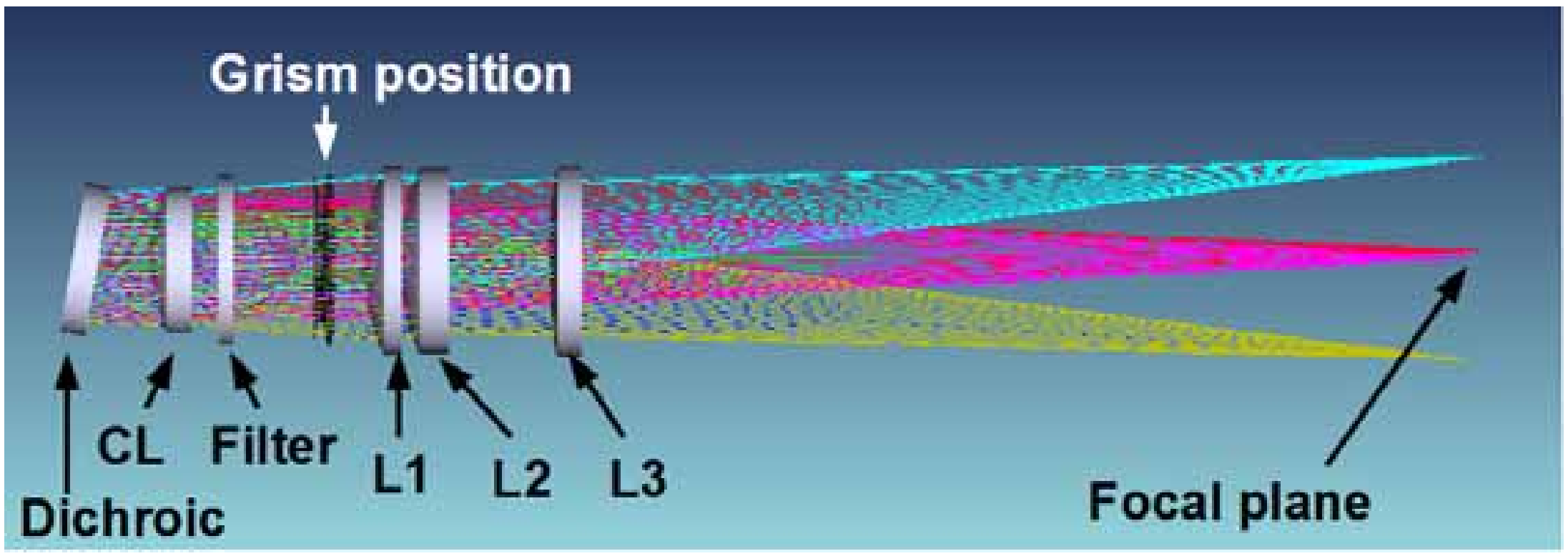

*Figure 4.11: sketch of the NISP optics. The dichroic is part of the telescope assembly.*

Thermo-mechanical effects in optical elements are compensated via adaptation rings, glued to the optical elements. These rings are required to introduce quasi zero forces and torques to elements, preserving the shape and position in compliance with optical tolerances.

The telescope provides the entrance beam through the dichroic. The detector system (NI-DS), controlled by the Detector Control Unit (NI-DCU), acquires the image and sends the data to the data processing unit (NI-DPU). The Instrument control Unit (NI-ICU) commands the functions of NI-OMA.

The *Filter Wheel Assembly* (FWA) and the *Grism Wheel Assembly* (GWA) are shown on Figure 4.10 and Figure 4.12. They provide the interchanging functionality between the photometric and spectroscopic modes. The FWA performs the filter switching function. The structure houses and positions three near-infrared filters in fused silica. Each filter is 10 mm thick and 120 mm in diameter. The wheel is configured with five slots in order to integrate in addition an open slot and a closed one.

The NI-GWA positions into the optical beam two kinds of grisms with different passband coatings. The first type (Blue) transmits between 1.1μm and 1.45μm, while the other (Red) is transparent between 1.45 μm and 2μm. The 4 grisms are mounted in two orientations: blue 0 deg, red 0 deg, blue 90 deg and red 90 deg.

A grism is a transmission grating ruled on the hypotenuse of a prism and thus at a particular wavelength the diffraction of the grating is compensated by the prism deviation. The NISP grism is a binary surface grating characterised by the combination of a shallow profile (2 degrees) with a low groove density (19 g/mm). The ruling technique will use photolithography to imprint saw tooth groove profile in a resin layer. The grisms provide a linear dispersion in main dispersion direction of 9.8Å/pix.

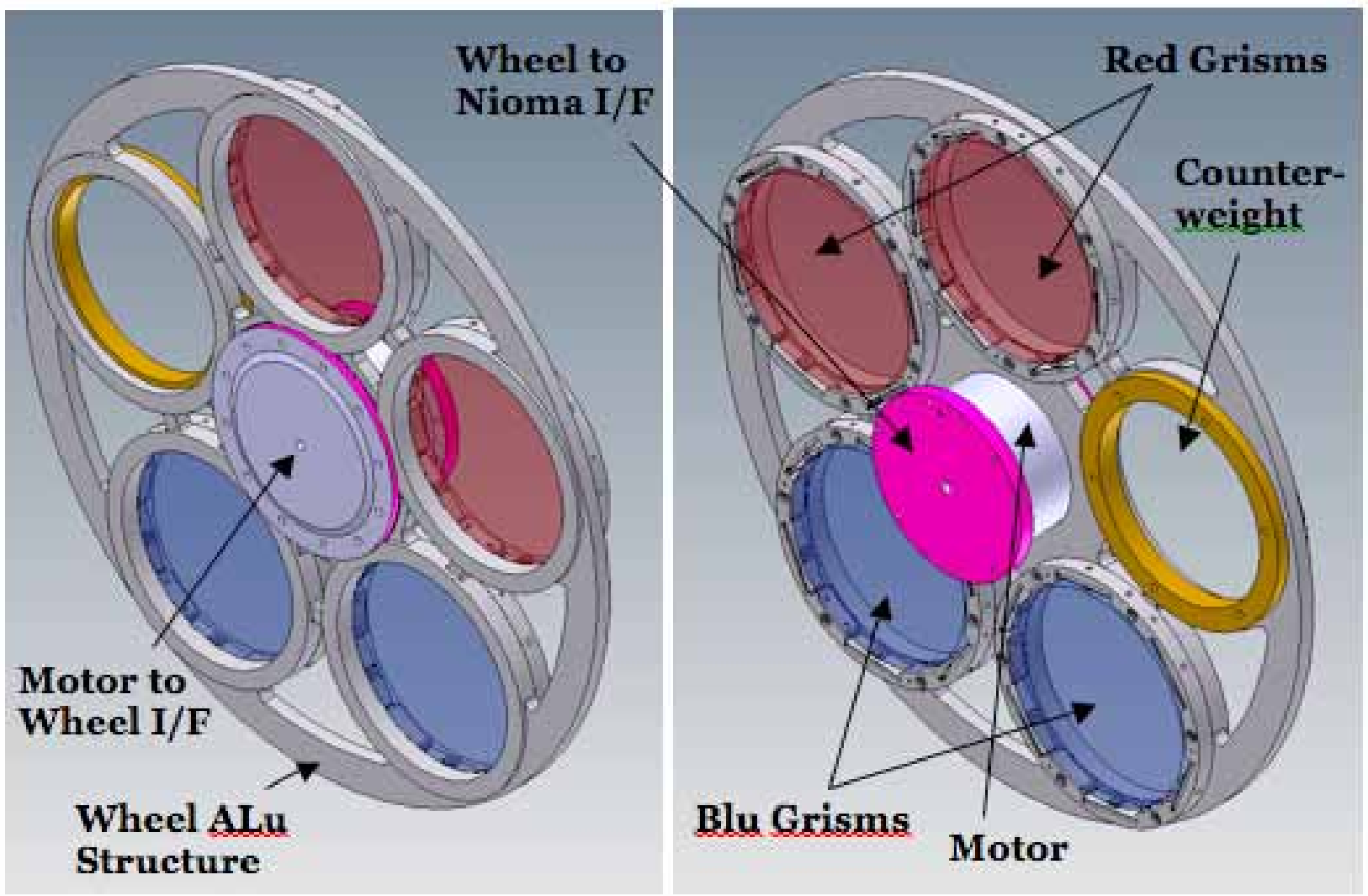

*Figure 4.12: Grism Wheel Configuration*

The cryogenic mechanism in NI-FWA and NI-GWA is a rotating actuator designed to operate from 300K down to 20K under vacuum. It is based on elements mounted in a stainless steel frame: a stepper space qualified motor, 360steps/rev; duplex angular ball bearings; a clutch system with Hirth gears (360 teeth/rev); a monostable electromagnet with bellows, which allows the clutch to be closed without power consumption (see Figure 4.13).

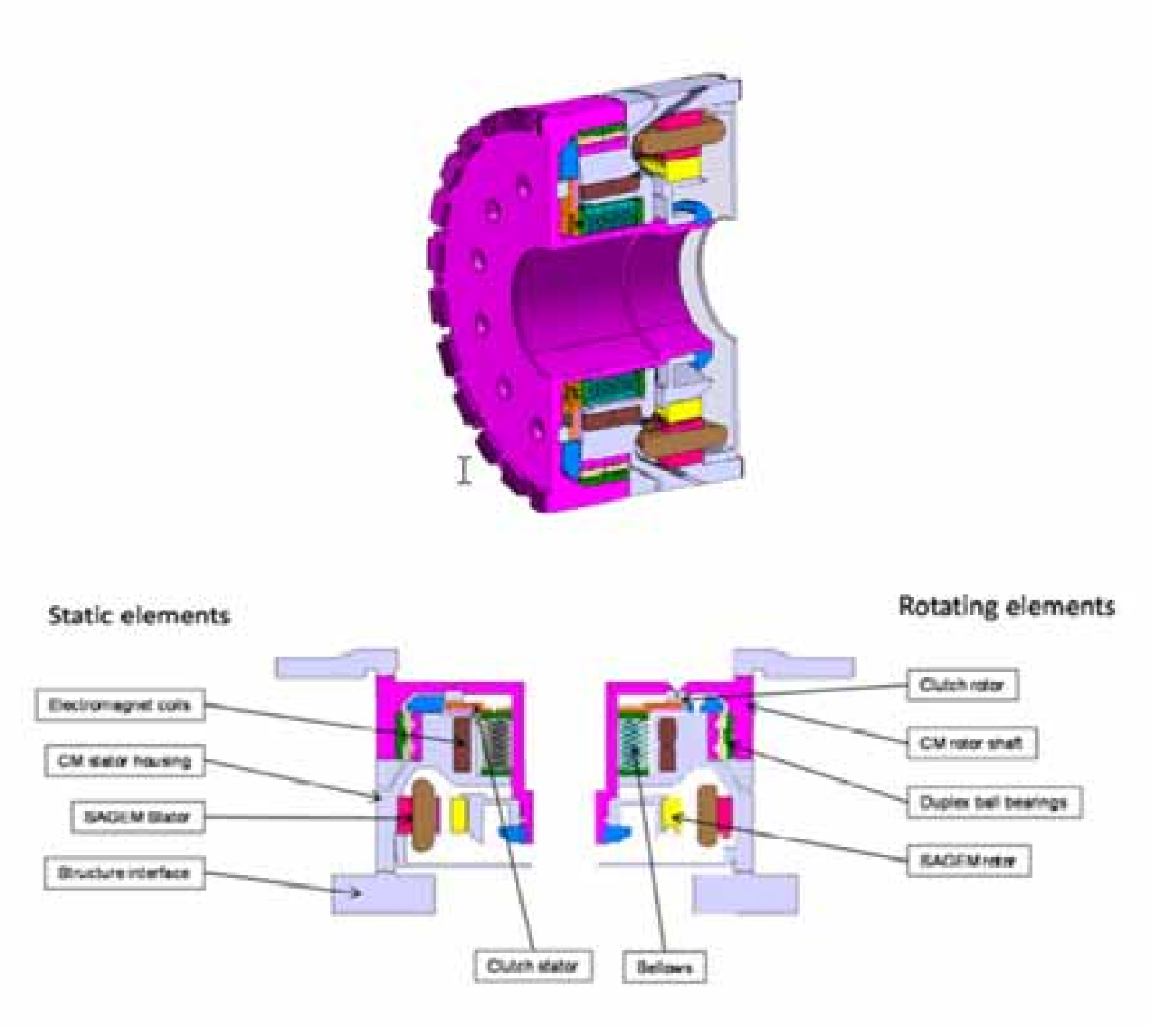

*Figure 4.13: Schematic cut view of cryo-mechanism*

The NISP *Calibration Unit* (NI-CU) is mounted on the NI-OMA and enables flat fields for the calibration of the pixel-to-pixel response of NISP detectors. It applies independently to the Y, J and H bands. The NISP detector is illuminated directly by the far-field of three optical fiber bundles (for Y, J and H band) mounted near the rim of the last lens (part of CaLA). The applied optical fibers are radiation resistant step-index fibers.

### 4.3.2 Detector system

At the present study level, the NISP detector system NI-DS includes the following electronic subsystems dedicated to the detectors control and data-handling:

- 16 H2RG detectors
- 16 H2RG controllers (SIDECAR ASIC)
- cryogenic flexi cables interconnecting each H2RG/SIDECAR pair
- a data, telemetry and command rooter based on SpaceWire link technology
- a set of wires interconnecting each SIDECAR to the SWL
- a temperature control and sensing system

The detector concept is the one developed by Teledyne, with HAWAII 2RG detectors having an interface mounting plate made of SiC (Figure 4.14). The baseline design uses 2 µm cut-off detectors. However, the NI-DS mechanical and thermal concept of this study can be easily adapted to cope with constraints imposed by 2.5µm detectors.

*Table 4.5: Summary of the thermal background for different configurations; the telescopes temperatures refer to those adopted by TAS (240 K) and Astrium (150 K).*

| **Flux in e/S** | | **Y** | **J** | **H** | **Blue** | **Red** |
|---|---|---|---|---|---|---|
| **2.05 micron Cut-Off** | 240 K telescope | $4.7\ 10^{-5}$ | $4.7\ 10^{-5}$ | 0.066 | $4.5\ 10^{-5}$ | 0.08 |
| | 150 K telescope | $4.7\ 10^{-5}$ | $4.7\ 10^{-5}$ | $4.7\ 10^{-5}$ | $4.7\ 10^{-5}$ | $4.7\ 10^{-5}$ |
| **2.5 micron Cut-Off** | 240 K telescope | 0.01 | 0.01 | 0.076 | 0.01 | 0.09 |
| | 150 K telescope | 0.01 | 0.01 | 0.01 | 0.01 | 0.01 |

The NI-DS Cryogenic Support Structure is a 3D monolithic structure made of SiC which holds the mosaic of detectors (Figure 4.14), while two perpendicular walls fix the SIDECARS spread over 2 floors (2×4 per wall).

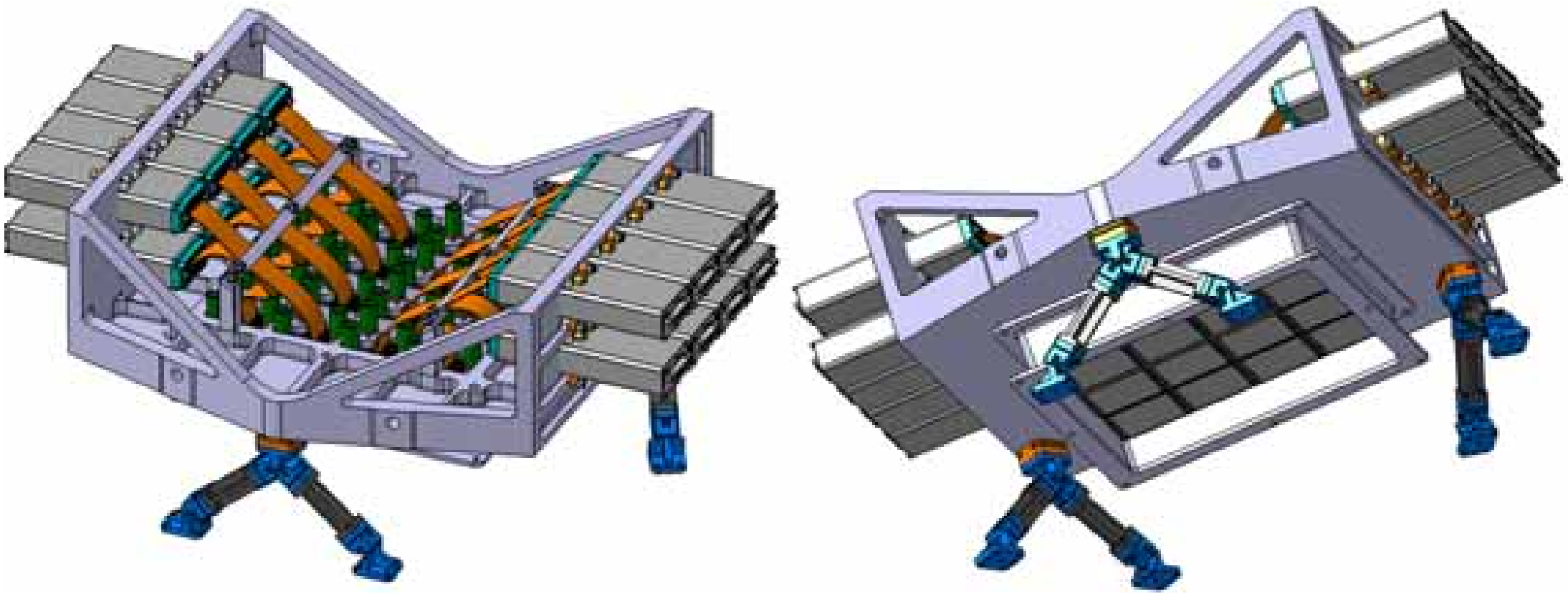

*Figure 4.14: NI-DS layout.*

The detector connection with its SIDECAR is made using a flexi cable (with gold plated EMC shielding). The bipods are made of titanium.

### 4.3.3 Thermal Architecture

The thermal design for the NISP instrument presented in this section is based on a double passive radiator system two radiators that provide the main temperature references for the NI-OMA and NI-DS.

The first radiator will provide a 135 K heat sink for the NI-OMA while working as a parasitic interception stage for the warm harness. A second colder radiator will set a reference temperature at 120K for the NI-DS.

The NI-OMA thermal architecture is based on the instrument cavity reference temperature ≤ 140K with a stability of ±2K over the full mission. The emissivity inside the instrument cavity is required to be 0.9.

Since the NISP thermal design has to be valid in the two thermal cases of at 150K and at 240K, two different requirements for the interface conductance between NI-OMA and optical bench are considered. The conducted leak through the interface hexapod between NI-OMA and telescope optical bench does not exceed 0.3W. The conducted power between NI-FPA and NI-OMA does not exceed 0.5 W. All harnesses from service module to NI-OMA are thermally linked to the 130K stage for maximum parasitic interception.

Detector stability is a key issue for instrumental performance and requires a careful design of the thermal control system. The NI-DS thermal control is achieved by a combination of passive and active systems. The passive component exploits thermal masses and resistances of the components (struts and flanges) to damp temperature oscillations during their propagation from the instability source (e.g. the cold radiator) to the detectors.

The active control is made by a double stage: a PID type controller on the SiC support plate provides a first coarse level of stabilisation, and then a finer control operated by the SIDECAR.

The requirement for peak-to-peak temperature fluctuations is 10 mK over a typical exposure time (500s).

## 4.3.4 Electronics Architecture

The NISP electronic chain includes the following functional elements (see Figure 4.15):

- NI-DS (Detector System)
  - NISP Detector Control Unit (NI-DCU)
  - NISP Data Processing Unit (NI-DPU)
  - NISP Instrument Control Unit (NI-ICU)

The NI-DCU, NI-DPU and NI-ICU are part of the NISP Warm Electronics (NI-WE, Figures 4.16 and 4.17), and are all located in the service module. Detectors and SIDECARs are part of the NI-DS and are located in the payload module.

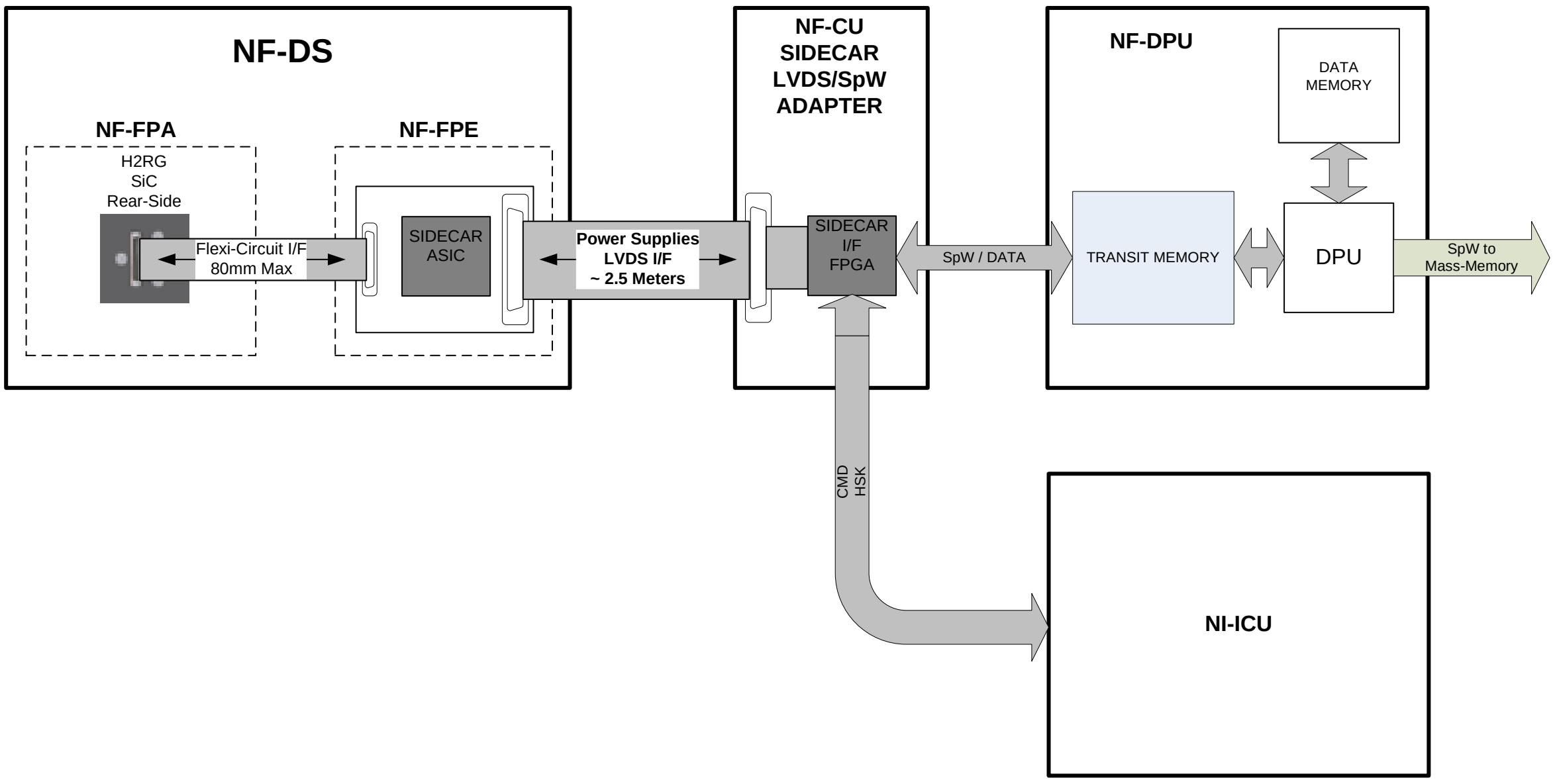

*Figure 4.15: Layout of a single NISP detection channel (One H2RG chain out of sixteen).*

The NI-DCU controls, configures and monitors the SIDECARs and accepts housekeeping from it. It carries out the pixel data pre-processing according to the Up-the Ramp or Fowler detector readout modes, computes the slopes, deglitches and makes linearity corrections. It also manages the detector readout mode for calibration and flat-field measurements and provides reference signals for reducing common mode errors.

The NI-DPU is mainly devoted to science data compression and transmission to the spacecraft Mass Memory. It handles communication with the NI-ICU, manages commands and housekeeping data for NI-DCU and the NI-DPU itself. It also maintains the file configuration for booting the SIDECARS.

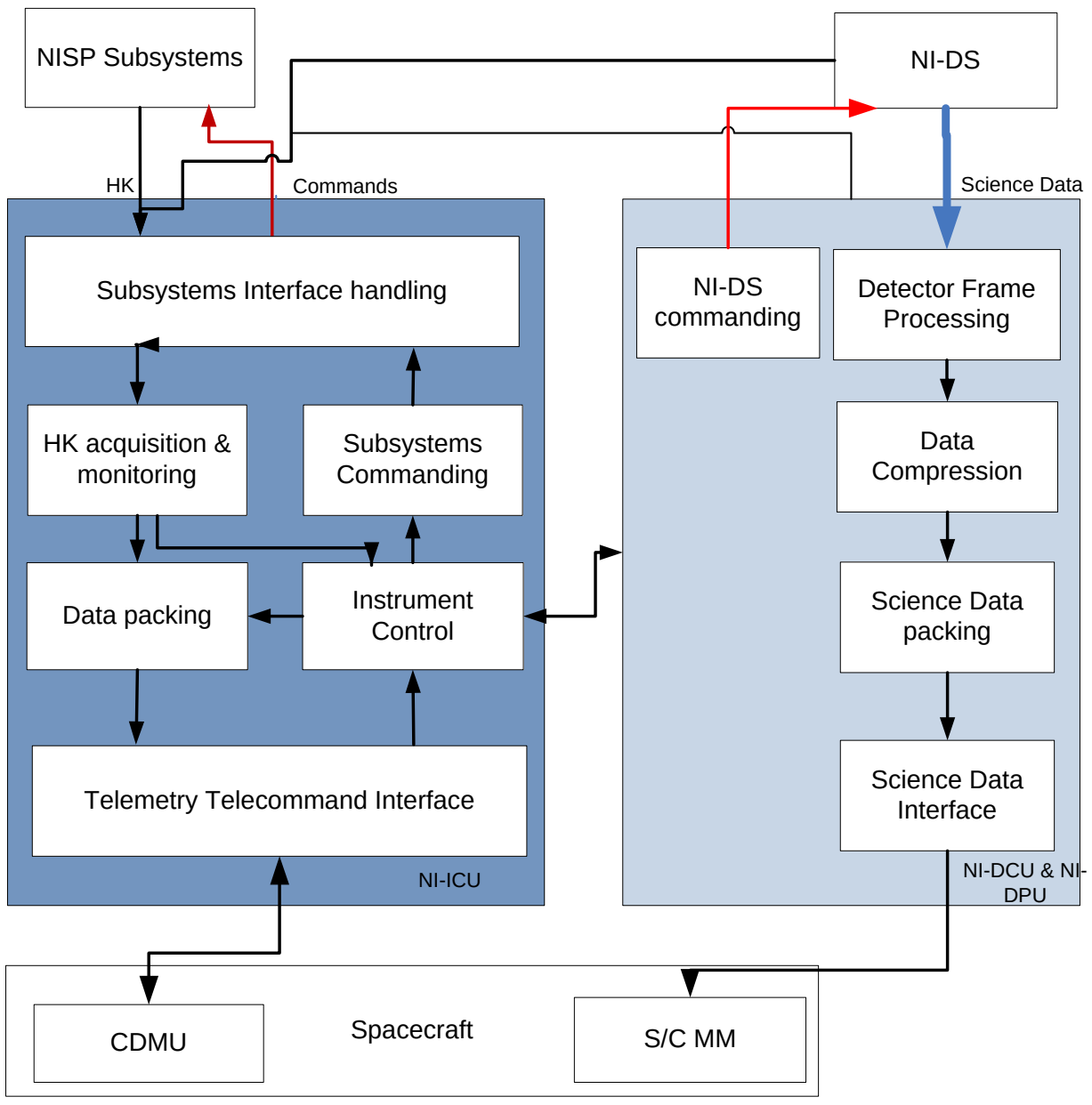

*Figure 4.16: High level functionality of the NI-WE OBS*

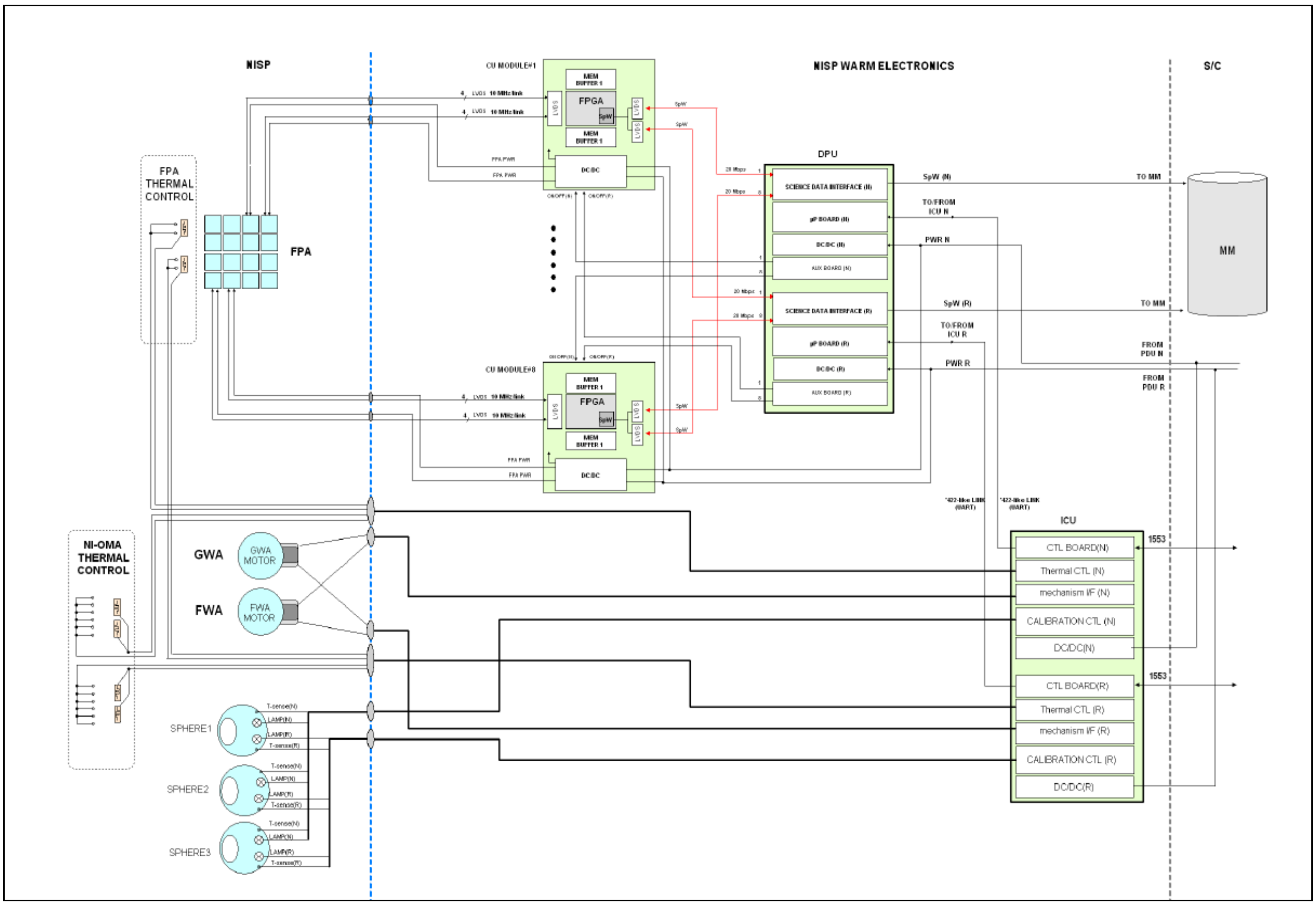

*Figure 4.17: NISP inserted in the on-board processing*

The NI-ICU handles all the NISP functionalities, and interfaces the NISP instrument to the S/C control system. It communicates with the NI-DPU. It provides the control electronics for the NI-FWA, the NI-GWA and the NI-CU. It also is the control electronics for the temperature control of the NI-OMA and NI-DS and maintains the housekeeping data of the NISP warm electronics (NI-WE).

The electronics concept has been driven by the possibility to keep individual H2RG channels as independent as possible, to reduce the possibility of introducing single-point-failure mechanisms.

The NISP software consists of the NI-DPU software, mainly dedicated to instrument data management, and of the NI-ICU software, devoted to the satellite platform interface and to all the functionalities related to instrument commanding and to provide ground visibility. This comprises dialogue with spacecraft and subsystems, and check of health status of the instrument (autonomous functions). On the basis of the housekeeping parameters values NI-ICU can ask the NI-CDMU to switch off part or the whole instrument.

### 4.3.6 Mass and power budget

*Table 4.6: NISP mass budget*

| Unit | Nominal (Kg) | Margin | Total (Kg) |
|---|---|---|---|
| NI-OMA | 38.4 | 20% | 46.0 |
| NI-GWA | 7.4 | 20% | 8.9 |
| NI-FWA | 7.4 | 20% | 8.9 |
| NI-CU | 1.0 | 20% | 1.2 |
| NI-DS | 8.7 | 10%,20% | 10.4 |
| NI-DPU | 8.6 | 20% | 10.3 |
| NI-DCU (8 modules) | 13.9 | 20% | 16.7 |
| NI-ICU | 13.4 | 20% | 16.1 |
| Total | 98.8 | 20% | 118.5 |

*Table 4.7: NISP Power Budget*

| Sub-system | Stand-by (W) | Observations (W) | FWA/GWA latch/unlatch (W) | FWA/GWA wheel actuation (W) | | Flat-field Calibration (W) |
|---|---|---|---|---|---|---|
| | | | (50 ms TBC) | Min | Max | |
| NI-DCU | 52 | 52 | 52 | 52 | 52 | 52 |
| NI-DPU | 20 | 20 | 20 | 20 | 20 | 20 |
| NI-ICU | 28.5 | 28.5 | 57.1 | 30.8 | 37.5 | 30.5 |
| Total Primary | 100.5 | 100.5 | 129.1 | 102.8 | 109.5 | 102.5 |

### 4.3.7 NISP Critical Items

***Instrument alignment***: The Euclid telescope is designed to provide a focal plane for the VIS instrument. The NISP instrument, therefore, is designed and integrated to cope with this requirement. The NISP alignment, both stand-alone and at satellite level, has to be carefully studied and controlled.

***Calibration***: the high accuracy required by the Euclid scientific objectives implies an accurate instrument calibration both on ground and during the mission operation. Periodic performance verification will be implemented to understand the NISP instrument evolution with time.

***Mechanism-induced perturbation***: the wheel rotation and the clutch release induce a significant perturbation on the telescope pointing. In principle, this can be compensated by the satellite, at the expense of reduced survey efficiency and/or an increased attitude system activation rate. A disturbance compensating mechanism may possibly have to be introduced in the payload design. This mechanism would require accurate mechanical modeling and a specific test activity to verify its performance.

***NI-DS***: The development of the Focal Plane Array is strictly related to the detector producer Teledyne activities. The trade-off between 2.0 and 2.5 μm cut-off detectors and their complete qualification is to be performed in the early implementation phase. An important activity will be detector characterisation, to ensure instrument performance and optimum control of systematic effects.

***Grism manufacturing and cryogenic aspherical lens systems***: These optical components are central to the NISP performance; their manufacture and the maintenance of their alignment require careful procedures and design.

***NI-WE***: Instrument performance is strongly related to detector sampling algorithms and Cosmic Ray identification and non-linearity correction. The NISP electronics needs to be able to sample detectors at high frequency and requires high computing power to perform on-board data reduction and compression. Detailed architecture and implementation, within the mission requirements and budget, implies a dedicated design.

***Contamination control***: The presence of CFRP in the payload cavity may imply water release throughout the mission lifetime, impacting on NISP overall performance. Dedicated procedures will be applied during instrument development and test phases to control this effect.

## 4.4 Field of View gap-filling evaluation

While the detectors in the VIS and NISP arrays are close-butted, there are still spaces between their active surfaces. Also, because the two detector types have different effective fields of view, and the array formats are different, these gaps do not coincide. This is the main reason for the four exposure frames to be displaced by the dithers (the other reasons are to allow for cosmic ray subtraction, to counter radiation damage effects and to allow sub-pixel spatial resolution images to be reconstituted). The pointing displacements during the dithers have been optimised to the following:

- Dither 1: ΔX: 100", ΔY: 50"
- Dither 2: ΔX: 100", ΔY: 0"
- Dither 3: ΔX: 100", ΔY: 0"

With these dithers, in the case of no overlap between fields, the pattern of coverage obtained is shown in Figure 4.18 Deep red colours indicate the fraction of sky within a field covered with 4 frames, orange with three frames, yellow-green with two frames and blue with only one.

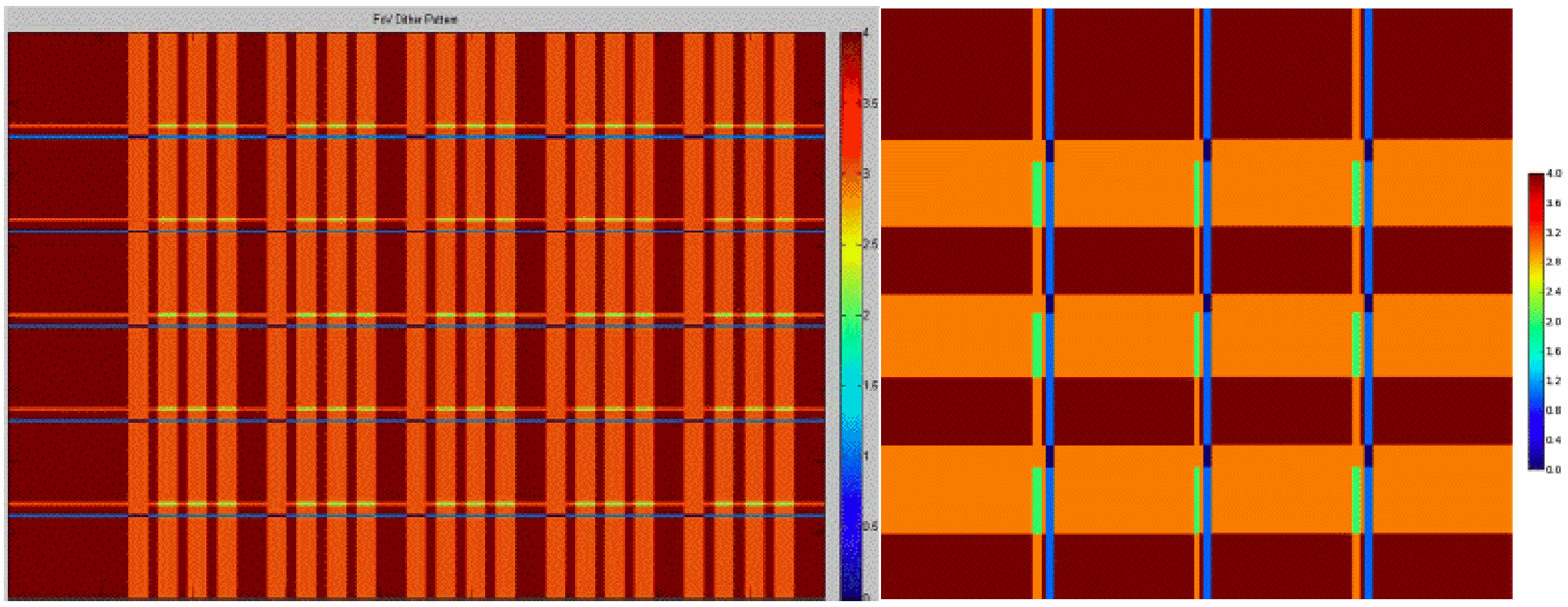

*Figure 4.18: Field coverage provided by the dither pattern for VIS (left) and NISP (right)*

The cumulative fraction of pixels covered by three and four frames is shown in Table 4.8. This shows that half of the sky is covered by 4 frames in both instruments, and that the fraction of the field covered in two or fewer frames is <7% (NISP) and <4% (VIS).

*Table 4.8: The cumulative fraction of pixels covered by three or four frames.*

| Instrument | % of pixels with >3 frames | % of pixels with 4 frames |
|---|---|---|
| VIS | 96.5% | 49.5% |
| NISP | 93.4 % | 50.9% |

# 5 Mission Design

The system design concept of Euclid is driven by the required properties of the sky survey and by programmatic constraints. Major issues of the sky survey are its speed, depth, precision, and imaging quality. The main programmatic constraints are technology readiness, schedule, and cost ceiling.

The survey speed is guaranteed by the combination of a large field of view, about 0.54 $deg^2$, and an optimised survey strategy. Ensuring the high image quality leads to demanding requirements on the pointing and thermo-elastic stability. The survey depth leads to a minimum telescope aperture, dedicated baffling design, low temperature optics and detectors, a cold telescope for low near-infrared background, and on-board data processing for the noise reduction of the near-infrared detectors.

In this section the mission analysis is presented with a description of a possible implementation of the Euclid surveys based on the constraints imposed by the design drivers. The most important spacecraft subsystems are discussed in the second part of this section.

## 5.1 Mission analysis

A large amplitude orbit around the Sun-Earth Lagrange point 2 (SEL2) has been selected because it imposes minimum constraints on the observations and allows scanning of the sky outside the galactic latitude *b* ±30 degree band around the Milky Way within the mission duration.

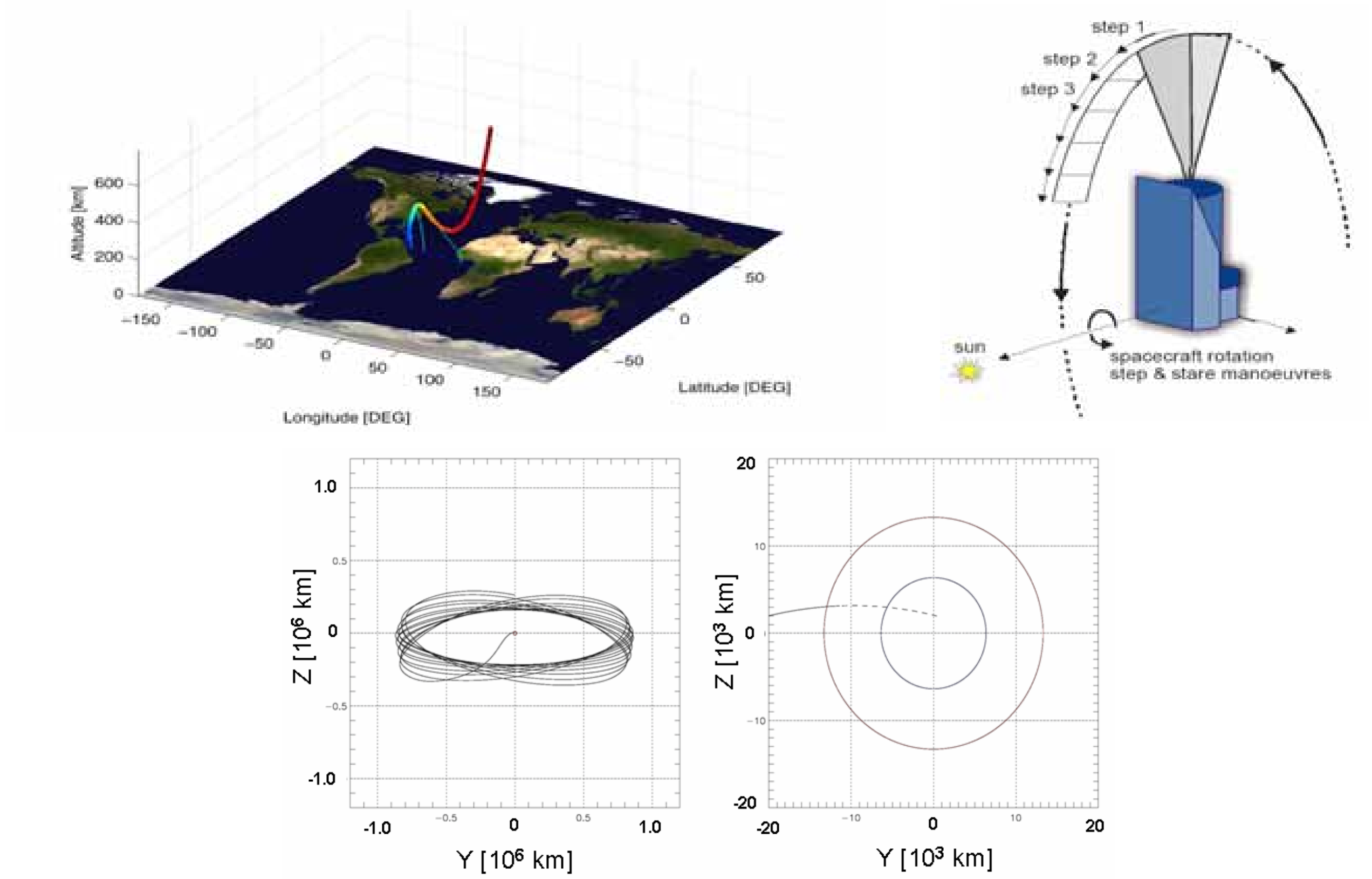

*Figure 5.1: upper left panel: Soyuz 2.1B + Fregat ascent trajectory for a direct L2 transfer. Bottom panels: reference orbit projected in the y-x plane of the co-rotating frame following a launch on 2 Nov 2017 at time 04:34:48 (upper plots) in large (left) and small (right) scales. The outer circle represents the size of the Earth's penumbra at L2 and the inner circle is the earth's shape. Dashed is the day side part of the trajectory. Seen from the Earth, the orbit describes a Lissajous figure around the anti-Sun direction on the sky with maximum distance from the anti-Sun direction of ~ 33 deg. Upper right panel: the step-and-stare scanning of the sky, the line of sight is kept perpendicular to the Sun direction (illustration kindly provided by Astrium).*

The Euclid spacecraft will be launched from the Guiana Space Centre, Kourou, on board a Soyuz ST 2.1-B carrying an increased tank for the Fregat. Euclid will be directly inserted into the specific transfer targeted to a large amplitude orbit at SEL2 (Figure 5.1). After a number of correction manoeuvres for launcher dispersion and fine-targeting during transfer, the spacecraft will be freely inserted into the final SEL2 orbit, i.e. no specific insertion manoeuvres are required. The transfer lasts about 30 days. The maximum Sun-Spacecraft-Earth (SSE) angle variation of the operational orbit is constrained to less or equal than 33 deg (Figure 5.1). The launch date and the launch conditions determine the ellipticity and size of the operational orbit, and influence the SSE angle plus the daily visibility from the ground station. The launch is possible at almost any day of the year with minor restrictions to avoid eclipses during transfer and in the operational orbit. The in- and out-of plane orbital periods are both close to 180 days. The frequency of station-keeping manoeuvres is ~30 days to correct for the instability inherent in the motion about L2.

The payload module (PLM) cools down passively to its operational temperature, shortly after separation from Fregat some 1500 s after launch. A heating and cool-down scheme for PLM and instruments is required to control the out-gassing of volatiles without contaminating the focal planes. After out-gassing and decontamination, spacecraft and instrument commissioning can start. This takes place already before the spacecraft reaches it operational orbit in SEL2. The commissioning and scientific performance verification phases are completed within three to six months after launch.

## 5.2 Mission operations concept

### 5.2.1 Survey Field Observation

The elementary observation sequence of a field is composed of four frames of the 0.54 $deg^2$ common area, observed with a dither step in-between. During each frame the visual instrument VIS and the near-infrared spectroscopy and imaging photometry instrument NISP carry out exposures of the sky simultaneously. At the end of the last frame, a slew towards the next field is performed.

For each frame the nominal integration time in the VIS and NISP are (see Figure 5.2): VIS integration time = 590s, NISP Spectroscopy integration time = 590s, followed by NISP photometric measurements with integration time: Y = 88s, J = 90s, H = 54s. In the first 610s both VIS and NISP take data for WL and spectra. Then, because of image disturbing vibration from filter wheel rotation, VIS has its shutter closed during the remaining exposures taken for NIR imaging.

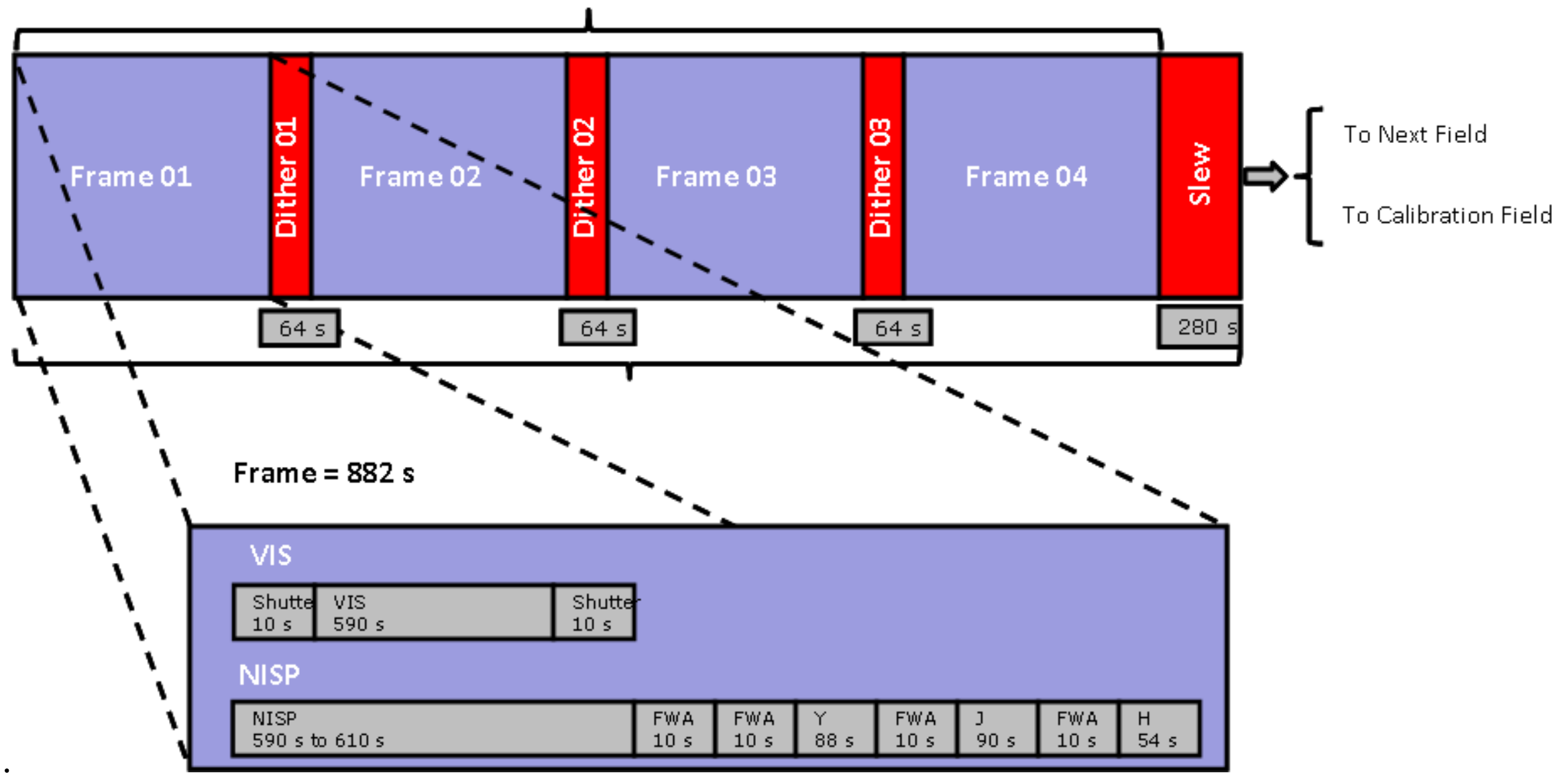

*Figure 5.2: Schematic timeline of an observation sequence of one field. Each frame (blue) starts with a simultaneous VIS and NIS exposure, followed by three NISP photometric exposures. The three dither-to-dither slews and the field-to-field slew are marked in red. The time allocations for the VIS shutter and the filter wheel movements of the filter wheel assembly (FWA) are indicated.*

The dithering strategy (see also Section 4.4) covers the gaps between detectors. It also mitigates the impact of cosmetics defects and cosmic rays on science data, and improves the sampling of the images. In the case of NISP spectroscopy, the four dithers are used to obtain slitless spectra in two adjacent wavelength intervals with two grism rotations to minimise source confusion due to overlapping spectra.

The time necessary to carry out a dither or a field-to-field slew depends on the attitude and orbit control system design, which is different for the different industrial contractors. Based on initial AOCS performance estimates, the following times have been used to construct a realistic reference survey: 64 sec for a dither, and 280 s for a field-to-field slew. These values are subject to change depending on the industrial evaluations.

### 5.2.2 Reference survey strategy

The main survey is the Wide Survey (WS), which covers ~15,000 $deg^2$ of extragalactic sky (|b|>30 deg) at a depth to observe ~30 galaxies per $arcmin^2$ useful for weak lensing, corresponding to a detection limit of 10 sigma extended source at $m_{AB}$=24.5, to reach 3.5 sigma for emission lines at $3\times10^{-16}$ erg $cm^{-1}s^{-1}$, and have NIR imaging in Y, J and H at $m_{AB}$= 24 (5 sigma point source). This is achieved by the exposure times considered in the previous section.

The survey strategy is determined by the following elements:

***Stability***: the image quality depends on the thermo-elastic deformations of the payload. These deformations are induced by variations in the illumination of the spacecraft by the Sun. We define the solar aspect angle (SAA) as the angle between the satellite boresight and the satellite-Sun direction. By design, the SAA is allowed to vary between 90 and 120 deg. SAA variations can cause significant variations in distance between the primary and secondary mirror thereby degrading the image quality. SAA variations may therefore require dedicated image stabilisation and characterisation times. The survey strategy aims at minimising SAA variations over the entire survey in order to secure the quality of the input raw data (before any corrections)

***Sky background***: a strong source of background light is scattering by zodiacal dust particles. This emission has a smooth distribution and is highest in the ecliptic plane and decreases towards the ecliptic poles. The shortest Euclid wavelengths are affected the most by zodiacal light. The survey strategy gives a higher priority to regions at higher ecliptic latitudes. Individual stars with m<17 mag would saturate a few pixels and create ghost images in the FoV; therefore regions with a high density of bright stars are also avoided.

***Galactic Extinction***: dust in the Galaxy causes extinction of the light of galaxies. Galactic extinction has been mapped down to arcminute scales using infrared data from IRAS, WMAP and Planck. These maps will be used to select the best areas on the sky in terms of low extinction.

***Specific pointed calibration observations***: dedicated pointings to high stellar density regions for PSF calibrations and observations of white dwarfs and planetary nebulae as spectral line calibration sources must be performed regularly over the mission lifetime. In addition, areas must be repeatedly observed with similar settings as for the Wide Survey, to search for systematic variations on long timescale and for scientific calibrations of the spectroscopic sample.

***Deep Survey field observations***: The Deep Survey is built by repeatedly observing the same area on the sky in the wide survey observing mode. The Deep Survey covers at least two separate fields in the northern and southern celestial hemisphere, each of 20 $deg^2$. Due to viewing constraints, sky areas can only be visited regularly by Euclid if they are situated at high ecliptic latitude, where visibility is highest (see Figure 5.1). The field orientation will be different at different epochs due to the satellite's annual motion around the Sun. This is an advantage because different orientations are needed to mitigate confusion due to overlapping spectra and to achieve accurate photometric inter-calibrations over large areas.

The reference survey described in the next section is based on a step and stare survey where strips of constant SAA are mapped, as illustrated in the right-upper panel of Figure 5.1. Due to Euclid's motion around the Sun, the great circle with SAA=90 deg moves by one degree per day and eventually the strips add up to large contiguous areas of the sky. High priority areas are the galactic poles and high ecliptic latitude regions. Around the equinoxes (21 March and 21 September), the great circles with SAA=90 deg mostly cover sky with |b|<30 deg2. These periods are used to carry out wide survey observations along arcs of small circles with SAA>90 deg, after taking into account stabilisation times and the different PSF calibrations for these SAA. At given periods, the Wide Survey is interrupted for pointed calibration observations. The fraction of observing time dedicated to these targeted observations, however, is likely to diminish in time due to the increased knowledge of the instruments and the system stability.

### 5.2.3 Euclid Reference Survey

The survey strategy outlined in the previous section has been worked out to a reference survey. This serves to demonstrate that Euclid can perform the required Wide Survey of 15,000 deg$^2$, the necessary calibrations, and the Deep Survey (~5 months) within 6 years.

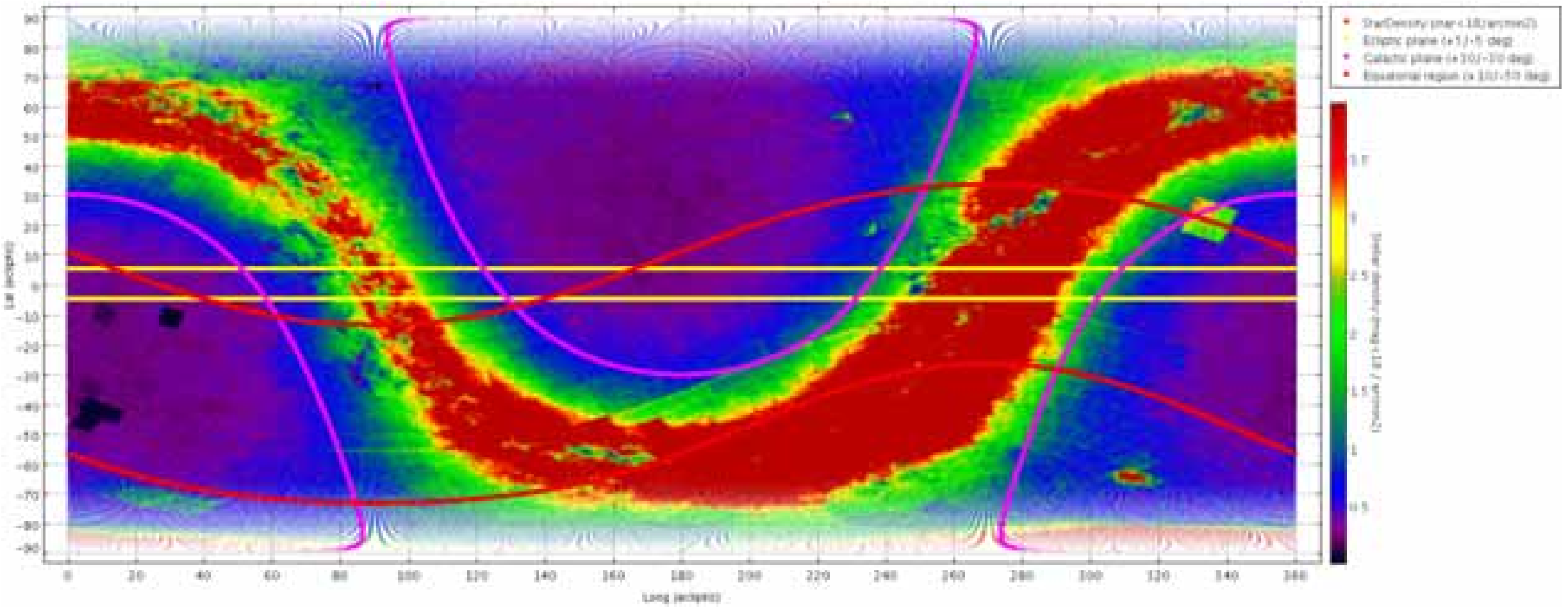

*Figure 5.3: Sky map of stellar densities projected in ecliptic coordinates indicating the regions best suitable for the Euclid sky survey: regions with galactic latitude |b|>30 deg, and regions with ecliptic latitude |beta| > 5 deg. Also indicated are regions with equatorial +10 < delta < -50 deg, presently best suited for ground based access by southern telescopes.*

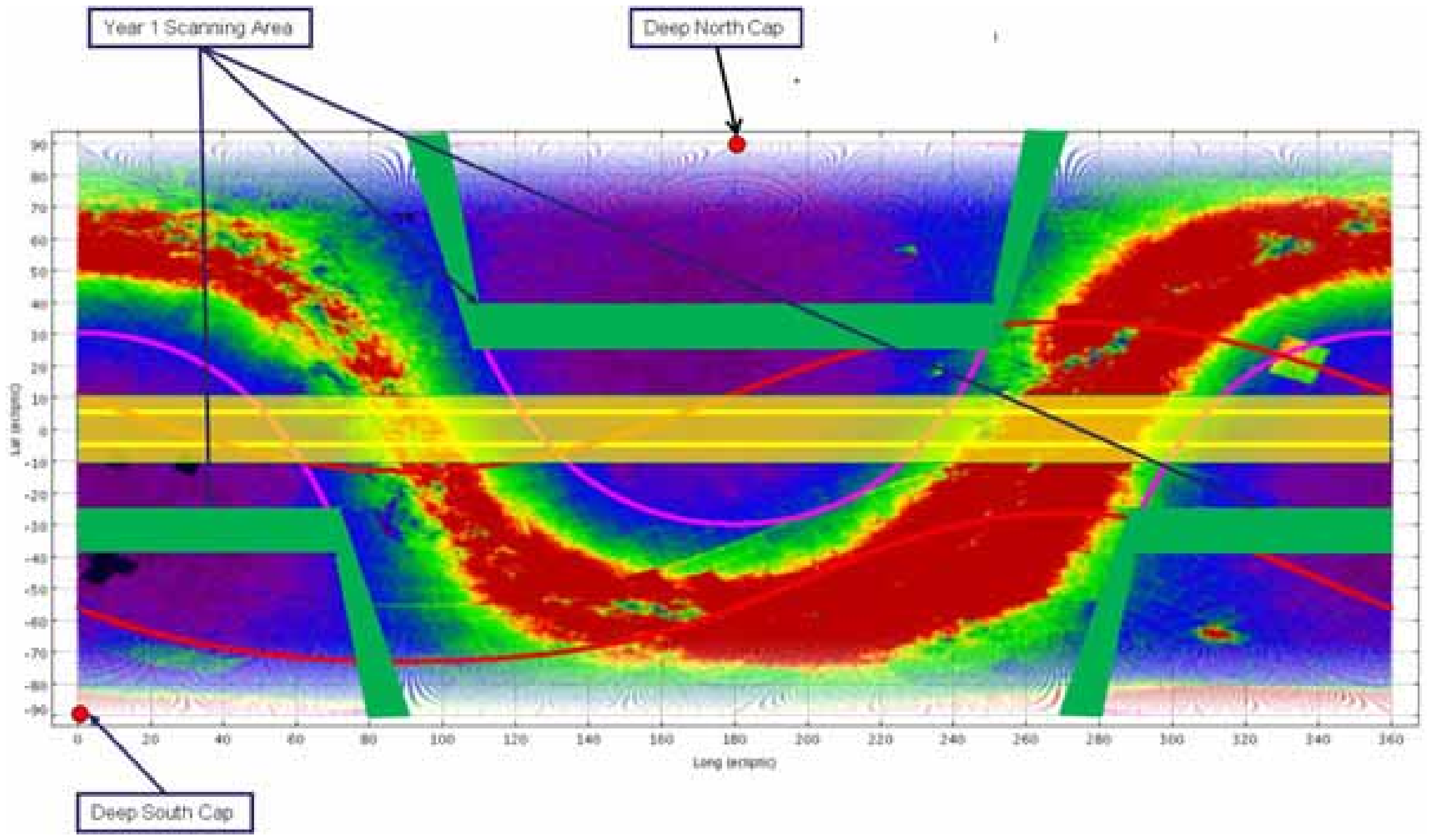

*Figure 5.4: Targeted area (green) for year 1 routine phase.*

**Wide Survey**: The first year is the most demanding in terms of amount and frequency of calibrations, which will be reduced during the rest of the mission. Besides pointings to specific calibration targets (white dwarfs, planetary nebulae, dense stellar fields), time has to be allocated for the Deep Survey. In the first year the Deep Survey concentrates on repeated visits of the Northern deep field. The first year targeted area for the Wide Survey is shown in Figure 5.4.

The efficiency in covering new areas decreases as the mission progresses, because the yet unobserved parts of the sky have lower visibility. At the latest stages of the survey, this introduces idle periods while waiting for specific regions to become visible. The ability to use this time for other purposes depends on the amount of propellant available for slewing. The amount of propellant is sized by design to carry out only the required Wide and Deep Surveys.

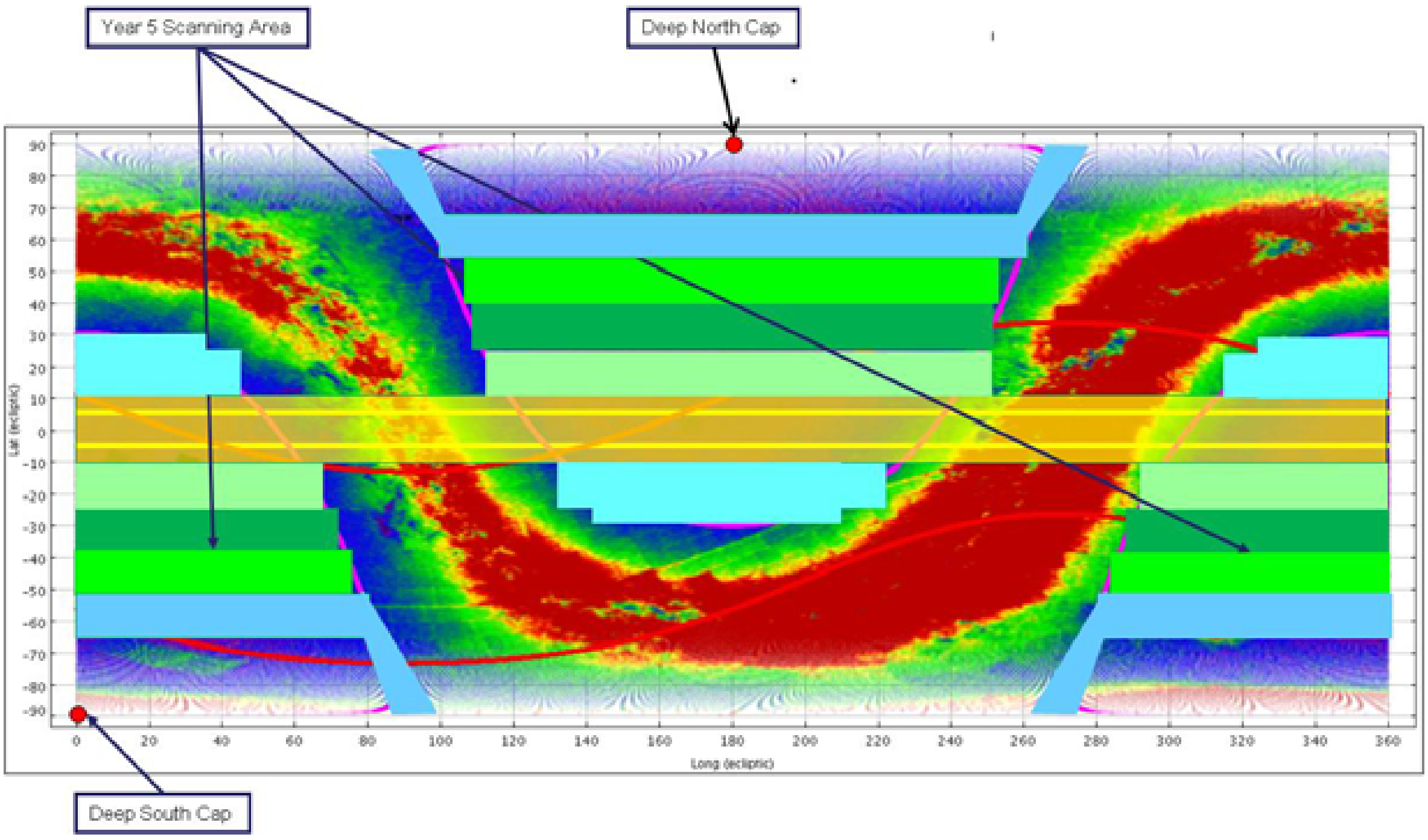

*Figure 5.5: Targeted area for year 5.*

**Deep Survey:** For the Deep Survey, the preferred observational sequence is to point close to the Ecliptic Poles in order to have the maximum visibility throughout the year and to be able to map the area with a wide range of rotation angles as required for slitless spectroscopy calibrations.

Two fields have been selected: one sits on top of the North Ecliptic Pole (NEP), and the second one lies as close as possible to the South Ecliptic Pole (SEP) while avoiding regions of high extinction (Figure 5.6).

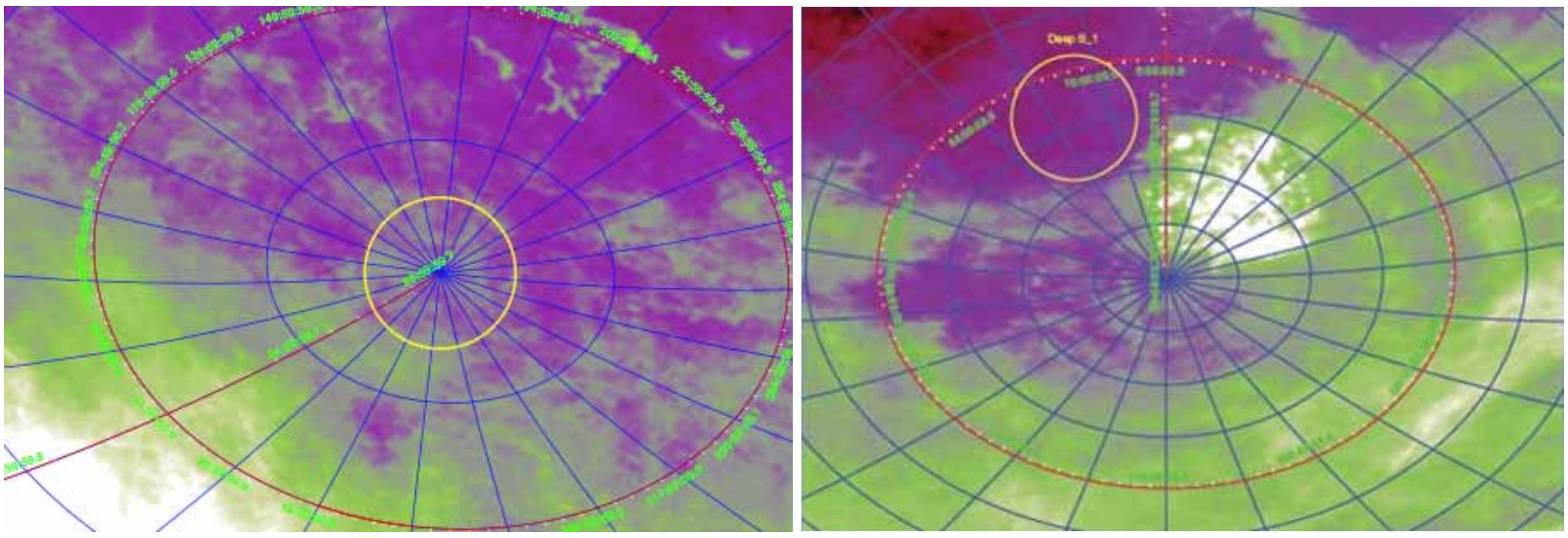

*Figure 5.6: Left panel: Northern Deep Field projected on a sky extinction map. Right panel: Southern Deep Field*

The centres of the Euclid Deep Field (EDF) areas are: Euclid Northern Deep Field (EDFN): ecliptic longitude: N/A, ecliptic latitude: +90.0 deg; Euclid Southern Deep Field (EDFS): ecliptic longitude: 24.6 deg, ecliptic latitude: −82.0 deg.

The precise location for the EDFS is subject to optimisation, which includes trade-offs between extinction versus visibility and SAA stability requirements. Both EDFs will have a number of adjacent FoVs to satisfy the requirement of total area. The EDF centres and an equivalent circular area of radius 2.5 degs are shown in Figure 5.6, superimposed on extinction maps derived from IRAS. To reach the required deep survey depth, a stack of ~40 wide survey field exposures is needed.

One full EDF can be covered in about two days, with the same exposure and dithering strategy as used in the Euclid Wide Survey and with a resulting maximum SAA variation of ±2 degs.

The pointing centres of the FoVs in repeated passes will be collectively shifted by some amounts of random amplitude and direction to satisfy the requirement from NIR imaging. For NIR Imaging to have accurate photometric relative calibrations repeated observations must be obtained of same objects detected at different positions in the focal plane. This can be achieved by staggered exposures with shifts of appropriate size.

Spectroscopy calibration requires observing the EDFs at different times to obtain slitless spectra with different dispersion orientations for the same object to minimise spectral confusion. This is accomplished by observing the EDFN every month over the first year, yielding 12 exposures with rotations spaced on average by 30 degrees in roll. The other 28 passes needed to reach the two magnitudes deeper are collected by visits every other month in the further 5 years of the Euclid Wide Survey. In contrast, the full area of the EDFS is planned to be done at once during the second year of observations, for both risk and scientific considerations.

Examples of possible implementations for the EDF are given in Figure 5.7.

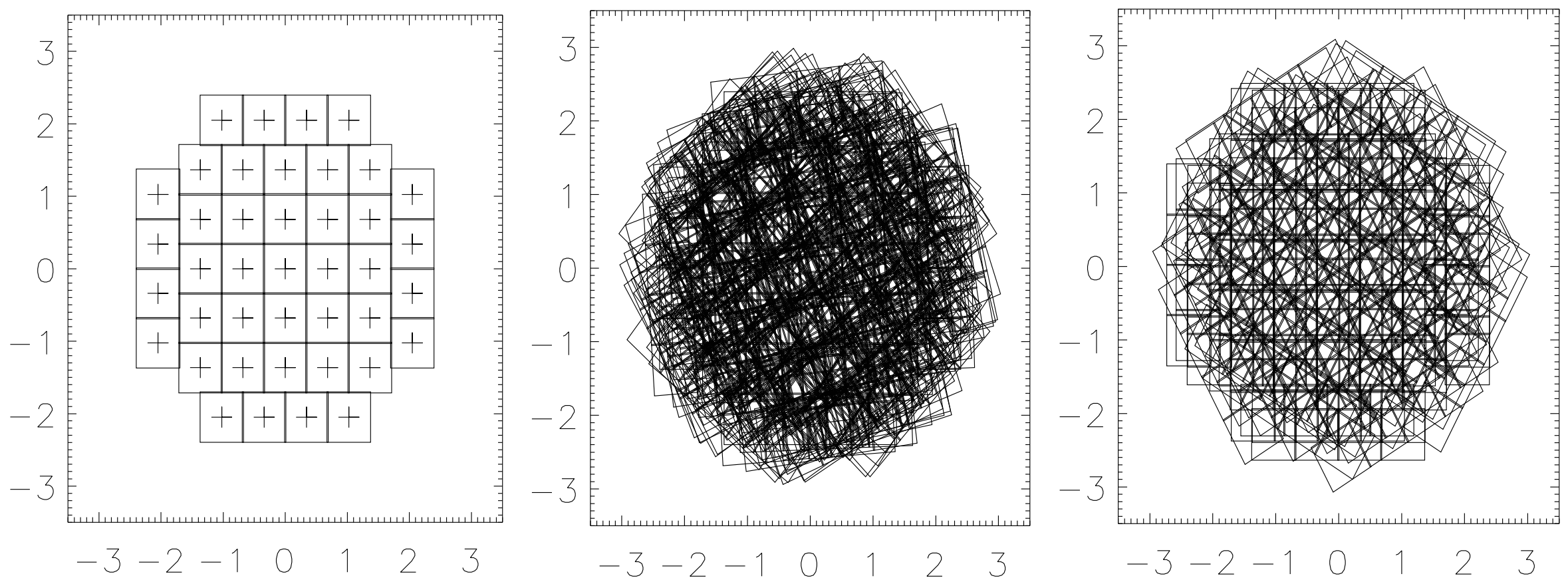

*Figure 5.7: Northern Deep Field example: basic pattern of 41 FoV (left); 28 exposures with random shifts over 2mo time (middle; field rotates of ~60 degs); 12 passes with equispaced rotations and random shifts (right)*

Alternative options could be explored, like different cadence or exposure sequence strategies that would be optimised toward high-z SN discovery, or selection of different EDFS fields, located at lower declination and better suited for follow-up observations with ground based facilities.

The mission time needed to complete the Euclid Wide and Deep Surveys is ~ 6 years. The spacecraft and operations are designed for an additional year as a maturity margin on the survey implementation, to absorb the possible impact of increased overheads and to accommodate a < 6-months commissioning & performance verification phase. This leaves open the possibility of increasing the Wide Survey area or carrying out additional surveys (see Section 2.4.7).

# 5.4 Spacecraft

## 5.4.1 Attitude and orbit control system

The weak lensing experiment places stringent requirements on the pointing stability to ensure optimum width and ellipticity stability of the point spread function (PSF). This demands a relative pointing error (RPE) of ~25 mas (1 sigma) in spacecraft $x$ and $y$ direction, and a fraction of arcsec (1 sigma) in $z$ (roll) over the image accumulation time of ~700 s. The dominating error source on the RPE is actuator thrust noise and the dominating control effort is the attitude knowledge noise. The absolute pointing error (APE) must also be controlled to minimise overlap between adjacent fields.

The pointing acquisition of each field can be performed with a standard high accuracy star tracker. To meet the RPE requirements the AOCS include a fine guidance sensor (FGS) for accurate attitude measurement. The FGS uses the same CCD type as the VIS instrument and share the same optical field of view. The FGS CCDs are placed on both sides of the VIS detector array.

### 5.4.2 Propulsion and actuators

Chemical (monopropellant hydrazine) propulsion is used for the transfer corrections, monthly station keeping manoeuvres, and large (180 deg) slew manoeuvres.

Conventional reaction wheels are discounted as the actuator to control RPE, because their noise budget is too high. Instead, cold gas micro-propulsion is selected enabling fine attitude control. Astrium also uses cold gas for the field steps 0.5 deg slews and dither step manoeuvres (~100 arcsec), whereas TAS uses reaction wheels for that purpose.

### 5.4.3 Communications

Low gain antennas in the X-band are used to support the telecommanding of the satellite and the transfer of instrument and spacecraft real time housekeeping data.

The nominal downlink science data is performed by K-band (26 GHz) together with the stored housekeeping data at a maximum data volume of 850 Gbit/day. A steerable K band high gain antenna is used to downlink the science data. To avoid antenna pointing mechanism induced perturbations, antenna re-pointing is allowed in between steps or scans. Due to the strong dependency of the K band link margin on ground station elevation, two ground stations are envisaged, one on the Northern and one on the Southern hemisphere with seasonal operations to ensure high elevation.

Since Euclid is likely to be the first ESA science mission to require K band capability at SEL2, it is likely to bear the costs of ground station upgrades.

### 5.4.4 Data handling

The solid state mass memory (SSMM) stores and encodes the compressed instrument data. SpaceWire links are used for reception of the instrument science and housekeeping telemetry from the on-board computers. The SSMM provides re-transmission capabilities for data that have not been received properly on ground. Taking a maximum data volume of 850 Gbit/day and a three-day storage capability, the SSMM capacity has been sized to ~4 Tbits with margins.

### 5.4.5 Thermal

The Astrium design foresees passive cooling of the payload, while that of TAS relies on active thermal control. The two different thermal concepts are driven by the different operating temperature of the primary mirrors. The structural/thermal architecture is designed with the aim to provide a high degree of thermal isolation to the payload module (PLM) as well as high thermal stability. The sunshield provides the main thermal barrier with respect to the solar heat load. The thermal isolation between the back of the sunshield and the top of the service module (SVM) with the PLM is performed with high performance multi-layer insulation (MLI). The telescope is protected by a thermal baffle, which is also used to mitigate straylight.

The mechanical mounting of the PLM to the SVM is performed via a set of bipods that minimise thermal conduction. The SVM is equipped with thermal hardware (i.e. radiators, blankets) to ensure a proper range of temperatures (operating and non-operating) for the different SVM units as well as the PLM warm units located in the SVM. There are several areas in the PLM where radiators can be accommodated in order to dissipate excess heat.

### 5.4.6 Power

Power is generated by a solar array body-mounted on the sunshield panels and stored by a Lithium-Ion battery. Different power regulation methods (MPPT or S3Q) are selected by Astrium and Thales, although both implement a 28V bus regulated voltage.

### 5.4.7 Overall configuration and budgets

The SVM architecture is derived from that of Herschel with a 6 panels and a central cone, in which the propellant tanks are mounted. The size of the SVM is scaled to fit the Soyuz launcher but also driven by the PLM overall dimensions.

## 5.5 Payload interfaces

**Mechanical interfaces:** The instruments VIS and NISP are located in the payload cavity of the payload module. They are attached to the optical bench which also supports on its other side the telescope assembly. The optical bench is connected to the service module by bipods.

**Thermal interfaces:** The payload module needs to be kept at a low temperature (<240 K) in order to minimise the amount of thermal radiation in the NISP instrument. The instruments themselves are kept at a maximum temperature of 150 K by use of black coated MLI and the overall PLM thermal design, while dedicated conductive interfaces with corresponding radiators provide sink temperatures for the structures, motors, and focal plane arrays of the two instruments, referred to as the “cold units”. Warm electronic units are located in the SVM.

**Electrical interfaces:** Main electrical interfaces between the payload module and the service module are:

- 28 Volt power bus for the different cold units;
- dedicated data buses based on SpaceWire interfaces to allow the transfer of the very high data rate of the VIS instrument;
- parallel data interfaces between the NISP focal plane array and the NISP Data Processing Unit ensuring the real time processing

For the instrument warm units, the electrical interfaces are based on:

- 28 Volt power bus;
- MIL std 1553 bus;
- SpaceWire interface to access central mass memory.

**Optical interfaces:** For the VIS instrument, the optical interface is at focal plane level. For the NISP instrument, the optical interface is the exit pupil of the telescope, more precisely the dichroic which is under the responsibility of industry. The dichroic insures separation of the input beam between the VIS and the NISP channel.

## 5.6 AIV and Development Issues

The Euclid schedule is driven by the M2 launch date at the end of 2018, a margin of 6 months between the flight acceptance review (FAR) and the start of the launch campaign and the time margins (three months) in the need dates of the instruments provided by the EMC. Besides these ESA requirements, the project schedule is driven by the payload module (PLM) development and assembly, integration, and verification (AIV). Procurement of the telescope primary mirror, including its polishing and verification, takes about 36 months. The payload module structure requires about 24 months to procure and has to be available in the early phases of the STM (see below) development.

The integration of the space segment comprising PLM and service module (SVM) is under the prime contractor's responsibility. The prime is also responsible for all the tests required to verify the full functionality of the system under environmental conditions, including end-to-end tests of the functional chains. Since the instruments are provided by the EMC, an overall PLM development approach is required, compatible with the overall development requirements.

Technology development activities (TDAs) are in progress on the dichroic mirror, cold gas thrusters, cryogenic optics, CCD detector charge transfer inefficiencies, and the telemetry K band in order to bring these subsystems to TRL 5 or higher by the end of 2011. There is still substantial potential of changes at both mission level and spacecraft level resulting from insights from TDAs.

### 5.6.1 System Level

To ensure a robust schedule, the developments of the payload module, the satellite and the SVM are widely decoupled and executed in parallel. In principle, the qualification and verification testing is done at the lowest level possible, ensuring that only qualified and verified assemblies are used for system level integration. This applies to the instruments as well. Mechanical and environmental tests are also made at satellite level. The system level verifications are performed end-to-end and, to the maximum extent possible, in flight representative conditions. A development approach has been selected which aims at maximising flexibility and minimising risks. It comprises the following development models:

- A structural thermal model (STM) on payload and spacecraft level for early mechanical qualification;
- An avionics model (AVM) for early electrical and functional verification;
- A proto-flight model (PFM) on payload and spacecraft level.

The STM programme is executed in parallel to the development of the PLM platform equipment and payload. In the case of Astrium, the spacecraft STM provides only a full mechanical qualification without thermal, since the optical performances and thermo-elastic stability of their all-SiC telescope with monolithic mirrors can be predicted with sufficient accuracy. The STM programme is carried out to verify the structure stability, strength, and stiffness, interface loads for platform and PLM equipment, mechanical properties and mechanical interfaces. The STM is used to validate the mechanical and thermo-elastic engineering models, mechanical integration, and handling procedures. In addition, it provides launcher compatibility tests and determinations of transfer functions of micro-vibrations. The built standard for the STM is:

- payload module primary structure;
- structural models of the instruments;
- the SVM structure;
- propulsion systems;
- mass dummies for platform equipment;
- sun shield;
- dummy harness bundles

The AVM is used for command, control and electrical interface verification between the platform avionics and instruments, for on board software validation, for AOCS performance verification by closed loop testing, for development and debugging of checkout software, and for initial validation of on-board flight procedures. The AVM is a full scale electrical and functional model of the satellite, consisting of the electronics of nearly all subsystems. It is used for:

- electrical interfaces verification;
- preparation of integration procedures, definition generation and debugging of test modules;
- verification of failure detection, isolation, and recovery (FDIR) and attitude and orbit control system (AOCS) procedures;
- validation of the detection chain with the engineering models of the instruments

This approach anticipates and mitigates the risk inherent to the electrical integration of the main satellite equipment and to the execution of functional tests at system levels. Test sequences and procedures can be tested and debugged first at AVM level before they are used to validate the PFM.

Availability of the AVM during PFM level tests also opens the possibility of performing detailed investigations and diagnostics of failures encountered with the PFM. The electrical ground support equipment (EGSE) can be used as spares for the PFM EGSE.

After completion of the STM AIT programme, the PLM is dismounted. All primary structures are refurbished from STM to the proto-flight model (PFM) standard. The PFM is developed for acceptance test campaign, spacecraft functional verification, flight acceptance and pre-launch certification. The spacecraft PFM is the model that will be launched.

### 5.6.2 Payload Level

At PLM level, testing activities are foreseen on two main models: the structural thermal model (STM) and the proto flight model (PFM).

The PLM STM undergoes mechanical testing for validation at subsystem level. Mechanical tests are repeated at satellite level, after coupling with the SVM STM for interface verification and better characterisation of modal frequencies. The built standard of the PLM STM is:

- PLM primary structure, (optical bench, blank mirrors, telescope support, outer baffle, bipods to the SVM, VIS interface structure);
- Structural models provided by the VIS and NISP instruments;
- Samples of thermal hardware are fitted to the SVM for qualification of the mechanical attachment;
- Selected major instrument harness runs are represented by dummy harness bundles.

An electrical functional model is used to validate the electrical and software interfaces to the spacecraft. However, no engineering model (EM) of the overall PLM is foreseen since most of the electronic units are located in the SVM and are best validated at spacecraft level. Engineering models at subsystem level of the few electrical functions are directly integrated to the spacecraft AVM

The PLM-PFM is subject to a full proto-flight qualification programme covering functional, optical performance, mechanical, electrical, and thermal vacuum testing at qualification levels with acceptance duration.

The payload consortium delivers an STM EM for system AVM and a PFM to the prime. The instruments providers mitigate the risks associated with the system proto-flight philosophy by following a specific model approach, compatible with mission level requirements:

- NISP develops an optical structural thermal model (OSTM) as STM. This model is used for thermal qualification (thermal cycling, thermal balancing) and vibration qualification (vibration of assembly at qualification level). The OSTM provides geometrical and alignment verification at ambient and cold temperatures at the optics interfaces. The NISP EM is a subset of the NISP EQM, which consists of a complete reduced instrument to qualify the design of the instrument.
- VIS develops the STM-O, which is used for thermal qualification (thermal cycling, thermal balancing) and vibration qualification (vibration of assembly at qualification level) after feedback from STM environmental tests at spacecraft level. Specific mechanical and optical tests are also included (thermo-elastic properties measurement). A VIS verification model (VM) is developed to verify critical instrument performances and to release criticality of the PFM. The VIS-VM is used to demonstrate the full function and qualification of the FPA and its subsystems.

# 6. Performance

In the previous sections we described how the scientific objectives of Euclid can be realised in practice, through a specific implementation of a payload, mission concept and data analysis system. In this section we describe the techniques used to verify the performances of the Euclid mission and show the results of the verification that demonstrate that the scientific goals can actually be met.

The performance verifications have been carried out through simulation pipelines specifically developed for each of the two main cosmological probes of Euclid. This approach is particularly well suited for Euclid, given the demanding requirements and overall complexity of the mission. For example, because of the confusion due to overlapping spectra and the background produced by unresolved faint sources, slitless spectroscopy carries an intrinsic risk of failing to measure galaxy redshifts. The reliability and success rate of any redshift recovery method can only be tested through extensive simulations, which have to be as realistic as possible. These specificities require the level of realism and detail of the simulations to be taken to a significantly more advanced stage compared to what is normally achieved at this phase of a mission.

In the following sections we discuss how the simulations were implemented and used to demonstrate that the required level of precision on the cosmological parameters can indeed be reached with the proposed configuration. Particular emphasis is put on verifying that the impact of potential systematic errors on the measured quantities is kept under strict control, making Euclid a most powerful mission, and the best understood next-generation dark energy experiment in terms of precision *and* accuracy.

## 6.1 Instrument simulation and performance

### 6.1.1 Visible imaging

The main role of the visible imaging is to measure galaxy shapes and hence, the shear induced by weak gravitational lensing. This is achieved through the analysis of a large sample of galaxies, from which the (luminous plus dark) matter maps can be reconstructed as a function of distance, to a high level of precision. From these maps, the growth of the density perturbations, and consequently the constraints on the cosmological parameters, can be derived. The small and stable point spread function, and the uniformity afforded by wide field imaging using a stable platform in space, above the atmosphere, is the critical advance that the Euclid visible imager (VIS) provides.

In order to achieve these goals, the visible imaging must:

1. observe a large area of the sky in order to build a sufficiently large sample of galaxies;
2. have sufficient sensitivity to reach faint galaxies at large distances (redshift);
3. provide images of excellent quality, the characteristics of which are exceptionally well understood

These are the top-level requirements for the weak lensing listed in section 3. The first two are classical, driving the field of view of the detectors and the size of the telescope. The third requirement is the most critical one, and consists of two parts: not only must the point spread function (PSF) be acceptable, but its characteristics must also be known very precisely at any point over the field of view and for any observation.

In Section 2 it is shown how, if the PSF is well known and controlled, the shear measurements can be used to derive unbiased cosmological parameters to the precision required by the science objectives of Euclid. The main requirements on the acceptability of the PSF are on its size, its ellipticity, as well as a constraint on the extension of its wings. Then, the *knowledge* of this PSF can be expressed in terms of residuals of these quantities (as determined by a model) from their true values. In orbit, the technique adopted by Euclid relies on the availability of a number of relatively bright stars across the focal plane, which allows to calibrate the PSF at these points, and from which a model of the PSF at the location of each galaxy can be constructed. Additional subsidiary calibrations, for example for characterising the linearity, are also used.

The main contributions to the imperfection of PSF knowledge are:

a. the model of the measured PSF (optics+pointing jitter+detector), together with pixelisation, at any point in the focal plane;

b. the effectiveness of modelling the charge smearing caused by the imperfect transfer of charge to the readout node in the CCD detectors (Charge Transfer Inefficiency or CTI) before it reaches the detection chain; CTI increases with time due to the cumulative damage by Solar protons to the CCD lattice. The CTI also suffers from non-linear effects which depend on the source characteristics (size, intensity).

These imperfections are allocated equal headroom in their effect on performance, and each of them is checked separately against their allocation in the discussion below. At the level of accuracy reached by Euclid, almost every contributor to the performance has to be assessed, so it necessary to account for many other limitations as well, for example imperfect calibration. These however, have been reduced by design, to the point that the allocations made to them are secondary.

For the performance evaluation, both large full field-of-view simulations, and more detailed simulated explorations of the PSF behaviour over the focal plane have been undertaken. These are shown schematically in Figure 6.1.

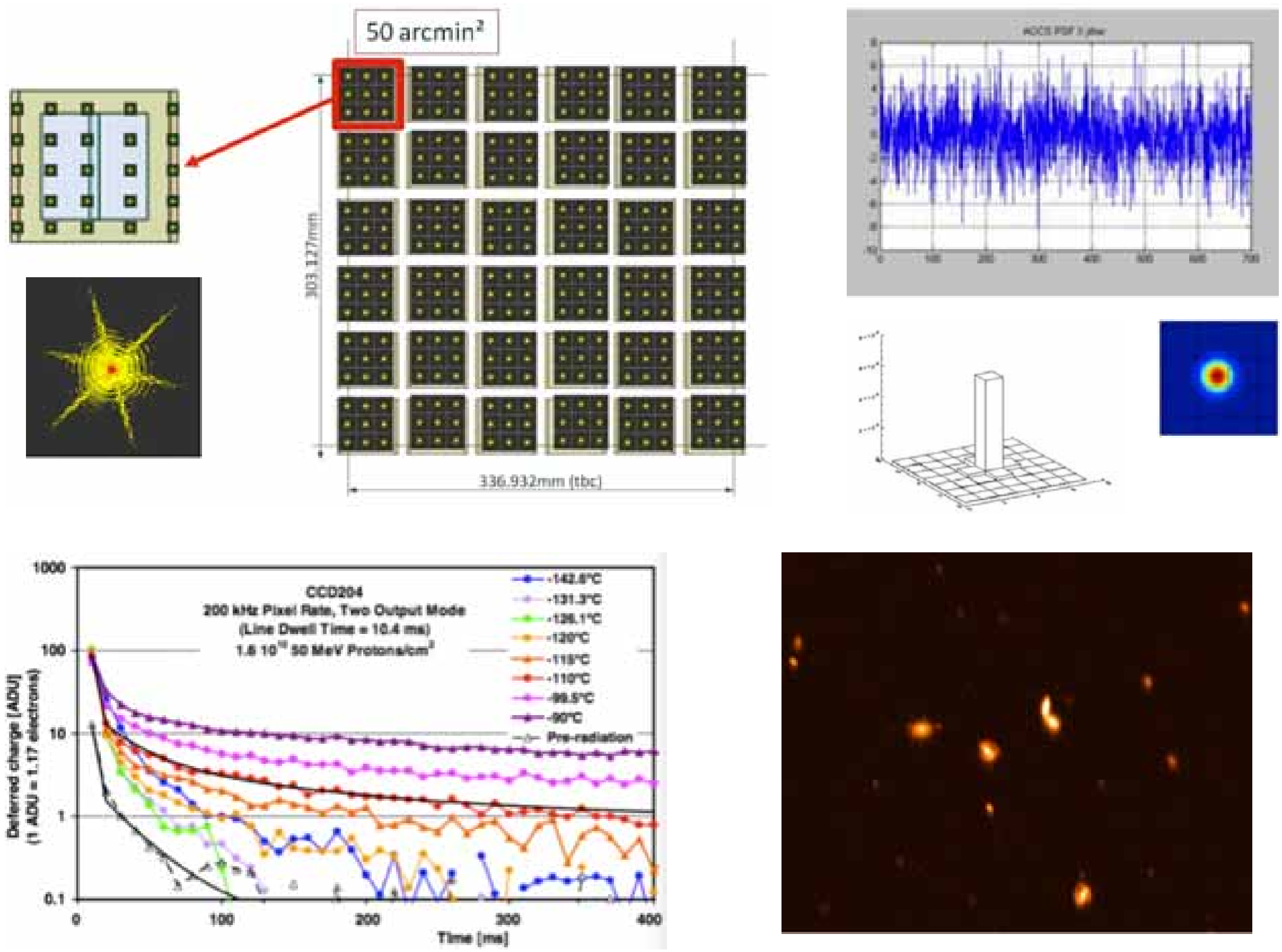

*Figure 6.1: The simulation process for VIS: (top left) optical PSF, (top right) pointing jitter, (middle right) CCD charge diffusion, (bottom left) CCD radiation-impacted CTI measurements, and (bottom right) simulated HST UDF fields.*

The simulation process is as follows (see Fig 6.1):

(a) The Euclid telescope model is perturbed in terms of as-manufactured optical quality and optical alignment to produce up to 60 different instances of the PSF at up to 324 locations in the field-of-view. For one CCD, a finer sampling of 25 field-of-view points for 5 instances is produced, together with additional PSFs to explore wavelength dependence. All of these are sufficient to sample the instrument states and the spatial and spectral variation of the PSF at many points on the field of view in a representative fashion;

(b) industry-supplied pointing jitter time series predicted from attitude control system simulations are used to build up jittered PSFs from the displaced optical PSFs integrated over the exposure;

(c) charge spreading within the CCD from charge diffusion measurements on the Euclid-like CCD204 is then added: the result at this stage is the "system PSFs" for this jitter run;

(d) images are then pixelised interpolating the PSF over the detector area. Internal and external background light, cosmic rays, noise and instrumental cosmetic effects are injected in the model;
(e) the CTI models developed for Euclid from both the *Gaia* (CDM03) and *HST* programmes (Massey, 2010) using real test data with parameters derived from radiation testing of Euclid-like CCD204 devices are then applied;
(f) galaxy images are produced using a galaxy model, or real data (such as the *HST* ultra-deep field), by scaling/rotating each galaxy individually and convolving it with the system PSF derived from steps a-c above before convolution,; weak lensing shears are also added at this stage as required;
(g) full-scale 24k × 24kpix mosaics are produced using a simulated CCD metrology for the full array and 144 detection chains (4 per CCD)

In addition to, and as part of the PSF simulations, the instrumental throughput is also simulated, as well as the survey area using dedicated survey planning tools presented in Section 6.2. The simulation chain for the VIS instrument is therefore quite complete.

Having produced simulated images, the next step is the VIS performance evaluation. Examining first the throughput, VIS is required to reach AB=24.5 for extended sources at 10 σ in 3 exposures lasting 540 sec, the duration set by spectroscopic exposures and the requirement for stable pointing. The calculation includes the telescope and fold mirrors, dichroic beam splitter and detectors (all with their end-of-life performances); it uses a careful definition of extended sources to allow for aperture losses, and different cases of Zodiacal light background. The conclusion is that there is sufficient throughput to reach the required depth. There is additional margin in that the measured readout noise is lower than that used in the calculation, and because half of the sky will be covered with 4 exposures, instead of 3.

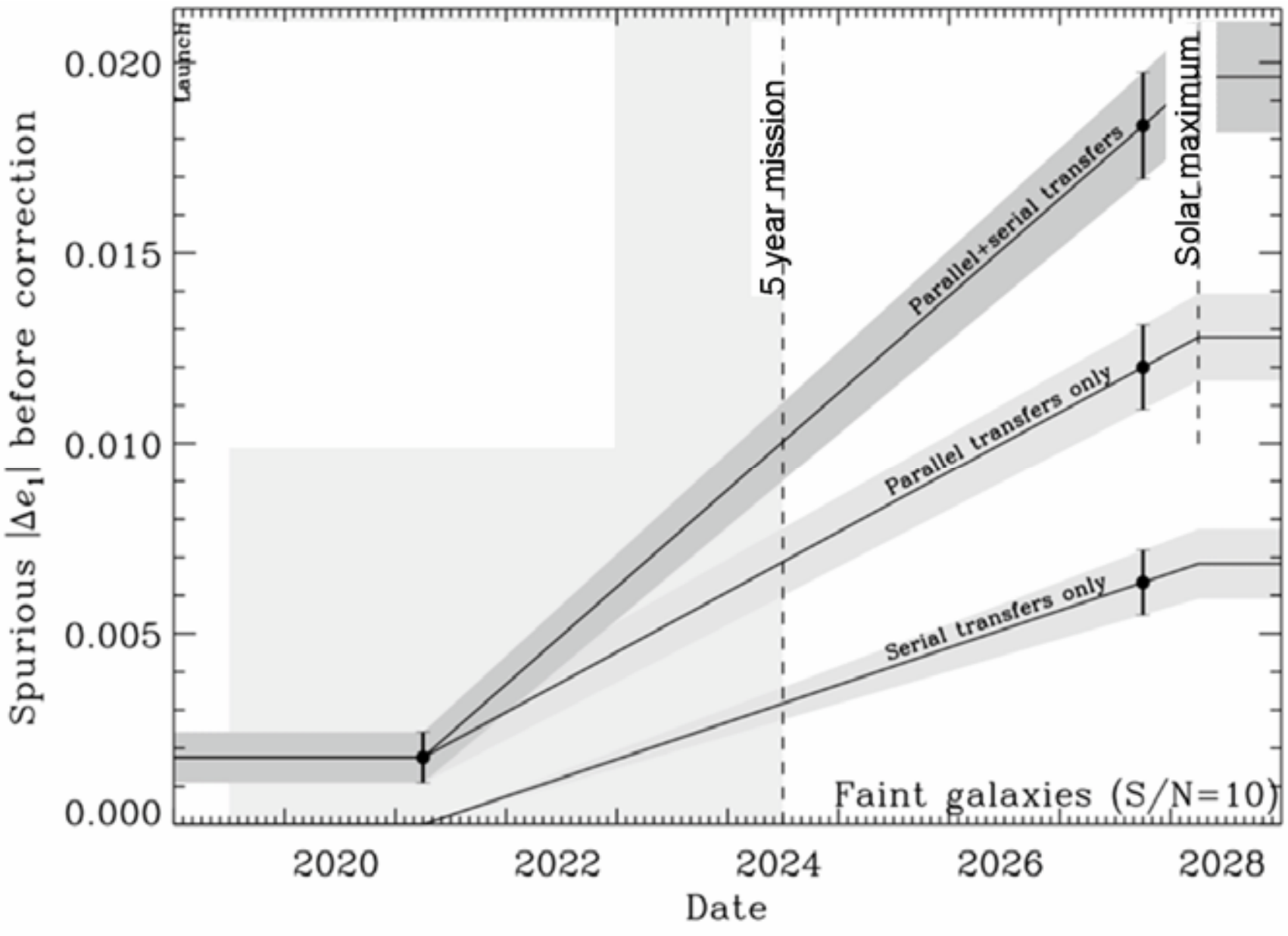

*Figure 6.2: Spurious ellipticity induced by Charge Transfer Inefficiency (CTI) in the shapes of bright stars from which the PSF is modelled. This contributes little to the overall absolute ellipticity of the PSF, and fits easily within the allocated budget. The CTI is enhanced by solar proton radiation damage, which is predicted at 95% confidence level using the standard ESA SPENVIS radiation model. To be conservative, the stars are placed at the maximum distance from the readout node, which maximises CTI.*

The PSF performance can now be evaluated, having in mind that if it can be shown that the PSF has the appropriate size and ellipticity, and more stringently, if the PSF properties are known sufficiently accurately, it will then be possible to determine the cosmological parameters with the required accuracy.

The simulations of the PSF covering the focal plane generated in steps a-d above are used for this part of the assessment. The first check is to assess whether the simulated PSF meets the requirements in absolute terms. As noted in (a) and (b) above, the requirements provide separate allocations in the telescope and focal plane. Using the Euclid optical model, with sensible ranges in manufacturing and alignment errors, the PSF size, ellipticity and wings for the different field positions and instrument states have been explored and found to be compliant. The PSF size and ellipticity at the end of mission were then estimated taking into account the amount of CTI expected from the solar proton radiation dose. This is illustrated in Figure 6.2. As can be seen, the absolute level of spurious ellipticity easily fits within the 20% allocation for this effect.

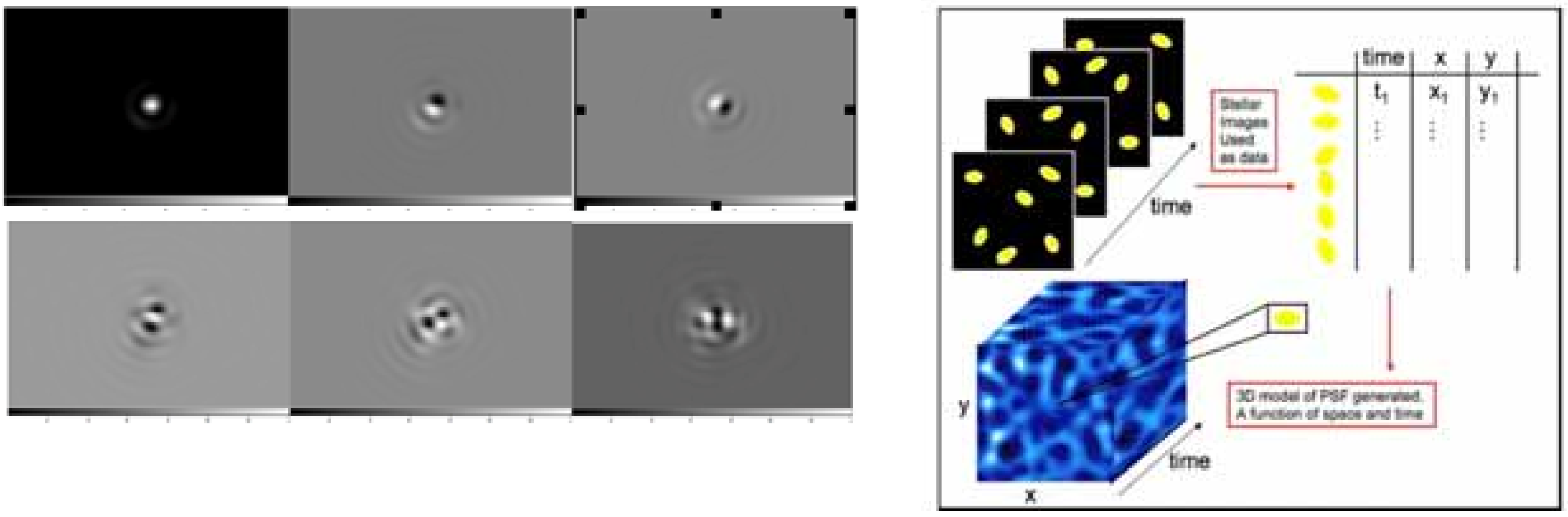

*Figure 6.3: The PCA components, which are to be combined in different quantities to make up a PSF at any point in the field of view (left). Right panel shows the extension to the time (or instrument state) domain.*

Having assessed the characteristics of the PSF, the *knowledge* requirements are addressed next. Starting again with the system PSF, mainly dominated by the optics, the PSF is then decomposed into its main contributor using Principle Components Analysis (PCA, Figure 6.3). By construction, the PCA components can be added in combinations to reproduce a high fidelity model of the PSF anywhere in the field of view. The combinations are functions of position. The concept can be extended into the time domain, so that PCA decomposition can be used to describe the PSF for any state of the instrument (for example primary-secondary mirror separation) at any time and for point in the field of view.

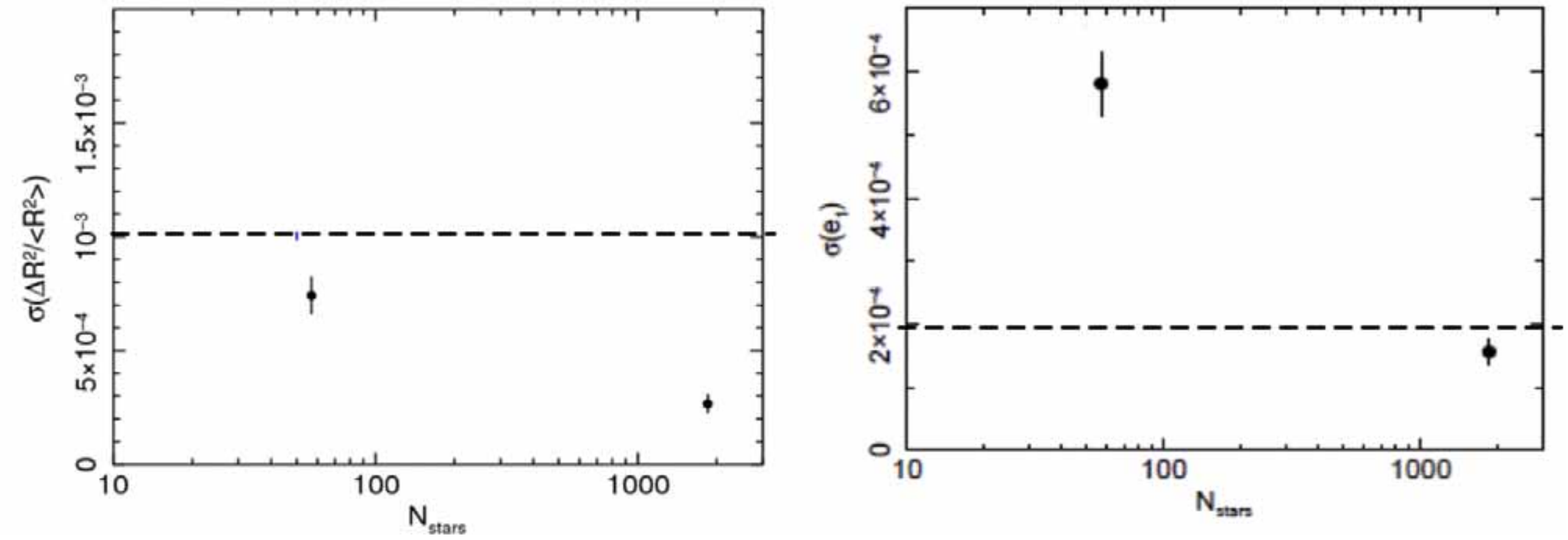

*Figure 6.4: Residuals in the knowledge of the PSF in terms of size (left) and ellipticity (right) as a function of the number of suitable stars used to model the PSF, derived using the PCA modelling. The levels, below which the requirements allocated to this contribution are met, are shown as dotted lines. A typical full focal plane has ~1800 unsaturated stars with signal-to-noise ratio >100. This shows that the PSF knowledge (excluding CTI effects, which are dealt with separately) can be met for each exposure individually, even before sequences of exposures are combined.*

The number of components required and the fidelity with which the PSF can be modelled have been examined using the simulated, pixelised noisy data from our simulations (steps a-d above). The results are shown in Figure 6.4 which demonstrates that, provided that there are ~1800 suitable stars (S/N ratio>100) in every field, the knowledge requirement is met by each single exposure. It is met without even imposing requirements on PSF stability. Margin is available since one can use PCA decomposition in the time domain and use successive exposures to further improve the PSF characterisation.

Finally, analysis of the knowledge of the PSF deformation induced by the detector CTI is performed using a proven technique developed for *HST* (Massey et al 2010, Massey, 2010) and now part of the processing chain. A copy of the CTI-damaged image is passed through a CTI model to inject a level of radiation damage, and linear combinations of this image are subtracted from the original data. By repeating this process a number of times, the PSF deformation can be reduced by a factor ≥ 30, allowing the knowledge requirements to be met

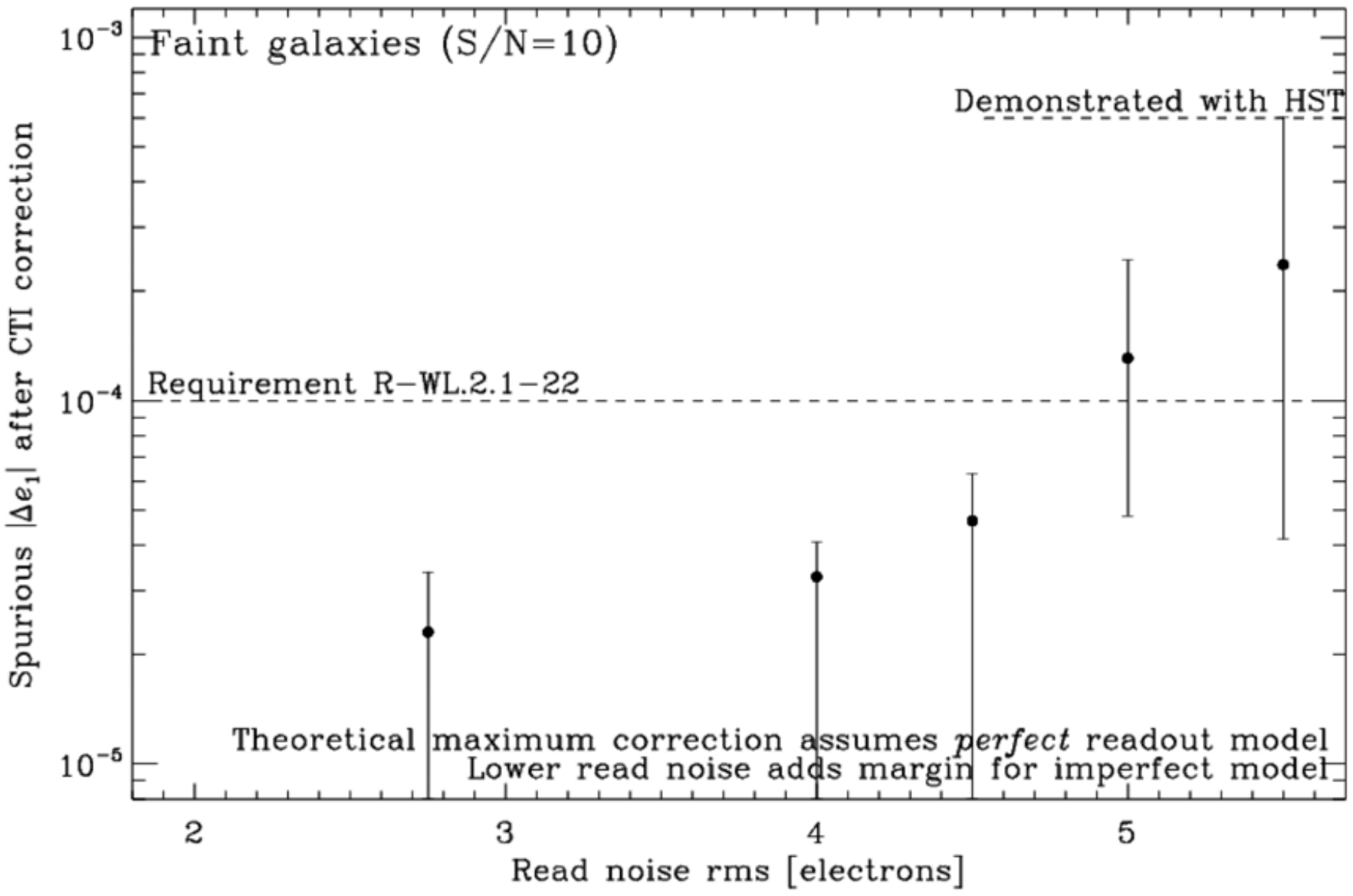

*Figure 6.5: The CTI-induced residual ellipticity in faint galaxies furthest from the readout node after correction during image post-processing as a function of read-out noise. The correction algorithm assumes a model of the CCD readout mechanism and its accuracy is limited by readout noise. The error budget includes a 1% allocation for inaccuracies in the readout measurement parameters. Again an end-of-mission dose is used, calculated with the ESA SPENVIS model.*

The CTI correction technique relies on a low level of readout noise, since this noise arises after charge transfers and therefore limits the accuracy with which the original image can be reconstructed. It also relies on a model of the radiation damage. The effect of readout noise on the residual ellipticity knowledge is illustrated in Figure 6.5. This shows that if the read-out noise can be kept ≤ 4.5 $e^-$ - and the slow readout of the VIS CCD guarantees this - the knowledge requirement can be met. The procedure is "self calibrating" in the sense that residual effects (after the parameters derived from the calibrations are applied) can be adjusted until the effect is nulled out. The above calculations carry margins since the benefit of multiple exposures (where the star images are located at slightly different positions and suffer from slightly different CTI) is not taken into account and their location far from the readout node is a worst case.

### 6.1.2 Near-Infrared photometry

The NISP imaging photometry mode provides photometric measurements in three near-infrared bands (Y, J and H). The photometric measurements supplemented with ground-based multiband measurements are used to estimate the photometric redshifts of the weak lensing galaxies (see section 3.2.3 and 3.2.4). To meet the science requirements, the imaging mode of the NISP instrument is required to have a depth of $Y_{AB}$, $J_{AB}$ and $H_{AB}$ = 24 mag (5σ) with a high image quality defined in terms of radii of encircled energy (EE50, EE80) of (<0.30, <0.62) arcsec, (<0.30, <0.63) arcsec and (<0.33, <0.70) arcsec at the centre of the Y, J and H bands respectively. To reduce crowding, the pixel scale is set to 0.3 ± 0.03 arcsec and therefore 16 NIR detectors are needed to cover the 0.5 $deg^2$ instrument field-of-view. The NISP imaging mode is thus under-sampled but implementation of the NISP instrument meets all the currently verifiable scientific imaging mode requirements. In this section, we summarise the instrument's performance compared to two critical requirements: (i) the imaging quality requirement and (ii) the depth requirement.

As for the visible instrument, the optical performance of the NISP imaging mode has been evaluated by constructing system PSF which combine contributions from:

- the optical system PSF, generated from the telescope and instrument optical design perturbed according to realistic manufacturing and element misalignment errors;
- the satellite's pointing jitter provided by industry as time series from their attitude control system simulations;
- a NISP PSF, which accounts for detector effects, such as cross talk and inte-rpixel capacitance.

The system PSFs generated from the worst performing focal plane positions, defined as those with the largest radii of encircled energy (EE50 and EE80 values), are shown in Figure 6.6. Even at these positions, the imaging quality requirements are met.

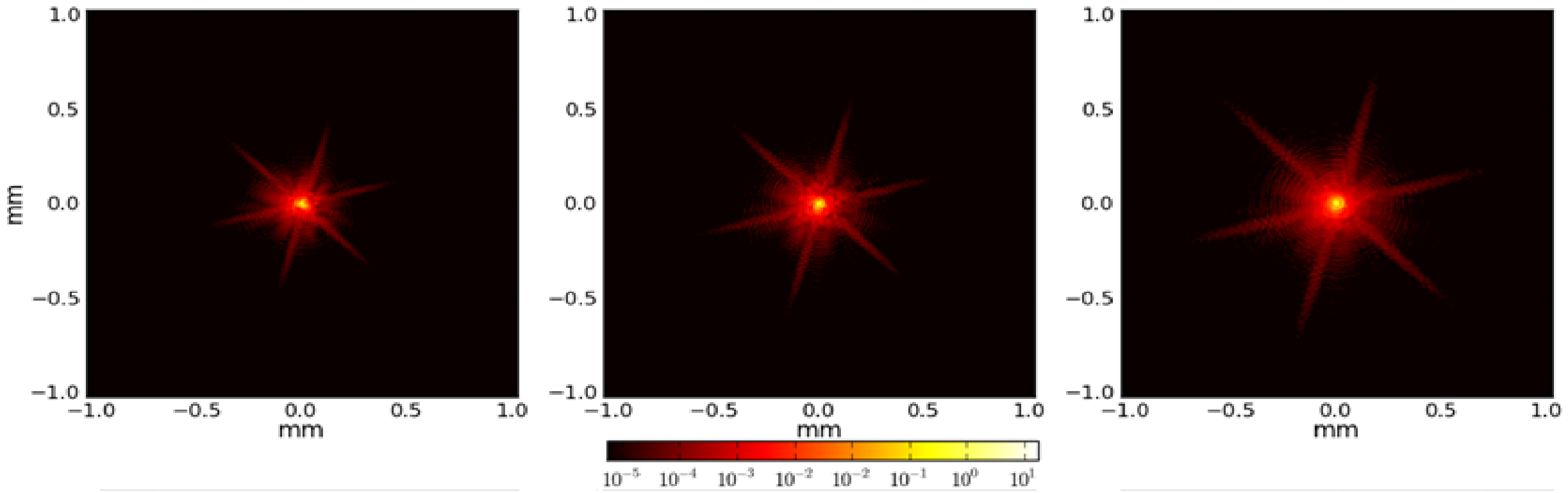

*Figure 6.6: The point spread functions (left = Y band, middle = J band & right = H band) on a log scale generated at the focal plane positions where the image quality is worst (largest EE50 & EE80). These PSF are constructed from accurate optical, pointing stability and detector models and are used as reference in other performance evaluations*

The imaging mode limiting point-source magnitudes have been calculated based on this worst case system PSF, and by combining only 3 individual dithered exposures. This configuration corresponds to over 90% of the survey data, while half of the survey is imaged with 4 dithers exposures and will therefore be deeper.

The imaging mode limiting performances have been assessed by taking the end-of-life telescope and instrument throughput profiles and with representative source, background, thermal, scattered light, dark current and detector readout noise contributions. With this conservative model, the NISP instrument's imaging mode is capable of reaching the limiting sensitivity within the time allocated to it during a 6-year mission. Figure 6.7 shows a simulated, single chip H-band NISP photometry mode image.

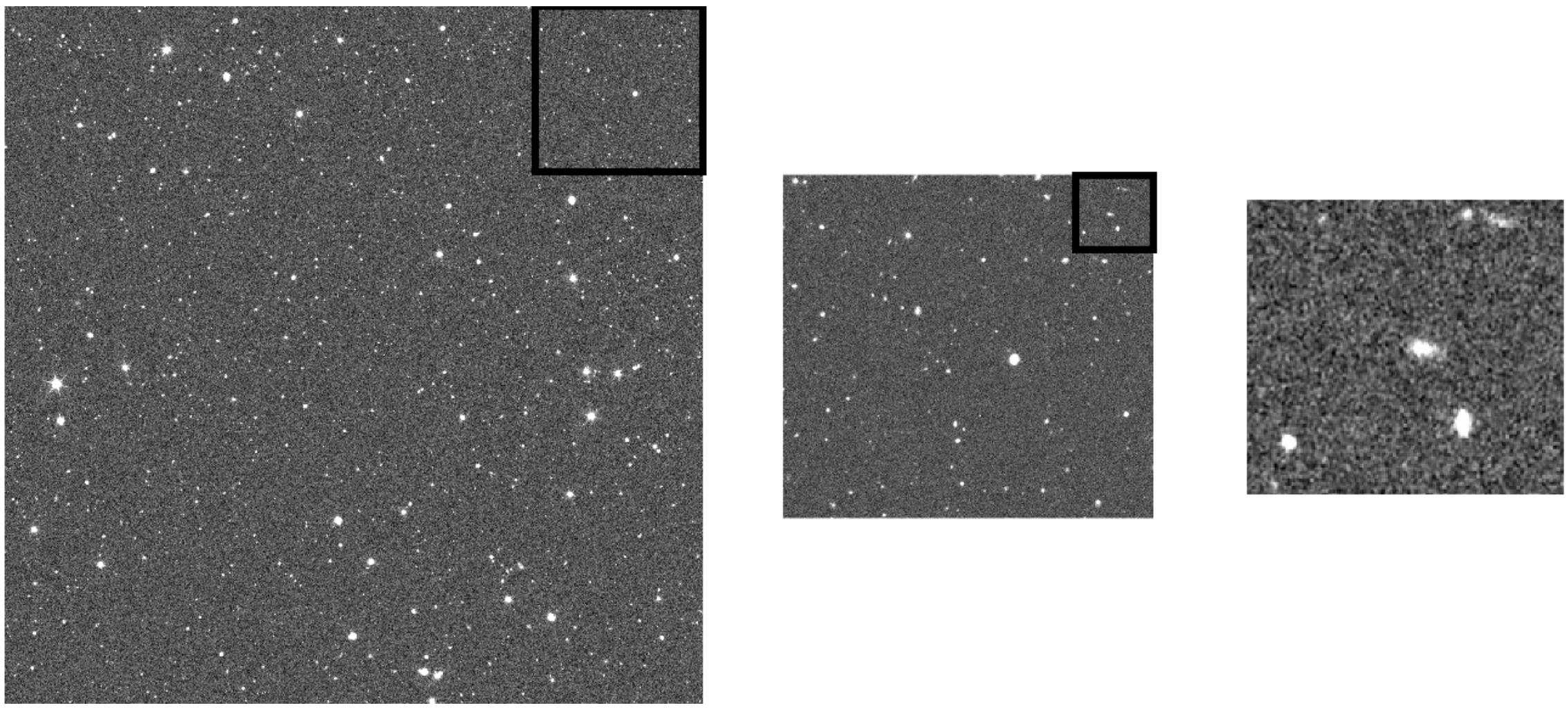

*Figure 6.7: A single chip, H-band image simulated with the NISP Photometry Simulator (Left). Simple drizzling has been used to combine three dithered measurements of the same field. (Middle) and (Right) show a progressively more zoomed in section of the top right corner of the chip.*

## 6.1.3 Near-infrared Spectroscopy

The NIR spectroscopic channel is designed to carry out redshift measurements with $\sigma_z<0.001(1+z)$, of more than 50 million galaxies in several redshift bins up to a redshift of 2. The redshift measurement relies on the detection of emission lines, mainly the Hα line in the near infrared range. For a given Hα line flux, the precision and accuracy of the measurements depend on the intrinsic NISP instrumental parameters (PSF size, resolution, and instrumental background) and on the observing strategy adopted to mitigate the specific limitations of the slitless technique, namely confusion and a higher astrophysical background. The performance verification has to demonstrate (i) that single redshifts are unbiased and measured with sufficient precision in order to reach the needed flux limit of 3 $10^{-16}$ erg $cm^{-2}s^{-1}$ (at 3.5 $\sigma$ for a 1 arcsec diameter size source) and (ii) that the completeness is sufficient to get galaxy number densities per redshift bins that meet the scientific goals. The survey selection function, both on the sky and in redshift, has also to be understood to a level sufficient to push global systematic errors on the derived cosmological quantities well below statistical errors.

The overall Euclid observational strategy in spectroscopic mode has been devised to optimise all these aspects. A space mission is in principle most suitable for slitless spectroscopy, due to the lower background emission (Glazebrook et al., 2005). This is even more important in the infrared, where airglow emission lines fully contaminate ground-based spectra. Still, by its very nature, slitless spectroscopy is affected by the confusion arising from the superposition of spectra from adjacent objects, while on the other hand the background is enhanced owing to light from unresolved spectra of faint background sources, and *zodiacal light*.

Reducing confusion produced by overlapping spectra is the first concern. At the depth of Euclid spectroscopic observations, essentially every spectrum is at least partially superimposed on another. In such conditions, contamination is the main cause of redshift measurement failures. To overcome this difficulty, the spectroscopic observing strategy entails first splitting the total wavelength coverage into two separate observations, using red and blue grisms as described in Section 4. The resulting shortening of each spectrum already reduces the percentage of overlaps. In addition, for each band two independent exposures are taken, with the dispersion rotated by 90 deg, as to separate spectra that may overlap along one direction. These four corresponding frames (0 deg blue, 0 deg red, 90 deg blue, 90 deg red) are also dithered around the reference position, with the strategy described in Section 4, to fill detector gaps and accommodate the combined VIS-NISP tiling pattern on the sky. Finally, through the combination of the Y, J and H direct images, a deep (AB=24) NIR image of the field is built and cross-referenced with the dispersed image to obtain an identification and accurate zero point for the bulk of the observed spectra.

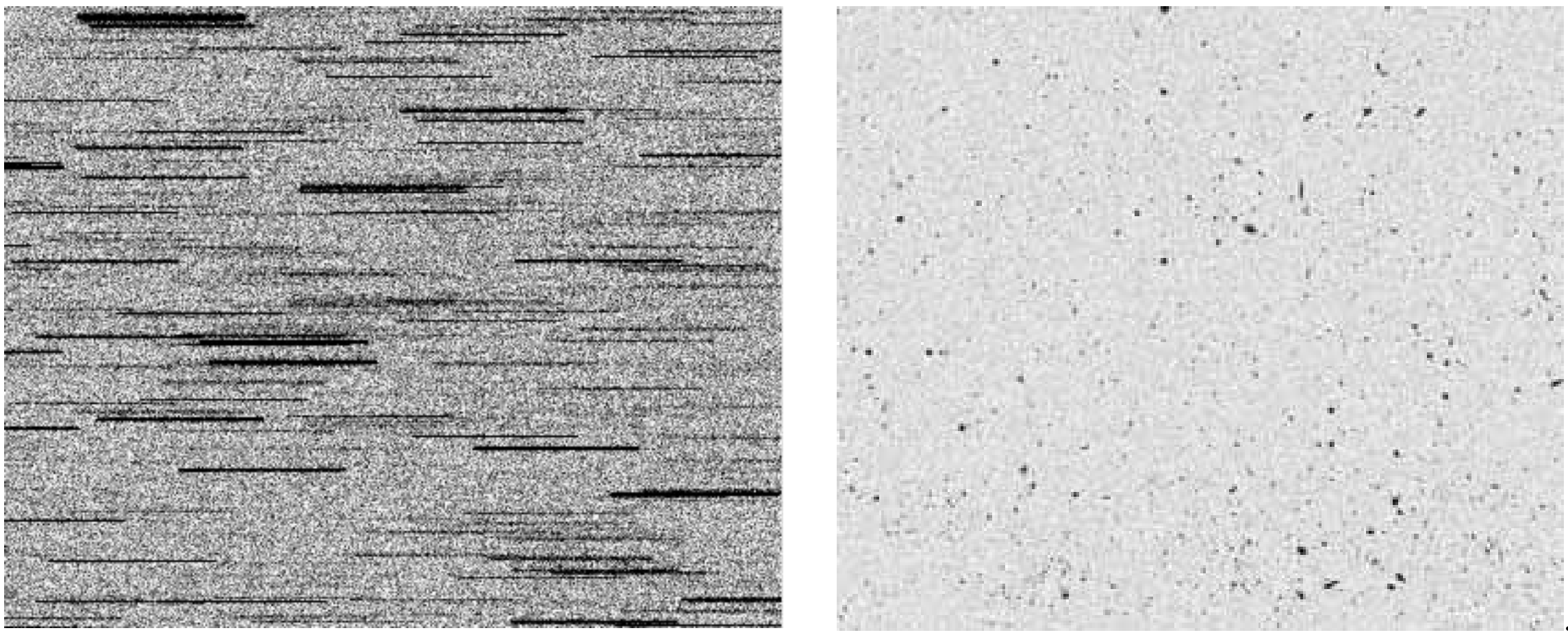

*Figure 6.8: Simulation of a typical Euclid slitless observation (left: single array, 10.2 arcmin side); (right) and its corresponding direct image*

To verify that the observing strategy and instrument parameters allow meeting the scientific requirements, the flux sensitivity is verified using the NISP instrument characteristics described in Section 4 and for a nominal exposure time of 540 s. An advanced end-to-end simulation pipeline has been developed to evaluate the expected completeness and purity of the resulting selected spectroscopic sample. The output is also used to compute cosmological quantities and quantify the impact on the various scientific figures of merit. The simulation pipeline consists of three components: i) a realistic input source catalogue with spectro-photometric information; ii) a module simulating 2D dispersed images based on an end-to-end radiometric model of Euclid for a given instrumental and observational set-up; iii) an automated analysis of the extracted spectra to classify the sources and derive galaxy redshifts.

Another fundamental ingredient for the realism of the simulation is the galaxy angular size. In slitless spectroscopy this is a crucial parameter, as it sets the effective spectral resolution and the level of confusion between adjacent spectra. An intrinsic and realistic angular size was thus assigned to each galaxy based either on the observed half-light radii $r_e$ measured from the HST images of the COSMOS field, or on the size-luminosity relation when HST images were not available. Finally, stars with the appropriate magnitude distribution are added to the galaxy catalogue, experimenting with surface densities corresponding to galactic latitudes |b|=60° and |b|=30°. Spectra for contaminating stars are created from the Pickles stellar templates, corresponding to the spectral and luminosity type, rescaled for each object to its magnitude. The Hα counts and redshift distribution of galaxies in the catalogue obtained this way are in fair agreement with independent predictions derived from observations (Geach et al., 2010).

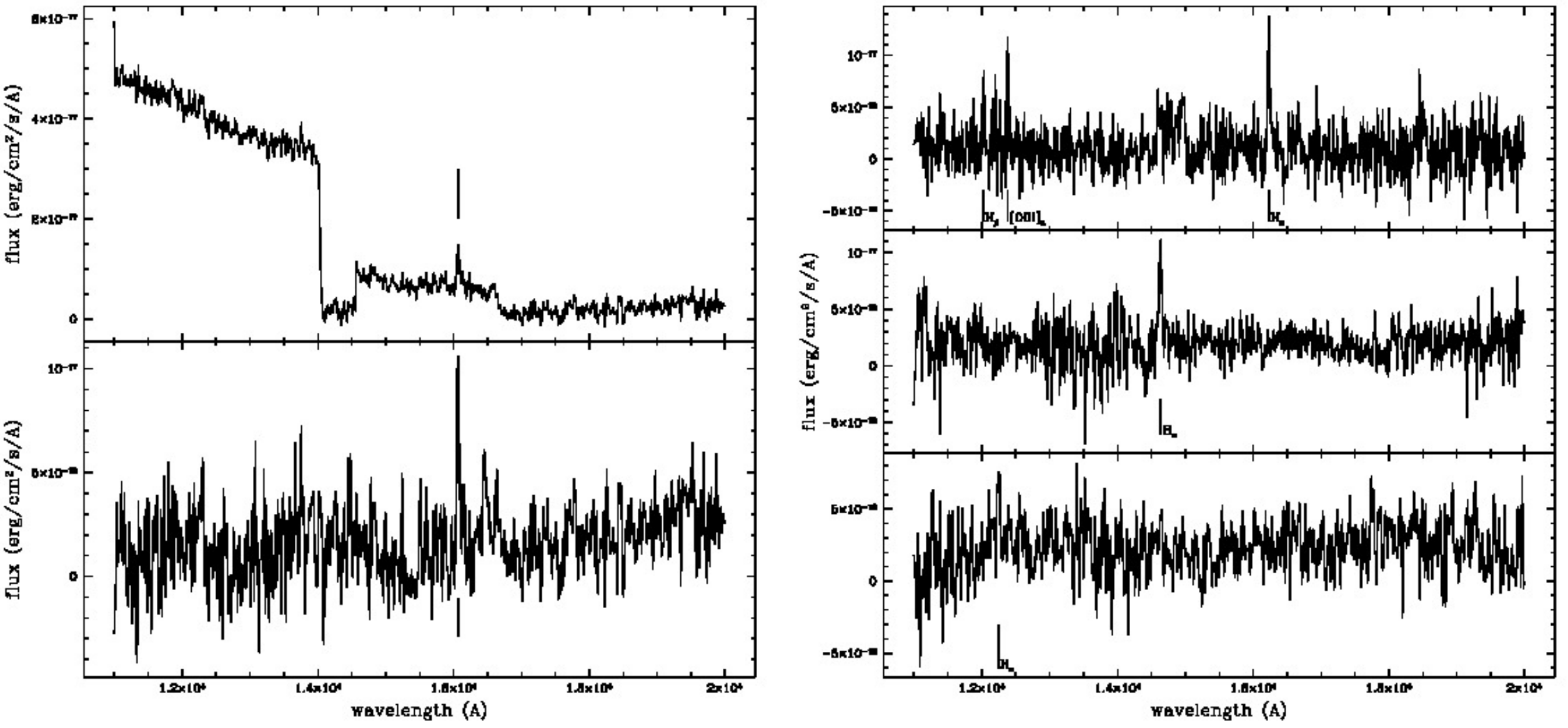

*Figure 6.9: Simulated Euclid spectra of Hα emitting galaxies, before / after decontamination (left) and for three galaxies with different Hα flux (right), respectively $5.9x10^{-16}$, $4.7x10^{-16}$ and $3.6x10^{-16}$ erg $cm^{-2}$ $s^{-1}$ from top to bottom. The positions of detected emission lines are marked.*

Simulated spectra are generated using the package aXeSIM3[3], which produces a direct image and a dispersed image of the simulated field, based on the input instrumental parameters (PSF, transmission curve, etc.). The code also allows for inclusion of the zodiacal light background, estimated by integrating over the observed wavelength range. Figure 6.8 shows an example of a slitless image and its direct counterpart.

Once the 2D dispersed image has been simulated, 1D spectral extraction is performed by aXeSIM itself. Then, the red and blue sub-spectra for each roll angle are concatenated together, so that the results of this first part are two 1D spectra – one for each roll angle – each covering the full wavelength range 1.1-2.0 micron. Figure 6.9 shows examples of simulated spectra at different stages of extraction and reconstruction: on the left, a spectrum is shown before and after background decontamination; on the right, three final 1D spectra are shown for different Hα line fluxes. By construction, the simulation automatically takes into account any border effect or differences in exposure or confusion, which may occur in some areas as a result of the dithering pattern.

As contamination by adjacent spectra is a major concern, the first step is to check and flag spectra for spurious features by using, for each grism, the two observations at 0 and 90 degrees. Spectra can thus be cleaned and combined together. A blind search for emission lines is then carried out. All emission lines found are checked together to see whether they lead to a concordant redshift: if this is not the case, the strongest line is assumed to be Hα and the redshift is computed accordingly. The same happens in case only one feature is detected. If no emission lines are found, or if one of the sub spectra is missing (e.g. at the field border) a standard cross-correlation technique against galaxy templates is used. Finally, all spectra are processed through the EZ redshift measurement code (Garilli et al., 2010), which assigns a reliability flag to each measurement. The completeness (i.e. the fraction of spectra measured above a given line flux limit) and the purity (i.e. the fraction of measured spectra for which the redshift is correct) are then finally estimated on the final spectra. The results obtained with the reference instrumental configuration and survey strategy are shown in Figure 6.10, plotted as a function of redshift and Hα flux. These show a completeness that is well above the requirement value. The purity of the measurements (i.e. the complement of the fraction of catastrophic redshifts) is also compliant with the requirements, with the possible exception of the highest redshift bin. There is however ample margin for improving the data processing techniques so as to further enhance the purity. The extraction technique, the redshift measurement method and the reliability assessments techniques used here are based on existing tools. The methods that will be developed in the Science Ground Segment will be optimised toward Euclid specificities and will have much better performances.

---

[3] http://axe.stsci.edu/axesim/

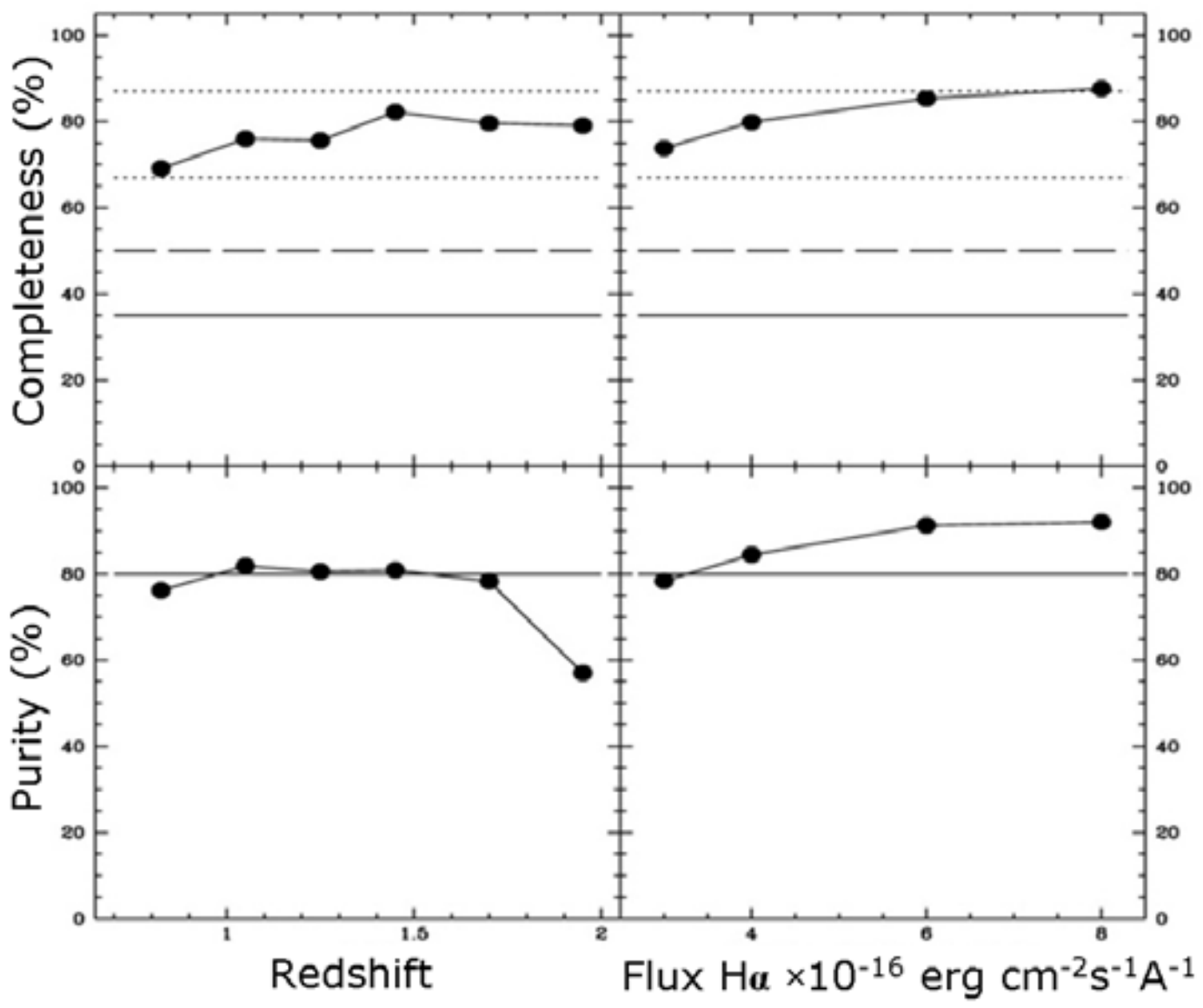

*Figure 6.10: Completeness and purity as a function of redshift and Hα flux, based on end-to-end simulations of the mission and survey strategy. Solid and long-dashed lines indicate the current requirement and goal, respectively.*

# 6.2 Wide survey implementation

## 6.2.1 Survey simulation

In this section it is shown how the Euclid wide survey will cover the required 15,000 deg$^2$ in 5.5 years, through a step-and-stare observation mode (4 dithered frames + overheads), given a joint VIS / NISP field-of-view of 0.54 deg$^2$ and an overlap between adjacent fields of 1%. Depending on the specific survey epoch and visibility, at the end of the sequence on one field the spacecraft can either slew to the next wide survey field, or to a calibration field. During the observation of a given field, the Solar aspect angle (SAA) varies slightly. The SAA has to be zeroed before slewing to the next field. To guarantee thermal stability of the optics, the maximum allowed SAA variation is set to 5 deg. The capability of de-pointing of the satellite is limited to ±5° in roll, and [0° – 30°] in pitch. The key parameters for each sequence are summarised in Table 6.1. The reference observation sequence for one field (integration time + overheads for dither and slew to next adjacent field) requires 4000 s. The slewing, pointing and stabilisation overheads provided by industry are based on the cold gas technology (see Section 5).

*Table 6.1: main parameters of the Euclid observational sequences (see Section 5.1 for observing sequence). These values include a 10% margin.*

| Parameter | Value |
|---|---|
| VIS frame Integration time | 590 s |
| NISP-P Y frame Integration time | 88 s |
| NISP-PJ frame Integration time | 90 s |
| NISP-P H frame Integration time | 54 s |
| NISP-S frame Integration time | 590 s |
| Shutter /rotation of wheel | 10 s |
| Global Frame time assumption | 882 s |
| Orbital Manoeuvre | 1 day / month |
| Flip Manoeuvre | 200 s / 6 months |
| Antenna Manoeuvre | During Slew |
| SAA acceptable variation range | Baseline 0° to 5°<br>Possible 0°-30° with stabilisation time |

### 6.2.2 Expected performance

A complete simulation of the wide survey has been carried out using the observing strategy presented in Section 5 and the above mission parameters. The result is shown in Figure 6.11 for 5.5 years survey duration. This is the result of a first optimisation of the observation strategy so as to cover the best area on the sky (in terms of minimum extinction and star density) while maintaining the solar aspect angle within an acceptable range. It shows that it is possible to complete the 15,000 deg² Wide Survey, perform the Deep Survey (6 months duration, executed partly during low visibility periods of the Wide Survey) and observe the calibration fields in 6 years. The additional year is reserved for commissioning, performance verification and initial calibration (up to 6 months), contingencies and maturity margin to cover for additional operational overheads which are still poorly determined at this stage. Should time be left at the end of the mission, it will be used to extend the survey area beyond 15,000 deg² so as to further increase the FoM

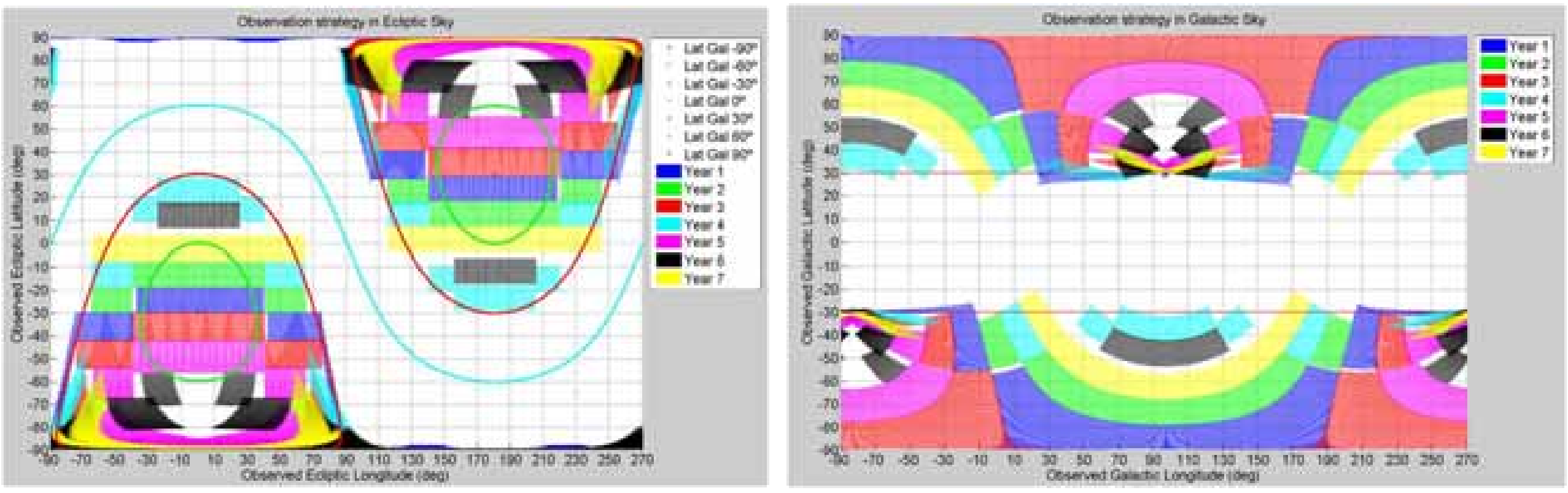

*Figure 6.11: the survey coverage simulation in 7 years in function of the sky fields in ecliptic coordinates (left) and galactic coordinates (right), colour-coded by year.*

Numerous effects, linked to detectors (e.g. dead pixels, gaps), instrument and telescope (e.g. ghosts) or specific areas of the sky (saturating stars, large galaxies) contribute to data loss. These effects can be considered on 3 typical scales which affect the survey in different ways: (1) small (<5 arcmin²); (2) medium (~20 arcmin²); (3) large (~50 arcmin²). Area losses spread over small holes have a negligible impact on the scientific performance (see Section 6.3), suggesting that the Euclid should maximise the sky coverage, rather than going back to fill small missing holes in an inefficient manner. Large and medium-size gaps remaining after dithering or because of possible detector losses, lead to an estimated 90% efficiency in area coverage. This has been obtained assuming that one CCD is lost after 2.5 years of mission, 1 NIR array is lost after 2.5 years and two CCD with filter are used for calibration but not for Weak Lensing measurements, thus creating holes in the survey.

## 6.3 Scientific performance

Having examined instrumental performances and survey coverage, the final step is to verify the overall end-to-end scientific performance of the mission.

### 6.3.1 Weak Lensing Performance

To verify the weak lensing performance, one must demonstrate that the number density and observable quantities, photometry and galaxy shapes, meet the requirements. Requirements have been set to control both random and systematic errors that would impair these cosmological measurements if left uncontrolled.

- The effective number density of galaxies used for weak lensing: an average of > 30 per square arcminute over the Wide Survey is required.

- Shape Measurement: the algorithms that measure the true shear from data are required to be unbiased with an accuracy of $10^{-3}$.
- Photometric Redshift Performance: the mean scatter between photometric and spectroscopic redshift is required to be < 0.05(1+z) with less than 10% catastrophic outliers and a 0.002 error on the mean redshift per tomographic redshift bin.
- Impact of unobserved sky regions in the survey (i.e. "holes" due to the dithering pattern, bright stars or other contaminating events): less than 3% of pixels are permitted to be lost due to glitches, cosmic rays, or any effect that generate masks on scales of less than 5 arcminutes square.
- Knowledge of the cosmic shear power spectrum: the cosmic shear power spectrum is required to be reconstructed with an integrated accuracy (over all scales) of $10^{-7}$.

Using simulations, it is shown here that each of these requirements is verified such that the final cosmological measurements can be made to the required precision.

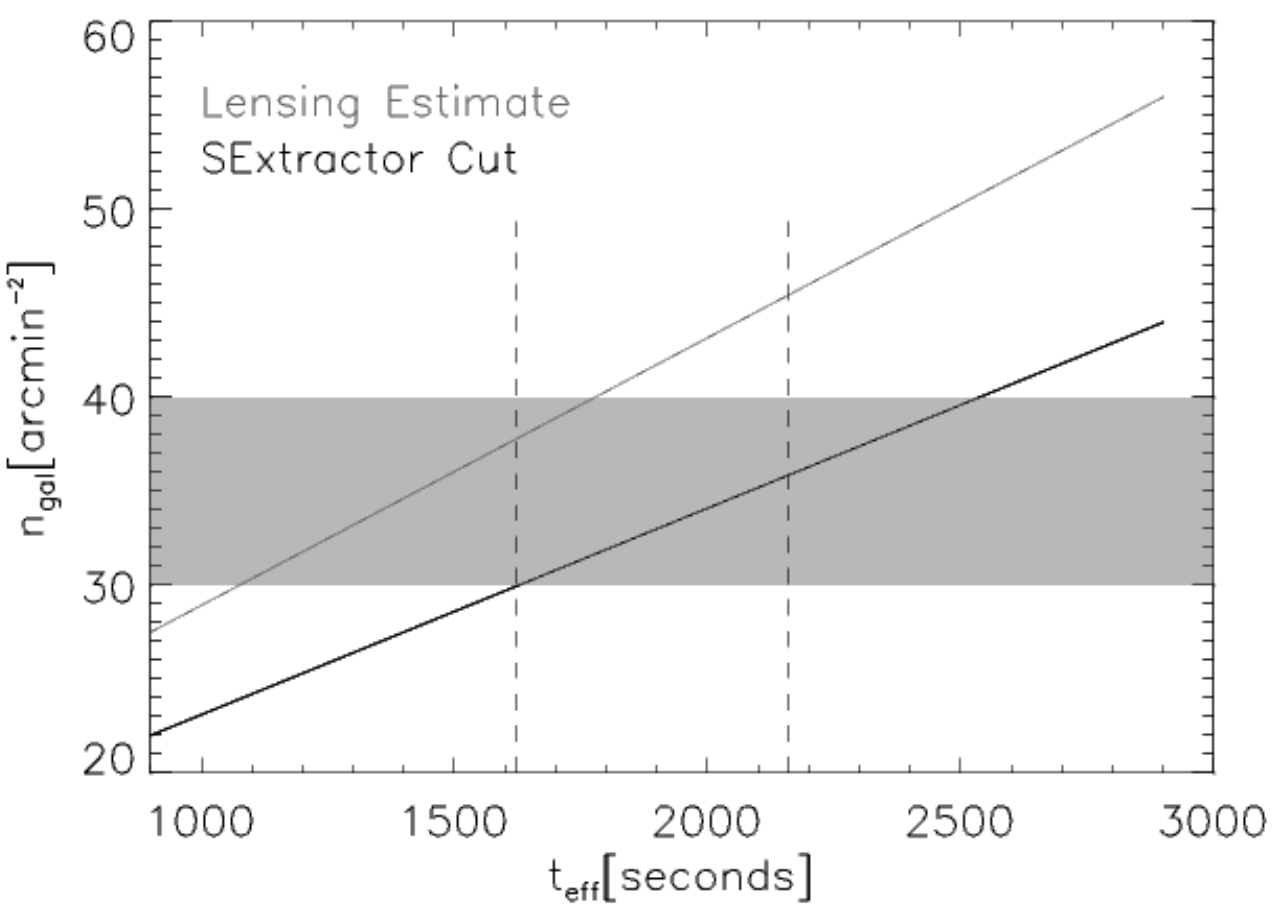

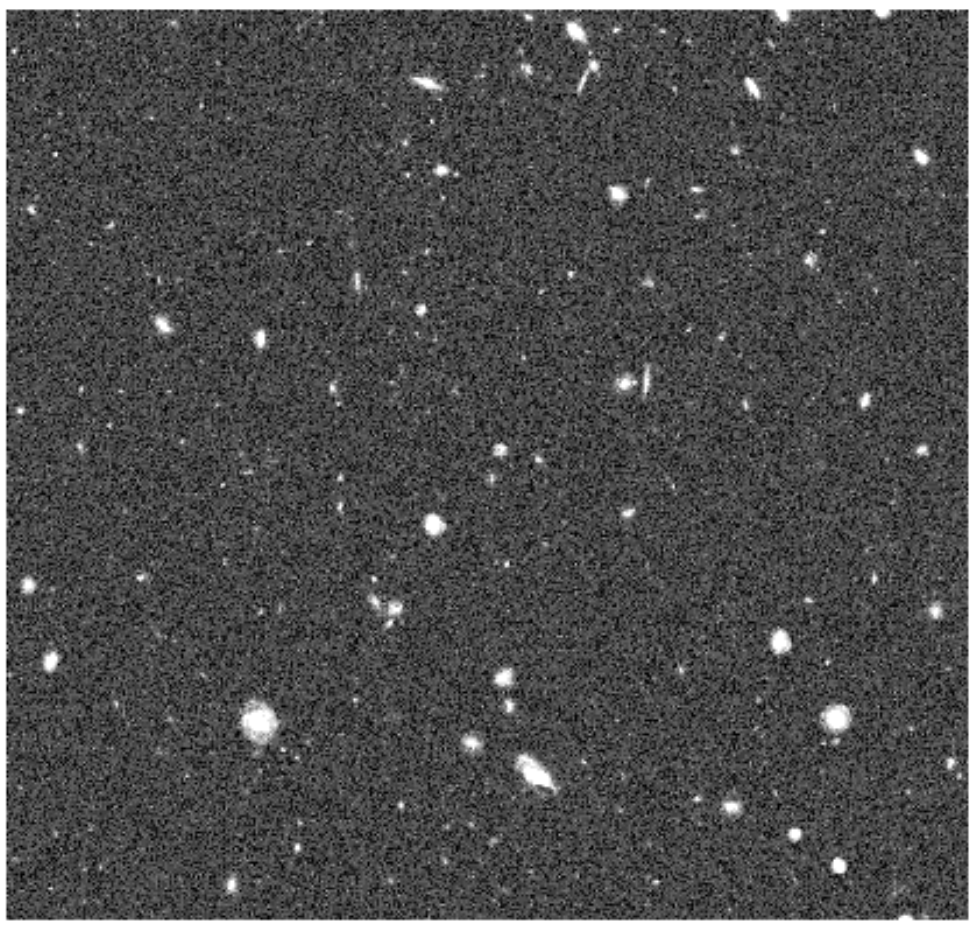

*Figure 6.12: a. Expected number counts of galaxies useful for lensing as a function of exposure time. The solid line is made using a simple cut on SExtractor detection with S/N>10 and FHWM[gal]>1.25FWHM[PSF], the dashed line is from the shape measurement pipelines that sum the lensing weight assigned to each galaxy, with a cut in ellipticity error of 0.1. The vertical dashed lines show the effective total exposure time for 3 dithers and 4 dithers (with an exposure time of 540 seconds). We see that we are able to reach our requirements of 30-40 gal/arcmin$^2$. b. (Right) A simulated image of the Euclid VIS (RIZ band-pass) with a size of 400 arcseconds.*

***Number Counts*:** Given the photometry and PSF characteristics, two independent image simulation pipelines have been used to evaluate the expected galaxy number counts for Euclid. Both image simulation pipelines produce realistic galaxies using a Shapelet decomposition (Massey et al 2004; Melchior et al., 2007) of the Hubble Space Telescope (HST) Ultra Deep Field (UDF; Beckwith et al., 2006). The two image simulations pipelines (described in Meneghetti et al., 2008 and Dobke et al 2010) have undergone detailed cross-comparisons with each other and the HST COSMOS survey (Meneghetti et al 2009 for details of these comparisons). For the Euclid specific configuration, the simulations use accurate models of the total Euclid throughput, noise characteristics, PSF shape and the zodiacal sky background. The image simulations have then been analysed using two independent shear measurement algorithms, all of which have been successfully applied to existing data (KSB and *lens*fit). The results are summarised in Figure 6.12 which shows the number of galaxies useful for lensing for two examples of cuts in the data: (i) simple cuts based on SExtractor parameters keeping galaxies with S/N >10 and FWHM size greater than 1.25 times that of the PSF, (ii) cuts based on shear measurement pipelines such that the error on measured ellipticity is $\sigma_e < 0.1$. For both of these, the galaxy number counts meet the Euclid requirements with 3 dithers totalling ~1600s exposure time.

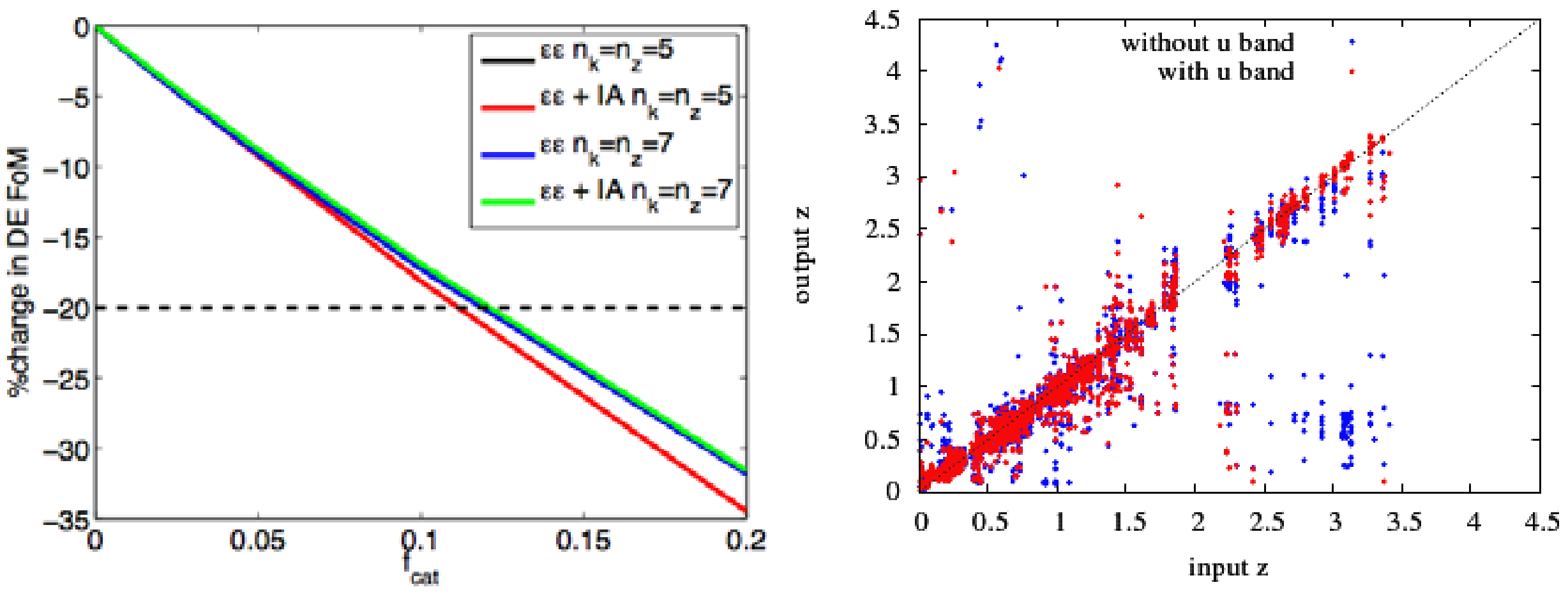

*Figure 6.13: a. (Left) Change in the FoM as a function of catastrophic photometric outlier fraction. The various lines show the change in the FoM for the ellipticity-ellipticity correlation function only (ee) and with the inclusion of Intrinsic Alignments (+IA) or not, with a binned parameterisation in scale and redshift with $n_k$ and $n_z$ respectively. The dotted horizontal line shows a FoM drop of 20% below which the choice of IA parameterisation becomes important From Kirk et al. (2011) b. (Right) A simulation of the Euclid photometric redshift estimates using the (u)griz from the ground and the infrared YJH bands*

***Photo-z Performance***: Photometric redshift estimates are needed for two key steps in the weak lensing analysis. The first is to divide galaxies into redshift slices and the second to characterise the distribution of galaxies in each slice. This is important for capturing the dark energy information in cosmic shear, and to characterise the systematic effects of intrinsic alignment. As demonstrated in Ilbert et al. (2006), based on a comparison between spectroscopic and photometric redshifts in the CFHTLS deep fields, and also verified with simulated Euclid data, the requirement of 0.05 (1+$z$) on the photometric redshift error is achievable with photometric depths in the *griz* bands described in Section 3, while the addition of *u*-band allows the goal of 0.03 (1+$z$) to be achieved. These depths should be reached by the planned Pan-STARRS 2, KiDS and DES surveys. Clearly surveys that go deeper (such as LSST) would further improve the photo-z performance. The fraction of catastrophic outliers, required to be less than 10%, is also met by the photometric performance; this is shown in Figure 6.13 (also discussed in Hearin et al., 2010).

Improvement is expected in algorithms that have been proposed to measure the flux from a galaxy and estimate its magnitude. The common practice consists of using top-hat apertures, whose size is determined from the light emitted by the brightest part of the object under investigation (e.g. Kron 1980, Petrosian 1976). However, this is not the optimal method to carry out photometric measurements. Indeed, magnitudes estimated from top-hat apertures are generally subject to biases, as they usually return a different fraction of the total light emission depending on the shape of light profile. When combining datasets obtained with different instruments and/or under different atmospheric conditions, as will need to be done for Euclid, further complications arise from the fact that fluxes should be measured within consistent apertures, wavelength dependent blurring effects should be taken into account, and pixelisation plays an important role. Optimally defining the aperture in a given band such to maximise the S/N ratio of the object leads to significant gains in the photometric accuracy and in the effective depth. An example of this is Kuijken (2006), where defined a flux measurement is defined that is independent of PSF and pixel scale called “Gaussian-aperture-and-PSF flux”. Applying this method on simulations, the systematic errors in the matched aperture fluxes are reduced to level of one percent. Maximising the S/N ratio in this way makes possible to measure magnitudes 0.2−0.3 fainter than reached in the above simulations.

***Shape Measurement***: The estimated shear to be measured from the data is required to be unbiased with respect to the true shear to $\Delta|\gamma|=|\gamma_{true}-\gamma_{measured}|\leq 10^{-3}$ accuracy**.** Over the past 10 years the accuracy of shape measurement has doubled every two years, and has been driven by methods developed within the Euclid Consortium, for example *lens*fit (Miller et al., 2007) already achieves an accuracy of $< 2\text{x}10^{-3}$ (Kitching et al., 2008) when tested on simulations (Heymans et al., 2006; Massey et al., 2007). As shown in Bridle et al. (2009) for Euclid galaxies (well sampled $R_{GAL}/R_{PSF}$>1.25 with a small stable PSF) the development needed to reach Euclid goals is *modest*, and recent advances (e.g. Bernstein, 2010; Melchior et al., 2011) show that the accuracy required by Euclid is already achieved for particular populations of galaxy. Members of the Euclid Consortium are involved in code development in two ways, (i) in-house simulations such the ones described in (Dobke et al., 2010; Meneghetti et al., 2009, 2011) and (ii) through the GRavitational lEnsing Accuracy Testing (GREAT) challenges (Bridle et al., 2009; Kitching et al, 2010).

***Effects of incomplete survey coverage*:** To verify the robustness of the requirements on the sky coverage and masking, simulations are used to generate realistic cosmological models and weak lensing maps are created from these models (Kiessling et al., 2011a, 2011b). These simulations cover a field-of-view of 100 square degrees and we simulate 100 independent lines of sight to generate a Monte Carlo suite of simulations, such that data covariances can be estimated. This analysis does not take into account methods that exist in the CMB (pseudo-Cl's for example) that can account for masking in the power spectrum analysis. We expect to implement these algorithms on Euclid data, and hence what is presented here is conservative.

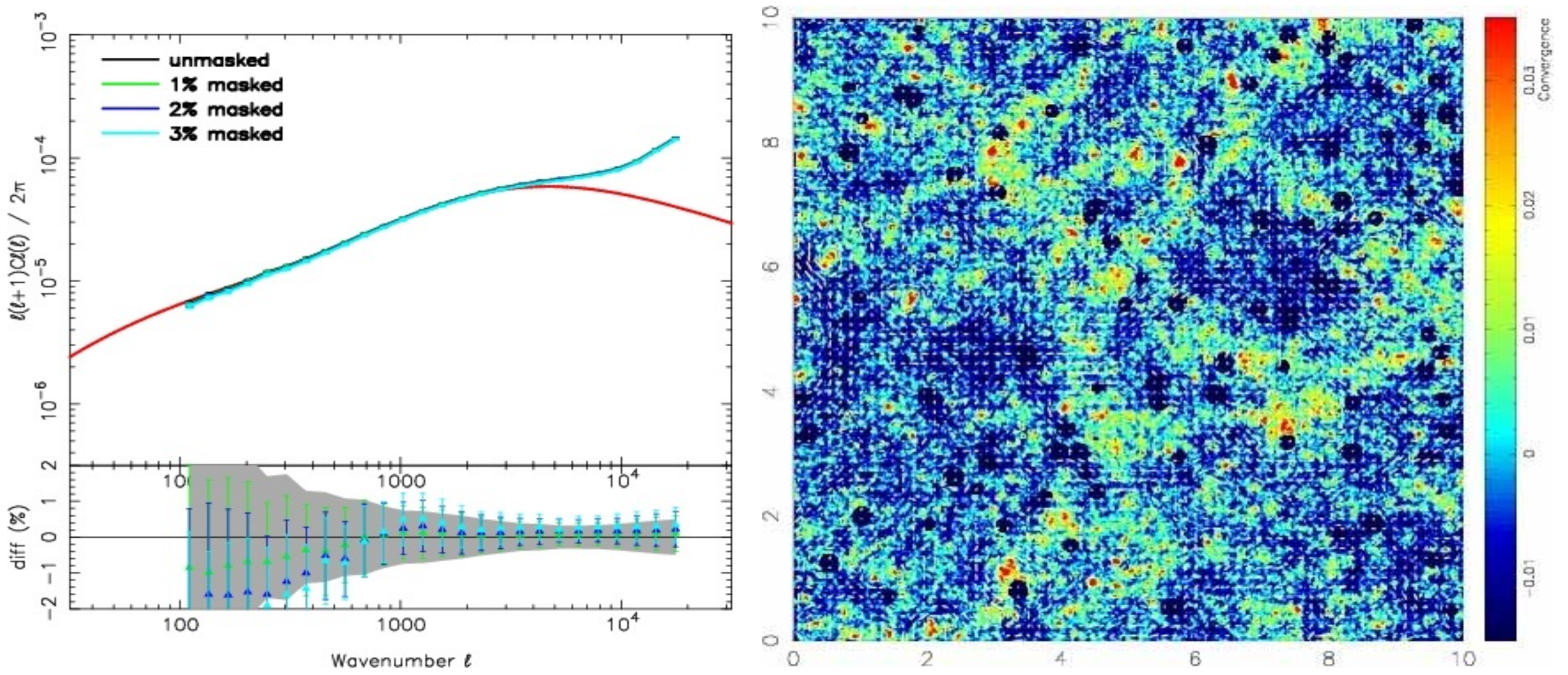

*Figure 6.14: Left: Unmasked shear power spectrum and masked power spectra for a survey with 1, 2 and 3% masking. The lower panel shows the percentage difference between the unmasked and masked power spectra as compared with the theoretically expected power spectrum. The grey region shows the 1-sigma error on the unmasked power spectrum Right: Simulated shear field where the colour represents the matter over density at a particular position (red is more dense) and the small whisker lines represent the shear amplitude and direction at each position. The masked regions can be seen as small black circular patches*

Figure 6.14 shows an example of the simulation. The simulations are used to determine the impact of small, star masks, glitches and cosmic rays, on cosmic shear power spectra. Star masks are simulated by placing circular masks across the shear field with a random distribution of sizes less than or equal to 5 $arcmin^2$, see Figure 6.14. Masking by up to 3% of the area does not bias the shear power spectrum by more than the error on the power spectrum. The amplitude of the difference between the masked and unmasked spectra is for 3% masked is always less than 2% over all scales which meets requirements for small masked areas less of less than 5 $arcmin^2$ (see previous sections). This is a test of the effects of small area-loss. Large area loss (for example entire fields or chips missing) in the data simply acts to decrease number counts (the inter-chip and inter-field cosmic shear signal has a subdominant contribution to the dark energy *FoM*) and the requirement of less than 15% of the survey lost in this manner is also met (see previous sections).

***Expected Performance*:** Figure 6.15 provides the final expected performance by plotting realistic power spectra with associated error bars as expected from Euclid. Figure 6.15 shows an example of the auto-power (within a redshift bin) and cross-power (between bins) weak lensing tomographic power spectrum for realistic mock galaxy shear catalogues generated from dark matter-only N-body simulations (Kiessling et al., 2011). The reconstructed power spectrum from simulations described above (including a realistic level of shot noise) is compared with the input power spectrum from theory. The figure shows that the signal is modelled accurately as a function of redshift. The power spectrum is recovered to sub-percent accuracy over signal-dominated scales and the integrated mean difference between the true and recovered power is $\sigma^2_{sys} < 10^{-7}$, which meets the requirements discussed in Section 3. At small scales the shot noise in the simulations begins to dominate – this is not a limitation of the technique and higher resolution simulations will enable the reconstruction over fully non-linear scales.

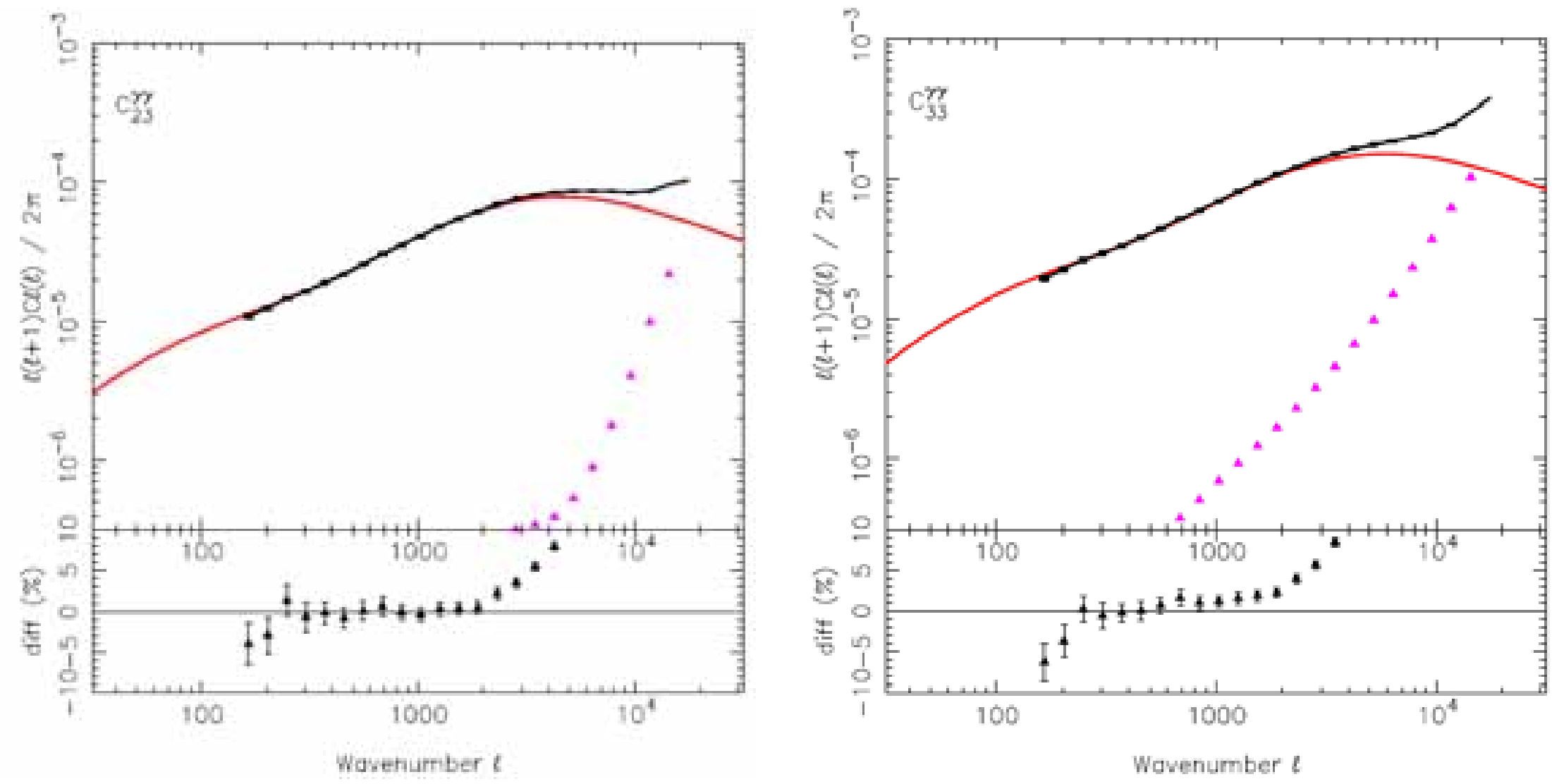

*Figure 6.15: Left panel: Tomographic weak lensing shear cross-power spectrum for redshifts 0.5 < z < 1.0 and 1.0 < z < 1.5, the red line shows the theoretically expected power spectrum and the black data shows the reconstructed power with error bars expected for Euclid, the pink points show the B-modes in the simulations which closely follows the shot-noise contribution. That this rises above the lensing signal at small scales that is not a limitation of the technique but a consequence of the simulation resolution. The lower panels show the percentage difference between the expected and measured power spectrum, which is recovered to 1% over all signal-dominated scales. Right panel: a similar analysis but for the tomographic weak lensing auto-power spectrum for 0.5 < z < 1.0.*

On small scales it is expected that baryonic effects will become important, however as shown by Semboloni et al (2011) Euclid will be able to correct for this effect, with negligible impact on cosmology performance by accounting for flexibility in model of baryonic feedback mechanisms.

### 6.3.2 Galaxy Clustering Performance

The primary requirement for galaxy clustering is to be able to measure the amplitude and anisotropy of the power spectrum (or correlation function) of the galaxy distribution within several redshift bins. From this, two fundamental pieces of information are obtained at each redshift: the first is the position of the BAO feature in the power spectrum. This is used as a standard ruler and associated to the reference mean redshift of the specific bin. A redshift-distance relation ("Hubble Diagram") is obtained mapping the expansion history of the Universe $H(z)$ and finally measuring $w(z)$. The second is the anisotropy of the power spectrum produced by linear motions, which is a measure of the growth rate of structure $f(z)$. The science requirements presented in Section 3 have been set to control both random and systematic errors that, if not limited, would impair these cosmological measurements, considering in particular the specific nature of the slitless redshift measurement technique. The sources of error can be identified within the following areas:

1. the precision and accuracy in the measurement of galaxy redshifts;
2. the impact of catastrophic redshift failures (i.e. wrong redshifts not identified as such);
3. the impact of unobserved sky regions in the survey (i.e. "holes" due to the dithering pattern, bright stars or other contaminating events);
4. the knowledge of the angular and redshift selection function.

These place requirements on both the Wide Survey and also on the Deep Survey, which is used to build a control sample of the selection function. The basic science performance for galaxy clustering as measured by Euclid was determined using the Fisher matrix approach described in Section 2, and was shown to meet the top-level scientific requirement when combined with the weak lensing data. This was based on the analysis described in Section 6.1.3, to estimate the expected redshift distribution of the Euclid galaxy sample, assuming that these galaxies were uniformly sampled based on angular position. In this section tests are outlined that have been performed beyond this basic analysis, confirming that Euclid will be able to match the required precision.

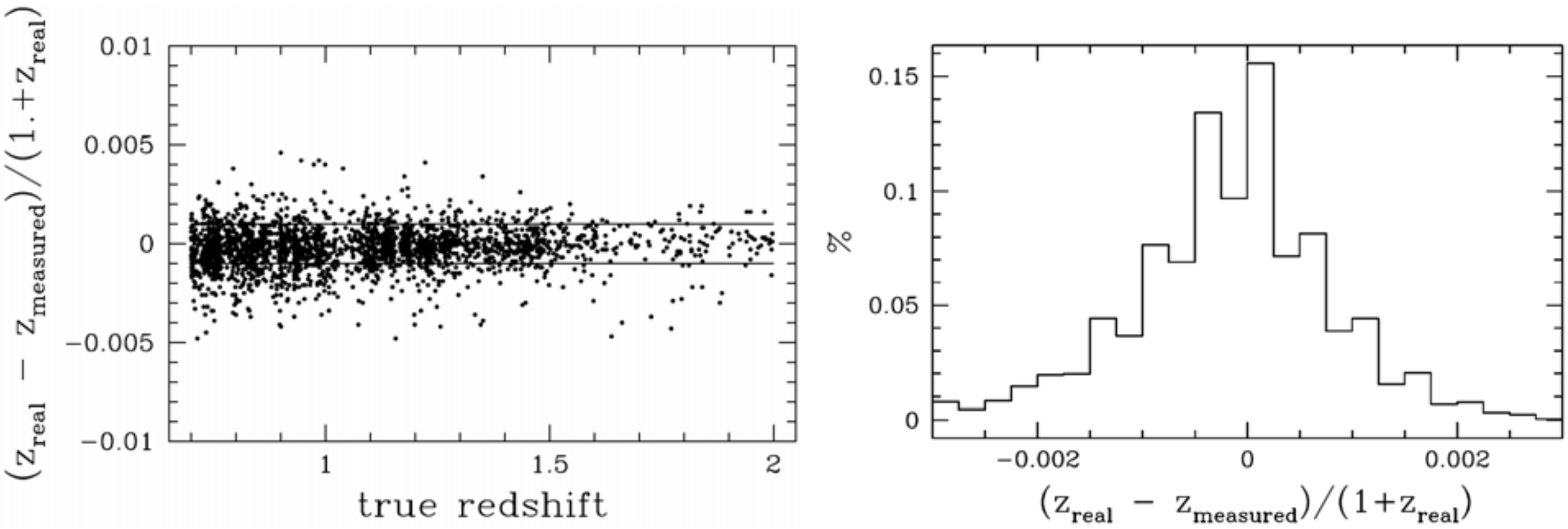

*Figure 6.16: Left: the difference between the measured and input redshifts in the end-to-end spectroscopic simulation, plotted as a function of the input redshift (excluding catastrophic measurements, which are accounted for in the systematic error budget, see text). Thin lines show the required 1-sigma error corridor. Right: corresponding histogram of the redshift errors normalised to the real redshift.*

***Precision and accuracy of redshift measurements:*** Measuring redshifts that are sufficiently precise (i.e. with low statistical uncertainty) and accurate (i.e. with negligible systematic error), lies at the basis of the whole Euclid spectroscopic experiment. The end-to-end simulations allow the direct comparison of the recovered redshifts with their original input values. Figure 6.16 plots their difference as a function of the true redshift, excluding catastrophic failures. The thin lines show the requirement of $\sigma(z)<0.001(1+z)$. 70% of the galaxies are within this limit. At the same time, as evident in the right panel of the same figure, the systematic offset of the measurements is very small: $6\times10^{-5}$, well below the requirement of $2\times10^{-4}$ excluding catastrophic failures. An accurate wavelength calibration is also required to ensure that this precision can be achieved without systematic effects.

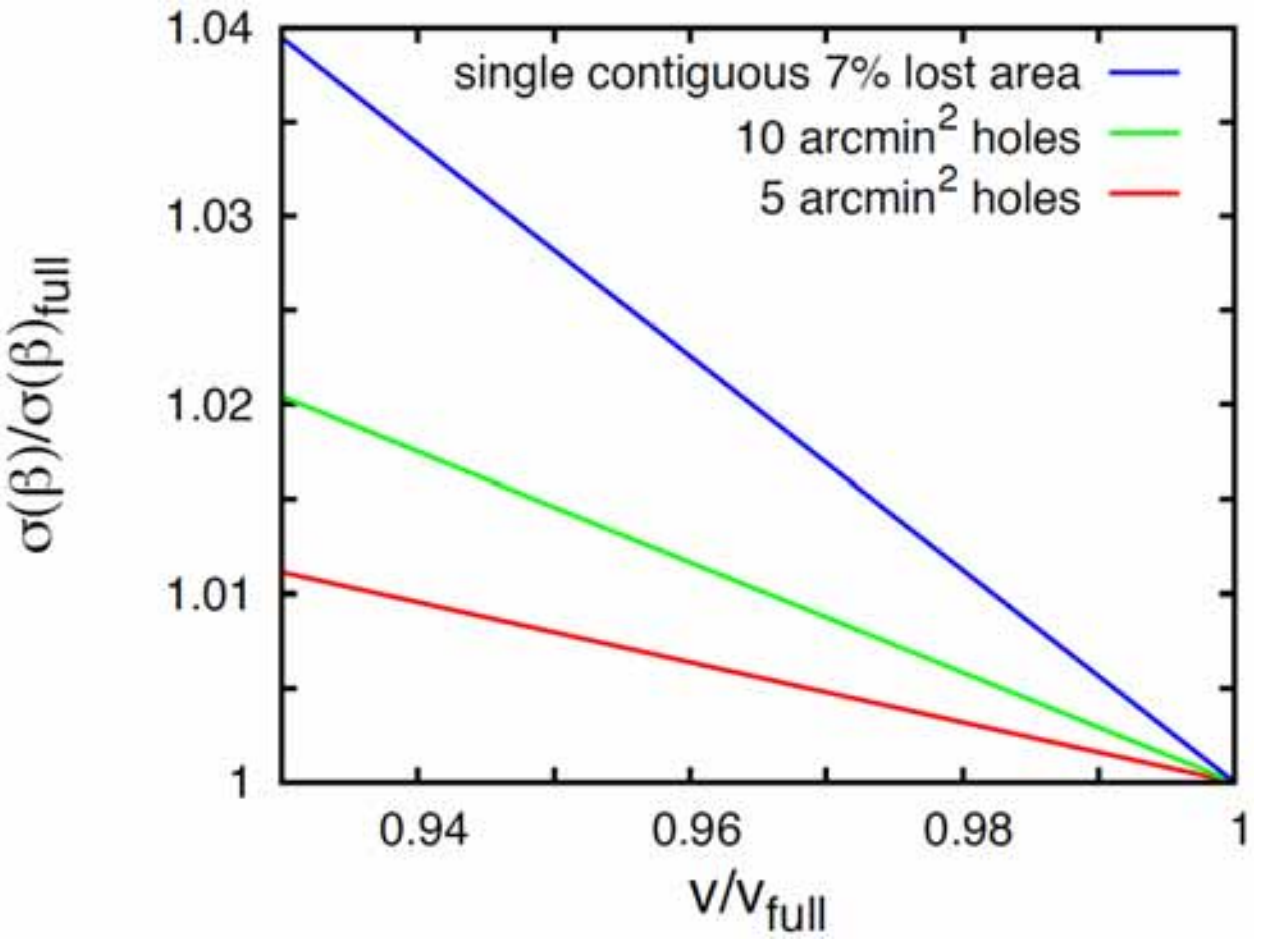

*Figure 6.17: Monte-Carlo estimate of the increase of the error on the growth rate of structure (in terms of the redshift distortion parameter β), one obtains when a given area of the survey is lost, either in a single continuous piece (blue curve), or through many small holes (as expected in the case of, for example, the presence of bright stars). The simulation shows the case of a lost area of 7% (i.e. ~1000 deg$^2$ for a wide survey of 15,000 deg$^2$), through to a complete survey. The impact of the lost area is significantly smaller if this is spread in small bits through the survey.*

***Impact of area losses throughout the wide survey*:** The impact on the computation of cosmological parameters – in particular the impact on measurements of the growth rate of structure, quantified in terms of the relative error on the redshift distortion parameter beta – from incompletely surveyed areas (in photometry, spectroscopy or both as described in Section 6.2.2), is examined next. The total area expected to be lost in this way may sum to 10% or more of the full survey (i.e. ~1500 deg$^2$ or more). The impact of this cannot be simulated with the standard Fisher Matrix forecasting technique (Section 2), in which only a global area can be specified, and a Monte Carlo approach is required. Figure 6.17 compares what happens if a total lost area of 7% is spread either in holes of 5 or 10 arcmin$^2$ (as expected) or in a single contiguous region. These measurements were made by "drilling" holes through a large simulation of a cosmological volume (the BASICC simulations, Angulo et al., 2009) and then estimating the anisotropy parameter $\beta$ using standard linear Kaiser modelling. The results show that when the area lost is spread in small disconnected regions, the loss of cosmological precision is significantly smaller than would be naively expected (i.e. proportional to the square root of the area – blue line) from the standard scaling of the Fisher matrix. The Fisher matrix error is matched by the Monte Carlo error only when the missing area is in a single piece (Bianchi et al., 2011, in preparation). A similar behaviour is seen when considering measurements of the *w(z)* using the BAO as standard ruler.

***Survey selection function: a) completeness and purity from the wide survey simulations.*** Figure 6.19 (left panel) shows the redshift distribution of all galaxies with good and reliable redshift expected in the Euclid wide survey. This distribution is obtained by convolving the Hα emitters galaxy counts (Geach et al., 2010) with the success rate obtained from the simulations.

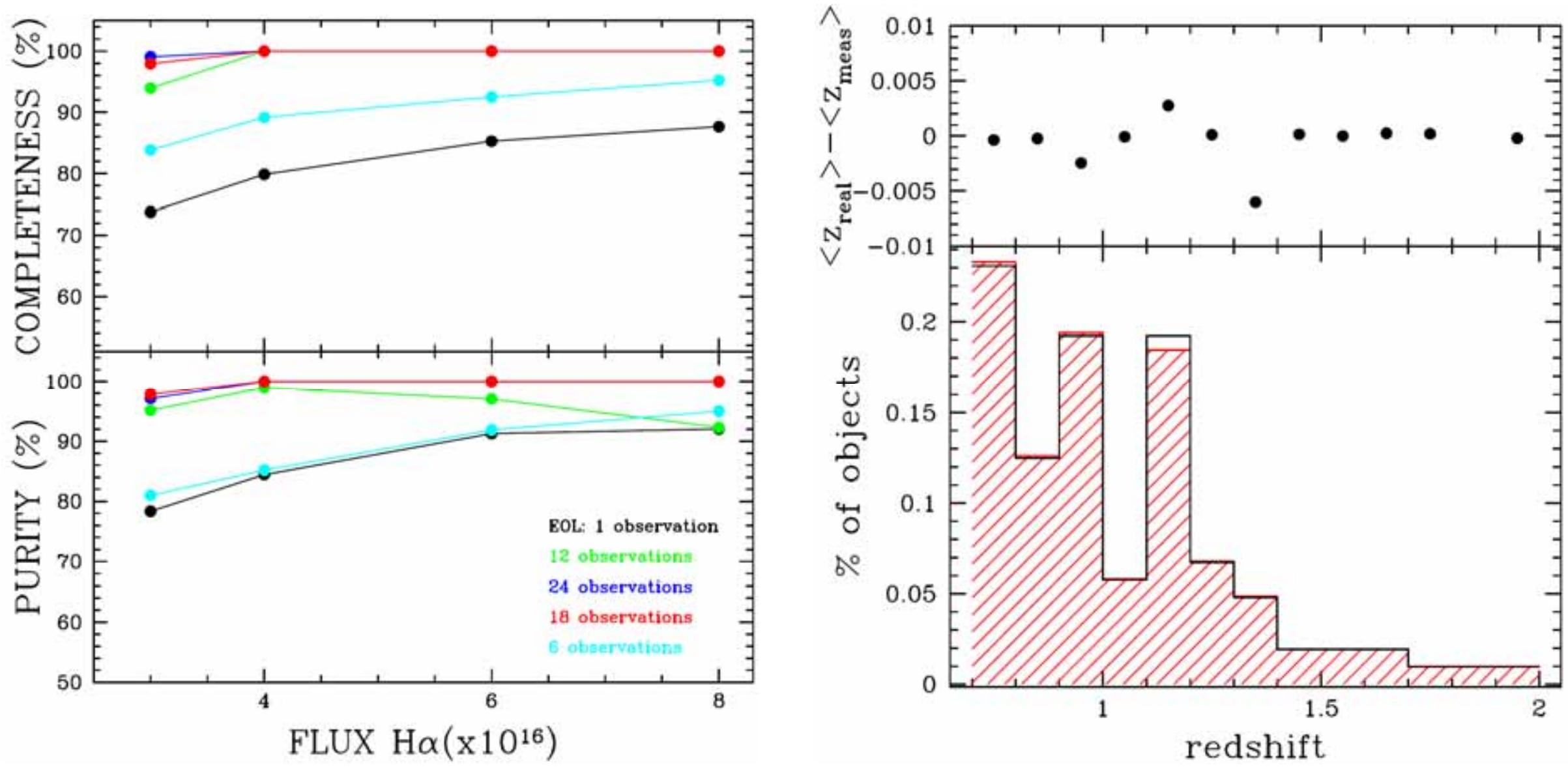

*Figure 6.18: Left: the recovered completeness (top panel) and purity (bottom panel) as a function of the Hα line flux, when increasing the number of repeated observations (with different position angles) of the deep survey. Right: real (black empty) and observed (red dashed) redshift distribution from the Deep survey. The top panel shows the difference between the mean redshift of the true distribution and the observed distribution*

***Survey selection function: b) completeness and purity in the reference deep survey.*** The Euclid deep spectroscopic survey will be fundamental, as to reach a complete characterisation of the wide survey selection function. To fulfil the requirements defined in Section 3.3.2, the deep survey has to allow the construction of a sub-sample with the same depth and redshift distribution as the wide survey, observed with the same strategy, but with a purity of 99% (see Section 5 for details). To verify these expectations, one specific simulation of the deep survey has been carried out over an area of 0.3 deg$^2$. The size and number of realisations of this simulation are limited by the large computational time involved. The very encouraging results of this simulation are shown in Figure 6.18, from which it is clear that the required level of purity for the control sample is reachable.

***Survey selection function: c) impact of density-dependent confusion on clustering measurements.*** Despite the observing strategy (two grisms plus two rotations) to reduce it, confusion caused by overlapping spectra still leads to a density-dependent galaxy selection function on the sky, which is potentially dangerous for the science goals. In order to assess the impact of this effect on clustering measurements, 100 deg$^2$ of the galaxy sample that will be delivered by Euclid has been simulated. To do this the density-dependent selection measured from the detailed 1deg$^2$ simulations of the wide survey described in Section 6.1.3 has been applied to a simulated "light cone", built from the Millennium simulation by applying a semi-analytic galaxy formation model (Bower et al., 2006). Available cosmological simulations currently limit this size, but the next generation of simulations will allow the simulation of the full Euclid area. This selection was contrasted to one based only on flux, without any density-dependence. The resulting correlation functions are compared in Figure 6.19 (right panel), which clearly indicates that the corrections required for the BAO peak position and the general correlation function are small and can be controlled. Note that, for the final cosmological measurements, it is not the size of the effect that is important, but how well it can be modelled. Larger simulations will be performed in future studies to verify and model these effects in detail.

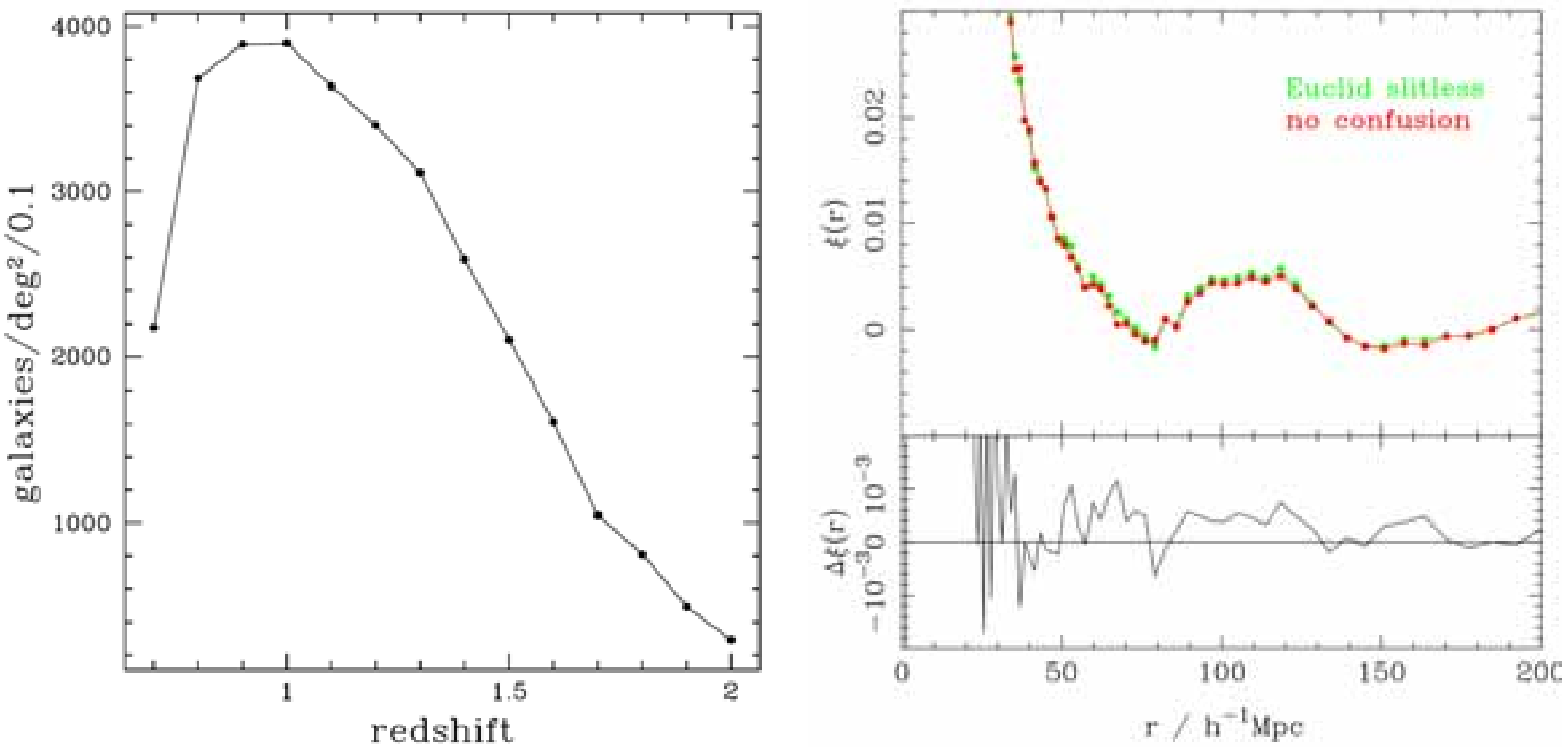

*Figure 6.19: Left: Predicted redshift distribution of Euclid galaxies with reliable redshift. The curve gives the differential dN/dz expected from the baseline Euclid wide spectroscopic survey, given the instrumental and survey configurations described in previous sections, the efficiency estimated f (Geach et al., 2010). Right***:** *Simulation of the two-point correlation function expected from a Euclid sky area of 100 deg$^2$, using spectroscopically confirmed galaxies with 0.9<z<1.1 (green line). To assess the impact of density-dependent confusion (thus, incompleteness), the plot compares this observed correlation function with (red line) that of the input sample, corresponding selecting a purely Hα flux limited sample. Larger simulations, undertaken prior to launch, will be used to accurately model and correct for this effect, which is shown here to be already small (lower panel).*

# 7 Ground Segment and Data Handling

## 7.1 Operations Ground Segment and Science Ground Segment

The Euclid ground segment consists of two blocks: the Operations Ground Segment (OGS), managed entirely by ESA, and the Science Ground Segment (SGS), for which the management is shared between ESA and the Euclid Mission Consortium (EMC).

The OGS covers the mission control components, which fall under ESA responsibility:

(a) At least one Ground Station, which supports a daily telemetry communications period (DTCP), expected to be 4 hours during nominal operations, and longer during the Commissioning and Performance Verification phases scheduled before the start of the nominal operations.

(b) The Mission Operations Centre (MOC), which is in charge of monitoring spacecraft health and safety, monitoring instrument safety, controlling the spacecraft attitude, and handling telemetry/-telecommands for both spacecraft and instruments. The MOC delivers the telemetry and flight dynamics products to the Science Ground Segment.

The Science Ground Segment (SGS) is responsible for data processing and archiving. The components of the SGS are:

(c) The Science Operations Centre (SOC), managed by ESA, which is responsible for executing the planned surveys, i.e. scheduling the spacecraft slews, scheduling the observations, monitoring the survey performance, rescheduling, and requesting MOC action via predefined procedures and sequences of telecommands. During routine operations, the SOC is the only SGS interface to the MOC. The SOC produces daily reports on the status of the payload and performs the Level 1 data (from raw to edited telemetry) the results of which are ingested into the Euclid Mission Archive together with first-level quality control (QC1), which is also produced by the SOC. The SOC produces quick release (Level Q) science products, which are released to the scientific community at large. The SOC manages the Euclid Mission Archive (EMA) and operates the Euclid Legacy Archive (ELA).

(d) The Instrument Operation Teams (IOTs), provided by the EMC and responsible for the maintenance of the instruments, monitoring of instruments health, instrument trend analysis, and the production of weekly instrument reports.

(e) The Science Data Centres (SDCs), also provided by the EMC.

1. Instrument-oriented SDCs host the IOTs and are in charge of instrument calibration activities and second-level quality control on calibrated data (QC2). They convert the edited telemetry frames into science frames and create the Level 2 data products such as calibrated mosaics of the survey data, and extracted spectra.
2. Data processing SDCs are in charge of science processing and of the creation of science-ready data products (Level 3) and corresponding quality controls (QC3).They link back to the IOTs mission critical issues, such as residuals deriving from systematic instrumental errors propagating into the errors of cosmological parameters (QC4).
3. Science support SDCs provide simulated data or reprocessed external data (Level S and Level E, respectively).

The SDCs are responsible for the selection of data items from Level 2 and Level 3 for public dissemination through the ELA after formal approval by the Euclid Science Team (EST, see also Section 8).

Despite the apparent complexity of the Euclid ground segment, ESA and the EMC have developed, and are committed to maintain, a tight collaboration in order to design and develop a *single*, *truly integrated SGS*.

## 7.2 Science Operations

The raw spacecraft data (both housekeeping and science telemetry data) are made available by the MOC to the SOC. SOC unpacks, decompresses, and formats the science telemetry to generate the Level 1 science data. MOC provides information on the performed pointing and the execution of the observations. The MOC further provides to the SOC on a regular basis auxiliary data like the station pass time information as well as products obtained by flight dynamics: orbit file, events file including information such as e.g. orbit maintenance manoeuvres, etc. The SOC puts this information into the Euclid Mission Archive; for retrieval by all other parties in the SGS.

The SOC performs the Level Q processing. The Level Q products are expected to be publicly distributed. A quick look system will use Level 1 and Level Q data to identify possible rescheduling cases and generate a daily quality report. The EMC retrieves the Level 1 data for processing to the higher data Levels 2 and 3; weekly instrument health reports are derived and sent to the SOC to support the planning and survey monitoring task. Derived calibration data are made available to the SOC via the EMA for use in the Level Q pipeline.

During the routine phase the SOC is formally the only party that has direct interaction with the MOC for nominal operations: it is the only interface *to* MOC. The SOC ensures that the instrument observation and commanding requests fit into the survey plan, to minimise the impact of these requests on the execution of the survey, and passes these requests to MOC.

The EMC provides the IOTs. Each IOT uses data from the SOC and relevant SDCs to monitor the instrument health and performance through, for example, trend analysis of the instrument calibration data. The IOTs produce weekly reports which are fed back to the SOC to be used to fine-tune the instruments on-board. IOTs are located at the SDC specialised in the corresponding instrument. Taking advantage of the experience gathered in previous missions, the IOTs are set-up early enough before launch in order to allow them to be involved in the end-to-end tests. They are composed mainly of members of the instrument development teams, ensuring proper transfer of knowledge obtained during the development of instruments and ground segment equipment (GSE). The IOTs play an important role before routine phase during the satellite commissioning and payload performance verification phases. During the routine phase the IOTs support the SOC.

Before the start of the routine phase the MOC offers the IOTs the possibility to monitor real time data available during the visibility periods via an Instrument Workstation (IWS), which is located within the local area network of the MOC. The IOTs use their quick look analysis (QLA) and real time assessment (RTA) facilities to monitor the science and housekeeping data, respectively. As mission operations switch from calibration activities to routine activities, the use of IWS should decrease. During routine phase the IWS is still expected to be useful, especially in the case of contingency, and is expected to be made available by MOC on a best-effort basis. Such a facility at MOC has been implemented successfully for other missions like e.g. Planck and Herschel.

For non-routine payload maintenance the SOC assists the IOTs in the delivery of the information and scheduling of the activity with MOC. The SOC needs full visibility of the instrument operations during periods of non-routine/contingency operations, to be able to maintain the survey schedule, in particular to prepare the observing plan immediately after a contingency period.

For the survey planning, SOC receives the survey strategy from the Project Scientist, who has been advised by the EST. The resulting long-term plan is provided to the MOC and EMC for their planning. The status of the survey as a whole, as assessed by the execution status of the individual fields, is maintained by the SOC using the daily and weekly reporting from SOC and IOTs. The survey status is reported to the Project Scientist.

## 7.3 SGS Organisation

Euclid delivers an unprecedented large volume of data for astronomical space missions: e.g. about 4 times more down-linked data than Gaia. Furthermore, a large volume of ground-based data from optical surveys like DES, Pan-STARRS or others is used for calibrations, quality control tasks and scientific data reduction, specifically for obtaining photometric redshifts. The ground based data have to undergo Euclid specific processing (such as the conversion to common astrometric and magnitude reference systems) in order to be consistently handled with Euclid data (“euclidisation”). The SGS components are connected to any other external data that might be needed via interfaces and cross matching tools widely available in the astrophysical community, such as ones currently provided by the Virtual Observatory.

### 7.4.1 Data processing levels and data production overview

The SGS data products, consequently processing levels, are subdivided in six levels, as summarised in Table 7.1. The description of the data processing levels concerns both input/output data products and the processes that consume and generate these data products:

- Level E: external data (images, spectra, catalogues, all relevant calibration and meta-data, observational data in science-usable format) derived from other missions and/or external survey projects, reformatted to be "euclidised". Level E provides final data products at the required level of accuracy. The data are delivered by the EMC.
- Level S: simulated pre-mission data, used before and during the mission, for calibration, validation, and modeling purposes (catalogues, satellite and mission modeling data, etc.). The data for this processing level are prepared before the mission and refined/updated during the mission. The data are delivered by the EMC.
- Level 1: is composed of three separate sub processing functions: telemetry checking and handling, including RTA on housekeeping; telemetry unpacking and decompression (edited telemetry); QLA on science telemetry and production of daily reports. The input data for this processing level come from the satellite via MOC and are used to perform quality control. The data are delivered by the SOC.
- Level Q: quick release data processing, i.e. basic removal of instruments signatures and production of calibrated images having "quick-look" quality for distribution to the scientific community. The data are delivered by the SOC.
- Level 2: instrumental data processing, including the calibrations as well as the removal of instrumental signatures in the data; trend analysis on instruments performance and production of weekly reports. The data processing at this level is under the responsibility of the SDCs in charge of the instruments monitoring. The data are delivered by the EMC.
- Level 3: data processing pipelines for the production of science-ready data. The Level 3 data are also produced by SDCs. The data are delivered by the EMC.

*Table 7.1: SGS processing Levels, indicating data products and responsibilities*

| Data Processing Levels / Data Levels | Data type after the processing (output) | S/W Development Responsibility | Data Production Responsibility |
|---|---|---|---|
| Level E | *Euclidised* External data | SDC | SDC |
| Level S | Simulations | SDC | SDC |
| Level 1 | Unpacked and edited telemetry | SDC | SOC |
| Level Q | Quick-release data | SDC | SOC |
| Level 2 | Instrument signature removed, calibrated data | SDC | SDC |
| Level 3 | Science-ready data | SDC | SDC |

As shown in Table 7.1, each level is split into software and data products. The responsibilities are shared between SOC and the SDCs. SOC will perform Level 1 activities, plus data processing leading to the production of Level Q data, but the software of Level 1 and Level Q are developed by the SDCs in charge of Level 2 and delivered to SOC for integration within its premises.

### 7.3.2 Quality Control

Detailed data quality control is essential for the management of the mission, the data processing and achieving the science goals. Quality control is performed at the different levels of data processing. Quick quality controls are done by the SOC, more elaborate quality controls involving full pipeline reductions and calibrations (e.g. effective tracking stability over different dithers of a single pointing field) are done by the IOTs. The IOTs report at least weekly to SOC. All these quality controls are critical for the success of the mission.

The results of the quality control are stored in the EMA so that each participant can be supplied with quality information. *All Quality Control information is shared over all participants*; there are 4 domains, summarised in Table 7.2.

*Table 7.2: Quality control domains*

| Name | Data processing level | Description | Product location |
|---|---|---|---|
| Quality Control 1 (QC1) | Level 1<br>Level Q | Data integrity (file level) and Quick Look Analysis | SOC |
| Quality Control 2 (QC2) | Level 2<br>Level E | Instrumental fingerprints removal (instrumental calibration) | SDCs |
| Quality Control 3 (QC3) | Level 3 | Data product quality check-up | SDCs |
| Quality Control 4 (QC4) | Level 3 | Final scientific product quality check-up | EST |

### 7.3.3 A data-centric approach

The key features of Euclid are the amount of data that the mission will generate, the heavy processing that is needed to go from the raw data to the science products, and the accuracy and quality control that are required at every step in the processing. This enforces a *data-centric* approach: all SGS operations revolve around the EMA a central storage and inventory of the data products and their metadata including quality control. The orchestration of data exchange and metadata update through SOC, SDCs and EMA is performed by a monitoring and control function. The ELA is a logical subset of the EMA and contains the data products to be distributed to the scientific community. The criteria for availability in the ELA are defined by the EST and the EMC, and are implemented in the EMA.

The EMA is a logical, rather than physical entity giving access to all mission-related analyses, wherever they have been carried out. The SOC and the EMC have the joint responsibility of guaranteeing homogeneity in data access, and of providing integrity, security and the appropriate level of quality control.

After approval of the data products for public release, the ELA shall be delivered to ESA as the unique distribution channel of Euclid data products to the scientific community.

## 7.4 Data Processing

### 7.4.1 Data processing functions

Since large and reliable computing resources are needed, the backbone of the science ground segment is formed by the SDCs, located in participating countries of the Euclid consortium. The SDCs are in charge of running different data processing functions of the Euclid pipeline. In most cases the infrastructure of these data centres already exists and provides services to the astrophysics or particle physics community. The SDCs are manned with information technology experts such that system maintenance is guaranteed from day one to the last days of the Euclid post-operations phase. As a customer of these data centres, the EMC contributes in kind so that CPU and storage capacities are available to suit the SGS computing needs, but maintenance of the SDC systems is not necessarily part of the Euclid SGS tasks.

Inside an SDC a further distinction can be made between the production part of the SDC, which runs and maintains the pipelines and is referred to as SDC-PROD, and the development part of the SDC, that participates in all the software development needed for the SGS and is referred to as SDC-DEV.

The SDC-PRODs have to be developed first. The aim is to design a system that can both achieve the construction of pipelines with a strict quality control system and benefit from the most advanced research in terms of signal processing for all aspects of the Euclid science. For this purpose, first the general data flow from raw data to science products was analysed and formalised; then a logical structure was designed through which this flow-chart can be replaced by actual high-performance codes.

The logical data processing flow is represented in Figure 7.1, where the different data processing levels are connected with logical data processing functions. These logical functions, or modules, represent self-contained processing units, i.e. they represent the highest-level break-down of a complete pipeline that can be achieved with parts that communicate only through the EMA. These processing units constitute a first step into the realisation of a distributed pipeline development. They are listed briefly hereafter:

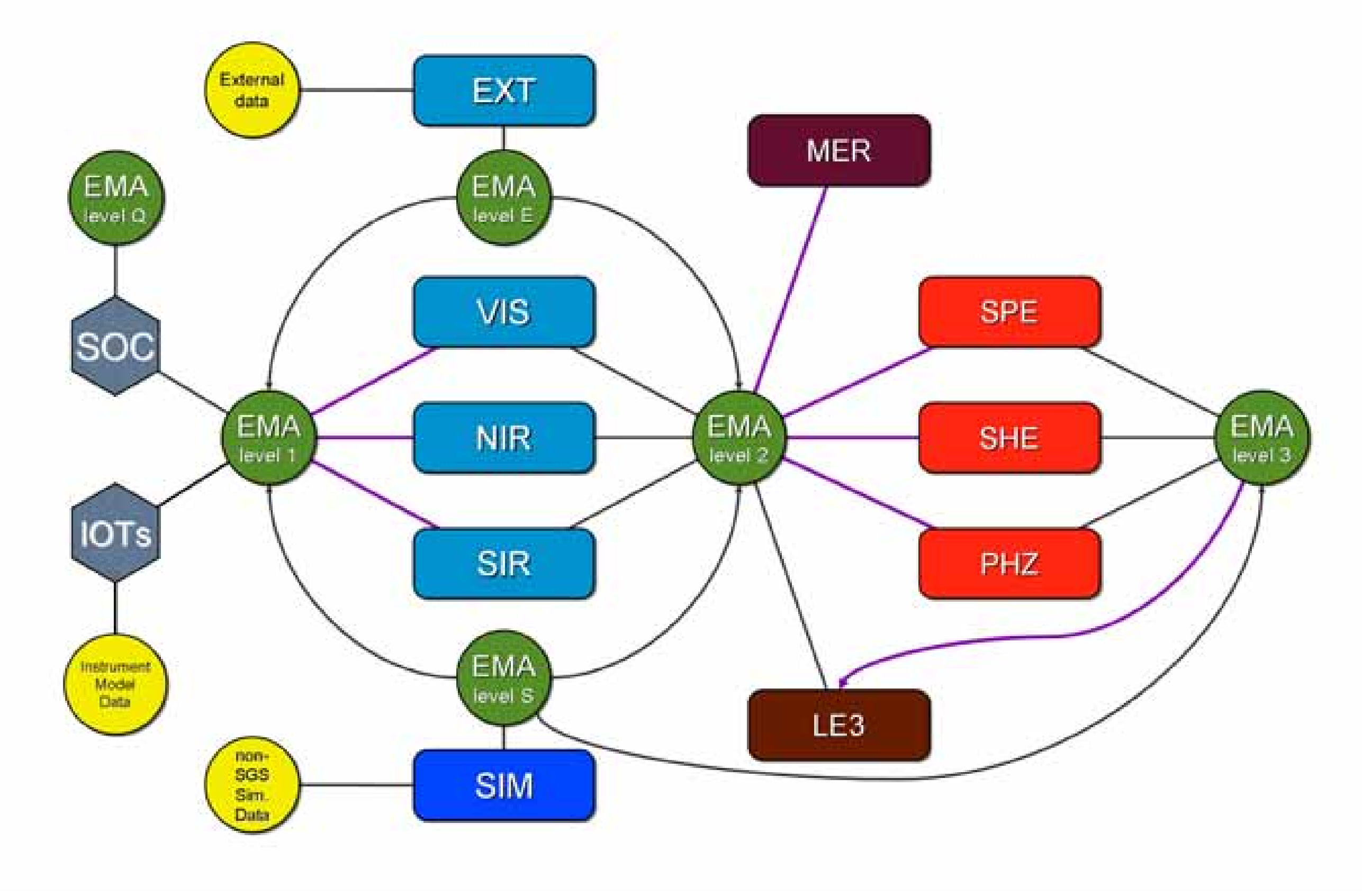

*Figure 7.1: The complete Euclid data processing pipeline all the way to the science data products. Each rectangular box defines a logical function where the operations can be considered as self-contained, i.e. interactions with the other functions occur through the EMA. Actual data are represented by the green circles and reside in the EMA. The grey arrows indicate relations where information is provided for the destination, the purple arrows signify more complex relations between data and processing units. See text for details.*

- VIS: is in charge or processing the visible imaging data from level 1 to level 2. It produces fully calibrated images, as well as source lists for quality check purposes only.
- NIR: is in charge of processing the near-infrared imaging data from level 1 to 2. It produces fully calibrated images as well as source lists for quality check purposes and to allow spectra extraction.
- SIR: is in charge of processing the near-infrared spectral imaging data from level 1 to 2. It produces fully calibrated spectral images and extracts spectra in the slitless spectroscopic frames taken by the NISP.
- EXT: is in charge of ingesting in the EMA all external data. This is mostly multi-wavelength data for photo-z estimation, but also spectroscopic data to validate the spectrometric redshift measurement tools.
- SIM: realises the simulations needed to test, validate, and qualify the whole pipeline, science products included.
- MER: realises the merging of all the level 2 information. It is in charge of providing stacked images and source catalogues where all the multi-wavelength data (photometric and spectroscopic) are aggregated.
- SPE: extracts spectroscopic redshifts from the level 2 spectra.
- PHZ: computes photometric redshifts from the multi-wavelength imaging data.
- SHE: computes shape measurements on the visible imaging data.
- LE3: computes the high-level science data products (Level 3), using the fully processed shape and redshift measurements.

A system is needed to turn the flow into an actual pipeline. A strong prerequisite from the EMC scientists is to keep the highest flexibility in the development of the most critical paths of the data processing. This includes allowing for different methods, different processing environments, and different experts to tackle the problems raised by the Euclid pipeline. Typical high flexibility areas are shape measurements and photo-metric redshifts. As the best European experts are residing in different countries of the consortium, the SGS activities must integrate this geographical and technical diversity into the Euclid ground segment activity.

These conditions are accommodated by the SGS, provided that research and development for the Euclid data processing pipelines are detached from the production-based SDCs and managed at a high level in the SGS.

Any new pipeline element is developed according to the following scheme:

- Scientific requirements are placed on the method to be used, or objective to be reached, for the pipeline element under consideration. The requirements are accompanied with descriptions of key tests to check the validity of the pipeline element. Responsibility of providing the requirements rests with the Euclid Science Working Groups (SWGs), which are specialised teams of scientists built around the Euclid science objectives (both core and legacy). Once requirements are agreed on in the SWGs, they can be turned into prototypes.
- Algorithmic research proceeds by designing prototypes, performing numerical tests, and comparing the results with the original requirements. Success/failure of the pipeline elements is gauged in collaboration with the SWG that issued the requirement. There are no formal requirements on the choice of infrastructure and language to be used, in order to maximise creativity. The algorithmic development is carried out by the Organisation Units (OUs), which are teams of EMC scientists with code-development know-how, grouped along their interest/competences for each of the individual data processing functions (Figure 7.1).
- Once validated by SWG and OU, the prototype is passed to an SDC, along with a test harness. The SDC-DEV turns this prototype into a Euclid pipeline element, which abides common coding standards, and uses pre-defined input and output mechanisms. The homogenisation and configuration control of the Euclid pipeline are the responsibility of the SDCs. The pipeline element is further tested and validated for inclusion in the SDC production chain. These decisions have to involve the SGS management.

The above description gives the formal steps to be followed to produce new pipeline elements. Rather than defining teams in the sense of groups of people, the steps define functions inside the SGS:

- the functional role of the SWGs to provide scientific requirements for the pipeline development;
- the functional role of the OUs to turn the requirements into code prototypes and to assess whether the requirements can be met;
- the functional role of the SDC-DEVs to turn the prototypes into pipeline modules.

It is perfectly acceptable, and it is likely to occur, that a single scientist in the EMC takes each of these functions in turn in the making of a particular element.

### 7.4.2 Data processing architecture and organisation

Figure 7.2 provides a visual estimation of the Euclid SGS data flow. The amount of data is estimated from the data provided by a source or a processing function, and is ingested by another data processing function for further analysis. The thickness of the arrows gives an indication of the data volume. These are order-of-magnitude estimates with large uncertainties. The factors 5 (or 2) and 3 in the data processing function boxes mean a multiplicative factor 5 (or 2) for data processing, and a factor 3 for reprocessing.

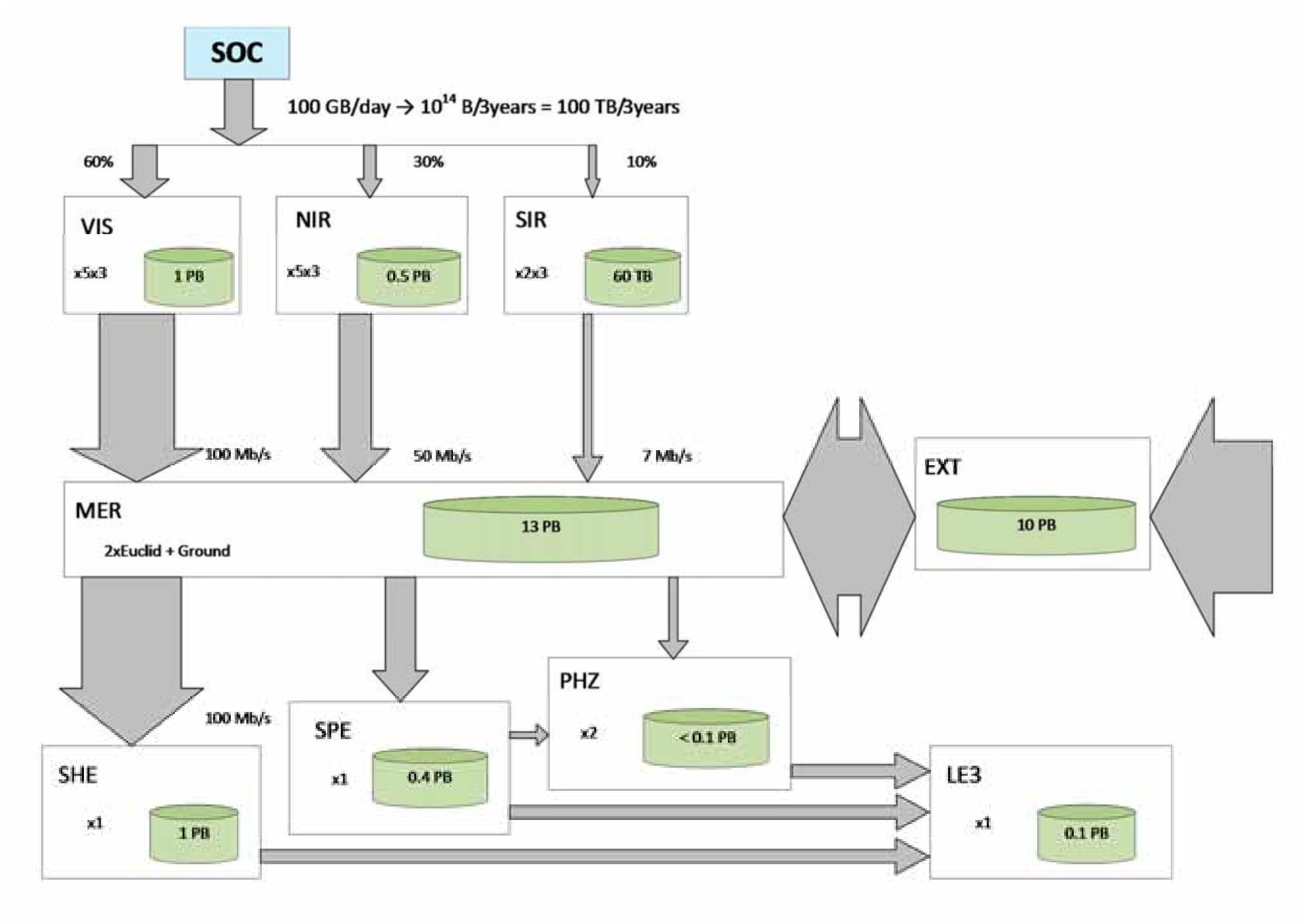

*Figure 7.2: Visual estimation of the Euclid SGS data flow, given by the amount of data that, provided by a data processing function, are ingested by another data processing function for further analysis. The arrows representing the data flow are not to scale.*

Analysis of simulation results shows that data access both by network and from disk is the main difficulty due to the large data volume. The SGS data processing architecture and organisation are set on solutions to:

- decrease as much as possible data access;
- decrease as much as possible data exchange;
- keep a very *thin* infrastructure between software modules and data access to avoid overheads while accessing data inside each SDC infrastructure;
- allow the inclusion of new national SDCs, as needed;
- simplify as much as possible the system design.

These considerations lead to the reference architecture as depicted in Figure 7.3. The proposed logical architecture relies on:

- a concept of software layers inside the SGS: metadata access layer (query/retrieve), data product access layer (open, read/write, get info, ...), data processing layer;
- a concept of distributed data products storage (bulk data products are stored at least twice among SDCs and metadata are indexed inside EMA) avoiding the unnecessary movement of huge amounts of data between SDCs and EMA/SDC;
- a single EMA metadata repository which inventories and indexes all metadata (and corresponding data locations).

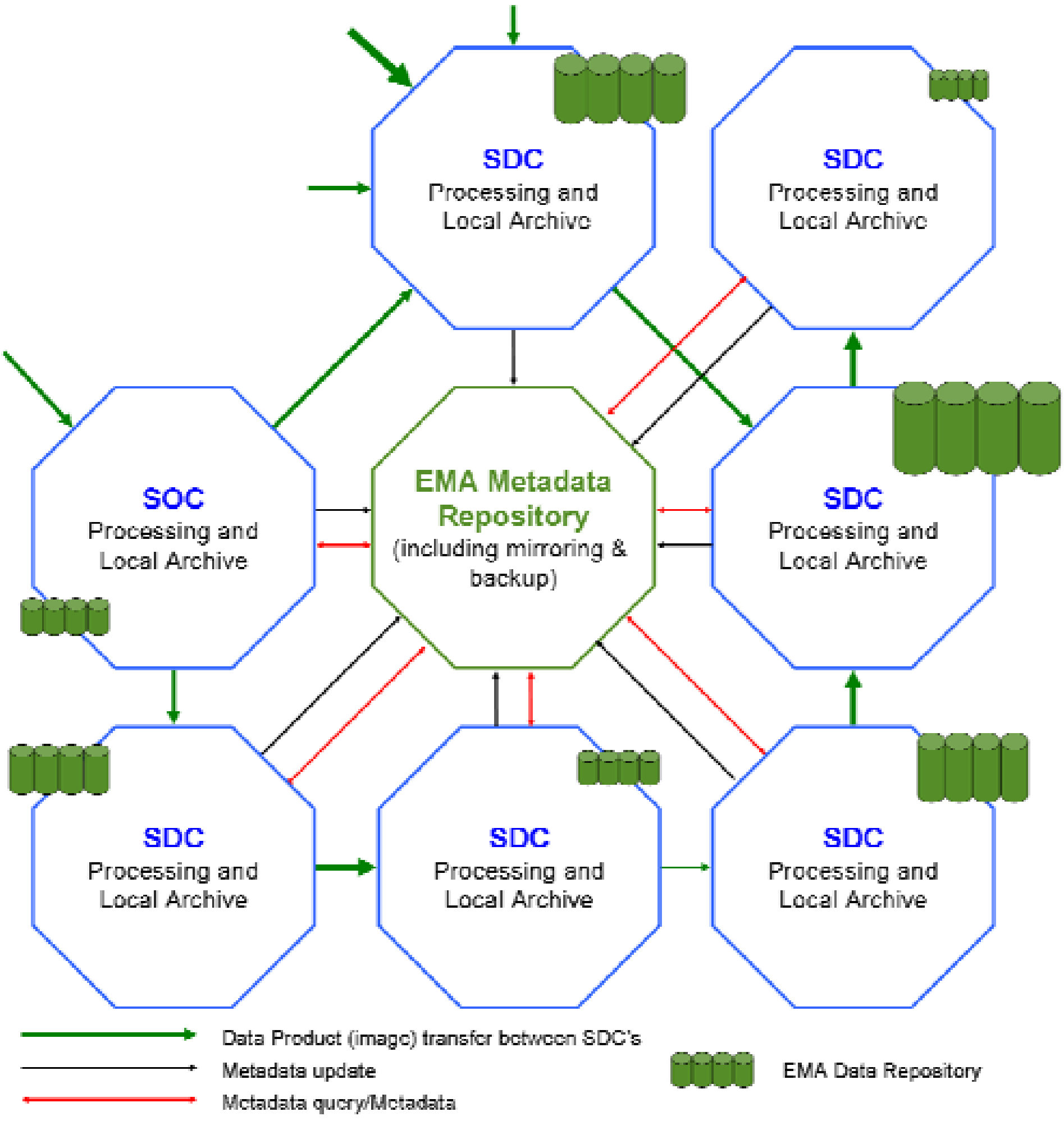

*Figure 7.3: Logical reference architecture*

The decisions on the mapping of pipeline modules onto SDCs, i.e. the deployment of functions on SDC infrastructures, are based upon:

- technical scientific requirements (input data necessary for pipeline processing);
- constraints set forth by each country, e.g. available expertise at specific geographic sites;
- architecture concepts derived from an architecture study conducted within the EMC SGS System Team: redundancy of huge data storage, redundancy of pipeline hosting infrastructure, distributed huge data storage among SDCs, single EMA repository of metadata and inventory of data products location.

### 7.4.3 Data processing responsibilities and expertise

Figure 7.4 shows the geographical distribution of data processing responsibilities within the EMC. Inclusion of new national SDCs is possible, whenever needed.

Relying on long-standing and in-depth experience gained from large ground based and space based all-sky surveys, the EMC in collaboration with ESA can present a feasible framework for the end-to-end handling and processing of the Euclid data.

The dataflow rate of Euclid is high, but similar to a number of currently operating and future large imaging surveys on the ground (e.g., ESO-VISTA, ESO-VST-OmegaCAM, CFHT-MegaCAM), which provide extensive data handling experience at various European institutes participating in the consortium. The *data centric* design of the data processing builds both on this ground-based astronomy expertise and on the expertise from the Gaia, Herschel and Planck missions, and elaborates on the following more Euclid-specific issues:

- Optimal hierarchical data handling infrastructure from the SOC to the science communities, involving quality control at each stage, and capitalizing as much as possible on the experience of the European scientific community in the development of data processing systems for ESA missions (e.g. XMM-Newton, INTEGRAL, Herschel, Planck, and Gaia).

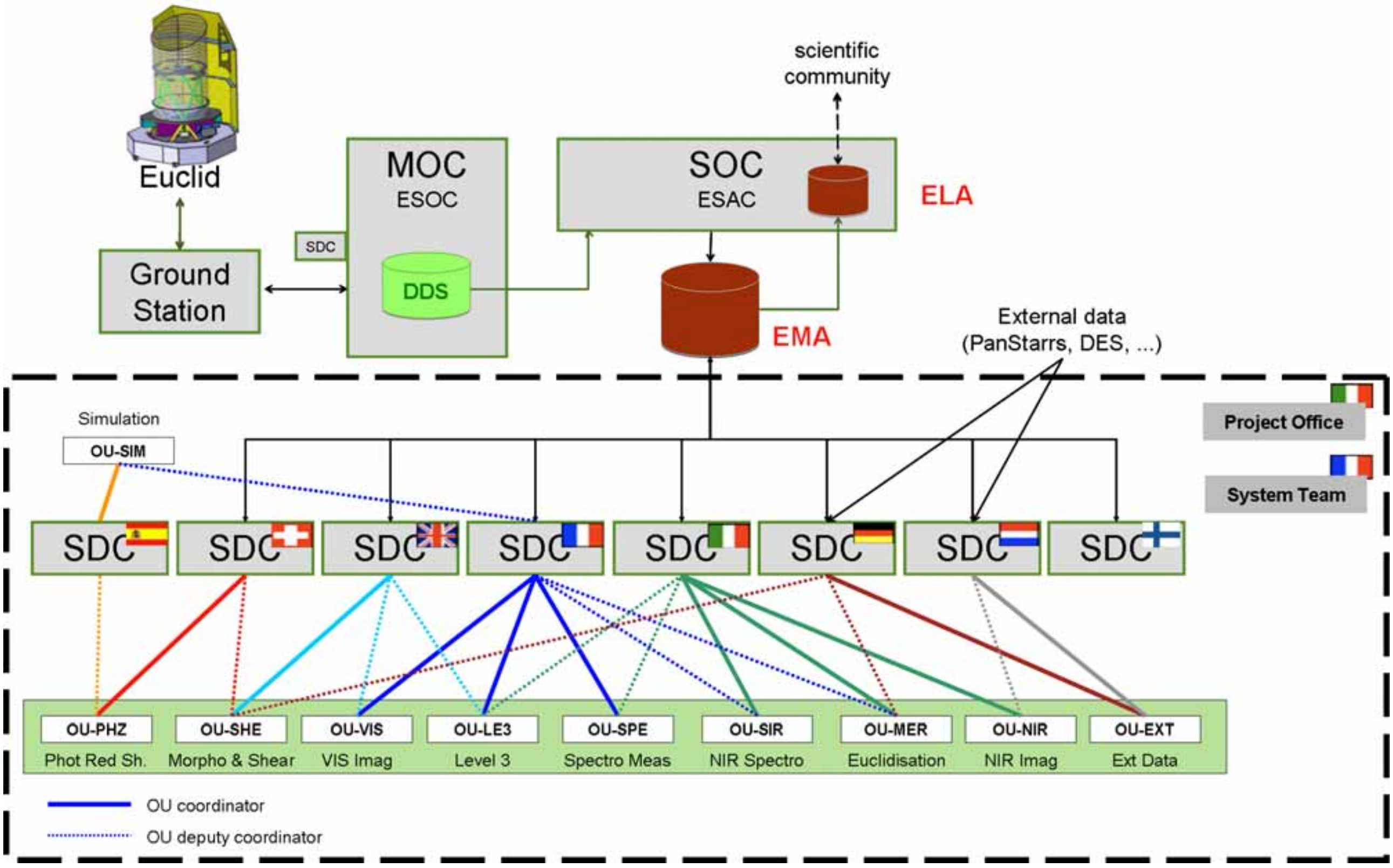

*Figure 7.4: The overall set of tasks for the EMC SGS and the national data processing responsibilities*

- Publication of all relevant Euclid data items into the ELA ready for additional studies. The ELA includes processed spectra and images, source catalogs, etc. provided in a format compatible with all of the international wide-spread standards (e.g. currently, FITS and the Virtual Observatory) and accessible through public e-infrastructures and web services.
- Optimal involvement of the EMC in the quality controls and calibration via the SDCs in charge of instrumental operations, together with additional SDCs connected in a distributed European-wide network. This approach is made possible by exploiting the extensive experience in European collaboration networks such as Euro-VO, and also the expertise on operating large surveys and particularly lensing surveys.
- Provision of the infrastructure for the EMC and partners to exchange and share data, also facilitating redundant processing as a verification of the key science results.
- Provision of a long-term (at least 10 years) and cost efficient solution for the data processing, storage, archiving and dissemination, jointly designed and implemented in close collaboration between the EMC and ESA.

# 8 Management

## 8.1 Introduction

In this section the roles and responsibilities of ESA, the industrial organisation, the Euclid Mission Consortium (EMC), and the scientific community at large are described. The overarching responsibility for all aspects of the Euclid mission rests with ESA's Directorate of Science and Robotic Exploration and its director.

The EMC is funded by the Member States and has been selected via an announcement of opportunity. The EMC provides the two instruments VIS and NISP, elements to the Euclid science ground segment, the science data products to the mission archive and members of the Euclid Science Team.

## 8.2 Industrial organisation

After a possible down-selection of Euclid in October 2011 and related endorsement in February 2012, ESA will release the Invitation to Tender (ITT) for the Implementation Phase (B2/C/D/E1) early 2012. The scope of this contract is to implement all industrial activities leading to a launch and commissioning of Euclid in the requested timeframe. The successful bidder will be appointed as Prime Contractor in charge, amongst other, of system engineering and management of the sub-contractors.

In the Implementation Phase and following the selection of the prime contractor, each subsystem is nominally procured through open competition in accordance with ESA best practice rules. The subsystem contractors are in charge of the procurement activities at lower levels. The industrial team of the prime contractor and the selected contractors for the first subsystem layers constitute the project industrial Core Team. The implementation phase schedule requires having the Core Team in place within six months after the selection of the prime contractor.

## 8.3 Payload Procurement

The Euclid telescope opto-mechanical assembly which includes the 1.2 m primary mirror, secondary, tertiary and folding mirrors are provided by the Agency under industrial contract. ESA procures the near-infrared detectors with their proximity electronics for the NISP and the CCDs and their proximity electronics for the VIS. EMC provides the necessary support to ESA in the procurement of the detectors to ensure that the science requirements can be met. The detectors are delivered to the EMC who is responsible for the delivery of the NISP and VIS instruments to ESA (see Figure 8.1).

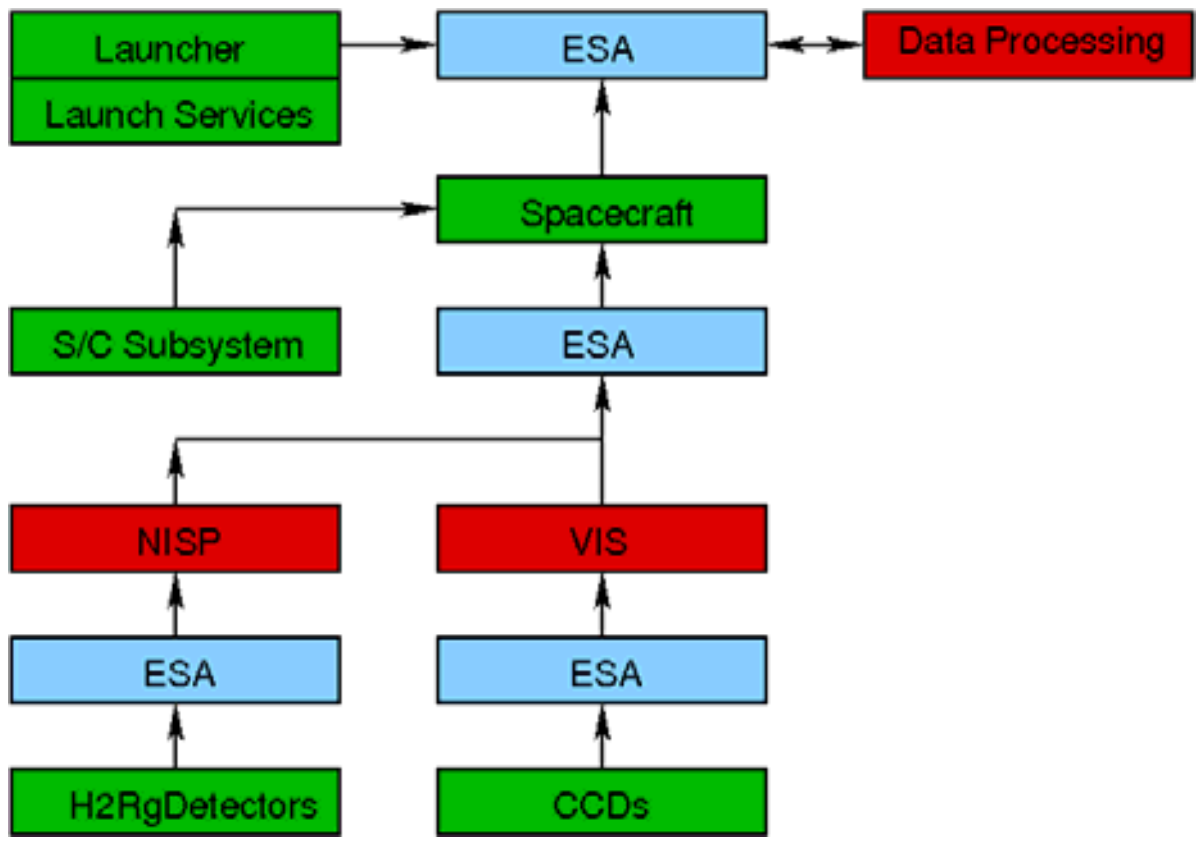

*Figure 8.1: Euclid hardware responsibilities. The green boxes indicate the elements to be delivered by industry to ESA. The red boxes are the elements to be delivered by the consortium to ESA. The data processing box is also explained in detail in Section 7.*

After Phase A/B1, the industrial prime contractor is responsible for the development, procurement, manufacturing, assembly, integration, test, verification and timely delivery of a fully integrated spacecraft capable of accommodating the defined payload elements, fulfilling the mission requirements and achieving the mission objectives. In particular, the prime contractor is responsible for the provision of the telescope. The VIS instrument, the NISP instrument, and the satellite subsystems to be provided by the EMC are considered by the prime contractor as customer furnished equipment. While the procurement of the telescope and instruments are managed separately, they constitute a single high precision cosmology experiment dedicated to a well-focused objective. As a consequence, their technical design must be carried out and optimised in a coherent fashion. For this reason close cooperation between the EMC and ESA is required to finalise the design and development of the payload.

## 8.4 Euclid Schedule

As a Cosmic Vision M-class mission, the Euclid schedule after the end of Phase A (July 2011) is as follows:

- October 2011: Down-selection for CV M1/M2 missions
- December 2011: Completion of the Definition Phase (A/B1)
- February 2012: Adoption for the Implementation Phase (B2/C/D/E1)
- September 2012: Start of the Implementation Phase
- End 2018: Launch (L)
- L+~days: start Satellite Commissioning & Payload Performance Verification Phases
- L+≤6 months: start Routine Phase
- L+7 years: end of mission.
- L+9 years: end of Active Archive Phase

The implementation phase (B2/C/D/E1) system study is expected to start in 2012 with the objective of launch in 2018. It will include 6 major reviews:

- System Requirements Review (SRR)
- Preliminary Design Review (PDR)
- Critical Design Review (CDR)
- Qualification Review (QR)
- Flight Acceptance Review (FAR)
- In-orbit Commissioning Review (IOCR)

The FAR is part of the satellite formal acceptance procedure. The IOCR is at system level, and marks the start of the routine science operations.

## 8.5 Science Management

After the spacecraft commissioning phase the ESA Science Operations Department (SOD) assumes responsibility of the Euclid Mission. ESA's Space Operations Centre (ESOC) implements the Mission Operations Centre (MOC), operates the spacecraft, and delivers telemetry and attitude data to the Euclid Science Ground Segment via the ESA Science Operations Centre (SOC). ESA's Space Astronomy Centre (ESAC) implements the SOC, which acts as the central node for the mission planning, performs an initial quality check and processing of the data and makes the telemetry available to the remainder of the SGS. The EMC manages the Science Data Centres (SDCs) responsible for the instrument specific data processing. The SDCs together with the SOC constitute the SGS. The SOC is also responsible for the development and operations of the Euclid Legacy Archive (ELA). The SOC populates and maintains the ELA and delivers the data products to the general scientific community. Details on the SGS management for both the EMC and SOC and the organisation of the uplink and downlink data processing are described in Section 7.

The Euclid Science Team (EST) oversees the preparations and execution of scientific operations, and endorses distribution of the data products to the community via the ELA. The EST is formed after selection of the EMC and is composed of nine scientists representing the EMC, two Independent Legacy Scientists (ILSs), and the ESA Project Scientist (PS) as chair. The ILSs are selected from the scientific community at large via a separate Announcement of Opportunity. The EST gives advice on all aspects of Euclid which affect its scientific performance. EST members participate in major project reviews and perform specific tasks as needed to support the development and operations of the mission.

# 8.6 Instrument procurement and EMC organisation

## 8.6.1 Hardware activities by the EMC

The hardware activities under the responsibility of the Euclid Consortium are split into two main systems organised into two management structures, the visible imaging (VIS) instrument and the near infrared imaging and spectrograph (NISP) instrument (see Figure 8.2a and 8.2b). The Consortium is responsible for the design, implementation, integration and tests, and timely delivery of the two scientific instruments to ESA. The characterisation of the detectors and the assemblies of the detector systems are under the EMC responsibility. The EMC carries out the normal system/subsystem engineering and interface engineering tasks for the ESA provided subsystems. The consortium participates in all formal subsystem reviews and supports the related progress meetings. The Euclid Consortium Lead (ECL) ensures that the complete instrument programme is implemented and executed within programmatic constraints.

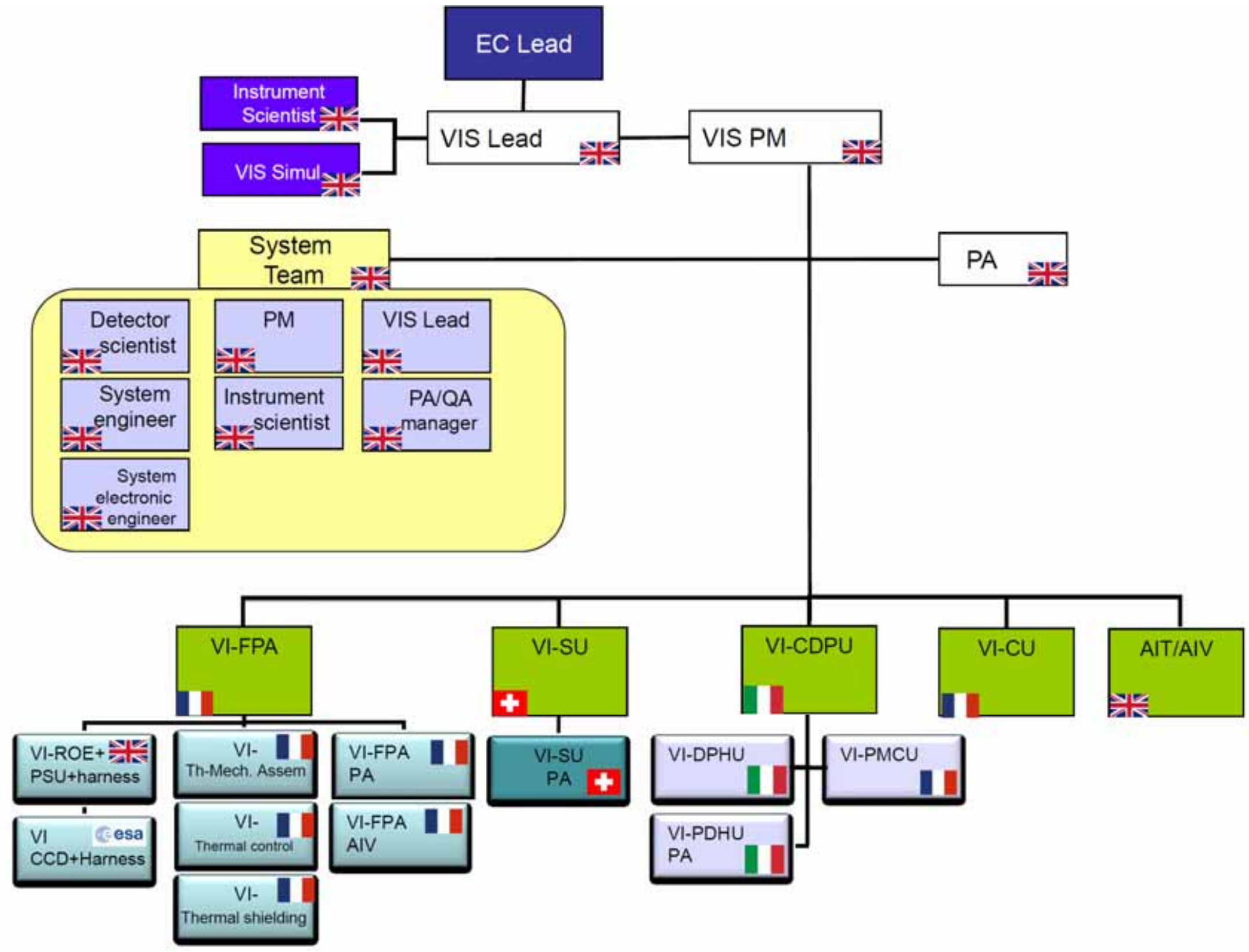

*Figure 8.2a: The VIS management structure and WBS.*

The VIS and NISP activitites are under the responsibility of their project offices led by the respective project managers (PM). The two project offices are responsible for the design, construction and the delivery of the two instruments to ESA. Due to the extremely challenging image quality requirements for weak lensing, the top level organisation of the VIS is consolidated by the VIS Instrument Lead. The VIS instrument lead is in charge of leading extensive and very detailed instrument performance simulations strongly coupled with the performance of the telescope, the VIS imaging calibration plan, and the survey implementation. This key position has to interface with the ECL, the VIS instrument scientist and with a specific VIS simulation lead. The interactions with ESA on payload requirements and interfaces or cross-cutting activities, and the coordination between VIS and NISP, are under the responsibility of the VIS and NISP project managers and the ECL System Engineer Support. The VIS and NISP instrument scientists are responsible for the performance evaluations and the compliance of the technical solutions with the scientific requirements of each instrument. The main main tasks are organised slightly differently, as the VIS is a single visible imaging instrument while NISP is more complex, comprising a NIR imaging-photometry and spectroscopy channel.

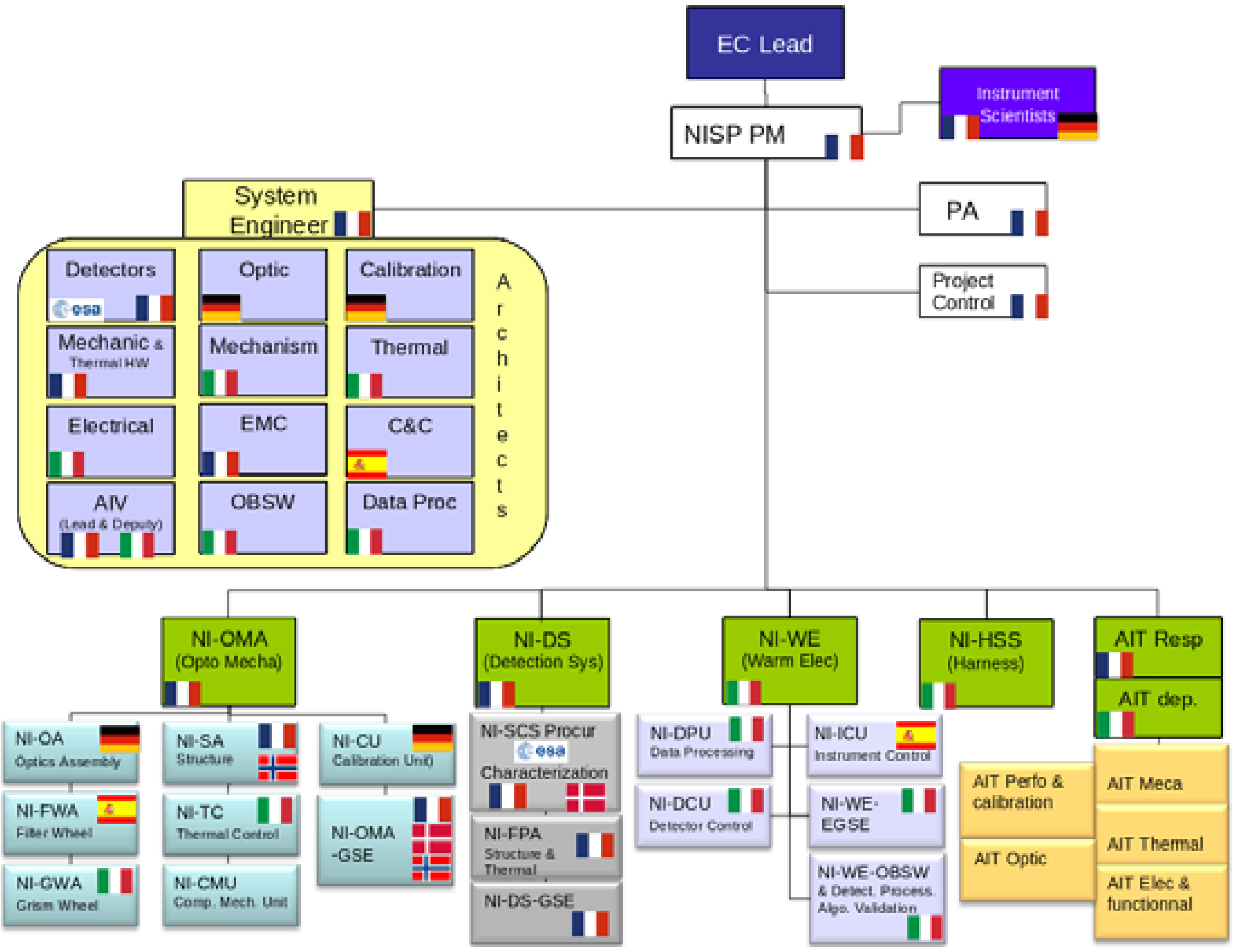

*Figure 8.2b: The NISP management structure and WBS.*

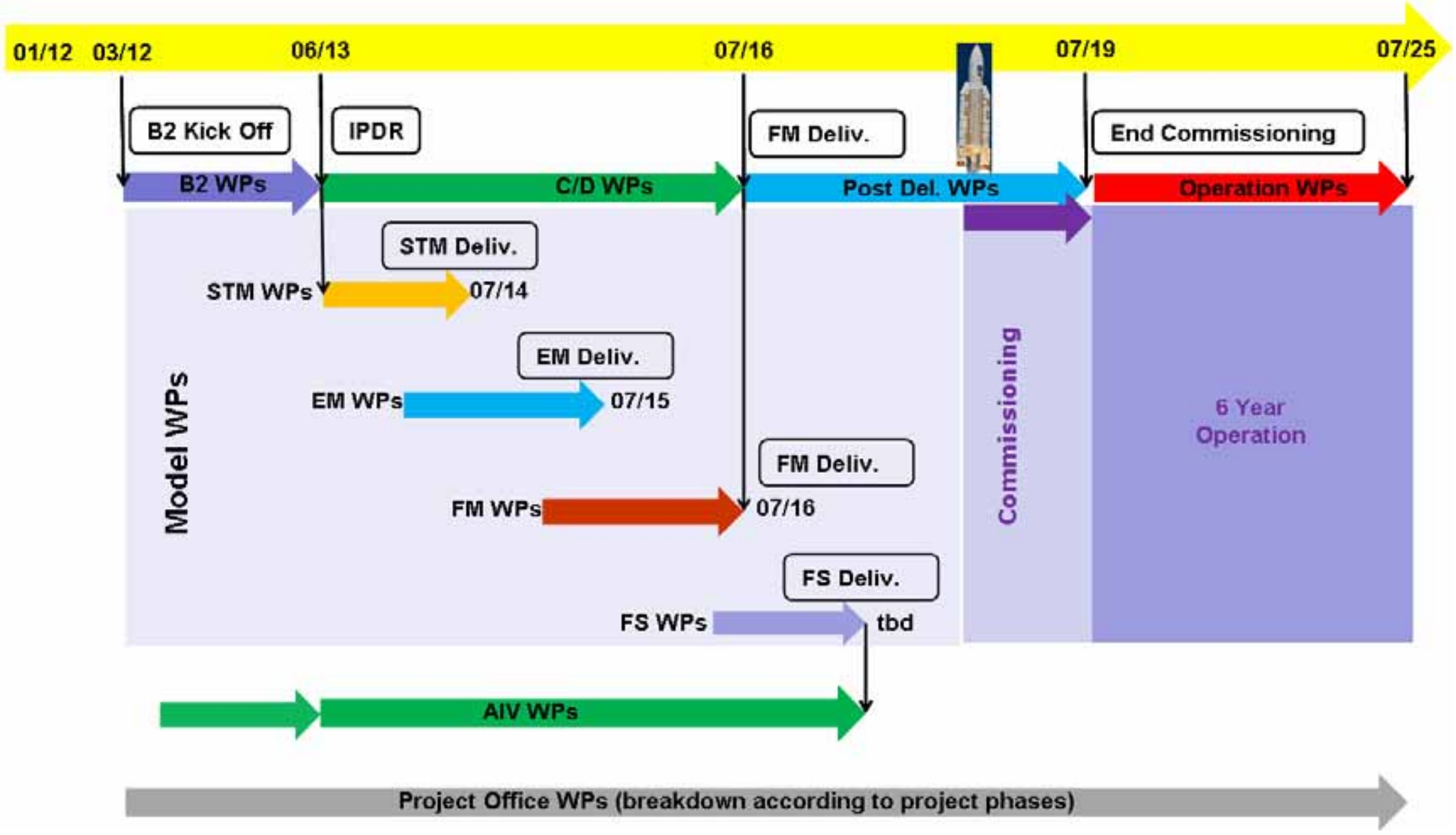

*Figure 8.3: Proposed instruments development schedule, assuming a launch date towards the end of the year 2018.*

## 8.6.2 Consortium organisation

Before the adoption of the implementation phase, based on the science management principles, a Multi-Lateral Agreement is established between ESA and the EMC funding agencies to formalise the commitments and deliverables of all parties.

The EMC is led by the EC Consortium Board (ECB). The role of the ECB is to define the EC policy with respect to the EC management and the scientific objectives. It is composed of 1-2 representative members per contributing countries. The ECB members are the points of the contact with their respective national agencies. The ECB provides reports as required to the Steering Committee composed of representatives of the supporting national agencies.

The Euclid Consortium is composed of about 800 members, of which about 450 researchers, spread over 15 countries, most of them being ESA member states, and 107 institutes. The membership to the Euclid Consortium is under the responsibility of the ECB. The ECB examines and endorses all requests for registration on a case by case basis, based on the added value for the Consortium and the affiliation of applicants. Institutes being in ESA member countries, or contributing significantly to the mission are favoured.

The organisation of the consortium is shown in Figure 8.4. The ECB steers the activities of the EC, and delegates the management and the coordination of the consortium, and the final decisions on trade-offs to the Euclid Consortium Lead (ECL), who is also the chair of the ECB. The ECL is the single formal interface of the consortium with ESA.

The ECL is assisted in his daily tasks by the ECL local office for administration, coordination and management, and system engineering support. The ECL and its local office work in very strong interactions with the instrument leads and the project managers.

The ECL is also assisted by an EMC Mission Survey Scientist. This scientist leads the EMC activities that needs a global views and understanding of the survey planned with Euclid, of the VIS and NISP science drivers, and of the performance of the telescope and the instruments. He is closely involved in the optimisation of the Euclid survey in order to maximise the scientific return of the mission. He also has a leading role in the end-to-end simulations. This pivot position aims at strengthening the day-to-day communication between the science working groups and the instrument and ground segment scientists, as well as the coordination of transverse scientific activities (mission definition, mission performances, calibrations, end-to-end simulations).

The consortium activities are organised in 5 groups: the Science Working Group (SWG), the VIS and NISP instrument groups, the SGS group, and the communication (COM) group. These activities are coordinated and monitored by the EC Coordination Group (ECCG). The ECCG advises the ECB on all consortium activities common and transverse to the instrument, ground segment, communication and science. It is led by the ECL who delegates the day-to-day activity to the ECL Support System Engineer. The ECCG is composed of the ECL, the ECL Support System Engineer, the ECL Support Coordinator and Manager, the Euclid Mission Survey Scientist, the Instrument, VIS, NISP and SGS managers, the instrument leads, the VIS, NISP and SGS scientists, the communication lead, the coordinators of the science WGs, and other coordinators, as needed (*e.g*. Calibration, Simulation).

The ECB delegates the day-to-day management and coordination of the instrument, ground segment, communication, and science activities to the Instrument Lead, instrument Project Managers (PMs), Ground Segment Project Manager, Communication lead, and Science WG coordinators. In particular, the technical interactions with ESA are delegated to the instrument Project Managers, the SGS manager, and the EST members from the EMC. These managers, leads, coordinators and scientists report to the ECB which is the ultimate authority for decisions regarding the EMC.

The science activities of the consortium are performed by the Science Working Groups (SWG). The working groups are organised into Cosmology, Legacy and Simulation panels which are led by working group coordinators. The SWGs are involved in the development of the science case for the mission, the definition of requirements and their translation into mission, instrument and SGS requirements. The SWGs also monitor the science performance of the instrument and SGS and supports trade off decisions. Together with the ECB, the working groups are responsible of the EMC science publications activities during the mission development and after launch using the SGS data products.

The scientific interface with ESA is done by the EST members nominated by the EMC and by the ECL who is invited ex-officio to all EST meetings.

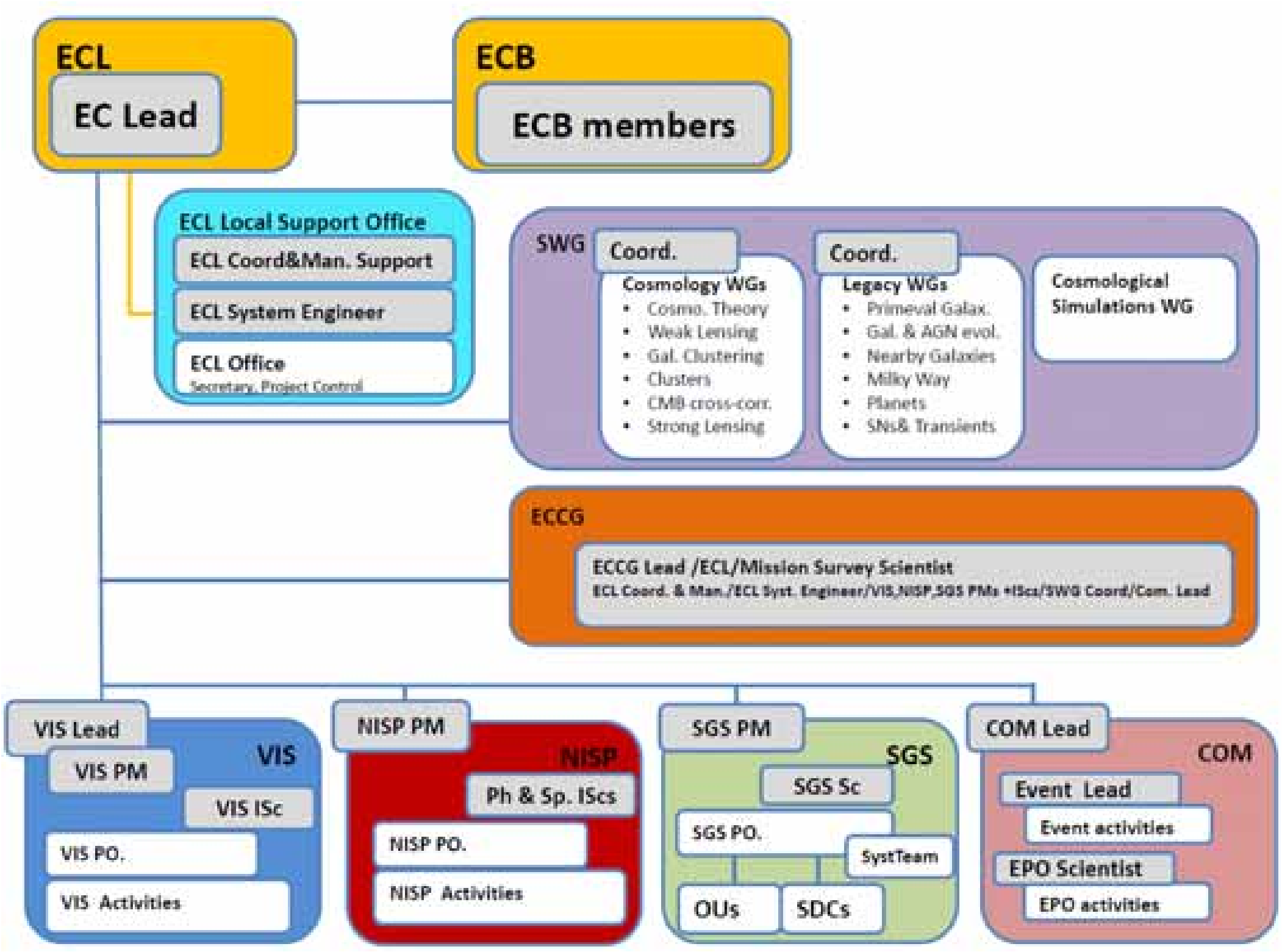

*Figure 8.4: the top level organisation of the Euclid Consortium. The Euclid Consortium Board (ECB) is made up of representative members of contributing countries and is led by the Euclid Consortium Lead (ECL). The ECB members are the points of the contact with their respective national agencies.*

# References

Abdalla, F.B., et al. 2008, arXiv:0812.3831
Albrecht, A., et al. 2009, arXiv:0901.0721
Albrecht, A., et al. 2006, arXiv:0609.591
Amara, A., Refregier, A. 2007, MNRAS, 381, 1018
Amendola, L., et al. 2010, arXiv:1008.1183
Amendola, L., Kunz, M., & Sapone, D. 2008, MEMSAI, 80, 22
Angulo, R.E., et al. 2009, MNRAS, 399, 983
Auger, M. W., et al. 2010, ApJ, 724, 511
Baldwin, J.A., Phillips, M.M., Terlevich, R. 1981, PASP, 93, 5
Balestra, I. et al. 2010, A&A, 512, 12
Beckwith, S. et al. 2010, Astron.J.132:1729-1755
Bernstein, G. 2010, MNRAS, 401, 1399
Bernstein, G., & Cai, Y.-C. 2011, arXiv:1104.3862
Blake, C., et al. 2011, arXiv:1105.2862
Blanton, M.R., Moustakas, J. 2009, ARA&A, 47, 159
Bordoloi, R., Lilly, S.J. & Amara, A. 2010, MNRAS, 406, 881
Bouwens, R.J., et al. 2010, ApJ, 709, L133
Bower, R., et al. 2006, MNRAS, 370, 645
Bridle, S., et al. 2010, MNRAS, 405, 2044
Bridle, S., et al. 2009, AnApS, 3, 6.
Burningham, B., et al.,2010, MNRAS, 406, 1885
Cimatti, A., et al. 2008, A&A, 482, 21
Cole, S., et al. 2005, MNRAS, 362, 505
Conselice, C.J., Rajgor, S., Myers, R. 2008, MNRAS, 386, 909
Conselice, C.J., Yang, C., Bluck, A.F.L. 2009, MNRAS, 394, 1956
Cypriano, E.S., et al. 2010, MNRAS, 405, 494
Daniel, S.F., et al. 2010, PRD, 81, 123508
de Ravel, L., et al. 2009, A&A, 498, 379
Deacon, N. R., Hambly, N. C. 2006, MNRAS, 371, 1722
Di Porto, C., & Amendola, A. 2007, PRD, 77, 083508
Dobke et al.2010, arXiv/1008.4112
Douspis, M., et al. 2008, A&A, 485, 395
Dupe, F.-X., et al. 2011, A&A, in press, arXiv:1010.2192
Dvali, G., Gabadadze, G., & Porrati, M. 2000, PLB, 485, 208
Eisenstein, D. et al. 2011, preprint (arXiv:1101.1529 [astro-ph.CO])
Eisenstein, D. et al. 2005, ApJ, 633, 560
Fan, X., et al. 2006, AJ, 132, 117
Fedeli, C., et al. 2011, MNRAS, 414, 1545
Ferreira, P.G, & Skordis, C. 2010, PRD, 81, 104020
Freedman, W.L., et al. 2001, ApJ 553, 47
Freedman, W.L., et al. 2009, ApJ, 704, 1036
Fu, L., et al. 2008, A&A, 479, 9
Garilli, B., et al. 2010, PASP 122, 827
Gavazzi, R., et al. 2007, ApJ, 667, 176
Geach, J.E., et al. ,2010, MNRAS 402,1330
Glazebrook, K., et al. 2005, New Astr. Rev. , 49, 374
Guzzo, L., et al. 2008, Nature, 451, 7178, 541
Hawkins, E., et al. 2003, MNRAS, 346, 78
Hearin, et al. 2010, ApJ 720 1351
Heavens, A.F., Kitching, T.D., & Verde, L. 2007, MNRAS, 380, 1029
Heymans, C.E. et al. 2006, MNRAS, 368, 1323
Hikage et al. 2010, MNRAS,412,65
Hirata, C. M., et al. 2007, MNRAS, 381, 1197
Hoekstra, H., Yee, H. K. C., Gladders, M. D. 2004, ApJ, 606, 67
Hopkins, A. M., Beacom, J. F. 2006, ApJ, 651, 142
Hopkins, P. F., et al. 2010, ApJ, 724, 915
Hu, W. & White 2004, Sci. Am. 290N2 44
Hu, W. 1999, ApJ, 522, 21
Ida, S., Lin, D.N.C. 2004, ApJ, 604, 388
Ilbert, O., et al. 2010, ApJ, 709, 644
Jain, B., & Taylor A.N., 2003, PRL, 91, 141
Jarosik, N., et al. 2011, ApJS, 192, 14
Joachimi, B., Bridle, S.L. 2010, AAP, 523, A1
Joachimi, B., et al. 2011, A&A, 527, 26
Jullo, E., & Kneib, J.-P. 2009, MNRAS, 395, 1319
Kaiser, N. 1987, MNRAS, 227, 1
Kiessling, A., et al. 2011a, MNRAS, 414, 2235
Kiessling, A., et al. 2011b, MNRAS, 416, 1045
Kirk, D., et al. 2011, arXiv:1109.4536
Kitching, T.D., et al. 2010, MNRAS, 1564
Kitching, T.D., et al. 2008, MNRAS,389, 173
Kitching, T.D., Taylor, A.N. 2011, arXiv:1012.3479
Kitching, T.D., et al. 2008, MNRAS, 390, 149
Koester, B., et al. 2007, ApJ, 660, 239
Komatsu, E., et al. 2011, ApJS, 192, 18
Koopmans, L. V. E. 2005, MNRAS, 363, 1136
Kron R.C., 1980, ApJS, 43, 305
Kuijken, K. 2006, A&A, 456, 827
Kümmel, M., et al. 2010, The aXe Manual, ST-ECF
Laureijs R., et al. 2009, ESA/SRE(2009)2; arXiv :0912,0914
Le Fevre, O., et al. 2005, A&A, 439, 845
Leauthaud, A., et al. 2010, ApJ, 709, 97
Leauthaud, A., et al. 2011, arXiv:1104.0928
Lidz, A., et al. 2009, ApJ, 690, 252
Lilly, S., et al. 2007, ApJS, 172, 70
Lima, M., Hu, W. 2005, PRD, 72, 043006
Lin, L., et al. 2008, ASPC, 399, 298
Lombriser, L., et al. 2010, arXiv:1003.3009
Mandelbaum, R., et al. 2005, MNRAS, 361, 1287
Mandelbaum, R., et al. 2011, MNRAS, 410, 844
Mannucci, F., et al. 2010, MNRAS, 404, 1355
Mannucci, F., et al. 2010, MNRAS, 408, 2115
Massey, R., et al 2004, MNRAS 348, 214
Massey, R., et al 2010, MNRAS, 409, 109
Massey, R., et al. 2007, ApJS, 172, 239
McCarthy P.J., et al. 2004, ApJ, 614, L9
McLure, R.J., et al. 2010, MNRAS, 403, 960
McQuinn, M., et al. 2007, MNRAS, 381, 75
Melchior, et al. 2007, A&A 484, 189-193
Melchior, et al. 2011,MNRAS, 412,1552
Meneghetti, M., et al. 2008, MNRAS, 385, 728
Meneghetti, M., et al. 2008, A&A, 482, 403
Meneghetti, M., et al. 2009, A&A, 514,A93
Meneghetti, M., et al. 2010, A&A, 514, 93
Meneghetti, M., et al. 2011. A&A, 530,A17
Mesinger, A., Furlanetto, S.R. 2008, MNRAS, 385, 1348
Miller, L., et al. 2007, MNRAS, 382, 315
Miralda-Escude, J. 1998, ApJ, 501, 15
Paulin-Henriksson, S., et al. 2008, A&A, 484, 67
Peacock, J., et al. 2006, ESA-ESO Working Group on Fundamental Cosmology
Percival, W., White, M. 2009, MNRAS, 393, 297
Petrosian V., 1976, ApJ, 209
Rapetti, D., et al. 2008, MNRAS, 388, 1265
Rapetti, D., et al. 2009, MNRAS, 400, 699
Rassat, A., et al. 2008; arXiv:0810.0003
Reyes, R., et al. 2010, Nature, 464, 7286, 256
Roche, N., et al. 2010,MNRAS,407,1231
Schlegel, et al. 2009, arXiv:0904.0468
Semboloni, E., et al. 2011, MNRAS, 410, 143
Semboloni, E., et al. 2011, arXiv:1105.1075
Seo, H.-J., Eisenstein, D. 2007, ApJ, 665, 14
Shim, H., et al. 2009, ApJ, 696, 785
Smith, G.P., et al. 2009, ApJ, 707, L163
Springel, V., et al. 2005, Nature, 435, 629
Sumi, T., et al. 2011, Nature, 473, 349
Suzuki, N., et al. 2011, arXiv:1105.3470
Taylor, A.N., et al. 2006, MNRAS, 374, 1377
Tegmark, M., Taylor, A.N., & Heavens, A. 1997, ApJ, 480, 22
Tinker, J.L., et al. 2008, ApJ, 688, 709
Treu, T., et al. 2011, arXiv:1104.5663
Treu, T., Koopmans, L.V.E. 2004, ApJ, 611, 739
van Daalen, M.P., et al. 2011, arXiv:1104.1174
Vanzella, E., et al. 2008, A&A, 478, 83
Vanzella, E., et al. 2011, ApJL, 730, 35
Voigt, L.M., et al. 2011, arXiv:1105.5595
Wang, Y., et al. 2010, MNRAS, 409, 737
Willott, C.J., et al. 2010, AJ, 139, 906

# Acronyms

| | |
|---|---|
| AGB | Asymptotic Giant Branch |
| AGN | Active Galactic Nucleus |
| AIT | Assembly, Integration and Testing |
| ALMA | Atacama Large Millimetre/Submillimetre Array |
| AOCS | Attitude and Orbit Control System |
| APE | Absolute Pointing Error |
| ASIC | Application Specific Integrated Circuit |
| AVM | Avionics Model |
| aXe | slitless spectroscopy data extraction software |
| BAO | Baryonic Acoustic Oscillations |
| BASICC | Baryonic Acoustic Oscillation Simulations at the Institute for Computational Cosmology |
| BOSS | Baryon Oscillation Spectroscopic Survey |
| CCD | Charge Coupled Device |
| CDM | Cold Dark Matter |
| CDR | Critical Design Review |
| CDPU | Control & Data Processing Unit |
| CEA | Commissariat a L'énergie Atomique |
| CERN | European Organisation for Nuclear Research |
| CFHTLS | Canada France Hawaii Telescope Legacy Survey |
| CFRP | Carbon Fiber Reinforced Plastic |
| CIB | Cosmic Infrared Background |
| CL | Corrector Lens |
| CMB | Cosmic Microwave Background |
| CoLA | Corrector Lens Assembly |
| COM | Communication |
| COSMOS | Cosmological Evolution Survey |
| CPU | Central Processing Unit |
| CR | Cosmic Ray |
| CSS | Cryogenic Support Structure |
| CTE | Charge Transfer Efficiency |
| CTE | Coefficient of Thermal Expansion |
| CTI | Charge Transfer Inefficiency |
| DS | Detector System |
| DDS | Digital Data Storage |
| DE | Dark Energy |
| DES | Dark Energy Survey |
| DETF | (NASA) Dark Energy Task Force |
| DGP | Dvali, Gabadadze, Porrati |
| DM | Development Model |
| DM | Dark Matter |
| DTCP | Daily Telemetry Communications Period |
| EC | Euclid Consortium (as EMC) |
| ECB | Euclid Consortium Board |
| ECCG | Euclid Consortium Coordination Group |
| ECL | Euclid Consortium Lead |
| EDF | Euclid Deep Field |
| EDFN | Euclid Northern Deep Field |
| EDFS | Euclid Southern Deep Field |
| EE | Encircled Energy |
| E-ELT | European Extremely Large Telescope |
| ELA | Euclid Legacy Archive |
| EM | Electro Magnetic Model |
| EMA | Euclid Mission Archive |
| EMC | Euclid Mission Consortium (as EC) |
| EOAT | Euclid Optimisation Advisory Team |
| EPO | Education & Public Outreach |
| eROSITA | Extended Röntgen Survey Imaging Telescope Array |
| ESA | European Space Agency |
| ESAC | European Space Astronomy Centre |
| ESO | European Southern Observatory |
| ESOC | European Space Operations Centre |
| EST | Euclid Science Team |
| EWS | Euclid Wide Survey |
| EZ | Easy-Z (Redshift Determination) |
| FAR | Flight Acceptance Review |
| Far-IR | Far Infrared |
| FGS | Fine Guidance Sensor |
| FH | Flight Harness |
| FM | Flight Model |
| FoM | Figure of Merit |
| FoV | Field of View |
| FPA | Focal Plane Array |
| FPA | Focal Plane Assembly |
| FS | Flight Spare |
| FWA | Filter Wheel Assembly |
| FWHM | Full Width Half Maximum |
| GOODS | Great Observatories Origins Deep Survey |
| GR | General Relativity |
| GREAT | Gravitational Lensing Accuracy Testing |
| GSE | Ground Segment Equipment |
| GWA | Grism Wheel Assembly |
| HETDEX | Hobby-Elberly Telescope Dark Energy Experiment |
| HGA | High Gain Antenna |
| HST | Hubble Space Telescope |
| IF | Interface |
| IGM | Inter Galactic Medium |
| IIP-SNe | Type II P SuperNovae |
| ILS | Independent Legacy Scientist |
| IM | Interface Module |
| IOCR | In-Orbit Commissioning Review |
| IOT | Instrument Operations Team |
| IPDR | Instrument Preliminary Design Review |
| IRAS | InfraRed Astronomical Satellite |
| ISW | Integrated Sachs Wolfe (effect) |
| ITT | Invitation to Tender |
| IWS | Instrument Work Station |
| JWST | James Webb Space Telescope |
| KiDS | Kilo Degree Survey |
| LED | Light Emitting Diode |
| LF | Luminosity Function |
| LHC | Large Hadron Collider |
| LOFAR | LOw Frequency Array for Radio Astronomy |
| LOS | Line of Sight |
| LRG | Luminous Red Galaxy |
| LSST | Large Synoptic Survey Telescope |
| Mid-IR | Mid Infrared |
| MLI | Multi Layer Insulation |
| MOC | Mission Operations Centre |
| MWA | Murchison Wide Field Array |
| NASA | National Aeronautics and Space Agency |
| Near-IR | Near Infrared |
| NEP | North Ecliptic Pole |
| NI-CaLA | NISP Camera Lens Assembly |
| NI-CoLA | NISP Corrector Lens Assembly |
| NI-DCU | NISP Detector Control Unit |
| NI-DS | NISP Detector System |
| NI-DPU | NISP Data Processing Unit |
| NI-FWA | NISP FWA |
| NI-GWA | NISP GWA |
| NI-HSS | NISP Harness |
| NI-ICU | NISP Instrument Control Unit |
| NI-OMA | NISP Opto-Mechanical Assembly |
| NI-WE | NISP Warm Electronics |
| NIP | Euclid Near-Infrared Imaging Photometer |
| NIR | Near Infrared |
| NISP | Near Infrared Spectrometer and Photometer |
| OB | Optical Bench |
| OGS | Operations Ground Segment |
| OU | Organisation Unit |
| PAH | Poly-Aromatic Hydrocarbons |
| Pan-STARRS | Panoramic Survey Telescope Rapid Response System |
| PCA | Principle Components Analysis |
| PCB | Printed Circuit Board |
| PDCU | Power Distribution Control Unit |
| PDR | Preliminary Design Review |
| PEM | Proximity Electronics Module |
| PLM | Payload Module |
| PM | Project Manager |
| PMCU | Power & Mechanism Control Unit |
| PO | Project Office |
| PS | Project Scientist |
| PSF | Point Spread Function |
| PSU | Power Supply Unit |
| PV | Performance Verification |
| QC | Quality Control |
| QLA | Quick Look Analysis |
| QR | Qualification Review |
| ROE | Read Out Electronics |
| RPE | Relative Pointing Error |
| RPSU | Power Supply Units |
| RSSD | (ESA) Research and Science Support Department |
| RTA | Real Time Assessment |
| S/C | Spacecraft |
| SAA | Solar Aspect Angle |
| SDC | Science Data Centre |
| SDSS | Sloan Digital Sky Survey |
| SEL2 | Sun-Earth Lagrange point 2 |
| SEP | South Ecliptic Pole |
| SFR | Star-Formation Rate |
| SGS | Science Ground Segment |
| SKA | Square Kilometre Array |
| SIDECAR | System for Image Digitisation, Enhancement, Control and Retrieval Application |
| SNe | SuperNovae |
| SNLS | SuperNovae Legacy Survey |
| SNR | Signal to Noise Ratio |
| SN | SuperNovae |
| SOC | Science Operations Centre |
| SOD | Science Operations Department |
| SRE | Science and Robotic Exploration |
| SRE-O | SRE-Operations Department |
| SRR | System Requirements Review |
| SSE | Sun-Spacecraft-Earth angle |
| STEP | Shear Testing Programme |
| STM | Structural End Thermal Model |
| SVM | Service Module |
| SWG | Science Working Group |
| SZ | Sunyaev-Zeldovich |
| TBC | To Be Confirmed |
| TBD | To Be Done / To Be Decided |
| TDA | Technology Development Activity |
| TRL | Technology Readiness Level |
| UDF | Ultra-Deep Field |
| UKIDSS | UKIRT Infrared Deep Sky Survey |
| UKIDSS-LAS | UKIDSS Large Area Survey |
| UKIRT | United Kingdom InfraRed Telescope |
| VI-CDPU | VIS Control and Data Processing Unit |
| VI-CU | VIS Calibration Unit |
| VI-FPA | VIS Focal Plane Assembly |
| VI-SU | VIS Shutter Unit |
| VIDEO | VISTA Deep Extra Galactic Observation Survey |
| VIKING | VISTA Kilo Degree Infrared Galaxy |
| VIPERS | VIMOS Public Extra-Galactic Redshift Survey |
| VIS | VISible Instrument |
| VISTA | Visible and Infrared Survey Telescope for Astronomy |
| VLT | Very Large Telescope |
| VO | Virtual Observatory |
| VST | VLT Survey Telescope |
| VVDS | VIMOS VLT Deep Survey |
| WBS | Work Breakdown Structure |
| WFC3 | HST White Field Camera 3 |
| WFE | Wave Front Error |
| WIMP | Weakly Interacting Massive Particle |
| WISE | Wide Field Infrared Survey Explorer |
| WL | Weak Lensing |
| WMAP | Wilkinson Microwave-Anisotropy Probe |
| WP | Work Package |
| WS | Wide Survey |
| XMM | X-Ray Multi-Mirror Mission |